\documentclass[12pt,a4paper,titlepage]{report}
\usepackage[toc]{appendix}
\usepackage[centertags]{amsmath}
\usepackage{amsfonts}
\usepackage{amsthm}
\usepackage{a4}
\usepackage{enumerate}
\usepackage{graphicx}
\usepackage{dsfont}
\usepackage{bm}
\usepackage{multirow}

\newcommand{\fig}[1]{Figure~\ref{#1}}
\newcommand{\tab}[1]{Table~\ref{#1}}
\newcommand{\be}{\begin{equation}}
\newcommand{\ee}{\end{equation}}
\newcommand{\eq}[1]{(\ref{#1})}

\def\nchapter#1{%
    \chapter*{#1}
    \addcontentsline{toc}{chapter}{#1}}
\def\nsection#1{%
    \section*{#1}
    \addcontentsline{toc}{section}{#1}}

\newlength{\defbaselineskip}
\newcommand{\setlinespacing}[1]%
           {\setlength{\baselineskip}{#1 \defbaselineskip}}

\theoremstyle{remark}

\begin{document}

\begin{titlepage}
\begin{center}
\Large{CZECH TECHNICAL UNIVERSITY IN PRAGUE} \\\normalsize Faculty
of Nuclear Sciences and Physical Engineering
\end{center}
\addvspace{150pt}
\begin{center}
\LARGE{\bf{DOCTORAL THESIS}}
\end{center}
\addvspace{50pt}
\begin{center}
\LARGE Interference Phenomena in Quantum Information
\end{center}
\addvspace{50pt}
\begin{center}
\Large Martin \v Stefa\v n\'ak \bigskip\bigskip\\ \normalsize Supervisor:
Prof. Ing. {I.} Jex, DrSc.
\end{center}
\addvspace{80pt}
\begin{center}
\large Prague, 2010
\end{center}
\end{titlepage}
\begin{titlepage}
\addvspace{350pt}
\bigskip
This thesis is the result of my own work, except where explicit
reference is made to the work of others and has not been submitted
for another qualification to this or any other university.
\begin{flushright}
\addvspace{30pt}
\bigskip
Martin \v Stefa\v n\'ak
\end{flushright}
\end{titlepage}

\setlinespacing{1.25}

\chapter*{Acknowledgement}

First of all, I would like to thank prof. Igor Jex for his kind supervision during the past years.

The first part of my thesis results from our longstanding and fruitful collaboration with Dr. Tamas Kiss from the Department of Nonlinear and Quantum Optics of the Research Institute for Solid State Physics and Optics belonging under the Hungarian Academy of Sciences. I would like to thank him in this way for numerous discussions.

The second part of my thesis follows from the results of my one year stay as a Marie Curie fellow in the group of prof. Schleich at the Department of Quantum Physics of the University of Ulm. I would like to thank him and the people from his group, in particular to Dr. Wolfgang Merkel, for stimulating discussions during my stay in Ulm. I also have to mention Dr. Daniel Haase from the Department of Number Theory and Probability Theory of the University of Ulm who contributed substantially to the discussions.

I would like to thank my fellow students and post-docs from the Department of Physics, especially to Dr. Jaroslav Novotn\'y, Dr. Hynek Lavi\v cka, Dr. Aur\'el G\'abris and Ing. V\'aclav Poto\v cek, for the very nice and stimulating atmosphere in our group.

The financial support from the Doppler Institute of the Faculty of Nuclear Sciences and Physical Engineering and the EU Marie Curie Research Network Training Project CONQUEST is gratefully acknowledged.

Last, but not least I would like to thank to my girlfriend, my family and friends for support during my studies.

\tableofcontents


\nchapter{Foreword}

One of the key features of quantum mechanics is the interference of probability amplitudes. The reason for the appearance of interference is mathematically very simple. It is the linear structure of the Hilbert space which is used for the description of quantum systems. In terms of physics we usually talk about the superposition principle valid for individual and composed quantum objects. So, while the source of interference is understandable it leads in fact to many counter-intuitive physical phenomena which puzzle physicists for almost hundred years.

The present thesis studies interference in two seemingly disjoint fields of physics. However, both have strong links to quantum information processing and hence are related. In the first part we study the intriguing properties of quantum walks. In the second part we analyze a sophisticated application of wave packet dynamics in atoms and molecules for factorization of integers.

The main body of the thesis is based on the original contributions listed separately at the end of the thesis. The more technical aspects and brief summaries of used methods are left for appendices.


\part{Recurrences in Quantum Walks}
\label{part:1}


\nchapter{Introduction}
\label{chap:1}


\nsection{From random walks to quantum walks}
\label{chap:1a}

The term random walk was first introduced by Pearson \cite{pearson} in 1905, is a mathematical formalization of a trajectory that consists of successive random steps. Shortly after that a paradigmatic application of a random walk - the explanation of Brownian motion \cite{brown} and diffusive processes, was found by Einstein \cite{einstein} and Smoluchowski \cite{smoluchowski}. Since then random walks have been used in many branches of science \cite{overview}, ranging from physics, economy, ecology to social sciences. Among others, the random walk is one of the cornerstones of theoretical computer science \cite{rw:compsc1,rw:compsc2}. Indeed, it can be employed for algorithmic purposes to solve problems such as graph connectivity \cite{graph:connect}, 3-SAT \cite{3-sat} or approximating the permanent of a matrix \cite{matrix:perm}.

Quantum walks have been proposed by Aharonov, Davidovich and Zagury \cite{aharonov} as a generalization of classical random walks to quantum domain. The unitary time evolution governing the walk can be either discrete as introduced by Meyer \cite{meyer1,meyer2} and Watrous \cite{watrous} leading to coined quantum walks or continuous as introduced by Farhi and Gutman \cite{farhi,childs}. It is interesting to note that similar ideas can be found already in the works of Feynman \cite{feynman} and Bialynicki-Birula \cite{birula} in the context of discretization of the Dirac equation. Scattering quantum walks \cite{hillery:2003,hillery:2004,kosik:2005,hillery:2007} were proposed by Hillery, Bergou and Feldman as a natural generalization of coined quantum walks based on an interferometric analogy. The connection between the coined quantum walks and the continuous time quantum walks has been established \cite{strauch,chandra:08}. Recently, it has been shown that both continuous \cite{childs:09} and discrete time \cite{lovett:09} quantum walks can be regarded as a universal computational primitive. By now, quantum walks form a well established part of quantum information theory \cite{bruss:leuchs}. For a review see e.g. the article by Kempe \cite{kempe:ovw} or books by Venegas-Andraca \cite{Venegas-Andraca} or Konno \cite{konno:book}.

Continuous-time quantum walks are suitable for the description of coherent transport of excitation in networks \cite{muelken:prl,muelken:pre1}. Recently, a coherent energy transfer in photosynthetic systems was observed \cite{engel}. This long-lived coherence which can be described by a generalized continuous-time quantum walk \cite{mohseni} together with the environmental noise leads to a substantial increase in energy transfer efficiency \cite{caruso}.

Coined quantum walk is well suited as an algorithmic tool \cite{kempe,ambainis}. Several algorithms based on coined quantum walks showing speed up over classical algorithms have been proposed \cite{shenvi:2003,ambainis:2003,childs:04,kendon:2006,aurel:2007,magniez,vasek}. Various properties of coined quantum walks have been analyzed, e.g. the effects of the coin and the initial state \cite{2dw1,chandrashekar:2007,miyazaki}, absorbing barriers \cite{bach:2004}, the hitting times \cite{kempe:2005,krovi:2006a,krovi:2006b} or the effect of decoherence \cite{aurel:2007,kendon:2006b}. Hitting times for continuous quantum walks related to the quantum Zeno effect were considered in \cite{varbanov:2008}. Great attention has been paid to the asymptotics of quantum walks \cite{nayak,carteret,Grimmett,konno:2002,konno:2005b}. In particular, localization was found in 2-D quantum walks \cite{2dqw,2dw1,localization} and in 1-D for a generalized quantum walk \cite{1dloc,sato:2008}. Several experimental schemes have been proposed to realize coined quantum walks including cavity QED \cite{sanders}, linear optics \cite{jeong,pathak}, optical lattices \cite{eckert,dur}, Bose-Einstein condensate \cite{chandrashekar:2006} and quantum rings \cite{Orsolya}. Recently, as proof of principle, experiments with neutral atoms \cite{karski}, ions \cite{schmitz} and photons \cite{Schreiber} have been performed.

In comparison to classical random walks coined quantum walks are considerably more flexible. The coin operator can be in principle an arbitrary unitary matrix. Moreover, one can choose the initial coin state. All of these influence the dynamics of the quantum walk. The diversity of quantum walks asks for a classification. Indeed, in order to exploit the full potential of quantum walks for algorithmic purposes one needs to know in which regimes they can be operated in.

The present thesis focuses mainly on one particular quantity which is suitable for the classification of both classical as well as quantum walks, namely the probability to return to the origin. The recurrence probability is known as the P\'olya number, after G. P\'olya who as the first discussed this property in the context of classical random walks on infinite lattices in 1921 \cite{polya}. P\'olya pointed out the fundamental difference between walks in different dimensions. In three or higher dimensions the recurrence probability is less than one and depends exclusively on the dimension \cite{montroll:1956}, whereas for walks in one or two dimensions the P\'olya number equals unity. As a consequence, in three and higher dimensions the particle has a non-zero probability of escape \cite{domb:1954}. Recurrence in classical random walks is closely related to first passage times as pointed out in a number of classics papers of statistical mechanics \cite{montroll:1964,hughes}. A summary of the results on recurrence of classical random walks is left for Appendix~\ref{app:a}.

We extend the concept of recurrence and P\'olya number to quantum walks in Chapter~\ref{chap:2} based on \cite{stef:prl} where a particular measurement scheme was considered. Other possible definitions of the quantum P\'olya number are briefly discussed following \cite{kiss:recurrence}. As we show in Appendix~\ref{app:b}, within the framework of our measurement scheme the criterion for recurrence of a quantum walk is the same as for the classical random walk - it is determined by the asymptotic behaviour of the probability at the origin. To be able to analyze the probability at the origin we first solve the time evolution equations. Since the quantum walks in consideration are translationally invariant we make us of the Fourier transformation and find a simple solution in the momentum picture. Probability amplitudes in the position representation are then obtained by performing the inverse Fourier transformation. Hence, they have a form of an integral over momenta where the time enters only in the rapidly oscillating phase. This allows us to perform the asymptotic analysis of the probability at the origin in a straightforward way by means of the method of stationary phase. Basic concepts of this method are reviewed in Appendix~\ref{app:c}. We find that the asymptotic scaling of the probability at the origin is affected by the additional degrees of freedom offered by quantum mechanics. Hence, the recurrence probability of a quantum walk depends in general on the topology of the walk, choice of the coin and the initial state. This is in great contrast to classical random walks, where the P\'olya number is characteristic for the given dimension.

Recurrence of unbiased quantum walks on infinite $d$-dimensional lattices is analyzed in Chapter~\ref{chap:4} which is based on \cite{stef:pra}. First, we show that for the quantum walk driven by Hadamard tensor product coin, the P\'olya number is independent of the initial conditions, thus resembling the property of the classical walks. We provide an estimation of the P\'olya number for this quantum walk in dependence of the dimension of the lattice. Second, we examine the Grover walk on a plane, which exhibits localization and thus is recurrent, except for a particular initial state for which the walk is transient. We generalize the Grover walk to show that one can construct in arbitrary dimensions a quantum walk which is recurrent. This is in great contrast with classical random walks which are recurrent only for the dimensions $d=1,2$. Finally, we analyze the recurrence of the Fourier walk on a plane. This quantum walk is recurrent except for a two-dimensional subspace of initial states. We provide an estimation of the P\'olya number in dependence on the initial states.

In Chapter~\ref{chap:5} we extend our analysis of recurrence to biased quantum walks following \cite{stef:njp}. As we illustrate in Appendix~\ref{app:a2}, recurrence of a classical random walk on a line is extremely sensitive to the directional symmetry, any deviation from the equal probability to travel in each direction results in a change of the character of the walk from recurrent to transient. Applying our definition of the P\'olya number to quantum walks on a line we show that the recurrence character of quantum walks is more stable against bias. We determine the range of parameters for which biased quantum walks remain recurrent. We find that there exist recurrent genuine biased quantum walks which is a striking difference to classical random walks .

Quantum walks involving more than one particle opens up the possibility of having entangled initial states or the particles can be indistinguishable - either bosons or fermions. In Chapter~\ref{chap:6} which is based on \cite{stef:meeting} we study the motion of two non-interacting quantum particles performing a quantum walk on a line. We analyze the meeting problem, i.e. the probability that the two particles are detected at a particular position after a certain number of steps. The results are compared with the corresponding classical problem which we review in Appendix~\ref{app:d}. We derive analytical formulas for the meeting probability and find its asymptotic behaviour. We show that the decay of the meeting probability is faster than in the classical case, but not quadratically as one could expect from the ballistic nature of a quantum walk. The effect of non-classical features offered by quantum mechanics on the meeting probability is analyzed. We summarize our results and present an outlook in the Conclusions.


\nsection{Quantum walk on a line - an introductory example}
\label{chap:1b}

Before we turn to the presentation of our results we briefly introduce the basic notions of quantum walks. For a more comprehensive review we refer to the literature \cite{kempe:ovw}.

Let us begin with the classical random walk on a line. Random walk is a stochastic process where the particle moves on an integer lattice in discrete time steps. In each step the particle can move from its current location (say $m$) to the neighboring lattice points (i.e. $m\pm 1$) with equal probability. Suppose that the particle is at time $t=0$ at the origin of the lattice $m=0$. After the first step, we can find the particle at site $m=1$ or $m=-1$ with probability one-half. To calculate the probability that the particle is at position $m$ at a latter time $t$ we can use the following recurrence relations
\begin{equation}
\label{cl:walk:time:evol}
P(m,t) = \frac{1}{2} P(m-1,t-1) + \frac{1}{2} P(m+1,t-1),\qquad m\in\mathds{Z}.
\end{equation}
The solution of the equations (\ref{cl:walk:time:evol}) with the initial condition $P(0,0) = 1$ has the form
\begin{equation}
\label{chap:1:crw:dist}
P(m,t) = \frac{1}{2^{t}} {t\choose \frac{t+m}{2}}.
\end{equation}
Indeed, each random path has the same probability $2^{-t}$ and the number of paths leading to the lattice point $m$ is given by the well-known binomial distribution. It is straightforward to calculate various attributes of the random walk, e.g. the mean value and the variance of the particle's position. We find that the mean value vanishes, in agreement with the unbiasedness of the random walk we consider. On the other hand, the variance grows with the square root of the number of steps. Indeed, random walk is a diffusion process.

The quantum walk is a generalization of a classical random walk to a discrete unitary evolution of a quantum particle. Hence, there is no randomness in the time evolution itself in the quantum case. Nevertheless, the randomness enters through the measurement. Indeed, if we want to know the position of the particle we have to measure it and a particular result is found with the corresponding probability given by the standard quantum-mechanical formula. The particle can be found on any lattice point $m\in\mathds{Z}$. We denote the corresponding position eigenstates by $|m\rangle$. These vectors form an orthonormal basis of the {\it position space} $\mathcal{H}_P$
$$
\mathcal{H}_P = {\rm Span}\left\{|m\rangle|m\in\mathds{Z}\right\},\quad \langle m|n\rangle = \delta_{mn},\quad \sum_{m}|m\rangle\langle m| = I.
$$
As in the classical random walk, the particle moves from its current position to the neighboring lattice points, but instead of choosing the path randomly it travels all paths simultaneously, i.e. it evolves into a superposition
$$
|m\rangle \longrightarrow |m-1\rangle + |m+1\rangle.
$$
However, we can easily see that such a time evolution is not unitary. Indeed, two orthogonal vectors $|0\rangle$ and $|2\rangle$ evolves into the states
$$
|0\rangle\longrightarrow |-1\rangle + |1\rangle,\qquad |2\rangle\longrightarrow |1\rangle + |3\rangle,
$$
which have non-zero overlap. To make the time-evolution unitary we have to consider a particle which has an internal degree of freedom with two orthogonal states $|L\rangle$ and $|R\rangle$. This additional degree of freedom is usually referred to as {\it coin} and its two orthogonal states $|L\rangle$, $|R\rangle$ form a basis of the corresponding {\it coin space} $\mathcal{H}_C$
$$
\mathcal{H}_C = {\rm Span}\left\{|L\rangle,|R\rangle\right\}.
$$
The state of the coin determines the next move of the particle according to
$$
|m\rangle|L\rangle\longrightarrow|m-1\rangle|L\rangle,\qquad |m\rangle|R\rangle\longrightarrow|m+1\rangle|R\rangle.
$$
Such a transformation is performed by the {\it conditional displacement operator} $S$
$$
S = \sum_m\left(|m-1\rangle\langle m|\otimes|L\rangle\langle L| + |m+1\rangle\langle m|\otimes|R\rangle\langle R|\frac{}{}\right),
$$
which is indeed unitary. However, a time evolution according to $S$ itself would be rather trivial. Indeed, if the particle will start the quantum walk in a definite coin state, say $|L\rangle$, it will simply move on to the left. Hence, to obtain a non-trivial time evolution we first rotate the coin by the {\it coin operator} before the conditional displacement $S$ is applied. As the coin operator we can in principle choose an arbitrary unitary transformation on the coin space $\mathcal{H}_C$. Here, we consider a particular choice of the {\it Hadamard coin} $H$ which performs the following rotation
$$
H|L\rangle = \frac{1}{\sqrt{2}}\left(|L\rangle + |R\rangle\right),\qquad H|R\rangle = \frac{1}{\sqrt{2}}\left(|L\rangle - |R\rangle\right).
$$
Finally, we can write the {\it unitary propagator} $U$ which performs a single step of the quantum walk
\begin{equation}
\label{chap:1:U}
U = S\cdot\left(I\otimes H\right).
\end{equation}
Suppose that the particle is initially at the origin with the coin state $|L\rangle$, i.e.
\begin{equation}
\label{chap:1:init:state}
|\psi(0)\rangle  = |0\rangle|L\rangle.
\end{equation}
After the first step of the quantum walk it evolves into the state
\begin{equation}
\label{chap:1:state:1}
|\psi(1)\rangle = U|\psi(0)\rangle = \frac{1}{\sqrt{2}}\left(|-1\rangle|L\rangle + |1\rangle|R\rangle\frac{}{}\right).
\end{equation}
Note that if we perform the measurement of the particle's position, we find it with equal probability on the sites $\pm 1$. This is the same result as for the classical random walk. Moreover, after the measurement the state of the particle is projected onto the eigenstate corresponding to the measurement outcome. Hence, by performing position measurements after each step we obtain one classical random path. By making a statistics of such paths we recover a classical random walk. To obtain different dynamics we have to let the quantum particle evolve unperturbed, i.e. without measurements, for a desired number of steps $t$, and perform the position measurement afterwards. In this way, each path will not obtain probability but probability amplitude, which involves a phase. Different paths leading to the same lattice point will interfere. Hence, a quantum walk is an interference phenomenon.

As we have seen from (\ref{chap:1:state:1}) the probability distribution of the quantum walk after the first step does not differ from the probability distribution of the classical random walk. Indeed, if the quantum particle is initially localized at the origin no interference can occur. The same applies to the second step and the state of the particle is given by
$$
|\psi(2)\rangle = U|\psi(1)\rangle = \frac{1}{2}\left(|-2\rangle|L\rangle + |0\rangle(|L\rangle + |R\rangle) - |2\rangle|R\rangle\frac{}{}\right).
$$
The probability to find the particle at the position $m$ after two steps $P(m,2)$ is given by
\begin{eqnarray}
\nonumber P(-2,2) & = & |\langle -2|\langle L|\psi(2)\rangle|^2 + |\langle -2|\langle R|\psi(2)\rangle|^2 = \frac{1}{4},\\
\nonumber P(0,2) & = & |\langle 0|\langle L|\psi(2)\rangle|^2 + |\langle 0|\langle R|\psi(2)\rangle|^2 = \frac{1}{2},\\
\nonumber P(2,2) & = & |\langle 2|\langle L|\psi(2)\rangle|^2 + |\langle 2|\langle R|\psi(2)\rangle|^2 = \frac{1}{4},
\end{eqnarray}
which is the same as for the classical random walk. Finally, in the third step the interference occurs for the first time. The state of the particle after the third step has the form
$$
|\psi(3)\rangle = U|\psi(2)\rangle = \frac{1}{2\sqrt{2}}\left(|-3\rangle|L\rangle + |-1\rangle(2|L\rangle + |R\rangle) - |1\rangle|L\rangle + |3\rangle|R\rangle\frac{}{}\right),
$$
and we see that the probability distribution
\begin{eqnarray}
\nonumber P(-3,3) & = & |\langle -3|\langle L|\psi(3)\rangle|^2 + |\langle -3|\langle R|\psi(3)\rangle|^2 = \frac{1}{8},\\
\nonumber P(-1,3) & = & |\langle -1|\langle L|\psi(3)\rangle|^2 + |\langle -1|\langle R|\psi(3)\rangle|^2 = \frac{5}{8},\\
\nonumber P(1,3) & = & |\langle 1|\langle L|\psi(3)\rangle|^2 + |\langle 1|\langle R|\psi(3)\rangle|^2 = \frac{1}{8},\\
\nonumber P(3,3) & = & |\langle 3|\langle L|\psi(3)\rangle|^2 + |\langle 3|\langle R|\psi(3)\rangle|^2 = \frac{1}{8},
\end{eqnarray}
differs from the classical one. As a consequence of the choice of the initial coin state (\ref{chap:1:init:state}) it is biased towards the left.

\begin{figure}[h]
\begin{center}
\includegraphics[width=0.7\textwidth]{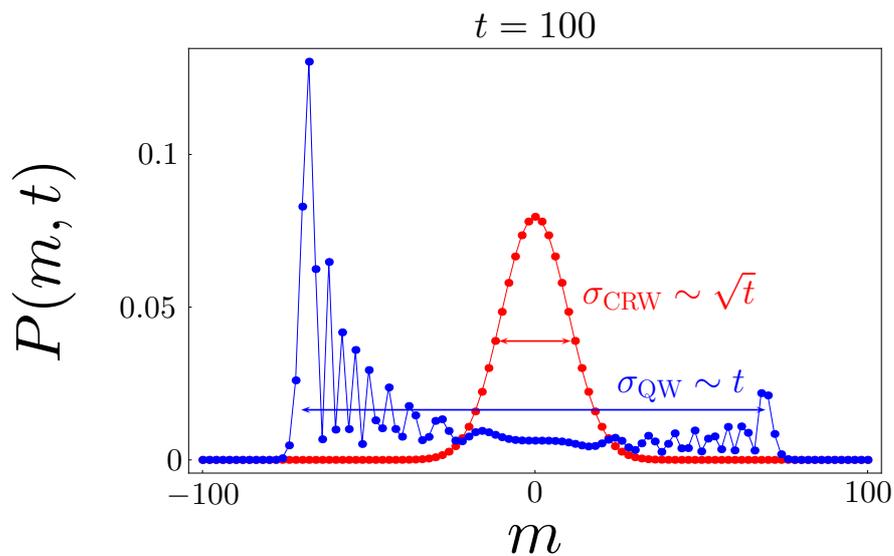}
\caption{The probability distribution of the classical and quantum walk on a line after 100 steps. For the classical random walk illustrated by the red points we find that the probability distribution is peaked at the origin and symmetric. Indeed, the mean value vanishes. The width of the distribution is rather small, since the variance of the classical random walk grows only with the square root of the number of steps. This is a typical signature of diffusion. In contrast, the probability distribution of the quantum walk described by the blue points shows striking differences. First, due to the choice of the initial coin state the distribution is biased towards left. Second, the width of the distribution is proportional to the number of steps. Indeed, the variance of the quantum walk grows linearly with time which is a typical signature of a ballistic process.}
\label{chap:1:fig2}
\end{center}
\end{figure}

In general, the state of the particle at a later time $t$ is given by the successive application of the propagator $U$ on the initial state $|\psi(0)\rangle$
\begin{equation}
\label{chap:1:state:t}
|\psi(t)\rangle = U^t|\psi(0)\rangle.
\end{equation}
Let us denote by $\psi_{L,(R)}(m,t)$ the probability amplitude of finding the particle at site $m$ with the coin state $|L(R)\rangle$ after $t$ steps of the quantum walk. These amplitudes are the coefficients of the decomposition of the state vector $|\psi(t)\rangle$ into the basis of the total Hilbert space $\mathcal{H} = \mathcal{H}_P\otimes\mathcal{H}_C$
\begin{equation}
|\psi(t)\rangle = \sum_m \left(\psi_{L}(m,t)|m\rangle|L\rangle + \psi_{R}(m,t)|m\rangle|R\rangle\frac{}{}\right).
\end{equation}
Using the form of the propagator $U$ (\ref{chap:1:U}) we find from the time evolution of the state vector (\ref{chap:1:state:t}) the equations of motions for the probability amplitudes
\begin{eqnarray}
\label{chap:1:amp:t}
\nonumber \psi_L(m,t) & = & \frac{1}{\sqrt{2}} \psi_L(m+1,t-1) + \frac{1}{\sqrt{2}} \psi_R(m+1,t-1),\\
\psi_R(m,t) & = & \frac{1}{\sqrt{2}} \psi_L(m-1,t-1) - \frac{1}{\sqrt{2}} \psi_R(m-1,t-1).
\end{eqnarray}
These equations are reminiscent of the time evolution equations of the classical random walk (\ref{cl:walk:time:evol}). However, in (\ref{chap:1:amp:t}) we transform probability amplitudes instead of probabilities. The probability to find the quantum particle at a particular position $m$ is given by the standard quantum-mechanical formula
$$
P(m,t) = \left|\langle m|\langle L|\psi(t)\rangle\right|^2 + \left|\langle m|\langle R|\psi(t)\rangle\right|^2 = \left|\psi_L(m,t)\right|^2 +\left|\psi_R(m,t)\right|^2.
$$
In Figure~\ref{chap:1:fig2} we display the probability distribution of the classical and quantum walk on a line obtained from the numerical simulation. Concerning the classical random walk depicted by the red points we observe a symmetric gaussian distribution with a rather small width. Indeed, the variance of the classical random walk grows with the square root of the number of steps, which is a typical signature of diffusion. The probability distribution of the quantum walk depicted by the blue points shows striking differences compared to the classical random walk. As we have already discussed, due to the choice of the initial coin state the distribution is biased to the left. More important observation is that the width of the distribution is proportional to the number of steps. Indeed, due to the interference of the probability amplitudes (\ref{chap:1:amp:t}) the growth of the variance is linear in time \cite{nayak}. Hence, the quantum walk is a ballistic process which is the key difference from the diffusive nature of the classical random walk. The quadratic speed-up of the variance is at the heart of the fast algorithms based on quantum walks \cite{shenvi:2003,ambainis:2003,childs:04,kendon:2006,aurel:2007,magniez,vasek}.

\chapter{Recurrence of Quantum Walks}
\label{chap:2}

\nsection{Introduction}

Classical random walks are defined as the probabilistic discrete time evolution of the position of a point-like particle on a discrete graph. Starting the walker from a well-defined graph point (the origin) one can ask whether the particle returns there at least once during the time evolution. The probability of this event is called the P\'olya number \cite{polya}. Classical random walks are said to be {\it recurrent} or {\it transient} depending on whether their P\'olya number equals to one, or is less than one, respectively.

The P\'olya number of a classical random walk can be defined in the following way \cite{revesz}
\begin{equation}
P\equiv\sum\limits_{t=1}^\infty q_0(t),
\label{polya:1:chap2}
\end{equation}
where $q_0(t)$ is the probability that the walker returns to the origin for the {\it first time} after $t$ steps. More practical expression of the P\'olya number is in terms of the probability $p_0(t)$ that the particle can be found at the origin at any given time instant $t$. It is straightforward to show that
\begin{equation}
P = 1-\frac{1}{\sum\limits_{t=0}^{+\infty}p_0(t)}.
\label{polya:def:crw}
\end{equation}
From (\ref{polya:def:crw}) we find that the recurrence behaviour of a random walk is determined solely by the infinite sum
\begin{equation}
{\cal S} \equiv \sum_{t=0}^{\infty}p_0(t).
\label{series}
\end{equation}
Indeed, $P$ equals unity if and only if the series ${\cal S}$ diverges \cite{revesz}. In such a case the random walk is recurrent. On the other hand, if the series $\cal S$ converges, the P\'olya number $P$ is strictly less than unity and the walk is transient. The well-known result found by P\'olya \cite{polya} is that unbiased random walks in one and two dimensions are recurrent while for higher dimensional lattices they are transient. For a more detailed review of recurrence of random walks see Appendix~\ref{app:a}.

We define the P\'olya number of a quantum walk in Section~\ref{chap:2a} by considering a specific measurement scheme. Other possible measurement schemes are briefly discussed. In accordance with the classical terminology we describe the quantum walk as recurrent or transient depending on the value of the P\'olya number. We find a condition for the recurrence of a quantum walk which is given by the asymptotic behaviour of the probability at the origin. A general description of a quantum walk on an infinite $d$-dimensional lattice is left for Section~\ref{chap:2b}. In particular, we find a simple form of the time evolution equation for probability amplitudes. In Section~\ref{chap:2c} we employ the translational invariance of the problem which allows us to solve the equations of motion easily in the momentum representation. We find that the amplitudes in the position representation can be written in the form of an integral over momenta where the time enters only in the oscillating phase. This form of the solution allows a straightforward analysis of the asymptotic behaviour of the amplitudes by means of the method of stationary phase. We perform this analysis in Section~\ref{chap:2d} and discuss the consequences on the recurrence nature of the quantum walk. In particular, we find that the latter is affected by the choice of the coin and the initial coin state.

\section{P\'olya number of a quantum walk}
\label{chap:2a}

For quantum walks we can keep the same definition of the P\'olya number (\ref{polya:1:chap2}) being the probability of returning to the origin at least once during the time evolution. However, to be able to talk about the position of a particle in quantum mechanics one must specify when and which type of measurement is performed. According to the definition (\ref{polya:1:chap2}) we would have to continuously measure whether the particle is at the origin. However, such a radical interruption of the system ultimately leads to a loss of coherence which is a vital ingredient of a quantum walk. It can be anticipated that within the continuous measurement scheme most of the quantum effects become rather weak. The analysis we have performed in \cite{kiss:recurrence} supports this conclusion.

In order to preserve the quantum interference we have considered different measurement scheme in \cite{stef:prl}. The recurrence is understood as a property of an ensemble of particles rather than an individual. The measurement scheme is the following: Prepare an ensemble of quantum walk systems in an identical initial state. Take one of such systems, let it evolve for one step, perform the measurement at the origin and then discard the system. Take a second, identically prepared system, let it evolve for two steps, make a position measurement at the origin and then discard it. Continue until a positive outcome is obtained. In the $t$-th trial we do not find the particle at the origin with the probability $1-p_0(t)$. Since the individual trials are independent the product
$$
\overline{P}_n = \prod_{t=1}^n(1-p_0(t))
$$
gives the probability that we have not found any particle at the origin in the first $n$ trials. In the complementary event, which occurs with the probability
\begin{equation}
P_n = 1-\prod_{t=1}^n(1-p_0(t)),
\label{polya:approx}
\end{equation}
we have found at least one particle at the origin. We define the P\'olya number of a quantum walk by extending $n$ to infinity
\begin{equation}
P = 1-\prod\limits_{t=1}^{+\infty}(1-p_0(t)).
\label{polya:def}
\end{equation}
This definition resembles the expression of the P\'olya number of a classical random walk in terms of the probability at the origin (\ref{polya:def:crw}). However, the inverted sum of $p_0(t)$ is replaced by the product of $1-p_0(t)$. Nevertheless, we show in Appendix~\ref{app:b} that definition (\ref{polya:def}) of the P\'olya number of a quantum walk leads to the same criterion for recurrence in terms of the probability at the origin $p_0(t)$. Indeed, the infinite product in (\ref{polya:def}) vanishes if and only if the series $\cal S$ (\ref{series}) diverges \cite{jarnik}. In such a case the P\'olya number of a quantum walk is unity and we call such quantum walks recurrent. If the series $\cal S$ converges, then the product in (\ref{polya:def}) does not vanish and the P\'olya number of a quantum walk is less than one. In accordance with the classical terminology we call such quantum walks transient.

The convergence of the series $\cal S$ (\ref{series}) is determined by the asymptotic behaviour of the probability at the origin. In the following Sections we find the means which allows us to perform this asymptotic analysis.


\section{Description of quantum walks on $\mathds{Z}^d$}
\label{chap:2b}

Let us first define quantum walks on an infinite $d$ dimensional lattice $\mathds{Z}^d$. The Hilbert space of the quantum walk can be written as a tensor product
$$
\mathcal{H} = \mathcal{H}_P\otimes\mathcal{H}_C
$$
of the position space
$$
\mathcal{H}_P=\ell^2(\mathds{Z}^d)
$$
and the coin space $\mathcal{H}_C$. The position space is spanned by the vectors $|\textbf{m}\rangle$ corresponding to the particle being at the lattice point $\textbf{m}$, i.e.
$$
\mathcal{H}_P=\text{Span}\left\{|\textbf{m}\rangle|\quad\textbf{m}=\left\{m_1,\ldots,m_d\right\}\in\mathds{Z}^d\right\}.
$$
The coin space $\mathcal{H}_C$ is determined by the topology of the walk. In particular, its dimension $n$ is given by the number of possible displacements in a single step. We denote the displacements by vectors
$$
\mathbf{e}_i\in\mathds{Z}^d,\quad i=1,\ldots,n.
$$
Hence, the particle can move from $\textbf{m}$ to any of the points $\textbf{m}+\textbf{e}_i, i=1,\ldots,n$ in a single step. We define an orthonormal basis in the coin space by assigning to every displacement $\mathbf{e}_i$ the basis vector $|\mathbf{e}_i\rangle$, i.e.
$$
\mathcal{H}_C = \textrm{Span}\left\{|\mathbf{e}_i\rangle|i=1,\ldots,n\right\}.
$$
A single step of the quantum walk is given by
\begin{equation}
U=S\cdot\left(I_P\otimes C\right).
\label{qw:time}
\end{equation}
Here $I_P$ denotes the unit operator acting on the position space $\mathcal{H}_P$. The coin flip operator $C$ is applied on the coin state before the displacement $S$ itself. The coin flip $C$ can be in general an arbitrary unitary operator acting on the coin space $\mathcal{H}_C$.

The displacement itself is represented by the conditional step operator $S$
$$
S = \sum\limits_{\mathbf{m},i}|\mathbf{m}+\mathbf{e}_i\rangle\langle\mathbf{m}|\otimes|\mathbf{e}_i\rangle\langle\mathbf{e}_i|,
$$
which moves the particle from the site $\mathbf{m}$ to $\mathbf{m}+\mathbf{e}_i$ if the state of the coin is $|\mathbf{e}_i\rangle$.

Let the initial state of the particle be
$$
|\psi(0)\rangle \equiv \sum\limits_{\mathbf{m},i}\psi_i(\mathbf{m},0)|\mathbf{m}\rangle\otimes|\mathbf{e}_i\rangle.
$$
Here $\psi_i(\mathbf{m},0)$ is the probability amplitude of finding the particle at time $t=0$ at the position $\mathbf{m}$ in the coin state $|\mathbf{e}_i\rangle$. The state of the particle after $t$ steps is given by successive application of the time evolution operator given by Eq. (\ref{qw:time}) on the initial state
\begin{equation}
|\psi(t)\rangle \equiv \sum\limits_{\mathbf{m},i}\psi_i(\mathbf{m},t)|\mathbf{m}\rangle\otimes|\mathbf{e}_i\rangle=U^t|\psi(0)\rangle.
\label{time:evol}
\end{equation}
The probability of finding the particle at the position $\textbf{m}$ at time $t$ is given by the summation over the coin state, i.e.
$$
p(\textbf{m},t) \equiv \sum_{i=1}^n|\langle\textbf{m}|\langle\mathbf{e}_i|\psi(t)\rangle|^2 = \sum_{i=1}^n|\psi_i(\mathbf{m},t)|^2 = ||\psi(\textbf{m},t)||^2.
$$
Here we have introduced $n$-component vectors
$$
\psi(\mathbf{m},t)\equiv{\left(\psi_1(\mathbf{m},t),\psi_2(\mathbf{m},t),\ldots,\psi_n(\mathbf{m},t)\right)}^T
$$
of probability amplitudes. We rewrite the time evolution equation (\ref{time:evol}) for the state vector $|\psi(t)\rangle$ into a set of difference equations
\begin{equation}
\psi(\mathbf{m},t) = \sum_l C_l\psi(\mathbf{m}-\mathbf{e}_l,t-1)
\label{time:evol2}
\end{equation}
for probability amplitudes $\psi(\mathbf{m},t)$. Here the matrices $C_l$ have all entries equal to zero except for the $l$-th row which follows from the coin-flip operator $C$, i.e.
$$
\langle\mathbf{e}_i\left|C_l\right|\mathbf{e}_j\rangle = \delta_{il}\langle\mathbf{e}_i\left|C\right|\mathbf{e}_j\rangle.
$$


\section{Time evolution of quantum walks}
\label{chap:2c}

The quantum walks we consider are translationally invariant which manifests itself in the fact that the matrices $C_l$ on the right-hand side of Eq. (\ref{time:evol2}) are independent of $\mathbf{m}$. Hence, the time evolution equations (\ref{time:evol2}) simplify considerably with the help of the Fourier transformation
\begin{equation}
\tilde{\psi}(\mathbf{k},t)\equiv\sum\limits_\mathbf{m}\psi(\mathbf{m},t) e^{i \mathbf{m}\cdot\mathbf{k}}, \quad \mathbf{k}\in\mathbb{K}^d.
\label{qw:ft}
\end{equation}
The Fourier transformation defined by Eq. (\ref{qw:ft}) is an isometry between $\ell^2(\mathds{Z}^d)$ and $L^2(\mathds{K}^d)$ where $\mathds{K}=(-\pi,\pi]$ can be thought of as the phase of a unit circle in $\mathds{R}^2$.

The time evolution in the Fourier picture turns into a single difference equation
\begin{equation}
\tilde{\psi}(\mathbf{k},t)=\widetilde{U}(\mathbf{k})\tilde{\psi}(\mathbf{k},t-1).
\label{qw:te:fourier}
\end{equation}
Here we have introduced the propagator in the momentum representation
\begin{equation}
\widetilde{U}(\mathbf{k}) \equiv D(\mathbf{k})\cdot C,\quad D(\mathbf{k}) \equiv \textrm{Diag}\left(e^{i\mathbf{e}_1\cdot\mathbf{k}},\ldots,e^{i\mathbf{e}_n\cdot\mathbf{k}}\right).
\label{teopF}
\end{equation}
We find that $\widetilde{U}(\mathbf{k})$ is determined both by the coin $C$ and the topology of the quantum walk through the diagonal matrix $D(\mathbf{k})$ containing the displacements $\mathbf{e}_i$.

We solve the difference equation (\ref{qw:te:fourier}) by formally diagonalising the matrix $\widetilde{U}(\mathbf{k})$. Since it is a unitary matrix its eigenvalues can be written in the exponential form
$$
\lambda_j(\mathbf{k})=\exp{\left(i\ \omega_j(\mathbf{k})\right)},
$$
where the phase is given by the eigenenergy $\omega_j(\mathbf{k})$. We denote the corresponding eigenvectors as $v_j(\mathbf{k})$. Using this notation the state of the particle in the Fourier picture at time $t$ reads
\begin{equation}
\tilde{\psi}(\mathbf{k},t) = \sum_j e^{i\ \omega_j(\mathbf{k})t}\left(v_j(\mathbf{k}),\tilde{\psi}(\mathbf{k},0)\right)v_j(\mathbf{k}),
\label{sol:k}
\end{equation}
where $\left(\ ,\ \right)$ denotes the scalar product in the $n$ dimensional coin space ${\mathcal H}_C$. Finally, we perform the inverse Fourier transformation and find the exact expression for the probability amplitudes
\begin{equation}
\psi(\mathbf{m},t) = \int_{\mathds{K}^d}\frac{d\mathbf{k}}{(2\pi)^d}\ \widetilde{\psi}(\mathbf{k},t)\ e^{-i \mathbf{m}\cdot\mathbf{k}}
\label{inv:f}
\end{equation}
in the position representation.

We are interested in the recurrence nature of quantum walks. As we have discussed in Section~\ref{chap:2a} the recurrence of a quantum walk is determined by the asymptotic behaviour of the probability at the origin
$$
p_0(t)\equiv p(\mathbf{0},t)=\left\|\psi(\mathbf{0},t)\right\|^2.
$$
as the number of steps approaches infinity. Hence, we set $\mathbf{m}=\mathbf{0}$ in Eq. (\ref{inv:f}). Moreover, in analogy with the classical problem of P\'olya we restrict ourselves to quantum walks which start at origin. Hence, the initial condition reads
\begin{equation}
\psi(\mathbf{m},0)=\delta_{\mathbf{m},\mathbf{0}}\psi, \quad \psi\equiv\psi(\mathbf{0},0)
\label{init:cond}
\end{equation}
and its Fourier transformation $\tilde{\psi}(\mathbf{k},0)$ entering Eq. (\ref{sol:k}) is identical to the initial state of the coin
$$
\tilde{\psi}(\mathbf{k},0)=\psi,
$$
which is a $n$-component vector. We note that due to the Kronecker delta in Eq. (\ref{init:cond}) the Fourier transformation $\tilde{\psi}(\mathbf{k},0)$ is independent of the momenta $\mathbf{k}$.

Using the above assumptions we find the exact expression for the probability at the origin
$$
p_0(t) = \left|\sum_{j=1}^c I_j(t)\right|^2
$$
where $I_j(t)$ are given by the integrals
\begin{equation}
I_j(t) = \int\limits_{\mathbb{K}^d}\frac{d\mathbf{k}}{(2\pi)^d}\ e^{i\ \omega_j(\mathbf{k})t}\ f_j(\mathbf{k}),\quad f_j(\mathbf{k}) = \left(v_j(\mathbf{k}),\psi\right)\ v_j(\mathbf{k}).
\label{psi:0}
\end{equation}


\section{Asymptotics of the probability at the origin}
\label{chap:2d}

Let us discuss how the additional freedom we have at hand for quantum walks influences the asymptotics of the probability at the origin $p_0(t)$. We suppose that the functions $\omega_j(\mathbf{k})$ and  $f_j(\mathbf{k})$ entering $I_j(t)$ are smooth. According to the method of stationary phase \cite{statphase} which we briefly review in Appendix~\ref{app:c} the major contribution to the integral $I_j(t)$ comes from the stationary points $\mathbf{k}^0$ of the eigenenergies $\omega_j(\mathbf{k})$, i.e. from the points where the gradient vanishes
$$
\left.\vec{\nabla}\omega_j(\mathbf{k})\right|_{\mathbf{k}=\mathbf{k}^0} = \mathbf{0}.
$$
The asymptotic behaviour of $I_j(t)$ is then determined by the stationary point with the greatest degeneracy given by the dimension of the kernel of the Hessian matrix
$$
H^{(j)}_{m,n}(\mathbf{k})\equiv \frac{\partial^2 \omega_j(\mathbf{k})}{\partial k_m\partial k_n}
$$
evaluated at the stationary point, i.e. by the flatness of $\omega_j(\mathbf{k})$. The function $f_j(\mathbf{k})$ entering the integral $I_j(t)$ determines only the pre-factor. We now discuss how the existence, configuration and number of stationary points affect the asymptotic behaviour of $I_j(t)$. As a rule of thumb, the decay of the probability at the origin $p_0(t)$ can slow down with the increase in the number of stationary points. Let us briefly discuss the results.

\subsection{No stationary points}
\label{chap:2d1}

If $\omega_j(\mathbf{k})$ has no stationary points then $I_j(t)$ decays faster than any inverse polynomial in $t$. Consequently, the decay of the probability at the origin is also exponential
$$
p_0(t) \sim e^{- \gamma t}
$$
with some positive rate $\gamma$. Indeed, quantum walks for which the probability at the origin decays so fast are transient. Such a situation occurs e.g. for extremely biased quantum walks which we analyze in Chapter~\ref{chap:5}.

\subsection{Finite number of stationary points}
\label{chap:2d2}

Suppose that $\omega_j(\mathbf{k})$ has a finite number of non-degenerate stationary points, i.e. the determinant of the Hessian matrix $H$ is non-zero for all stationary points. If the function $f_j(\mathbf{k})$ does not vanish at the stationary points then the contribution from all stationary points to the integral $I_j(t)$ is of the order $t^{-d/2}$. Consequently, the probability at the origin behave like
$$
p_0(t) \sim t^{-d}
$$
as $t$ approaches infinity. Clearly, the sum $\cal S$ defined in (\ref{series}) is convergent for $d>1$. Hence, the quantum walks for which the eigenenergies have only non-degenerate stationary points are recurrent only for the dimension $d=1$, i.e. on a line. This is e.g. the case of the Hadamard walk with tensor product coin studied in Chapter~\ref{chap:4b}.

\subsection{Continuum of stationary points}
\label{chap:2d3}

If $\omega_j(\mathbf{k})$ has a continuum of stationary points then the dimension of the continuum determines the decay of the integral $I_j(t)$. The case of 2-D integrals with curves of stationary points are treated in \cite{statphase}. It is shown that the contribution from the continuum of stationary points to the integral $I_j(t)$ is of the order $t^{-1/2}$. This is greater than the contribution arising from a discrete stationary point which is of the order $t^{-1}$. Hence, the continuum of stationary points has effectively slowed-down the decay of the integral $I_j(t)$. Consequently, the leading order term of the probability at the origin is
$$
p_0(t)\sim t^{-1},
$$
and we find that such a quantum walk is recurrent. We come across this situation in the case of the Fourier walk on a plane in Chapter~\ref{chap:4d}. Similar results can be expected for higher dimensional quantum walks where $\omega_j(\mathbf{k})$ have a continuum of stationary points.

A special case for a continuum of stationary points is when $\omega_j(\mathbf{k})$ does not depend on $n$ variables, say $k_1,\ldots k_n$, but has a finite number of stationary points with respect to the remaining $d-n$ variables $k_{n+1},\ldots, k_d$. Indeed, such an $\omega_j(k_{n+1},\ldots, k_d)$ has obviously a zero derivative with respect to $k_i,\ i=1,\ldots n$. Suppose that the function $f_j(\mathbf{k})$ factorizes
$$
f_j(\mathbf{k}) = g_j(k_1,\ldots,k_n)\cdot h_j(k_{n+1},\ldots,k_d).
$$
In such a case $I_j(t)$ is given by the product of time-independent and time-dependent integrals over $n$ and $d-n$ variables
$$
I_j(t) = \left[\ \int\limits_{\mathbb{K}^n} \frac{d\mathbf{k}}{(2\pi)^n} g_j(k_1,\ldots, k_n)\right]\cdot \left[\ \int\limits_{\mathbb{K}^{d-n}} \frac{d\mathbf{k}}{(2\pi)^{d-n}} e^{i\ \omega_j(k_{n+1},\ldots, k_d) t} h_j(k_{n+1},\ldots, k_d)\right].
$$
It is easy to find that if the time-independent integral does not vanish $I_j(t)$ behaves asymptotically like $t^{-(d-n)/2}$. Hence, the asymptotic behaviour of the probability at the origin is
$$
p_0(t)\sim{t^{-(d-n)}}.
$$
The quantum walks of this kind would be recurrent if the eigenenergy $\omega_j$ would depend only on a single component of the momenta $\mathbf{k}$. In the extreme case when $\omega_j(\mathbf{k})$ does not depend on $\mathbf{k}$ at all we can extract the time dependence out of the integral $I_j(t)$. If the remaining time independent integral does not vanish then $p_0(t)$ converges to a non-zero value and say that such a quantum walk exhibits {\it localization}. Note that since $p_0(t)$ has a non-vanishing limit the quantum walk is recurrent. Indeed, localization implies recurrence. We find localization in Chapter~\ref{chap:4c} for the Grover walk on a plane. Moreover, extending the 2-D Grover walk to $\mathds{Z}^d$ we find quantum walks where some of the eigenenergies are either constant or depend only on a single momentum component. As discussed above, such quantum walks are recurrent.

\subsection{Effect of the initial state}
\label{chap:2d4}

So far we have assumed that the function $f_j(\mathbf{k})$ is non-vanishing for $\mathbf{k}$ values corresponding to the stationary points. However, the initial state $\psi$ can be orthogonal to the eigenvector $v_j(\mathbf{k})$ for $\mathbf{k}=\mathbf{k}^0$ corresponding to the stationary point. In such a case the function $f_j(\mathbf{k})$ vanishes for $\mathbf{k}=\mathbf{k}^0$ and the stationary point $\mathbf{k}^0$ does not contribute to the integral $I_j(t)$. Consequently, the decay of $p_0(t)$ can speed up. Hence, for quantum walks we might change the recurrence behaviour and the actual value of the P\'olya number by altering the initial state $\psi$. Indeed, we find this non-trivial effect of the initial state for the Grover walk and the Fourier walk on a plane in Chapters~\ref{chap:4c} and \ref{chap:4d}.


\chapter{Recurrence of Unbiased Quantum Walks on Infinite Lattices}
\label{chap:4}


\nsection{Introduction}

In the present Chapter we determine the recurrence behaviour and the P\'olya number of several unbiased quantum walks. We concentrate on the effect of the coin operators and the initial states. For this purpose we fix the topology of the walks. We consider quantum walks where the displacements $\mathbf{e}_i$ have all entries equal to $\pm 1$
$$
\mathbf{e}_1 = \left(1,\ldots,1\right)^T,\ldots, \mathbf{e}_{2^d} = \left(-1,\ldots,-1\right)^T.
$$
In such a case the coin space has the dimension $n=2^d$ where $d$ is the dimension of the lattice. Moreover, the diagonal matrix $D(\mathbf{k})$ entering the propagator in the Fourier picture (\ref{teopF}) can be written as a tensor product
\begin{equation}
D(\textbf{k}) = D(k_1)\otimes\ldots\otimes D(k_d)
\label{dk2}
\end{equation}
of $2\times 2$ diagonal matrices
$$
D(k_j)=\textrm{Diag}\left(e^{-ik_j},e^{ik_j}\right).
$$
This fact allows us to extend some of the results for the quantum walks on a line or on a plane to quantum walks on a $d$-dimensional lattice.

First, in Section~\ref{chap:4b} we treat Hadamard walk on $\mathds{Z}^d$ with an independent coin for each spatial dimension. We find that for this quantum walk the probability at the origin is independent of the initial coin state. Hence, a unique P\'olya number can be assigned to this quantum walk for each dimension $d$. In contrast with the classical random walks the Hadamard walk is recurrent only for $d=1$. In Section~\ref{chap:4c} we analyze the recurrence of the Grover walk on a plane. This quantum walk exhibits localization \cite{localization} and therefore is recurrent. However, for a particular initial state localization disappears and the Grover walk becomes transient. We find an approximation of the P\'olya number for this particular initial state. We then employ the Grover walk on a plane to construct for arbitrary dimension $d$ a quantum walk which is recurrent. This is in great contrast with the classical random walks, which are recurrent only for the dimensions $d=1,2$. Finally, in Section~\ref{chap:4d} we analyze the Fourier walk on a plane. This quantum walk is recurrent except for a two-parameter family of initial states for which it is transient. For the latter case we find an approximation of the P\'olya number depending on the parameters of the initial state. We summarize our results in Section~\ref{chap:4e}.


\section{Hadamard walk on $\mathds{Z}^d$}
\label{chap:4b}

Let us start with the analysis of the recurrence behaviour of the Hadamard walk on a line which is driven by the coin
$$
H = \frac{1}{\sqrt{2}}\left(
            \begin{array}{cc}
              1 & 1 \\
              1 & -1 \\
            \end{array}
          \right).
$$
We find that the propagator in the Fourier picture
\begin{equation}
\widetilde{U}_H(k) = D(k)\cdot H = \frac{1}{\sqrt{2}}\left(
            \begin{array}{cc}
              e^{ik} & e^{ik} \\
              e^{-ik} & -e^{-ik} \\
            \end{array}
          \right)
\label{1d:Had}
\end{equation}
has eigenvalues $e^{i\ \omega_i(k)}$ where the phases $\omega_i(k)$ are given by
$$
\omega_1(k) = \arcsin\left(\frac{\sin k}{\sqrt{2}}\right),\quad \omega_2(k) = -\pi-\arcsin\left(\frac{\sin k}{\sqrt{2}}\right).
$$
Thus the derivatives of $\omega_i$ with respect to $k$ reads
\begin{equation}
\frac{d\omega_1(k)}{dk}=-\frac{d\omega_2(k)}{dk}=-\frac{\cos k}{\sqrt{2 - \sin^2 k}}
\label{der:1d}
\end{equation}
and we find that the phases $\omega_i(k)$ have common non-degenerate stationary points $k^0=\pm\pi/2$. It follows that the probability at the origin behaves asymptotically like $t^{-1}$. This asymptotic scaling is independent of the initial state. Indeed, no non-zero initial state $\psi$ exists which is orthogonal to both eigenvectors at the common stationary points $k^0=\pm\pi/2$. Hence, the Hadamard walk on a line is recurrent, i.e. the P\'olya number equals one, independent of the initial coin state.

We turn to the Hadamard walk on a $d$-dimensional lattice. The coin flip operator has the form of the tensor product of $d$ $2\times 2$ Hadamard matrices
\begin{equation}
H_d = H\otimes\ldots\otimes H.
\label{C:ind:x}
\end{equation}
Hence, we have an independent coin for each spatial dimension. It follows that also the propagator in the Fourier picture has the form of the tensor product
\begin{equation}
\widetilde{U}_{H_d}(\textbf{k}) = \widetilde{U}_H(k_1)\otimes\ldots\otimes \widetilde{U}_H(k_d)
\label{C:ind}
\end{equation}
of $d$ time evolution operators given by Eq. (\ref{1d:Had}) with different momenta $k_i$. Hence, the eigenenergies of the propagator (\ref{C:ind})  have the form of the sum
\begin{equation}
\omega_j(\textbf{k})=\sum_{l=1}^d \omega_{j_l}(k_l).
\label{eigenval:Cind}
\end{equation}
Therefore we find that the asymptotic behaviour of this quantum walk follows directly from the asymptotics of the Hadamard walk on a line. Indeed, the derivative of the phase $\omega_j(\textbf{k})$ with respect to $k_l$ reads
\begin{equation}
\frac{\partial \omega_j(\textbf{k})}{\partial k_l} = \frac{d \omega_{j_l}(k_l)}{d k_l},
\label{der:phase:Cind}
\end{equation}
and so $\omega_j(\textbf{k})$ has a stationary point $\textbf{k}^0=\left(k_1^0,k_2^0,\ldots,k_d^0\right)$ if and only if for all $l=1,\ldots,d$ the point $k_l^0$ is the stationary point of $\omega_{j_l}(k_l,\alpha_l)$. As we have found from Eq. (\ref{der:1d}) the stationary points of $\omega_{j_l}$ are $k^0_l=\pm\pi/2$. Hence, all phases $\omega_j(\textbf{k})$ have $2^d$ common stationary points $\textbf{k}^0=\left(\pm\pi/2,\ldots,\pm\pi/2\right)$. It follows that the asymptotic behaviour of the probability $p_0(t)$ is given by
\begin{equation}
p_0(t)\sim t^{-d}.
\label{asymp:Cind}
\end{equation}
As follows from the results for the Hadamard walk on a line the asymptotic behaviour given by Eq. (\ref{asymp:Cind}) is independent of the initial coin state. Compared to classical walks this is a quadratically faster decay of the probability at the origin which is due to the quadratically faster spreading of the probability distribution of the quantum walk.

\begin{figure}[p]
\begin{center}
\includegraphics[width=0.6\textwidth]{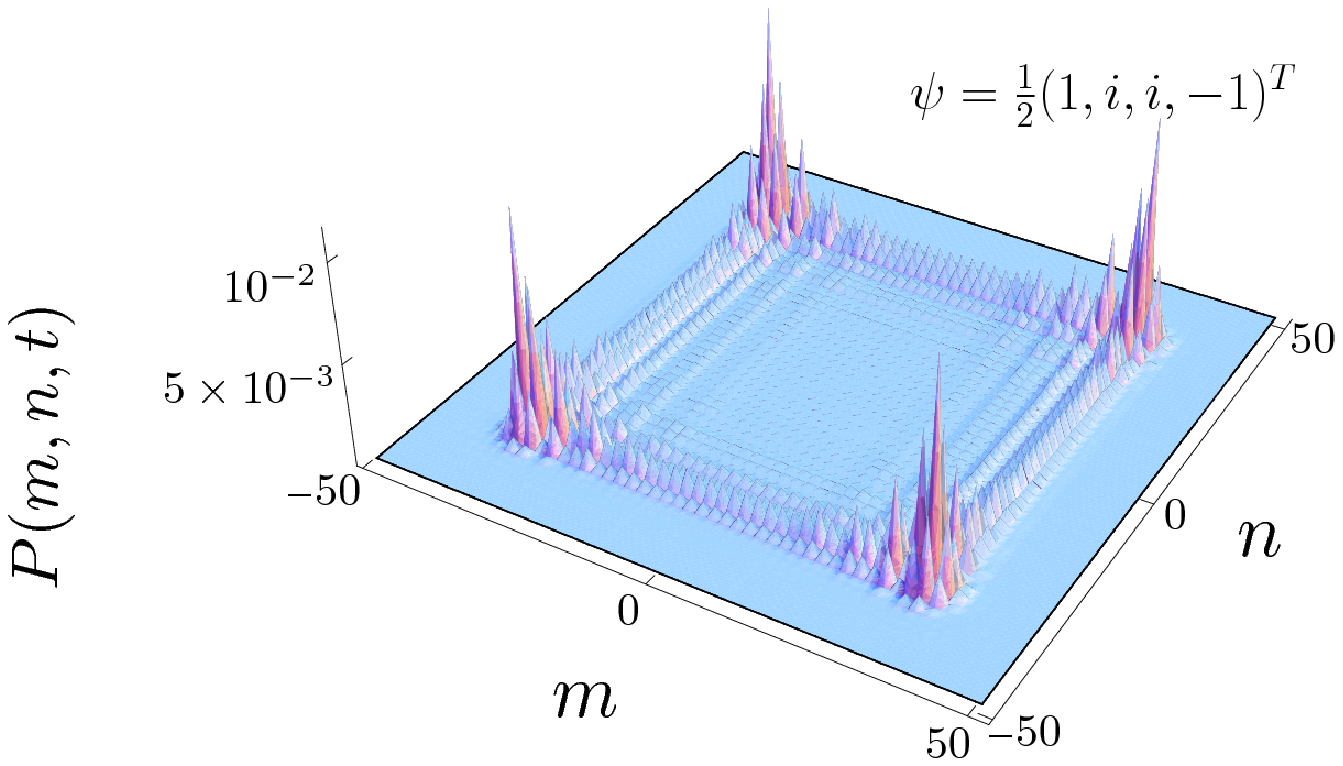}\vspace{24pt}
\includegraphics[width=0.6\textwidth]{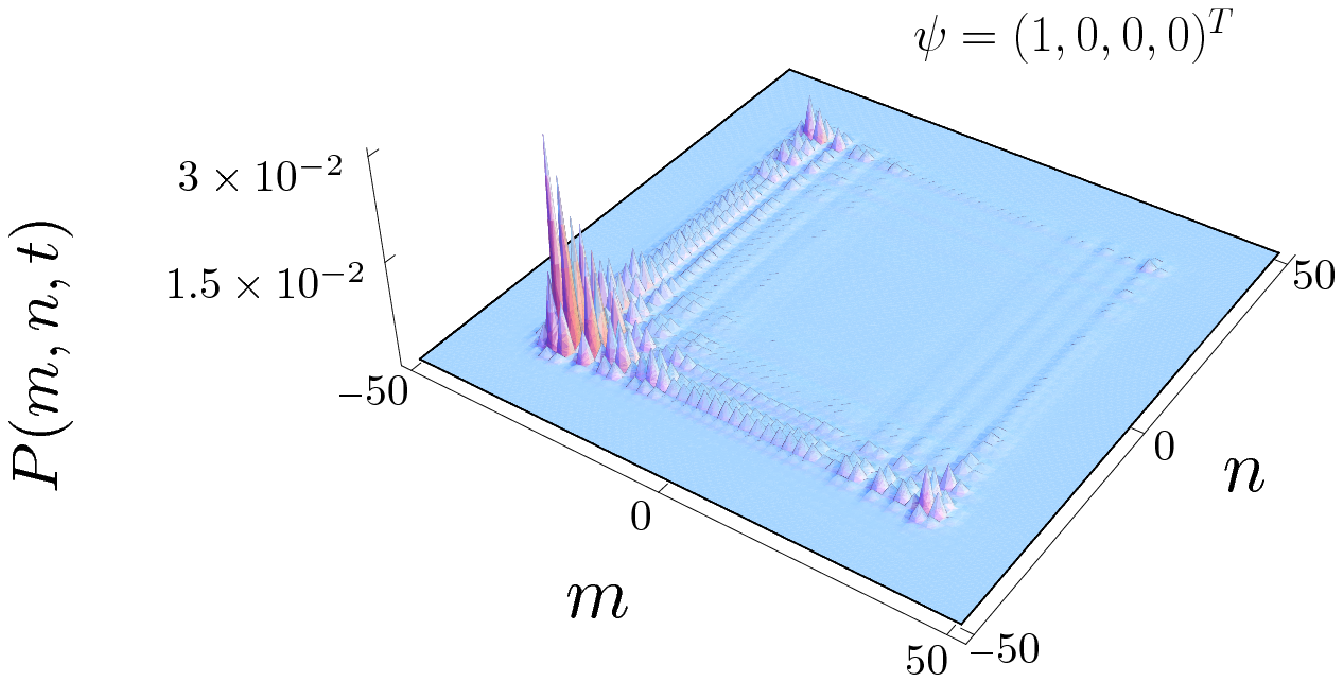}\vspace{24pt}
\includegraphics[width=0.45\textwidth]{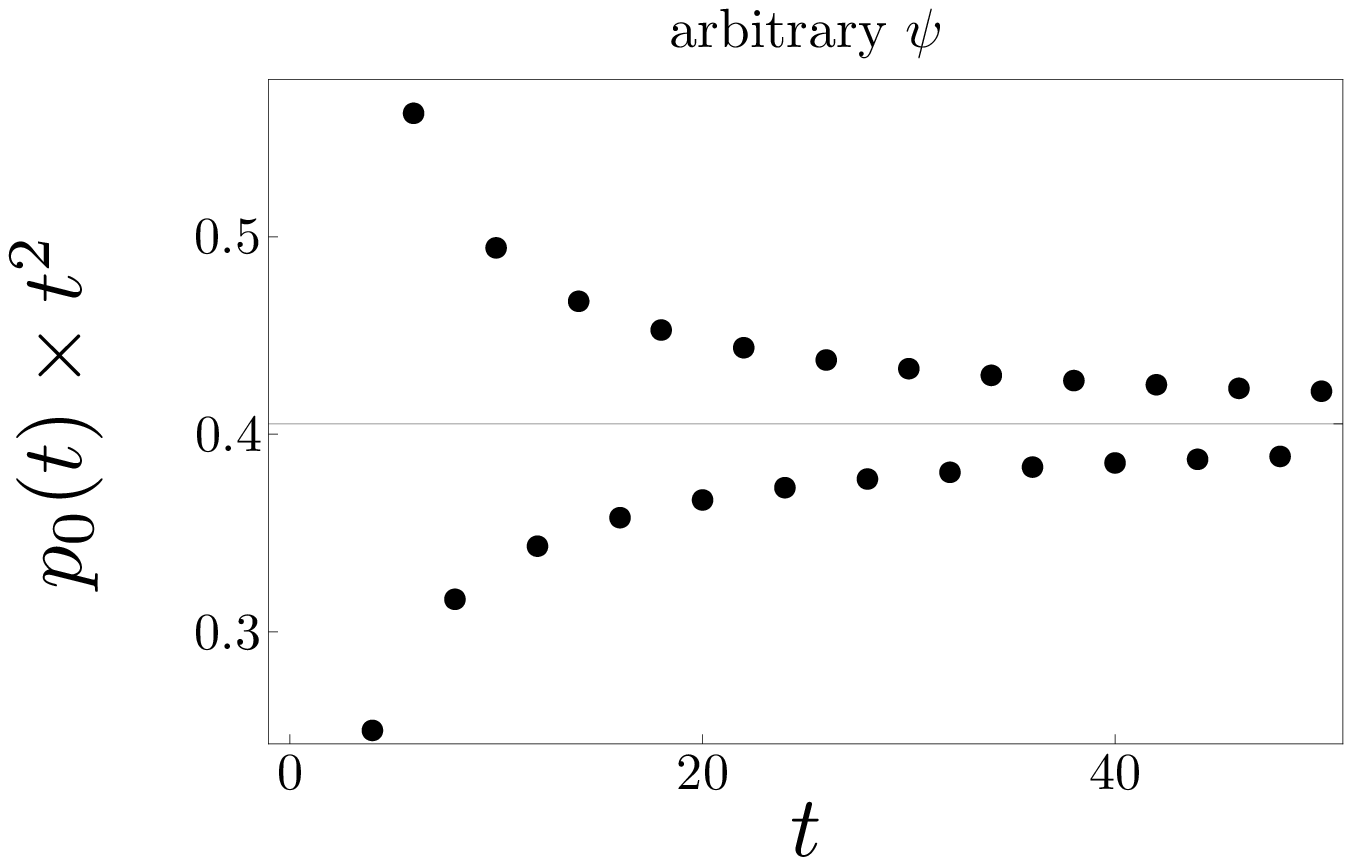}
\caption{Probability distribution of the Hadamard walk on a plane after 50 steps and the probability at the origin $p_0(t)$ for different choices of the initial state. In the upper plot we choose the initial state $\frac{1}{2}(1,i,i,-1)^T$ which leads to a symmetric probability distribution, whereas in the middle plot we choose the initial state $(1,0,0,0)^T$ resulting in a dominant peak of the probability distribution in the lower-left corner of the $(m,n)$ plane. However, the initial state influences the probability distribution only near the edges. The probability $p_0(t)$ is unaffected and is the same for all initial coin states. The lower plot confirms the asymptotic behaviour of the probability at the origin $p_0(t)\sim t^{-2}$ independent of the initial state.}
\label{had:2d}
\end{center}
\end{figure}

We illustrate the results for Hadamard walk on a plane driven by the coin
\begin{equation}
H_2 = \frac{1}{2}\left(
                      \begin{array}{rrrr}
                        1 & 1 & 1 & 1\\
                        1 & -1 & 1 & -1 \\
                        1 & 1 & -1 & -1\\
                        1 & -1 & -1 & 1\\
                      \end{array}
                    \right)
\end{equation}
in \fig{had:2d}. Here we show the probability distribution in dependence on the initial state and the probability at the origin $p_0(t)$. The first two plots  indicates that the initial state of the coin influences mainly the edges of the probability distribution. However, the probability $p_0(t)$ is unaffected and is exactly the same for all initial states. The lower plot confirms the asymptotic behaviour of the probability at the origin $p_0(t)\sim t^{-2}$.


Since the probability at the origin $p_0(t)$ decays like $t^{-d}$ we find that the Hadamard walk on $\mathds{Z}^d$ is recurrent only for dimension $d=1$ and is transient for all higher dimensions $d\geq 2$. Moreover, the whole sequence of probabilities $p_0(t)$ is independent of the initial state. Hence, the P\'olya number for this class of quantum walks depends only on the dimension of the walk $d$, thus resembling the property of the classical walks. On the other hand, this quantum walk is transient for the dimension $d=2$ and higher. This is a direct consequence of the faster decay of the probability at the origin which, in this case, cannot be compensated for by interference.

Let us estimate the value of the P\'olya number for the dimension $d\geq 2$. As depicted in the lowest plot of \fig{had:2d} the probability at the origin approaches quite rapidly its asymptotic form
\begin{equation}
p_0(t)\approx\frac{1}{(\pi t)^{d}}.
\end{equation}
Hence, already the first few terms of the product in Eq. (\ref{polya:approx}) are sufficient to estimate the value of the P\'olya number. Taking into account the first three terms of $p_0(t)$  which are found to be
\begin{equation}
p_0(2)=\frac{1}{2^d},\quad p_0(4)=p_0(6)=\frac{1}{8^d},
\end{equation}
we obtain the following approximation of the P\'olya number
\begin{equation}
P_{H_d}\approx 1 - \left(1-\frac{1}{2^d}\right)\left(1-\frac{1}{8^d}\right)^2.
\label{Polya:ind:est}
\end{equation}
We compare the estimation in Eq. (\ref{Polya:ind:est}) with the numerical results obtained from the simulation of the Hadamard walk with 1000 steps in the \tab{tab1} and find that they are in excellent agreement.

\begin{table}
\begin{center}
\begin{tabular}{|c|c|c|c|}
  \hline
  \multirow{2}{*}{Dimension} & \multirow{2}{*}{Simulation} & \multirow{2}{*}{Estimation (\ref{Polya:ind:est})} & \multirow{2}{*}{Error ($\%$)}\\
  & & & \\\hline
  \multirow{2}{*}{2} & \multirow{2}{*}{0.29325} & \multirow{2}{*}{0.27325} & \multirow{2}{*}{6.8}\\
  & & & \\\hline
  \multirow{2}{*}{3} & \multirow{2}{*}{0.12947} & \multirow{2}{*}{0.12841} & \multirow{2}{*}{0.82}\\
  & & & \\\hline
  \multirow{2}{*}{4} & \multirow{2}{*}{0.06302} & \multirow{2}{*}{0.06296} & \multirow{2}{*}{0.01}\\
  & & & \\\hline
  \multirow{2}{*}{5} & \multirow{2}{*}{0.031313} & \multirow{2}{*}{0.031309} & \multirow{2}{*}{0.01}\\
  & & & \\
  \hline
\end{tabular}
\caption{Comparison of the P\'olya number for the Hadamard walk on $\mathds{Z}^d$ obtained from the numerical simulation and the estimation of Eq. (\ref{Polya:ind:est}).}
\label{tab1}
\end{center}
\end{table}


\section{Grover walk on a plane}
\label{chap:4c}

We turn to the Grover walk on a plane which is driven by the coin
\begin{equation}
G = \frac{1}{2}\left(
                 \begin{array}{rrrr}
                   -1 & 1 & 1 & 1 \\
                   1 & -1 & 1 & 1 \\
                   1 & 1 & -1 & 1 \\
                   1 & 1 & 1 & -1 \\
                 \end{array}
               \right).
\label{grover:coin}
\end{equation}
It was identified numerically \cite{2dw1} and later proven analytically \cite{localization} that the Grover walk exhibits a localization effect, i.e. the probability $p_0(t)$ does not vanish but converges to a non-zero value except for a particular initial state
\begin{equation}
\psi_G\equiv\psi_G(0,0,0) = \frac{1}{2}\left(1,-1,-1,1\right)^T.
\label{grover:nospike:state}
\end{equation}

In order to explain the localization we analyze the eigenvalues of the propagator in the Fourier picture for the Grover walk
\begin{equation}
\widetilde{U}_G(k_1,k_2) = \left(D(k_1)\otimes D(k_2)\right) G.
\label{gkl}
\end{equation}
We find that they are given by
\begin{equation}
\label{eigenval:Grover}
\lambda_{1,2}  =  \pm 1,\qquad \lambda_{3,4}(k_1,k_2) = e^{\pm i\ \omega(k_1,k_2)}
\end{equation}
where the phase $\omega(k_1,k_2)$ reads
\begin{equation}
\cos(\omega(k_1,k_2)) = -\cos{k_1}\cos{k_2}.
\label{phase:Grover}
\end{equation}
The eigenvalues $\lambda_{1,2}$ are constant. As a consequence the probability at the origin is non-vanishing as discussed in detail in Chapter~\ref{chap:2d3}, unless the initial state is orthogonal to the eigenvectors corresponding to $\lambda_{1,2}$ at every point $(k_1,k_2)$. By explicitly calculating the eigenvectors of the matrix $\widetilde{U}_G(k_1,k_2)$ it is straightforward to see that such a vector is unique and equals that in Eq. (\ref{grover:nospike:state}), in agreement with the result derived in \cite{localization}.

\begin{figure}[p]
\begin{center}
\includegraphics[width=0.7\textwidth]{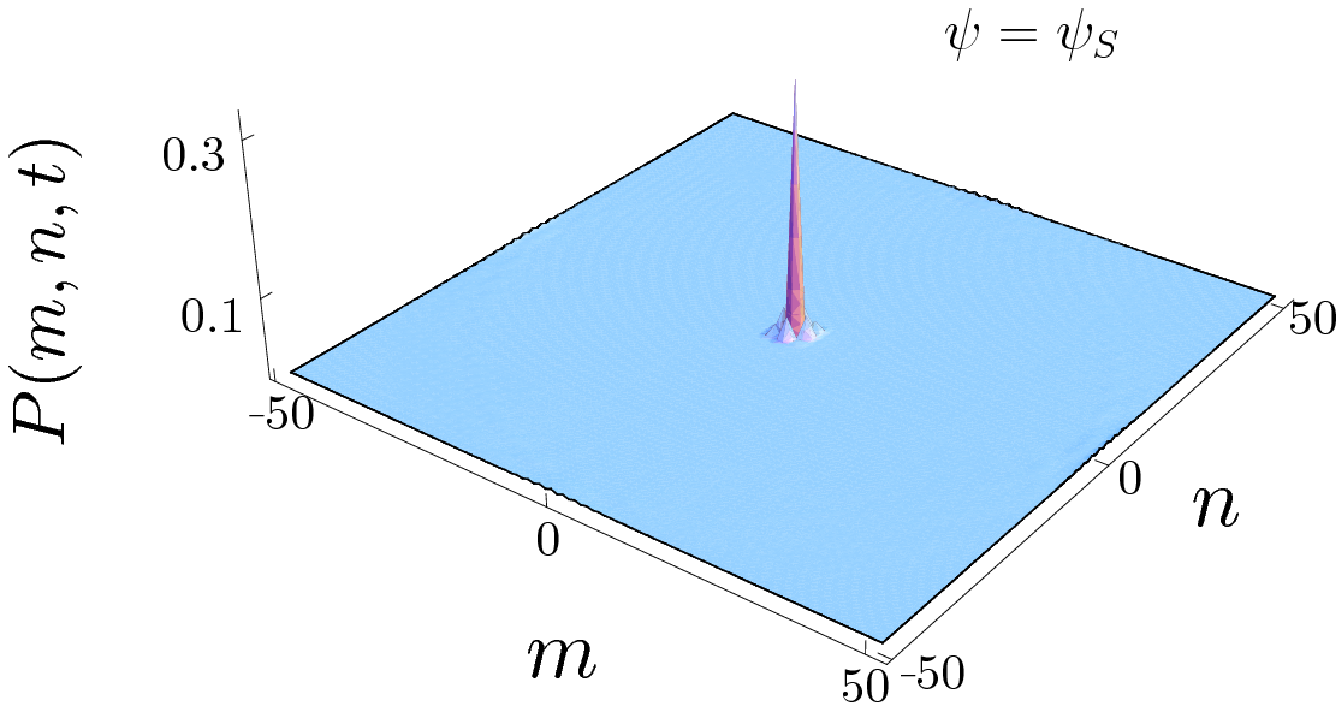}\vspace{36pt}
\includegraphics[width=0.6\textwidth]{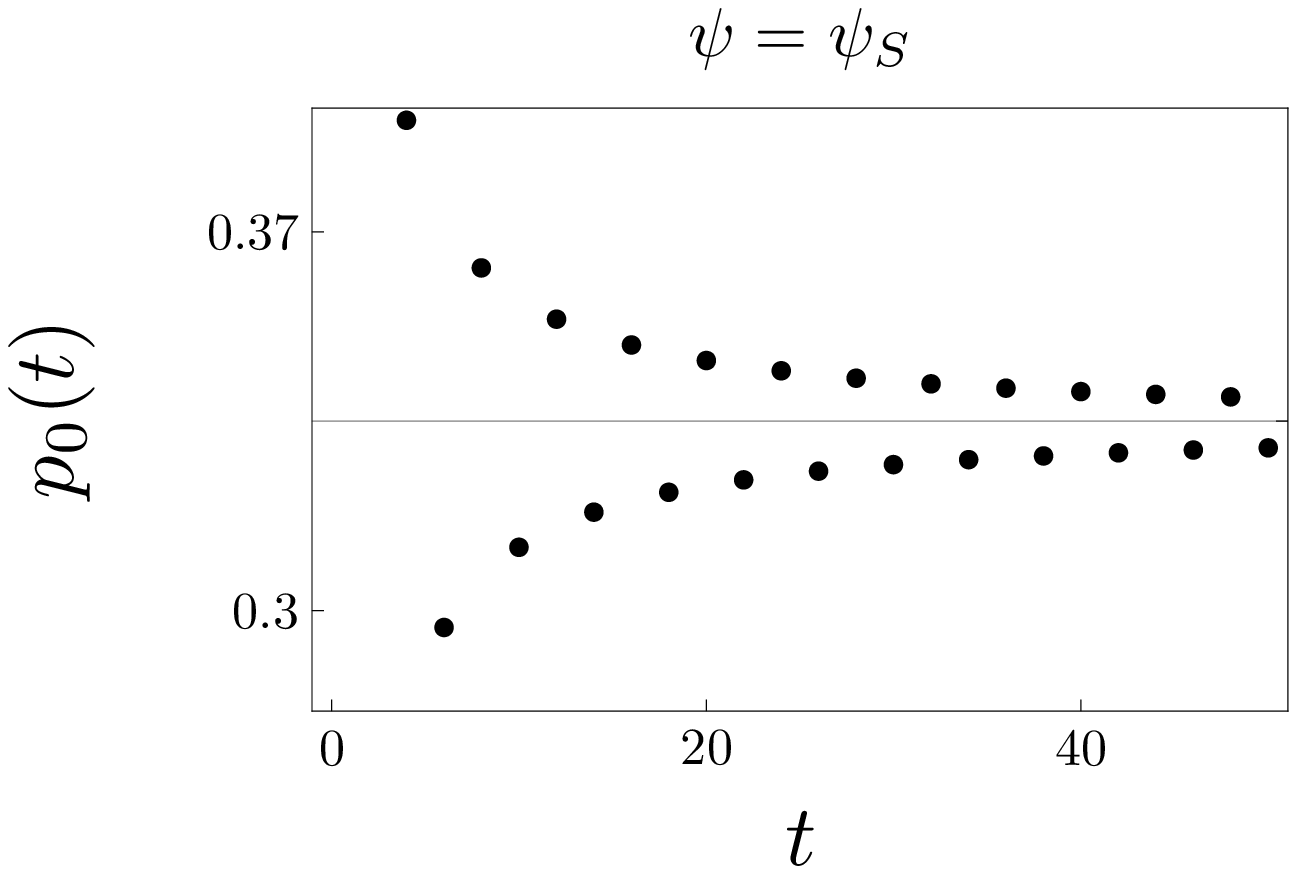}
\caption{Probability distribution of the Grover walk after 50 steps and the probability at the origin for a symmetric initial state (\ref{grover:psi:s}). This particular choice of the initial state leads to a symmetric probability distribution with a dominant central spike, as depicted in the upper plot. The lower plot indicates that the probability at the origin has a non-vanishing limit as $t$ approaches infinity. The results are qualitatively the same for all initial coin states except for $\psi_G$ given in (\ref{grover:nospike:state}), as we illustrate in \fig{grover:fig2}.}
\label{grover:fig1}
\end{center}
\end{figure}

\begin{figure}[p]
\begin{center}
\includegraphics[width=0.7\textwidth]{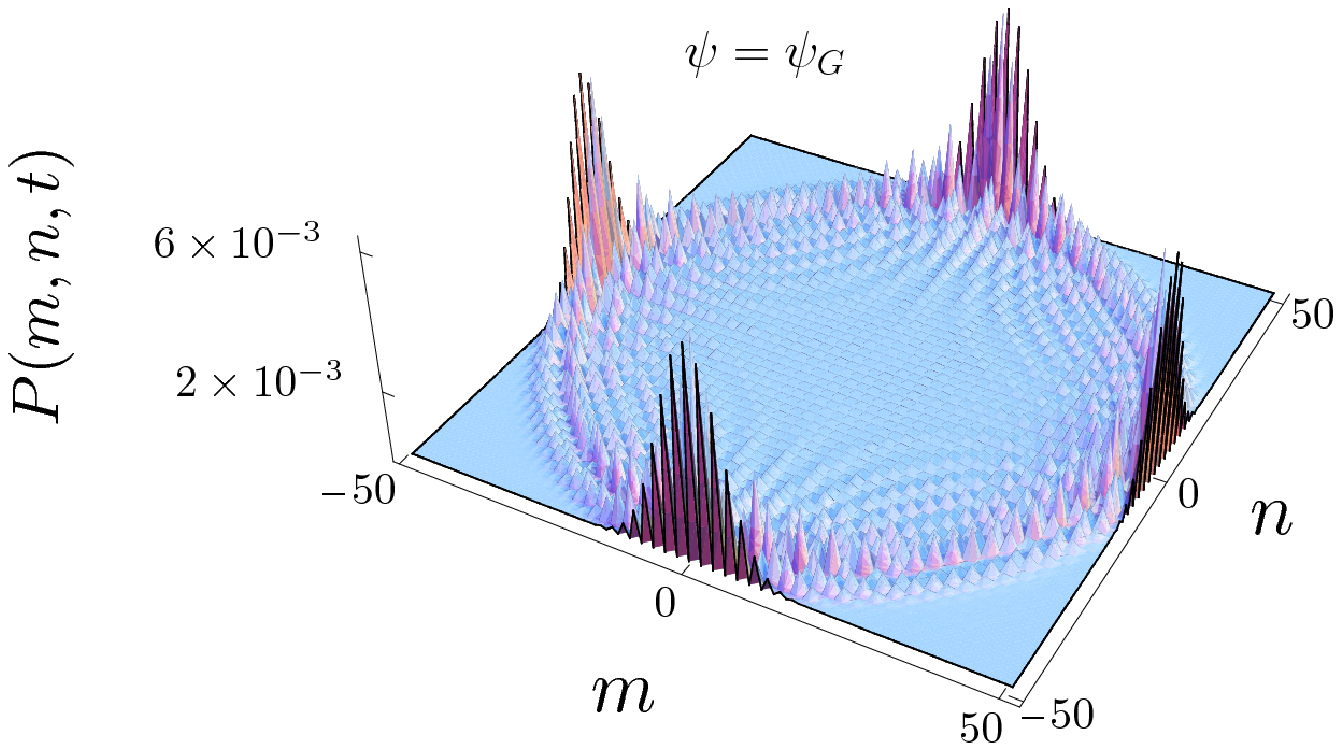}\vspace{36pt}
\includegraphics[width=0.6\textwidth]{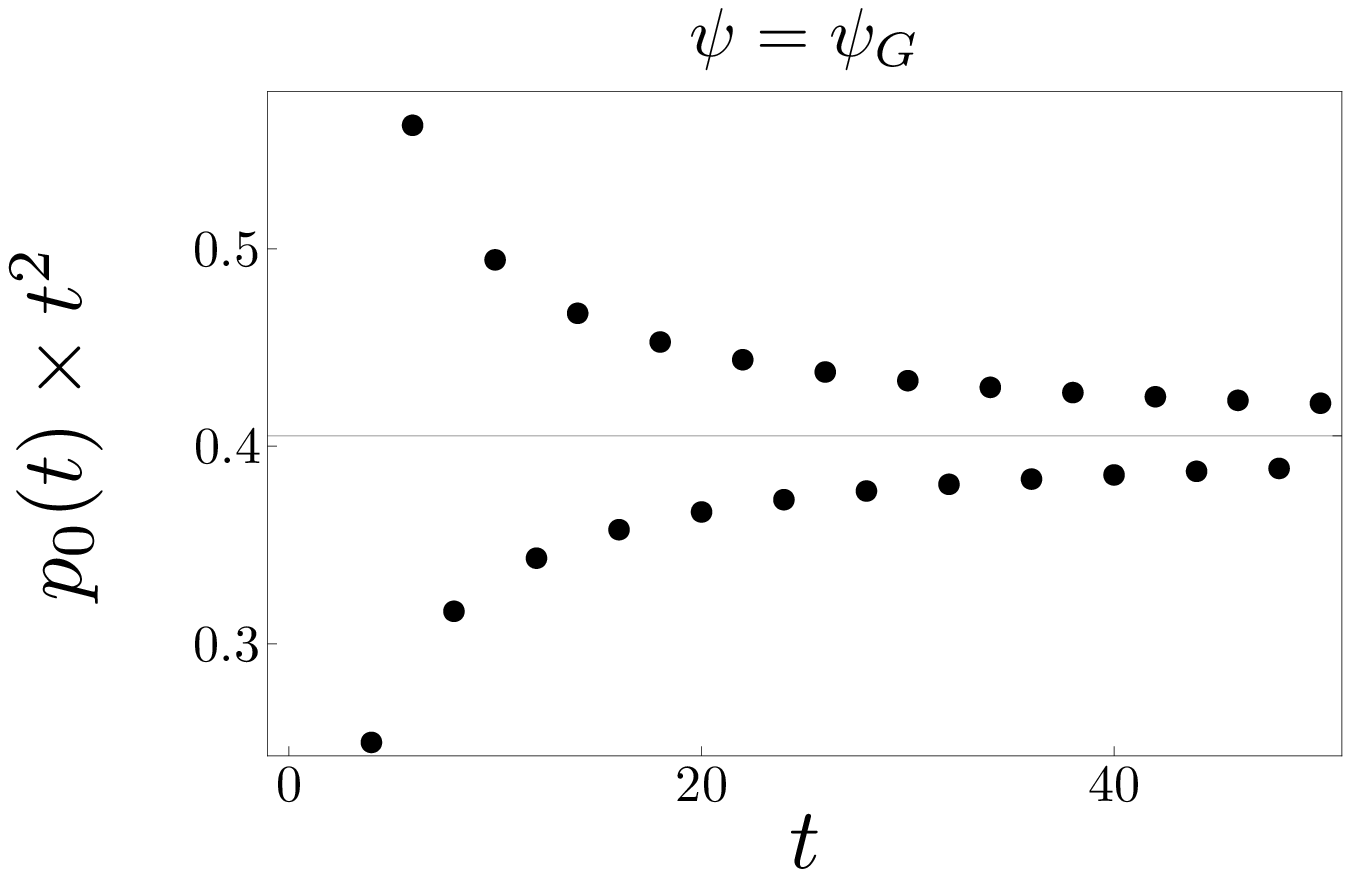}
\caption{Probability distribution of the Grover walk after 50 steps and the probability at the origin for a particular initial state $\psi_G$ given by Eq. (\ref{grover:nospike:state}). In contrast to \fig{grover:fig1} we find that the central spike vanishes and most of the probability is situated at the edges. Moreover, the probability at the origin vanishes as $t$ approaches infinity, as we illustrate in the lower figure. Here we plot the probability $p_0(t)$ multiplied by $t^2$ to unravel the asymptotic behavior of the probability at the origin. The plot confirms the analytic result of the scaling $p_0(t)\sim t^{-2}$.}
\label{grover:fig2}
\end{center}
\end{figure}

It is easy to show that for the particular initial state given by Eq. (\ref{grover:nospike:state}) the probability $p_0(t)$ decays like $t^{-2}$. Indeed, as the initial state of Eq. (\ref{grover:nospike:state}) is orthogonal to the eigenvectors corresponding to $\lambda_{1,2}$ the asymptotic behaviour is determined by the remaining eigenvalues $\lambda_{3,4}(k_1,k_2)$, or more precisely by the stationary points of $\omega(k_1,k_2)$. From Eq. (\ref{phase:Grover}) we find that it has only non-degenerate stationary points $k_1^0,\ k_2^0=\pm \pi/2$. For the initial state of Eq. (\ref{grover:nospike:state}) the probability that the Grover walk returns to the origin decays like $t^{-2}$. We conclude that the Grover walk on a 2-D lattice is recurrent and its P\'olya number equals one for all initial states except the one given in Eq. (\ref{grover:nospike:state}) for which the walk is transient. We illustrate these results in \fig{grover:fig1} and \fig{grover:fig2}.

In \fig{grover:fig1} we show the probability distribution generated by the Grover walk and the probability at the origin for a symmetric initial state
\begin{equation}
\psi_S=\frac{1}{2}(1,i,i,-1)^T.
\label{grover:psi:s}
\end{equation}
This particular choice of the initial state  results to a probability distribution with a dominant central spike, as depicted in the upper plot. The lower plot indicates that the probability at the origin has a non-vanishing limit.

In contrast for the initial state $\psi_G$ given by (\ref{grover:nospike:state}) the central spike in the probability distribution vanishes, as we illustrate in the upper plot of \fig{grover:fig2}. The lower plot indicates that the probability at the origin decays like $t^{-2}$.


Let us estimate the P\'olya number of the Grover walk for the initial state of Eq. (\ref{grover:nospike:state}). The numerical simulations indicate that the probability at the origin $p_0(t)$ for the initial state $\psi_G$ is the same as the probability at the origin of the 2-D Hadamard walk. Hence, their P\'olya numbers coincide. With the help of the relation (\ref{Polya:ind:est}) we can estimate the P\'olya number of the Grover walk with the initial state of $\psi_G$ by
\begin{equation}
P_G(\psi_G) \equiv P_{H_2}\approx 0.27325.
\end{equation}

The above derived results allow us to construct a quantum walk which is recurrent for an arbitrary dimension $d$, except for a subspace of initial states.
Let us first consider the case when the dimension of the walk is even and equals $2d$. We choose the coin as a tensor product
\begin{equation}
G_{2d} = \otimes^d G
\label{c2d}
\end{equation}
of $d$ Grover coins given by Eq. (\ref{grover:coin}). As follows from Eqs. (\ref{teopF}) and (\ref{dk2}) the time evolution operator in the Fourier picture is also a tensor product
\begin{equation}
\widetilde{U}_{G_{2d}}(\textbf{k}) = \widetilde{U}_G(k_1,k_2)\otimes\ldots\otimes \widetilde{U}_G(k_{2d-1},k_{2d})
\label{c2d:k}
\end{equation}
of the matrices $\widetilde{U}_G$ defined by Eq. (\ref{gkl}) with different Fourier variables $k_i$. Hence, the eigenvalues of $\widetilde{U}_{G_{2d}}(\textbf{k})$ are given by the product of the eigenvalues of $\widetilde{U}_G$. Since two eigenvalues of $\widetilde{U}_G$ are constant as we have found in Eq. (\ref{eigenval:Grover}) one half of the eigenvalues of $\widetilde{U}_{G_{2d}}(\textbf{k})$ are also independent of $\textbf{k}$. As we have discussed in Chapter~\ref{chap:2d3} the probability $p_0(t)$ converges to a non-zero value and therefore the quantum walk exhibits localization.

In the case of odd dimension $2d+1$ we augment the coin given by Eq. (\ref{c2d}) by the Hadamard coin for the extra spatial dimension
\begin{equation}
G_{2d+1} = G_{2d}\otimes H.
\label{c2d1}
\end{equation}
Performing a similar analysis as in the case of even dimensions we find that for the quantum walk driven by the coin $G_{2d+1}$ the probability that the walk returns to the origin decays like $t^{-1}$ due to the Hadamard walk in the extra spatial dimension. Hence, this quantum walk is recurrent.

We note that due the fact that the 2-D Grover walk is transient for the initial state $\psi_G$ the same statement holds for the above constructed quantum walks, supposed that the initial state contains $\psi_G$ in its tensor product decomposition. Such vectors form a subspace with dimension equal to $4^{d-1}$ for even dimensional walks and $2\times 4^{d-1}$ for odd dimensional walks.


\section{Fourier walk on a plane}
\label{chap:4d}

We turn to the 2-D Fourier walk driven by the coin
\begin{equation}
F = \frac{1}{2}\left(
                 \begin{array}{rrrr}
                   1 & 1 & 1 & 1 \\
                   1 & i & -1 & -i \\
                   1 & -1 & 1 & -1 \\
                   1 & -i & -1 & i \\
                 \end{array}
               \right).
\end{equation}
As we will see, the Fourier walk does not exhibit localization. However, the decay of the probability $p_0(t)$ is slowed down to $t^{-1}$ so the Fourier walk is recurrent, except for a subspace of states.

We start our analysis of the Fourier walk with the propagator
$$
\widetilde{U}_F(k_1,k_2) = \left(D(k_1)\otimes D(k_2)\right) F,
$$
which determines the time evolution in the Fourier picture. It seems to be hard to determine the eigenvalues of $\widetilde{U}_F(k_1,k_2)$ analytically. However, we only need to determine the stationary points of their phases $\omega_j(k_1,k_2)$. For this purpose we consider the eigenvalue equation
$$
\Phi(k_1,k_2,\omega)\equiv\det{\left(\widetilde{U}_F(k_1,k_2)-e^{i\ \omega} I\right)}=0.
$$
This equation gives us the eigenenergies $\omega_i(k_1,k_2)$ as the solutions of the implicit function
$$
\Phi(k_1,k_2,\omega) = 1 + \cos(2k_2)-2\cos(2\omega)+2\sin{2\omega} + 4\cos{k_2}\sin{\omega}\left(\sin{k_1}-\cos{k_1}\right) = 0.
$$
Using the implicit differentiation we find the derivatives of the phase $\omega$
\begin{eqnarray}
\nonumber \frac{\partial \omega}{\partial k_1} & = & -\frac{\cos{k_2}\sin{\omega}\left(\cos{k_1}+\sin{k_1}\right)}{\cos(2\omega)+\sin(2\omega)+\cos{k_2}\cos{\omega}\left(\sin{k_1}-\cos{k_2}\right)}\\
\frac{\partial \omega}{\partial k_2} & = & -\frac{2\sin{k_2}\sin{\omega}\left(\cos{k_1}-\sin{k_1}\right)-\sin(2k_2)}{2\left(\cos(2\omega)+\sin(2\omega)+\cos{k_2}\cos{\omega}\left(\sin{k_1}-\cos{k_2}\right)\right)}
\label{fourier:der}
\end{eqnarray}
with respect to $k_1$ and $k_2$. Though we cannot eliminate $\omega$ on the RHS of Eq. (\ref{fourier:der}), we can identify the stationary points $\textbf{k}^0=(k_1^0,k_2^0)$
$$
\left.\frac{\partial\omega(\textbf{k})}{\partial k_i}\right|_{\textbf{k}=\textbf{k}^0}=0,\quad i=1,2
$$
of $\omega(k_1,k_2)$ with the help of the implicit function $\Phi(k_1,k_2,\omega)$. We find the following:\\
({\it i}) $\omega_{1,2}(k_1,k_2)$ have stationary lines
$$\gamma_1=(k_1,0)\ \textrm{and}\ \gamma_2=(k_1,\pi)$$\\
({\it ii}) all four phases $\omega_{i}(k_1,k_2)$ have stationary points for
$$k_1^0=\frac{\pi}{4},\ -\frac{3\pi}{4}\quad \textrm{and}\quad k_2^0=\pm\frac{\pi}{2}$$

It follows from the discussion of Chapter~\ref{chap:2d3} that the two phases $\omega_{1,2}(k_1,k_2)$ with stationary lines $\gamma_{1,2}$ are responsible for the slow down of the decay of the probability $p_0(t)$ to $t^{-1}$ for the Fourier walk, unless the initial coin state is orthogonal to the corresponding eigenvectors $v_{1,2}(k_1,k_2)$ at the stationary lines. For such an initial state the probability $p_0(t)$ behaves like $t^{-2}$ as the asymptotics of the integral given by Eq. (\ref{psi:0}) is determined only by the stationary points ({\it ii}).

Let us determine the states $\psi_F$ which lead to the fast decay $t^{-2}$ of the probability that the Fourier walk returns to the origin. The states $\psi_F$ have to be constant vectors fulfilling the conditions
$$
\left(v_{1,2}(\mathbf{k}),\psi_F\right)=0 \quad \forall\ \mathbf{k}\in\gamma_{1,2},
$$
which implies that $\psi_F$ must be a linear combination of $v_{3,4}(\mathbf{k}\in\gamma_{1,2})$ forming a two-dimensional subspace in $\mathcal{H}_C$. For $k_2=0,\pi$ we can find the eigenvectors of the matrix $\widetilde{U}_F(k_1,k_2)$ explicitly
\begin{eqnarray}
\nonumber v_1(k_1,0) & = & v_2(k_1,\pi) = \frac{1}{2}\left(e^{-ik_1},1,-e^{-ik_1},1\right)^T\\
\nonumber v_1(k_1,\pi) & = & v_2(k_1,0) = \frac{1}{2}\left(-e^{-ik_1},1,e^{-ik_1},1\right)^T\\
\nonumber v_3(k_1,0) & = & v_3(k_1,\pi) = \frac{1}{\sqrt{2}}(1,0,1,0)^T \\
\nonumber v_4(k_1,0) & = & v_4(k_1,\pi) = \frac{1}{\sqrt{2}}(0,1,0,-1)^T.
\end{eqnarray}
The explicit form of $\psi_F$ reads
\begin{equation}
\psi_F(a,b) = \left(a,b,a,-b\right)^T,
\label{psi:F}
\end{equation}
where $a,b\in \mathds{C}$. We point out that the particular initial state
\begin{equation}
\psi_F\left(a=\frac{1}{2},b=\frac{1-i}{2\sqrt{2}}\right) = \frac{1}{2}\left(1,\frac{1-i}{\sqrt{2}},1,-\frac{1-i}{\sqrt{2}}\right)^T
\label{sym:F}
\end{equation}
which was identified in \cite{2dw1} as the state which leads to a symmetric probability distribution with no peak in the neighborhood of the origin belongs to the family described by Eq. (\ref{psi:F}).


\begin{figure}[p]
\begin{center}
\includegraphics[width=0.7\textwidth]{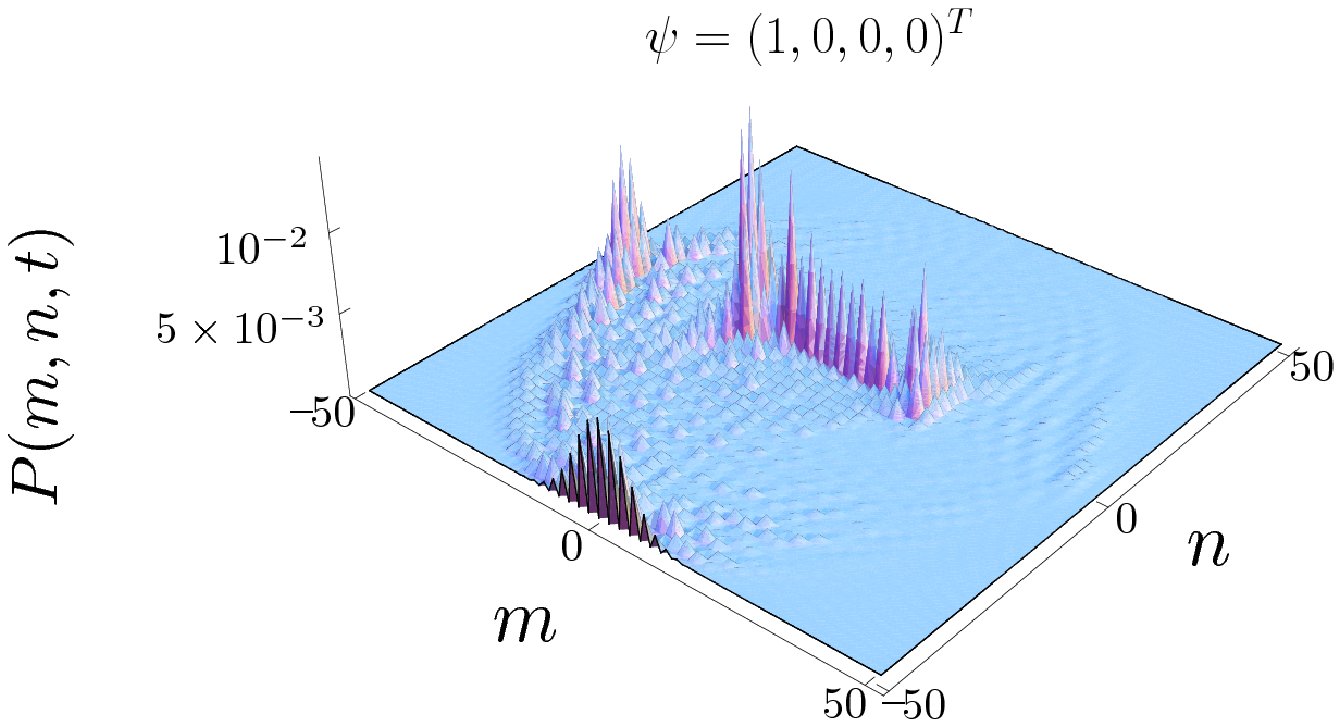}\vspace{36pt}
\includegraphics[width=0.6\textwidth]{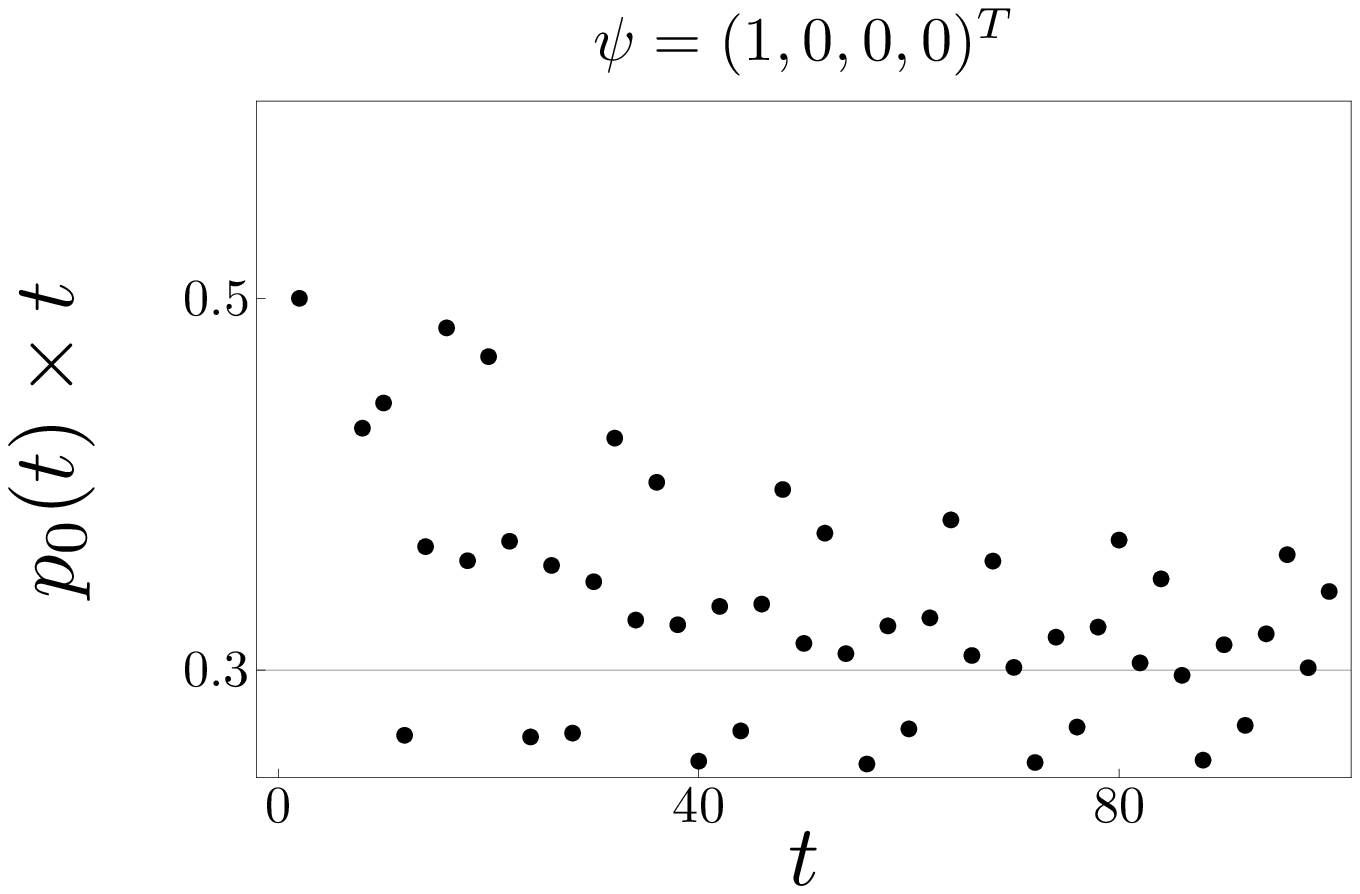}
\caption{Probability distribution after 50 steps and the time evolution of the probability $p_0(t)$ for the Fourier walk with the initial state $\psi=(1,0,0,0)^T$. The upper plot of the probability distribution reveals a presence of the central peak. Indeed, $\psi$ is not a member of the family $\psi_F(a,b)$. However, in contrast to the Grover walk the peak vanishes. In the lower plot we illustrate this by showing the probability $p_0(t)$ multiplied by $t$ to unravel the asymptotic behaviour $p_0(t)\sim t^{-1}$.}
\label{f3d1}
\end{center}
\end{figure}



\begin{figure}[p]
\begin{center}
\includegraphics[width=0.7\textwidth]{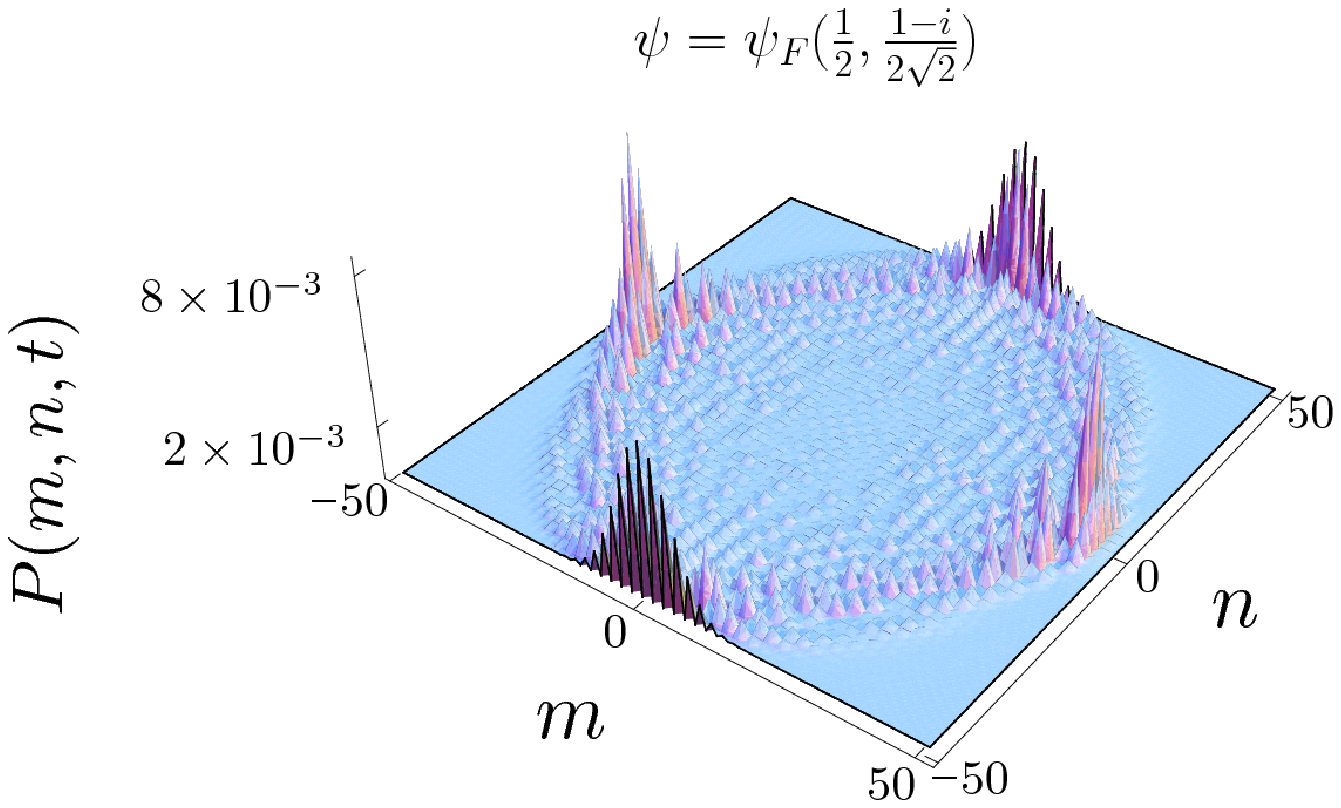}\vspace{36pt}
\includegraphics[width=0.6\textwidth]{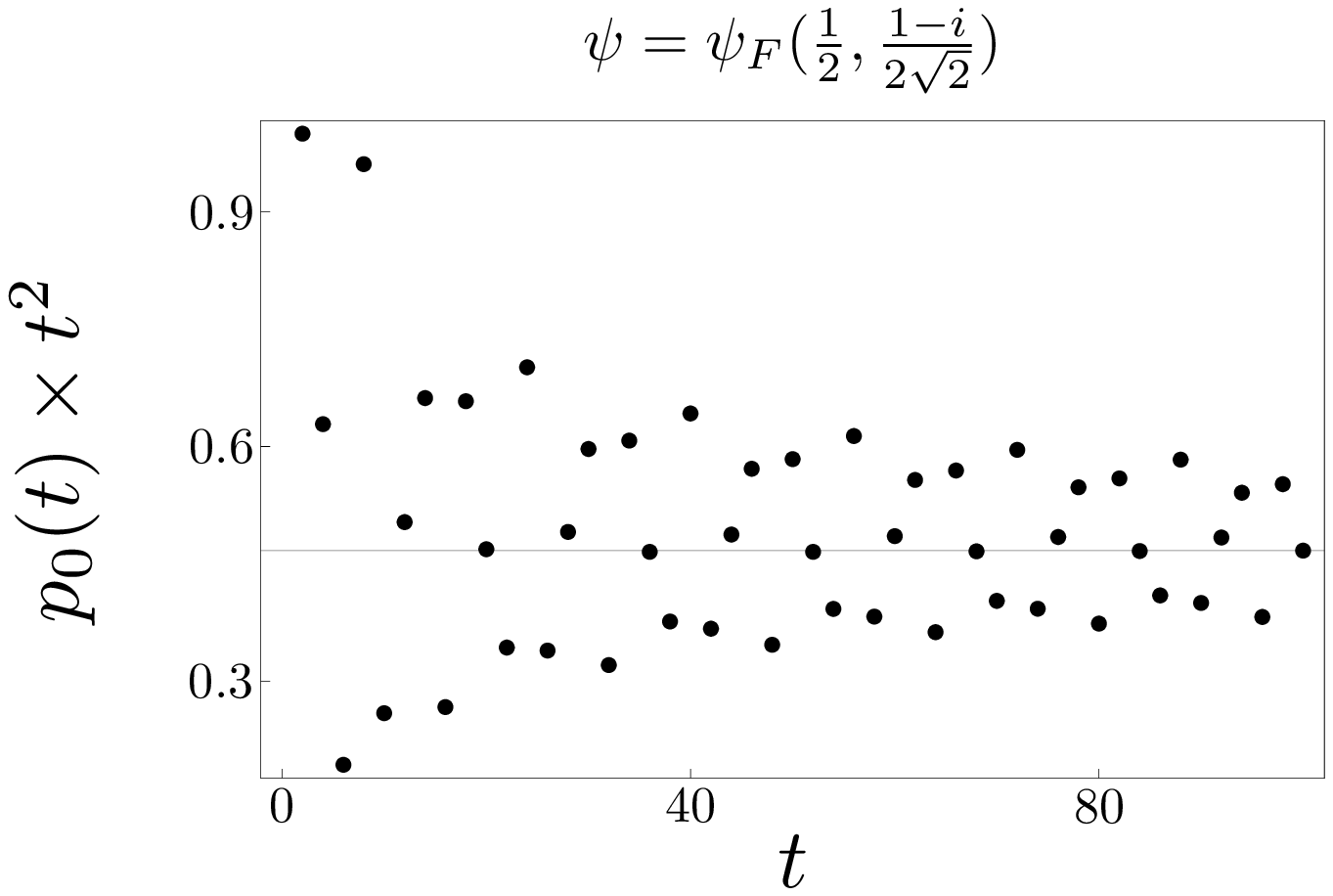}
\caption{Probability distribution after 50 steps and the time evolution of the probability $p_0(t)$ for the Fourier walk with the initial state given by Eq. (\ref{sym:F}). Since $\psi$ is a member of the family $\psi_F(a,b)$ the central peak in the probability distribution is not present, as depicted on the upper plot. The lower plot indicates that the probability $p_0(t)$ decays like $t^{-2}$.}
\label{f3d2}
\end{center}
\end{figure}


We illustrate the results in \fig{f3d1} and \fig{f3d2}. In \fig{f3d1} we plot the probability distribution and the probability $p_0(t)$ for the Fourier walk with the initial state $\psi=(1,0,0,0)^T$. This vector is not a member of the family $\psi_F(a,b)$ defined by Eq. (\ref{psi:F}). We find that a central peak is present, as depicted in \fig{f3d1}. However, in contrast to the Grover walk, the peak vanishes as shown by plotting the probability $p_0(t)$ multiplied by $t$ in \fig{f3d1} indicating a decay like $t^{-1}$, in agreement with the analytical result. In contrast, for \fig{f3d2} we have chosen the initial state given by Eq. (\ref{sym:F}) which is a member of the family $\psi_F(a,b)$. The upper plot shows highly symmetric probability distribution. However, the central peak is not present and as the lower plot indicates the probability $p_0(t)$ decays like $t^{-2}$.

We conclude that the Fourier walk is recurrent except for the two-dimensional subspace of initial states defined by Eq. (\ref{psi:F}) for which the walk is transient.

We turn to the estimation of the P\'olya numbers of the 2-D Fourier walk for the two-dimensional subspace of initial states given by Eq. (\ref{psi:F}). We make use of the normalization condition and the fact that the global phase of a state is irrelevant. Hence, we can choose $a$ to be non-negative real and $b$ is then given by the relation
$$
b = \sqrt{\frac{1}{2}-a^2}e^{i\phi}.
$$
Therefore, we parameterize the family of states defined by Eq. (\ref{psi:F}) by two real parameters --- $a$ ranging from $0$ to $\frac{1}{\sqrt{2}}$ and the mutual phase $\phi\in[0,2\pi)$. The exact expression for $p_0(a,\phi,t)$ can be written in the form
$$
p_0(a,\phi,t)=\frac{K_1(t)-K_2(t) a\sqrt{\frac{1}{2}-a^2}(\cos{\phi}-\sin{\phi})}{t^2},
$$
where $K_{1,2}$ has to be determined numerically. Nevertheless, the numerical simulation of $p_0(a,\phi,t)$ at two values of $(a,\phi)$ enables us to find the numerical values of $K_{1,2}(t)$ and we can evaluate $p_0(a,\phi,t)$ at any point $(a,\phi)$. The probability $p_0(a,\phi,t)$ shows the maximum at $a=\frac{1}{2}$, $\phi=\frac{3\pi}{4}$ and the minimum for the same value of $a$ and the phase $\phi=\frac{7\pi}{4}$. Consequently, these points also represent the maximum and the minimum of the P\'olya numbers.


\begin{figure}[p]
\begin{center}
\includegraphics[width=0.7\textwidth]{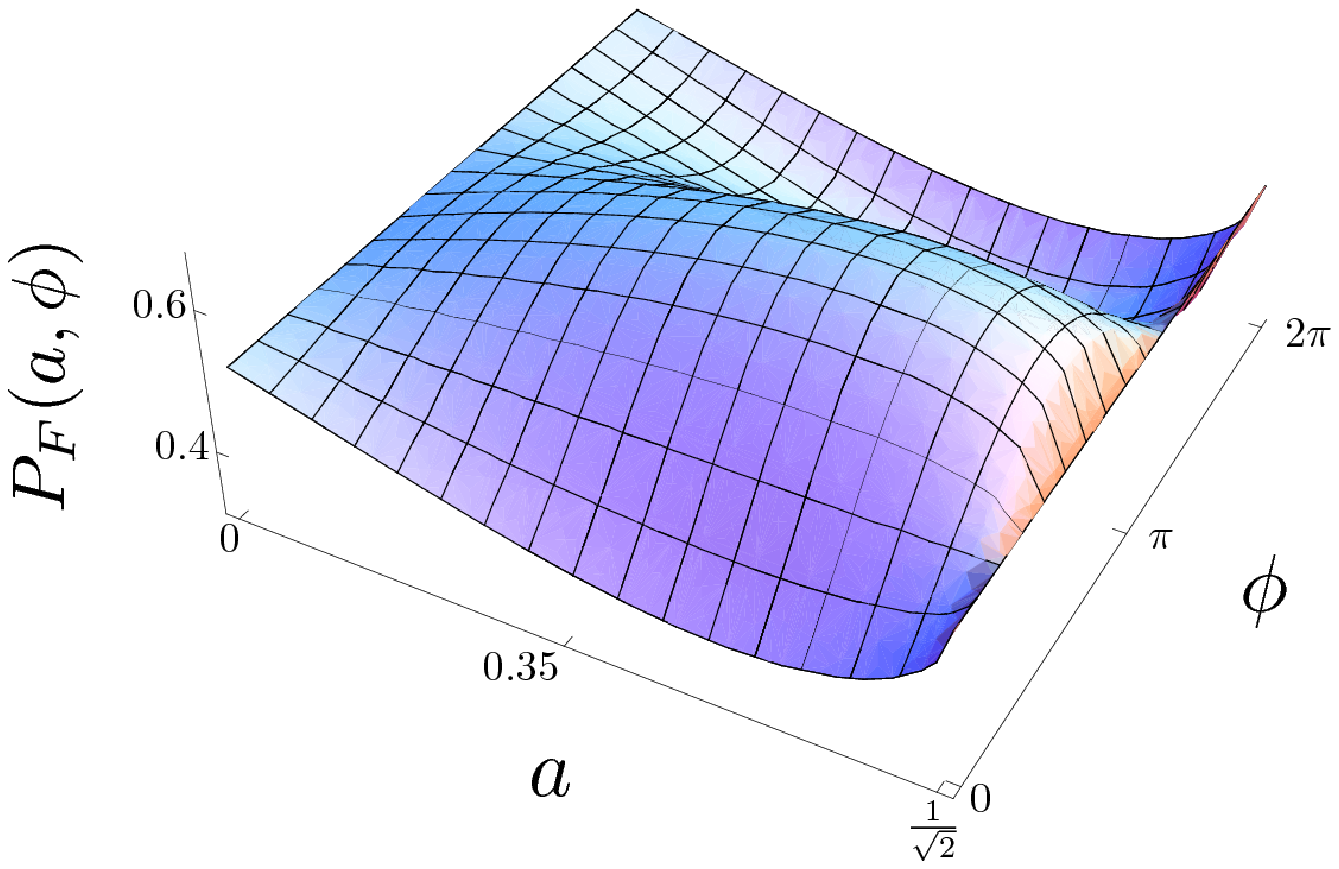}\vspace{36pt}
\includegraphics[width=0.6\textwidth]{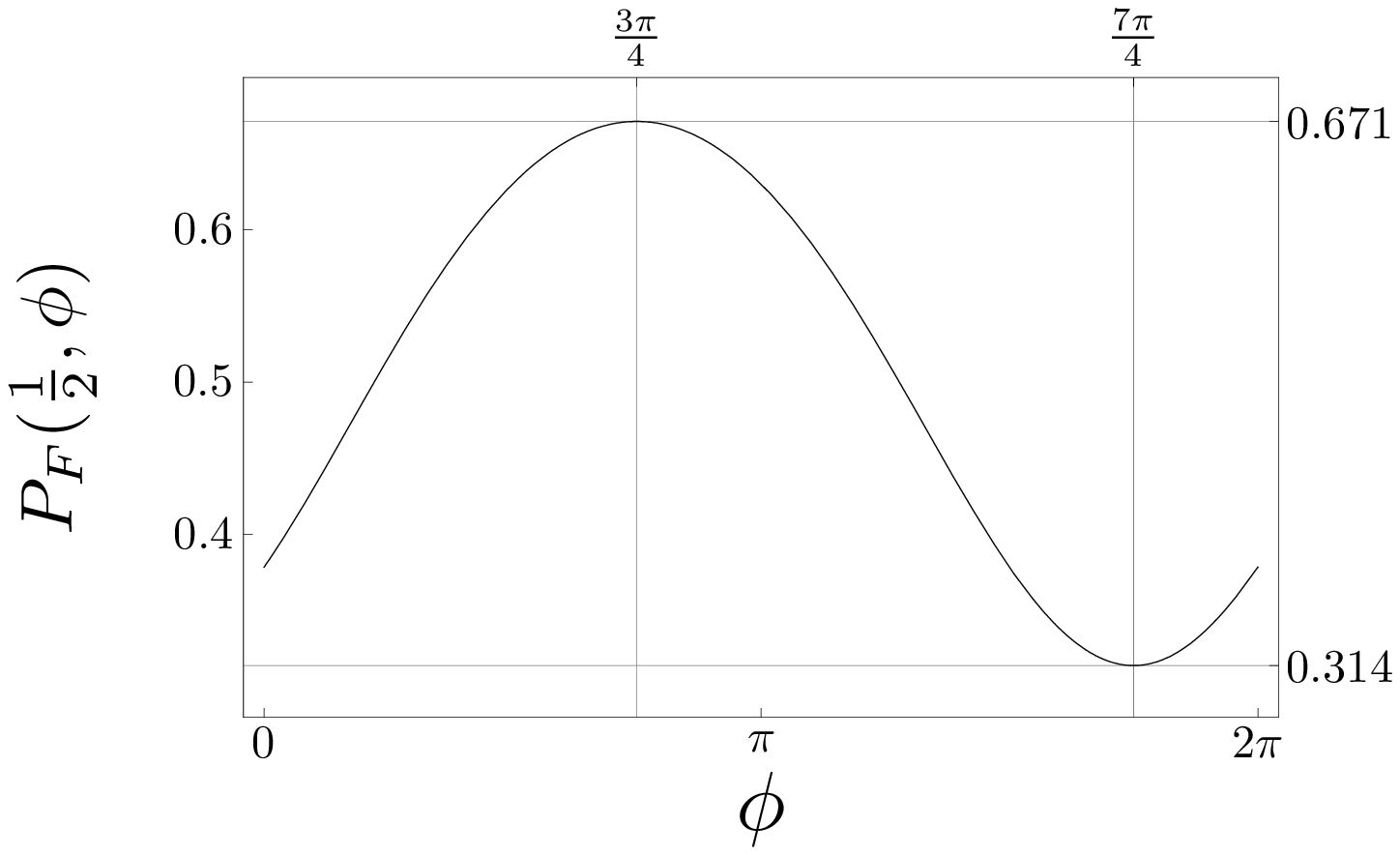}
\caption{Approximation of the P\'olya numbers for the 2-D Fourier walk and the initial states from the family of states defined by Eq. (\ref{psi:F}) in their dependence on the parameters of the initial state $a$ and $\phi$. Here we have evaluated the first 100 terms of $p_0(a,\phi,t)$ exactly. The P\'olya numbers cover the whole interval between the minimal value of $P_F^{min}\approx 0.314$ and the maximal value of $P_F^{max}\approx 0.671$. The extreme values are attained for $a=1/2$ and $\phi^{min}=7\pi/4$, respectively $\phi^{max}=3\pi/4$. On the lower plot we show the cut at the value $a=1/2$ containing both the maximum and the minimum.}
\label{polya:fourier}
\end{center}
\end{figure}


In \fig{polya:fourier} we present the approximation of the P\'olya number Eq. (\ref{polya:approx}) in its dependence on $a$ and $\phi$ and a cut through the plot at the value $a=1/2$ containing both the global minimum and the global maximum. Here we have evaluated the first 100 terms of $p_0(a,\phi,t)$ exactly. We see that the values of the P\'olya number vary from the minimum $P_F^{min}\approx 0.314$ to the maximal value of $P_F^{max}\approx 0.671$. We note that for the initial states that do not belong to the subspace defined by Eq. (\ref{psi:F}) the P\'olya number equals one.


\section{Conclusions}
\label{chap:4e}

Our results, summarized in \tab{tab2}, demonstrate that there is a remarkable freedom for the value of the P\'olya number for quantum walks, depending both on the initial state and the coin operator, in contrast to the classical random walk where the dimension of the lattice uniquely defines the recurrence probability. Hence, the quantum P\'olya number is able to indicate physically different regimes in which a quantum walk can be operated in.

\begin{table}[h]
\begin{center}
\begin{tabular}{|c|c|c|c|}
  \hline
  \multirow{2}{*}{Quantum walk} & \multirow{2}{*}{Section} & \multirow{2}{*}{Probability at the origin} & \multirow{2}{*}{P\'olya number}\\
  & & & \\\hline
  \multirow{3}{*}{$d$-D Hadamard} & \multirow{3}{*}{\ref{chap:4b}} & \multirow{3}{*}{$t^{-d}$ for any $\psi$} & $1$ for $d=1$\\
  & & & $<1$ for $d\geq 2$ \\
  & & & independent of $\psi$ \\\hline
  \multirow{4}{*}{2-D Grover} & \multirow{4}{*}{\ref{chap:4c}}  & \multirow{2}{*}{const. for $\psi\neq\psi_G$} & \multirow{2}{*}{1}\\
  & & & \\
  & & \multirow{2}{*}{$t^{-2}$ for $\psi=\psi_G$} & \multirow{2}{*}{$<1$}\\
  & & & \\\hline
  \multirow{4}{*}{2-D Fourier} & \multirow{4}{*}{\ref{chap:4d}}  & \multirow{2}{*}{$t^{-1}$ for $\psi\not\in\psi_F$} & \multirow{2}{*}{1}\\
  & & & \\
  & & \multirow{2}{*}{$t^{-2}$ for $\psi\in\psi_F$} & $<1$\\
  & & & dependent on $\psi$ \\\hline
\end{tabular}
\caption{Summary of the main results. We list the types of studied quantum walks, the asymptotic behaviour of the probability at the origin and the P\'olya number in the respective cases in its dependence on the initial state $\psi$.}
\label{tab2}
\end{center}
\end{table}


\chapter{Recurrence of Biased Quantum Walks on a Line}
\label{chap:5}

\nsection{Introduction}
\label{chap:5a}

Recurrence of classical random walks is a consequence of the walk's symmetry. As we briefly review in the Appendix~\ref{app:a2}, they are recurrent if and only if the mean value of the position of the particle vanishes. This is due to the fact that the spreading of the probability distribution of the position is diffusive while the mean value of the position propagates with a constant velocity. In contrast, for quantum walks both the spreading of the probability distribution and the propagation of the mean value are ballistic. In the present Chapter we show that this allows for maintaining recurrence even when the symmetry is broken.

The Chapter is organized as follows: In Section~\ref{chap:5b} we describe the biased quantum walk on a line, find the propagator in the momentum representation and solve the time evolution equation. The recurrence of the quantum walk is determined by the asymptotics of the probability at the origin. We perform this analysis in Section~\ref{chap:5c} and find the conditions under which the biased quantum walk on a line is recurrent. In Section~\ref{chap:5d} we analyze the recurrence of biased quantum walks from a different perspective. We find that the recurrence is related to the velocities of the peaks of the probability distribution of the quantum walk. The explicit form of the velocities leads us to the same condition derived in Section~\ref{chap:5c}. Finally, in Section~\ref{chap:5e} we derive the formula for the mean value of the position of the particle in dependence of the parameters of the walk and the initial state. We find that there exist genuine biased quantum walks which are recurrent. We summarize our results in the conclusions of Section~\ref{chap:5f}.

\section{Description of the walk}
\label{chap:5b}

\begin{figure}[h]
\begin{center}
\includegraphics[width=0.6\textwidth]{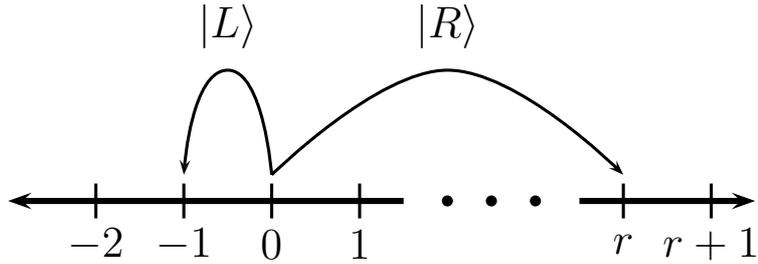}
\caption{Schematics of the biased quantum walk on a line. If the
coin is in the state $|R\rangle$ the particle moves to the right
to a point at distance $r$. With the coin state $|L\rangle$
the particle makes a unit length step to the left. Before the step
itself the coin state is rotated according to the coin operator
$C(\rho)$.}
\label{chap5:fig1}
\end{center}
\end{figure}

Let us consider biased quantum walks on a line where the particle has two possibilities --- jump to the right or to the left. Without loss of generality we restrict ourselves to biased quantum walks where the jump to the right is of the length $r$ and the jump to
the left has a unit size. We depict the biased quantum walk schematically in \fig{chap5:fig1}.

The Hilbert space of the particle has the form of the tensor product
$$
{\cal H} = {\cal H}_P\otimes{\cal H}_C
$$
of the position space
$$
{\cal H}_P = \ell^2(\mathds{Z}^d) = \textrm{Span}\left\{|m\rangle|\ m\in\mathds{Z}\right\},
$$
and the two dimensional coin space
$$
{\cal H}_C = \mathds{C}^2 = \textrm{Span}\left\{|R\rangle,|L\rangle\right\}.
$$
The propagator of the quantum walk in the position representation is
$$
U = S \left(I_P\otimes C\right),
$$
where the displacement operator $S$ has the form
$$
S = \sum\limits_{m=-\infty}^{+\infty}|m+r\rangle\langle m|\otimes|R\rangle\langle R|+\sum\limits_{m=-\infty}^{+\infty}|m-1\rangle\langle m|\otimes|L\rangle\langle L|.
$$
The coin flip $C$ can be in general an arbitrary unitary operator acting on the coin space $\mathcal{H}_C$. However, as has been discussed in \cite{2dw1} the probability distribution is not affected by the complex phases of the coin operator. Hence, it is sufficient to consider the one-parameter
family of coins
$$
C(\rho) = \left(
            \begin{array}{cc}
              \sqrt{\rho} & \sqrt{1-\rho} \\
              \sqrt{1-\rho} & -\sqrt{\rho} \\
            \end{array}
          \right).
$$
From now on we restrict ourselves to this family of coins. The value of $\rho=1/2$ corresponds to the well known case of the Hadamard walk.

In the momentum representation the propagator has the form
$$
\widetilde{U}(k) = \textrm{Diag}\left(e^{ikr},e^{-ik}\right)\cdot C(\rho)  = \left(
                                            \begin{array}{cc}
                                              \sqrt{\rho}e^{ikr} & \sqrt{1-\rho}e^{ikr} \\
                                              \sqrt{1-\rho}e^{-ik} & -\sqrt{\rho}e^{-ik} \\
                                            \end{array}
                                          \right).
$$
Since it is a unitary operator its eigenvalues are $e^{i\ \omega_{1,2}}$ where the phases are given by
\begin{eqnarray}
\nonumber \omega_1(k) & = & \frac{r-1}{2}k+\arcsin\left(\sqrt{\rho}\sin\left(\frac{r+1}{2}k\right)\right),\\
\omega_2(k) & = & \frac{r-1}{2}k -\pi-\arcsin\left(\sqrt{\rho}\sin\left(\frac{r+1}{2}k\right)\right).
\label{omega}
\end{eqnarray}
We denote the corresponding eigenvectors by $v_{1,2}(k)$. We give their explicit form in Section~\ref{chap:5f}. The solution of the time evolution equation in the Fourier picture has the standard form
$$
\widetilde{\psi}(k,t) = \sum_{j=1}^2 e^{i\ \omega_j(k)t}\left(v_j(k),\tilde{\psi}(k,0)\right)v_j(k),
$$
where $\tilde{\psi}(k,0)$ is the Fourier transformation of the initial state. We restrict ourselves to the situation where the particle is
initially localized at the origin as dictated by the nature of the problem we wish to study. Hence, the Fourier transformation of such an initial
condition is equal to the initial state of the coin which we denote by $\psi$. Since $\psi$ can be an arbitrary normalized complex two-component vector we parameterize it by two parameters $a\in[0,1]$ and $\varphi\in[0,2\pi)$ in the form
\begin{equation}
\psi = \left(
         \begin{array}{c}
           \sqrt{a} \\
           \sqrt{1-a}e^{i\varphi} \\
         \end{array}
       \right).
\label{psi:init}
\end{equation}
The solution in the position representation is obtained by performing the inverse Fourier transformation
\begin{equation}
\psi(m,t) = \int_{-\pi}^\pi\frac{dk}{2\pi}\ \widetilde{\psi}(k,t)\ e^{-imk} = \sum_{j=1}^2\int_{-\pi}^\pi\frac{dk}{2\pi}e^{i(\omega_j(k)t-mk)}\ \left(v_j(k),\psi\right)v_j(k).
\label{chap5:inv:f}
\end{equation}


\section{Asymptotics of the probability at the origin}
\label{chap:5c}

\begin{figure}[h]
\begin{center}
\includegraphics[width=0.5\textwidth]{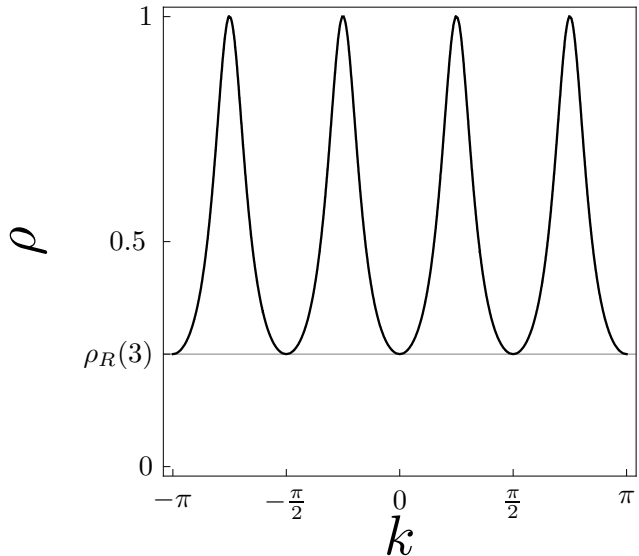}
\caption{The existence of stationary points of the phases
$\omega_{1,2}(k)$ in dependence on the parameter $\rho$ and
a fixed step length $r$. We plot the implicit functions $\omega_{1,2}'(k)\equiv
0$ for $r=3$. The plot indicates that for
$\rho<\rho_R(3)=\frac{1}{4}$ the phases $\omega_{1,2}(k)$ do
not have any stationary points. Consequently, the probability amplitude
at the origin decays fast and such biased quantum walk on a line is
transient. For $\rho\geq\rho_R(3)$ the stationary points exists and the
quantum walk is recurrent.}
\label{chap5:fig2}
\end{center}
\end{figure}

To determine the recurrence nature of the biased quantum walk we
have to analyze the asymptotic behaviour of the probability at the
origin. Exploiting (\ref{chap5:inv:f}) the amplitude at the origin
reads
\begin{equation}
\psi(0,t) = \sum_{j=1}^2\int_{-\pi}^\pi\frac{dk}{2\pi}e^{i\ \omega_j(k)t}\ \left(v_j(k),\psi\right)v_j(k),
\label{chap5:psi:0}
\end{equation}
which allows us to find the asymptotics of the probability at the origin
with the help of the method of stationary phase \cite{statphase}.
The important contributions to the integrals in (\ref{chap5:psi:0}) arise
from the stationary points of the phases (\ref{omega}). We find that
the derivatives of the phases $\omega_{1,2}(k)$ are
\begin{eqnarray}
\nonumber \omega_1'(k) & = & \frac{r-1}{2}+\frac{\sqrt{\rho}(r+1)\cos\left(k\frac{r+1}{2}\right)}{\sqrt{4+2\rho\left[\cos(k(r+1))-1\right]}},\\
\omega_2'(k) & = & \frac{r-1}{2}-\frac{\sqrt{\rho}(r+1)\cos\left(k\frac{r+1}{2}\right)}{\sqrt{4+2\rho\left[\cos(k(r+1))-1\right]}}.
\label{phase:der}
\end{eqnarray}
Using the method of stationary phase we find that the
amplitude will decay slowly - like $t^{-\frac{1}{2}}$, if at least
one of the phases has a vanishing derivative inside the integration
domain. Solving the equations $\omega_{1,2}'(k) = 0$ we find
that the possible stationary points are
\begin{equation}
k_0 = \pm\frac{2}{r+1}\arccos\left(\pm\sqrt{\frac{(1-\rho)(r-1)^2}{4\rho r}}\right).
\label{k0}
\end{equation}
The stationary points are real valued provided the argument of the arcus-cosine in (\ref{k0}) is less or equal to unity
$$
\frac{(1-\rho)(r-1)^2}{4\rho r} \leq 1.
$$
This inequality leads us to the condition for the biased quantum walk on a line to be recurrent
\begin{equation}
\rho_R(r) \geq \left(\frac{r-1}{r+1}\right)^2.
\label{crit:rec}
\end{equation}
We illustrate this result in \fig{chap5:fig2} for a particular choice of the walk parameter $r=3$.

Our simple result proves that there is an intimate nontrivial
link between the length of the step of the walk and the bias of the
coin. The parameter of the coin $\rho$ has to be at least equal to
a factor determined by the size of the step to the right $r$ for the
walk to be recurrent. We note that the recurrence nature of the
biased quantum walk on a line is determined only by the parameters
of walk itself, i.e. the coin and the step, not by the initial
conditions. The parameters of the initial state $a$ and $\varphi$
have no effect on the rate of decay of the probability at the
origin.


\section{Velocities of the peaks}
\label{chap:5d}

We can determine the recurrence nature of the biased quantum walk on
a line from a different point of view. This approach is based
on the following observation. The well known shape of the
probability distribution generated by the quantum walk consists of
two counter-propagating peaks. In between
the two dominant peaks the probability is roughly independent of $m$ and decays like $t^{-1}$.
On the other hand, outside the decay is exponential as we depart from the peaks. As it has been found in \cite{nayak} the positions of the peaks varies linearly with the number of steps. Hence, the peaks propagate with constant velocities, say $v_L$ and $v_R$. For the
biased quantum walk to be recurrent the origin of the walk has to
remain in between the two peaks for all times. In other words, the
biased quantum walk on a line is recurrent if and only if the
velocity of the left peak is negative and the velocity of the right
peak is positive.

\begin{figure}[h]
\begin{center}
\includegraphics[width=0.7\textwidth]{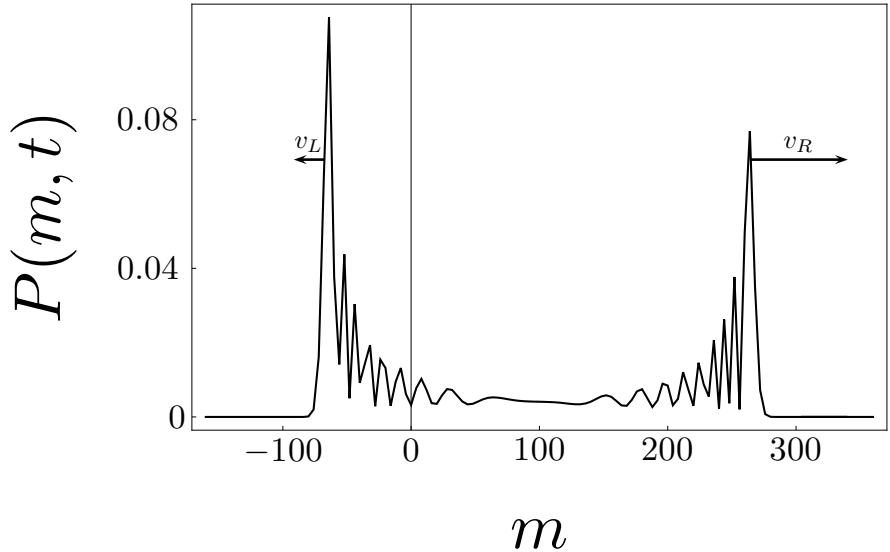}
\caption{Velocities of the left and right peak of the probability distribution generated by the biased quantum walk on a line and the recurrence.
We have chosen the parameters $r=3$, $a=\rho=\frac{1}{\sqrt{2}}$ and $\varphi = \pi$. The left peak propagates with the velocity $v_L\approx -0.68$,
the velocity of the right peak is $v_R\approx 2.68$. In between the two peaks the probability distribution behaves like $t^{-1}$ while outside the
decay is exponential. Since the velocity $v_L$ is negative the origin of the walk remains in between the left and right peak. Consequently, this quantum
walk is recurrent.}
\label{chap5:fig3}
\end{center}
\end{figure}

The velocities of the left and right peak are easily determined. We
rewrite the formula (\ref{chap5:inv:f}) for the probability amplitude
$\psi(m,t)$ into the form
$$
\psi(m,t) = \sum_{j=1}^2\int_{-\pi}^\pi\frac{dk}{2\pi}e^{i(\omega_j(k)-\alpha k)t}\ \left(v_j(k),\psi\right)v_j(k),
$$
where we have introduced $\alpha = \frac{m}{t}$. Due to the fact that we
concentrate on the amplitudes at the positions $m\sim t$ we have to
consider modified phases
$$
\widetilde{\omega}_j(k) = \omega_j(k)-\alpha k.
$$
The peak occurs at such a position
$m_0$ where both the first and the second derivatives of
$\widetilde{\omega}_j(k)$ vanishes. The velocity of the peak is
thus $\alpha_0 = \frac{m_0}{t}$. Hence, solving the equations
\begin{eqnarray}
\nonumber \widetilde{\omega}_1'(k) & = & \frac{r-1}{2}+\frac{\sqrt{\rho}(r+1)\cos\left(k\frac{r+1}{2}\right)}{\sqrt{4+2\rho\left[\cos(k(r+1))-1\right]}} - \alpha = 0 ,\\
\nonumber \widetilde{\omega}_2'(k) & = & \frac{r-1}{2}-\frac{\sqrt{\rho}(r+1)\cos\left(k\frac{r+1}{2}\right)}{\sqrt{4+2\rho\left[\cos(k(r+1))-1\right]}} - \alpha = 0,\\
\nonumber \widetilde{\omega}_1''(k) & = & -\widetilde{\omega}_2''(k) = \frac{(\rho-1)\sqrt{\rho}(r+1)^2\sin\left(k\frac{r+1}{2}\right)}{\sqrt{2}\left[2-\rho+\rho\cos(k(r+1))\right]^\frac{3}{2}} = 0,
\end{eqnarray}
for $\alpha$ determines the velocities of the left and right peak $v_{L,R}$. The third equation is independent of $\alpha$ and we easily find the
solution
$$
k_0 = \frac{4n\pi}{r+1},\ k_0=\frac{2\pi(2n+1)}{r+1},\ n\in\mathds{Z}.
$$
Inserting this $k_0$ into the first two equations we find the velocities of the left and right peak
\begin{equation}
v_L = \frac{r-1}{2}-\frac{r+1}{2}\sqrt{\rho},\quad v_R = \frac{r-1}{2}+\frac{r+1}{2}\sqrt{\rho}.
\label{velocities}
\end{equation}
We illustrate this result in \fig{chap5:fig3} where we show the probability distribution generated by the quantum walk for the particular choice of the parameters $r = 3,\ \rho = \frac{1}{\sqrt{2}}$. The initial state was chosen according to $a = \frac{1}{\sqrt{2}}$ and $\varphi = \pi$. Since the velocity of the left peak $v_L$ is negative this biased quantum walk is recurrent.

The peak velocities have two contributions. One is identical
and independent of $\rho$, the second is a product of $r$ and $\rho$
and differs in sign for the two velocities. The obtained results indicate that biasing the walk by having
the size of the step to the right equal to $r$ results in dragging
the whole probability distribution towards the direction of
the larger step. This is manifested by the term $\frac{r-1}{2}$
which appears in both velocities $v_{L,R}$ with the same sign. On
the other hand the parameter of the coin $\rho$ does not bias the
walk. As we can see from the second terms entering the velocities it
rather influences the rate at which the walk spreads.

As we have discussed above the biased quantum walk on a line is recurrent if and only if $v_L$ is negative and $v_R$ is positive. The form of the
velocities (\ref{velocities}) implies that this condition is satisfied if and only if the criterion (\ref{crit:rec}) is fulfilled.


\section{Mean value of the position}
\label{chap:5e}

As we discuss in the Appendix~\ref{app:a2} the classical random walks are recurrent if and only if the mean value of the position vanishes. We show that this is not true for biased quantum walks, i.e. there exist biased quantum walks on a line which are recurrent but cannot produce probability distribution with zero mean value. This is another unique feature of quantum walks compared to the classical ones.

Let us derive the formula for the mean value of the position of the particle for the biased quantum walk. With the help of the weak limit theorem
\cite{Grimmett} we express the mean value after $t$ steps in the form
$$
\left\langle \frac{x}{t}\right\rangle \approx \sum_{j=1}^2\int_{-\pi}^\pi\frac{dk}{2\pi}\ \omega_j'(k)\ \left|\left(v_j(k),\psi\right)\right|^2,
$$
up to the corrections of the order $O(t^{-1})$. Here $v_j(k)$ are eigenvectors of the
unitary propagator $\widetilde{U}(k)$, $\omega_j'(k)$ are the
derivatives of the eigenenergies and
$\psi$ is the initial state expressed in (\ref{psi:init}). The
derivatives of the phases are given in (\ref{phase:der}). We express
the eigenvectors in the form
\begin{eqnarray}
\nonumber v_1(k) & = & n_1(k)\left(\sqrt{1-\rho}, -\sqrt{\rho} + e^{i(\omega_1(k)-rk)}\right)^T,\\
\nonumber v_2(k) & = & n_2(k)\left(\sqrt{1-\rho}, -\sqrt{\rho} + e^{i(\omega_2(k)-rk)}\right)^T.
\end{eqnarray}
The normalizations are given by
\begin{eqnarray}
\nonumber n_1(u) & = & 2-2\sqrt{\rho}\cos\left(u-\arcsin\left[\sqrt{\rho}\sin u\right]\right),\\
\nonumber n_2(u) & = & 2+2\sqrt{\rho}\cos\left(u+\arcsin\left[\sqrt{\rho}\sin u\right]\right),
\end{eqnarray}
where we have introduced $u = \frac{k(r+1)}{2}$ to shorten the notation. The mean value is thus given by the following integral
$$
\left\langle \frac{x}{t}\right\rangle \approx \int\limits_0^{(r+1)\pi}\frac{f(a,\varphi,\rho,r,u)du}{2(r+1)\pi \left[1 +\sqrt{\rho}\cos u_1\right] \left[1-\sqrt{\rho} \sin u_2\right]}+O(t^{-1}),
$$
where
$$
u_1  =  u + \arcsin(\sqrt{\rho}\sin u),\quad u_2  =  u + \arccos(\sqrt{\rho}\sin u),
$$
and the numerator reads
\begin{eqnarray}
\nonumber f(a,\varphi,\rho,r,u) & = & (1-\rho) \left[r-1 + \rho \left(a + r (a-1)\right)\left(1+\cos(2u)\right)+\right.\\
\nonumber & & \left.+ \sqrt{a(1-a)}\sqrt{\rho(1-\rho)} (r+1) \left(\cos{\varphi}+\cos(\varphi+2u)\right)\right].
\end{eqnarray}
Performing the integrations we arrive at the following formula for the position mean value
\begin{eqnarray}
\nonumber \left\langle\frac{x}{t}\right\rangle & \approx & (1-\sqrt{1-\rho})(a(r+1)-1) + \frac{r-1}{2}\sqrt{1-\rho} + \\
& & + \frac{\sqrt{a(1-a)}(1-\sqrt{1-\rho})(1-\rho)(r+1)\cos\varphi}{\sqrt{\rho(1-\rho)}} + O(t^{-1}).
\label{mean}
\end{eqnarray}

We see that for quantum walks the mean value is affected by both the
fundamental walk parameters through  $r$ and $\rho$ and the
initial state parameters $a$ and $\varphi$. The mean value is
typically non-vanishing even for unbiased quantum walks (
with $r=1$ ). However, one easily finds \cite{2dw1} that the initial
state with the parameters $a=1/2$ and $\varphi=\pi/2$ results in a
symmetric probability distribution with zero mean independent of the
coin parameter $\rho$. Indeed, the quantum walks with $r=1$, i.e.
with equal steps to the right and left, does not intrinsically
distinguish left from right. On the other hand the quantum walks
with $r>1$ treat the left and right direction in a different way.
Nevertheless, one can always find for a given $r$ a coin parameter
$\rho_0$ such that for all $\rho\geq\rho_0$ the quantum walk can
produce a probability distribution with zero mean value. This is
impossible for quantum walks with $\rho<\rho_0$ and we will call
such quantum walks genuine biased.

Let us determine the minimal value of $\rho$ for a given $r$ for
which mean value vanishes. We first find the parameters of the
initial state $a$ and $\varphi$ which minimizes the mean value.
Clearly the term on the second line in (\ref{mean}) reaches the
minimal value for $\varphi_0=\pi$. Differentiating the resulting
expression with respect to $a$ and setting the derivative equal to
zero gives us the condition
$$
2 + \frac{(2a-1)\sqrt{\rho(1-\rho)}}{\rho\sqrt{a(1-a)}} = 0
$$
on the minimal mean value with respect to $a$. This relation is satisfied for
$a_0=\frac{1}{2}(1-\sqrt{\rho})$. The resulting formula for the mean
value reads
\begin{equation}
\left\langle\frac{x}{t}\right\rangle_{a_0,\varphi_0}
= \frac{r-1}{2}+\frac{\left(1-\sqrt{1-\rho}-\rho\right) (1+r)}{2
\sqrt{(1-\rho) \rho}}.
\label{mean:min}
\end{equation}
This expression vanishes for
\begin{equation}
\rho_0(r) = \left(\frac{r^2 - 1}{r^2 + 1}\right)^2.
\label{rho:0}
\end{equation}
Since (\ref{mean:min}) is a decreasing function of
$\rho$ the mean value is always positive for $\rho<\rho_0$
independent of the choice of the initial state. For $\rho>\rho_0$
one can achieve zero mean value for different combination of the
parameters $a$ and $\varphi$.

The formula (\ref{rho:0}) is reminiscent of the condition
(\ref{crit:rec}) for the biased quantum walk on a line to be
recurrent. However, $r$ is in (\ref{rho:0}) replaced by $r^2$.
Therefore we find the inequality $\rho_R<\rho_0$. Hence, the quantum
walks with the coin parameter $\rho_R<\rho<\rho_0$ are recurrent but
cannot produce a probability distribution with zero mean value. We
conclude that there are genuine biased quantum walks which are
recurrent in contrast to situations found for classical walks.

\section{Conclusions}
\label{chap:5f}

We have analyzed one dimensional biased quantum walks. Classically, the bias leading to a non-zero mean value of the particle's position can be introduced in two ways --- unequal step lengths or unfair coin. In contrast, for quantum walks on a line the initial state can introduce bias for any coin. On the other hand, for symmetric initial state modifying only the unitary coin operator while keeping the equal step lengths will not introduce bias. Finally, the bias due to unequal step lengths may be compensated for by the choice of the coin operator for some initial conditions. For this reason we have introduced the concept of the genuinely biased quantum walk for which there does not exists any initial state leading to vanishing mean value of the position.

\begin{figure}[h]
\begin{center}
\includegraphics[width=0.6\textwidth]{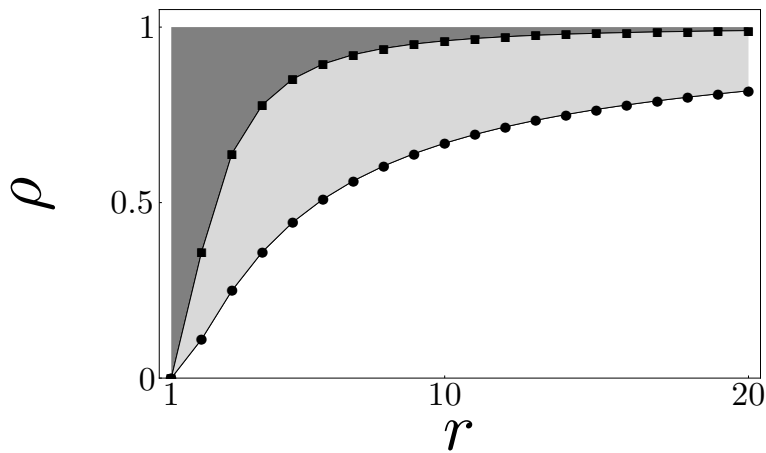}
\caption{"Phase diagram" of biased quantum walks on a line. The horizontal axis represents the length of the step to the right $r$ and the vertical axis shows the coin parameter $\rho$. The dotted line corresponds to the recurrence criterion (\ref{crit:rec}), while the squares represent the condition (\ref{rho:0}) on the zero mean value of the particle's position. The quantum walks in the white area are transient and genuine biased. In between the two curves (light gray area) we find quantum walks which are recurrent but still genuine biased. The quantum walks in the dark gray area are recurrent and for a particular choice of the initial state they can produce probability distribution with vanishing mean value.}
\label{chap5:fig4}
\end{center}
\end{figure}

We have determined the conditions under which one dimensional biased quantum walks are recurrent. This together with the condition of being genuinely biased give rise to three different regions in the parameter space which we depict as a "phase diagram" in \fig{chap5:fig4}.

\chapter{Meeting Problem in the Quantum Walk}
\label{chap:6}

\nsection{Introduction}
\label{chap:6a}

In this Chapter we study the evolution of two particles
performing a quantum walk. The evolution of each of the two
particles is subjected to the same rules. One of the interesting
questions, when two particles are involved, is to clarify how the
probability of the particles to meet changes with time (or number of
steps taken in walk). Because the behavior of a single particle performing a quantum walk
differs from its classical counterpart it has to be expected that
the same applies to the situation when two particles are involved.
Interference, responsible for the unusual behavior of the single
particle should play also a considerable role when two particles are involved. The possibility to change the input states (in
particular the possibility to choose entangled initial coin
states) adds another interesting point to the analysis. In the
following, we study the evolution of the meeting probability for
two particles. We point out the differences to the classical case
and discuss the influence of the input state on this probability.

Before we turn to the meeting problem we generalize in Sections~\ref{chap:6b} and \ref{chap:6c} the quantum walk to two distinguishable and indistinguishable particles. The meeting problem for two distinguishable particles with initially factorized coin states is analyzed in Section~\ref{chap:6d}. We derive the asymptotic behavior of the meeting probability and compare it with the results for the classical random walk which are summarized in Appendix~\ref{app:d}. The effect of entanglement on the meeting probability is considered in Section~\ref{chap:6e}. Finally, in Section~\ref{chap:6f} we analyze the meeting probability for two indistinguishable bosons and fermions. We summarize our results in the conclusions of Section~\ref{chap:6g}

\section{Quantum walk with two distinguishable particles}
\label{chap:6b}

The Hilbert space of the two particles is given by a tensor product of
the single particle spaces, i.e.
$$
\mathcal{H} = (\mathcal{H}_P\otimes\mathcal{H}_C)_1\otimes(\mathcal{H}_P\otimes\mathcal{H}_C)_2.
$$
Each particle has its own coin which determines his movement on the
line. Since we assume that there is no interaction between the two
particles they evolve independently and the time evolution of the
whole system is given by a tensor product of the single particle
time evolution operators. We describe the state of the
system by vectors
$$
\psi(m,n,t)= \left(%
\begin{array}{c}
  \psi_{LL}(m,n,t) \\
  \psi_{RL}(m,n,t) \\
  \psi_{LR}(m,n,t) \\
  \psi_{RR}(m,n,t) \\
\end{array}%
\right),
$$
where e.g. the component $\psi_{RL}(m,n,t)$ is the amplitude of
the state where the first particle is on $m$ with the internal state
$|R\rangle$ and the second particle is on $n$ with the internal
state $|L\rangle$. The state of the two particles at time $t$ is
then given by
\begin{equation}
\label{dist}
|\psi(t)\rangle = \sum_{m,n}\sum_{i,j=\ R,L}\psi_{ij}(m,n,t)|m,i\rangle_1|n,j\rangle_2.
\end{equation}
The conditional probability that the first particle is on a site $m$
at time $t$, provided that the second particle is at the same time
at site $n$, is defined by
\begin{equation}
\label{P2}
P(m,n,t) = \sum_{i,j=L,R}|\langle m,i|\langle n,j|\psi(t)\rangle|^2 = \sum_{i,j=L,R}|\psi_{ij}(m,n,t)|^2.
\end{equation}
Note that if we would consider a single quantum particle but on a two dimensional lattice, with two independent Hadamard coins for each spatial dimension, (\ref{P2}) will give the probability distribution generated by such a two dimensional walk. This shows the relation between a one dimensional walk with two particles and a two dimensional walk. The reduced probabilities for the first and the second particle are given by
$$
P_1(m,t) = \sum_n P(m,n,t),\quad P_2(n,t) = \sum_m P(m,n,t).
$$

The dynamics of the two particles is determined by the single particle motion. Since we can always decompose the initial state of the two
particles into a linear combination of a tensor product of a single particle states and because the time evolution is also given by a
tensor product of two unitary operators, the shape of the state will remain unchanged. Thus we can fully describe the time
evolution of the two quantum particles with the help of the single particle wave-functions. A similar relation holds for the
probability distribution (\ref{P2}). Moreover, in the particular case when the two particles are initially in a factorized state
\begin{equation}
\label{fact}
|\psi(0)\rangle = \left( \sum_{m,i}\psi_{1i}(m,0)|m,i\rangle_1\right)\otimes\left(\sum_{n,j}\psi_{2j}(n,0)|n,j\rangle_2\right),
\end{equation}
which translates into $\psi_{ij}(m,n,0)=\psi_{1i}(m,0)\psi_{2j}(n,0)$. Hence, the probability distribution remains a product of a single particle probability
distributions
\begin{eqnarray}
\label{fp}
\nonumber P(m,n,t) &=& (|\psi_{1L}(m,t)|^2+|\psi_{1R}(m,t)|^2)(|\psi_{2L}(n,t)|^2+|\psi_{2R}(n,t)|^2) \\
&=& P_1(m,t)P_2(n,t).
\end{eqnarray}
However, when the initial state of the two particles is entangled
\begin{equation}
\label{ent}
|\psi(0)\rangle = \sum_\alpha\left\{\left( \sum_{m,i}\psi^\alpha_{1i}(m,0)|m,i\rangle_1\right)\otimes\left(\sum_{n,j}\psi^\alpha_{2j}(n,0)|n,j\rangle_2\right)\right\},
\end{equation}
the probability distribution cannot be expressed in terms of single
particle distributions but probability amplitudes
\begin{equation}
\label{ment}
P(m,n,t) = \sum_{i,j=L,R} \left| \sum_\alpha\psi^\alpha_{1i}(m,t)\psi^\alpha_{2j}(n,t)\right|^2.
\end{equation}

Notice that the correlations are present also in the classical random walk with two particles, if we consider
initial conditions of the following form
\begin{equation}
\label{clcor}
P(m,n,0) = \sum_\alpha P_1^\alpha(m,0)P_2^\alpha(n,0).
\end{equation}
The difference between (\ref{ment}) and (\ref{clcor}) is that in the quantum case we have probability amplitudes not probabilities. The
effect of the quantum mechanical dynamics is the interference in (\ref{ment}).

Let us define the meeting problem. We ask for the probability that the two particles will be detected at the position $m$ after $t$
steps. This probability is given by the norm of the vector $\psi(m,m,t)$
\begin{equation}
\label{md}
M_D(m,t) = \sum_{i,j=L,R}|\psi_{ij}(m,m,t)|^2 =  P(m,m,t).
\end{equation}
As we have seen above for a particular case when the two particles are initially in a factorized state of the form (\ref{fact}) this
can be further simplified to the multiple of the probabilities that the individual particles will reach the site. However, this is
not possible in the situation when the particles are initially entangled (\ref{ent}). The entanglement introduced in the initial
state of the particles leads to the correlations between the particles position and thus the meeting probability is no longer a product
of the single particle probabilities.

\section{Quantum walk with two indistinguishable particles}
\label{chap:6c}

We analyze the situation when the two particles are indistinguishable. Because we work with indistinguishable particles
we use the Fock space and creation operators, we use symbols $a_{(m,i)}^\dagger$ for bosons and $b_{(n,j)}^\dagger$ for fermions,
e.g. $a_{(m,i)}^\dagger$ creates one bosonic particle at position $m$ with the internal state $|i\rangle$. The dynamics of the quantum walk with indistinguishable particles is defined on a one-particle level, i.e. a single step is given by the following transformation of the creation operators
$$
\hat{a}_{(m,L)}^\dagger \longrightarrow \frac{1}{\sqrt{2}}\left(\hat{a}_{(m-1,L)}^\dagger + \hat{a}_{(m+1,R)}^\dagger\right),\quad
\hat{a}_{(m,R)}^\dagger \longrightarrow \frac{1}{\sqrt{2}}\left(\hat{a}_{(m-1,L)}^\dagger - \hat{a}_{(m+1,R)}^\dagger\right),
$$
for bosonic particles, similarly for fermions. The difference is that the bosonic operators fulfill the commutation relations
\begin{equation}
\label{commut}
\left[\hat{a}_{(m,i)},\hat{a}_{(n,j)}\frac{}{}\right] = 0,\qquad \left[\hat{a}_{(m,i)},{\hat{a}_{(n,j)}}^\dagger\right] = \delta_{mn}\delta_{ij},
\end{equation}
while the fermionic operators satisfy the anticommutation relations
\begin{equation}
\label{anticommut}
\left\{\hat{b}_{(m,i)},\hat{b}_{(n,j)}\frac{}{}\right\} = 0,\qquad \left\{\hat{b}_{(m,i)},\hat{b}^\dagger_{(n,j)}\right\} = \delta_{mn}\delta_{ij}.
\end{equation}
We will describe the state of the system by the same vectors of amplitudes $\psi(m,n,t)$ as for the distinguishable particles.
The state of the two bosons and fermions analogous to (\ref{dist}) for two distinguishable particles is given by
\begin{eqnarray}
\label{psi:B:F}
\nonumber |\psi_B(t)\rangle & = & \sum_{m,n}\sum_{i,j=L,R}\psi_{ij}(m,n,t)\hat{a}_{(m,i)}^\dagger \hat{a}_{(n,j)}^\dagger|vac\rangle,\\
|\psi_F(t)\rangle & = & \sum_{m,n}\sum_{i,j=L,R}\psi_{ij}(m,n,t)\hat{b}_{(m,i)}^\dagger \hat{b}_{(n,j)}^\dagger|vac\rangle,
\end{eqnarray}
where $|vac\rangle$ is the vacuum state. Note that in (\ref{psi:B:F}) both summation indexes $m$ and $n$ run over all possible sites, even though e.g. the vectors $\hat{a}_{(m,i)}^\dagger \hat{a}_{(n,j)}^\dagger|vac\rangle$ and $\hat{a}_{(n,i)}^\dagger \hat{a}_{(m,j)}^\dagger|vac\rangle$ correspond to the same physical state. Using the commutation (\ref{commut}) and anticommutation (\ref{anticommut}) relations we can restrict the sums in (\ref{psi:B:F}) over an ordered pair $(m,n)$ with $m\geq n$. The resulting wave-function will be symmetric or antisymmetric.

The conditional probability distribution is given by
$$
P_{B,F}(m,n,t) = \sum_{i,j=L,R}\left|\langle 1_{(m,i)}1_{(n,j)}|\psi_{B,F}(t)\rangle\right|^2 =\sum_{i,j = L,R}\left|\psi_{ij}(m,n,t)\pm\psi_{ji}(n,m,t)\right|^2,
$$
for $m\neq n$, and for $m=n$
\begin{eqnarray}
\label{mbf}
\nonumber P_B(m,m,t) & = & \left|\langle 2_{(m,L)}|\psi_B(t)\rangle\right|^2 + \left|\langle 2_{(m,R)}|\psi_B(t)\rangle\right|^2 + \left|\langle 1_{(m,L)}1_{(m,R)}|\psi_B(t)\rangle\right|^2 \\
\nonumber & = & 2\left|\psi_{LL}(m,m,t)\right|^2+2\left|\psi_{RR}(m,m,t)\right|^2 + \left|\psi_{LR}(m,m,t)+\psi_{RL}(m,m,t)\right|^2\\
\nonumber & = & M_B(m,t),\\
\nonumber P_F(m,m,t) & = & \left|\langle 1_{(m,L)}1_{(m,R)}|\psi_F(t)\rangle\right|^2 = \left|\psi_{LR}(m,m,t)-\psi_{RL}(m,m,t)\right|^2\\
 & = & M_F(m,t).
\end{eqnarray}
The diagonal terms of the probability distribution (\ref{mbf}) define the meeting probability we wish to analyze.

Let us specify the meeting probability for the case when the
probability amplitudes can be written in a factorized form
$\psi_{ij}(m,n,t) = \psi_{1i}(m,t)\psi_{2j}(n,t)$, which for the
distinguishable particles corresponds to the situation when they are
initially not correlated. In this case the meeting probabilities are
given by
\begin{eqnarray}
\label{mb}
M_B(m,t) = 2\left|\psi_{1L}(m,t)\psi_{2L}(m,t)\right|^2+2\left|\psi_{1R}(m,t)\psi_{2R}(m,t)\right|^2 \nonumber \\
+\left|\psi_{1L}(m,t)\psi_{2R}(m,t)+\psi_{1R}(m,t)\psi_{2L}(m,t)\right|^2,
\end{eqnarray}
for bosons and
\begin{equation}
\label{mf}
M_F(m,t) = \left|\psi_{1L}(m,t)\psi_{2R}(m,t)-\psi_{1R}(m,t)\psi_{2L}(m,t)\right|^2,
\end{equation}
for fermions. We see that they differ from the formulas for the
distinguishable particles, except for a particular case when the two
bosons start in the same state, i. e. $\psi_1(m,0) = \psi_2(m,0) =
\psi(m,0) $ for all integers $m$. For this initial state we obtain
\begin{eqnarray}
\nonumber M_B(m,t) &=& |\psi_{L}(m,t)|^4+|\psi_{R}(m,t)|^4+2|\psi_{L}(m,t)\psi_{R}(m,t)|^2 \\
\nonumber &=& (|\psi_L(m,t)|^2+|\psi_R(m,t)|^2)^2  \\
\nonumber &=& P^2(m,t) ,
\end{eqnarray}
which is the same as for the case of distinguishable particles
starting at the same point with the same internal state.

\section{Meeting problem for distinguishable particles}
\label{chap:6d}

Let us compare the meeting problem in the classical and
quantum case. We study the two following probabilities: the total
meeting probability after $t$ step have been performed
\begin{equation}
\label{m2}
M(t) = \sum_{m}M(m,t),
\end{equation}
and the overall meeting probability during some period of steps
$T$ defined as
\begin{equation}
\label{ov}
{\cal M}(T) = 1-\prod_{t=1}^T\left(1-M(t)\right) .
\end{equation}
The total meeting probability $M(t)$ is the probability that the
two particles meet at time $t$ anywhere on the lattice, the overall
meeting probability ${\cal M}(T)$ is the probability that
they meet at least once anywhere on the lattice during the first $T$
steps.

We first concentrate on the influence of the initial state on the
meeting probability for the distinguishable particles. We consider
three situations, the particles start localized with some
initial distance $2d$ (for odd initial distance they can never
meet, without loss of generality we assume that the first starts
at the position zero and the second at the position $2d$), with
the coin states:

({\it i}) right for the first particle and left for
the second
$$
\psi_{RL}(0,2d,0) = 1 ,
$$

({\it ii}) symmetric initial conditions
$1/\sqrt{2}(|L\rangle+i|R\rangle)$ for both
$$
\psi(0,2d,0) = \frac{1}{2}\left( \begin{array}{c}
  1 \\
  i \\
  i \\
  -1
\end{array}\right) ,
$$

({\it iii}) left for the first particle and right for the second
$$
\psi_{LR}(0,2d,0) = 1.
$$
In the first case the probability distributions of the particles are
biased to the right for the first particle, respectively to the left
for the second, and thus the particles are moving towards each other.
In the second case the particles mean positions remain unchanged, as
for this initial condition the single particle probability
distribution is symmetric and unbiased. In the last case the particles
are moving away from each other as their probability distributions
are biased to the left for the first one and to the right for the
second.

Let us specify the meeting probabilities (\ref{m2}). Since the two particles are initially in a factorized state it follows from (\ref{fp}) and (\ref{md}) that the meeting probability is fully determined by the single particle probability distribution. Let
\begin{eqnarray}
|\psi^{(L)}(t)\rangle & = & \sum_m\left(\psi^{(L)}_L(m,t)|m,L\rangle+\psi^{(L)}_R(m,t)|m,R\rangle\right)\\
|\psi^{(R)}(t)\rangle & = & \sum_m\left(\psi^{(R)}_L(m,t)|m,L\rangle+\psi^{(R)}_R(m,t)|m,R\rangle\right)
\label{psi:LR}
\end{eqnarray}
be the state of a single quantum particle after $t$ steps, under the assumption that the initial condition was
$$
|\psi^{(L)}(0)\rangle=|0,L\rangle,\qquad |\psi^{(R)}(0)\rangle=|0,R\rangle.
$$
Let us denote by  $P^{(L,R)}(m,t)$ the corresponding single particle probability distributions. The meeting probabilities for the three situations ({\it i})-({\it iii}) are then given by
\begin{eqnarray}
\label{mq}
\nonumber M_{RL}(t,d) & = & \sum_m P^{(R)}(m,t)P^{(L)}(m-2d,t) \\
\nonumber M_{S}(t,d) & = & \sum_m\frac{P^{(L)}(m,t)+P^{(R)}(m,t)}{2}\frac{P^{(L)}(m-2d,t)+P^{(L)}(m-2d,t)}{2}\\
M_{LR}(t,d) & = & \sum_m P^{(L)}(m,t)P^{(R)}(m-2d,t).
\end{eqnarray}

\begin{figure}[p]
\begin{center}
\includegraphics[width=0.7\textwidth]{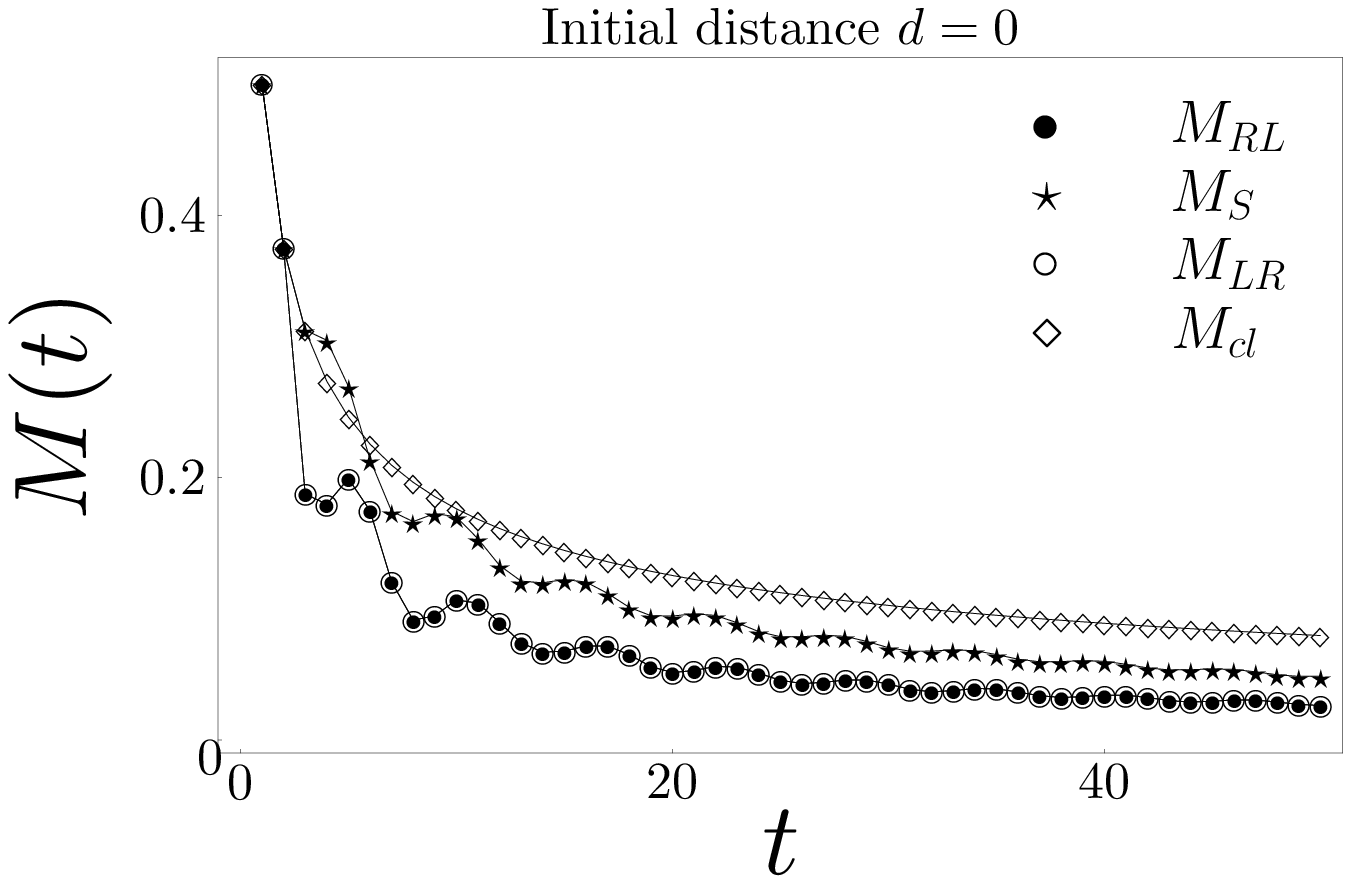}\vspace{24pt}
\includegraphics[width=0.7\textwidth]{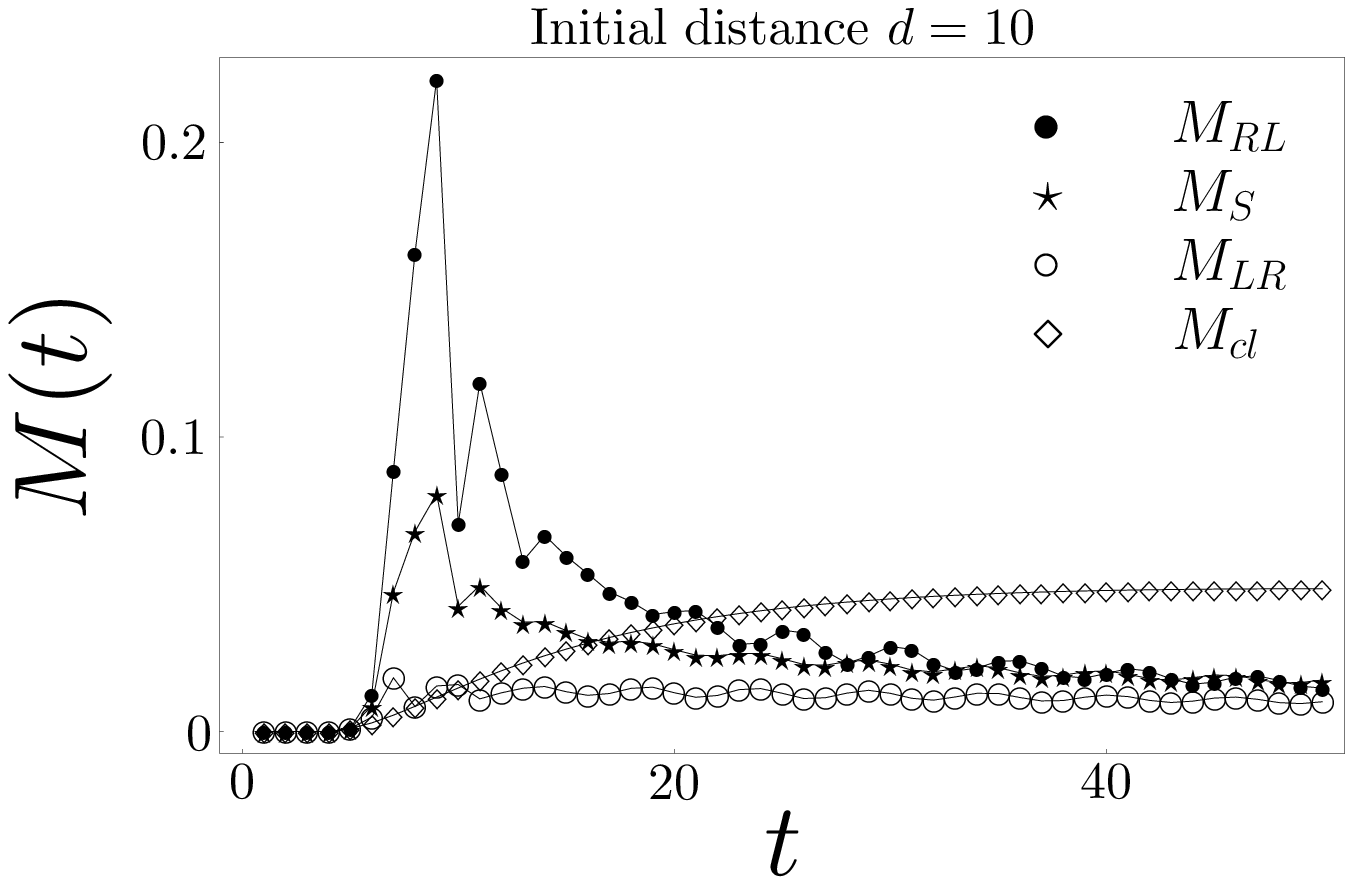}
\end{center}
\caption{Time evolution of the meeting probability for the three types of initial states and the classical random walk with two particles. The initial distance is set to 0 (upper plot) and 10 lattice points (lower plot). The upper plot shows a faster decay of the meeting probability when the two particles are initially at the same lattice point. Indeed, quantum walk spreads quadratically faster compared to the classical random walk. Since both particles start the walk from the origin the results for the initial states $|LR\rangle$ and $|RL\rangle$  are identical. In the lower plot, where the particles are initially separated, we observe an increase in the meeting probability for quantum walk. On the other hand, on a long-time scale the meeting probability decays faster in the quantum case.}
\label{fig:61}
\end{figure}

Figure~\ref{fig:61} shows the time evolution of the meeting probability for the three studied situations and compares it with the
classical case. The initial distance is set to 0 and 10 lattice points. The plot clearly shows the difference between the quantum and the
classical case.

In contrast to the classical walk, in the quantum case the meeting
probability is oscillatory. The oscillations arise from the single particle probability distribution. After
some rapid oscillations in the beginning we get a periodic function
with the characteristic period of about six steps, independent of
the initial state. In the quantum case the maximum of the meeting
probability is reached sooner than in the classical case - the
number of steps needed to hit the maximum is linear in the initial
distance $d$. This can be understood from the shape of the particles
probability distribution. The maximum of the meeting probability is
obtained when the peaks of the probability distribution of the first
and second particle overlap. If the initial distance between the two
particles is $2d$ then the peaks will overlap approximately after
$\sqrt{2}d$ steps. The value of the maximum depends on the choice of
the initial state.

We turn to the meeting probabilities on a long-time scale. We present the review of the results derived in Appendix~\ref{app:d}. For the classical random walk we find in Appendix~\ref{app:d1} that the meeting probability can by estimated by
\begin{equation}
M_{cl}(t,d)\approx\frac{1}{\sqrt{\pi t}} \exp(-\frac{d^2}{t})\sim \frac{1}{\sqrt{\pi t}}(1-\frac{d^2}{t})
\label{apcl}
\end{equation}
for large number of steps $t$. We see that the asymptotic behaviour of the meeting probability is determined by $t^{-\frac{1}{2}}$. Concerning the quantum walk, in Appendix~\ref{app:d2} we approximate the single particle probability distribution according to \cite{nayak} and replace the sum in (\ref{mq}) by integral. We find that within this approximation the meeting probability can be expressed in terms of the elliptic integrals. Finally, using the asymptotic expansion of the elliptic integrals we find the behaviour of the meeting probability for large number of steps
\begin{equation}
\label{apq}
M_D(t,d) \sim \frac{\ln{\left(\frac{2\sqrt{2}t}{d}\right)}}{t}.
\end{equation}
Hence, the meeting probability decays faster in the quantum case compared to the classical case (\ref{apcl}). However, the decay is not quadratically faster, as one could expect from the fact that the single particle probability distribution spreads quadratically faster in the quantum walk. The peaks in the probability distribution of the quantum walk slow down the decay.

Note that the estimation (\ref{apq}) holds for $d>0$, i.e. the initial distance has to be non-zero. As we mention in Appendix~\ref{app:d2}, the continuous approximation of the single particle probability distribution is not quadratically integrable, and therefore we cannot use this approach for the estimation of the meeting probability when the two particles are initially at the same lattice point. There does not seem to be an easy analytic approach to the problem. However, from the numerical results, the estimation
\begin{equation}
M_D(t) \sim \frac{\ln{t}}{t}
\label{mq:d0}
\end{equation}
fits the data the best.

\begin{figure}[p]
\begin{center}
\includegraphics[width=0.7\textwidth]{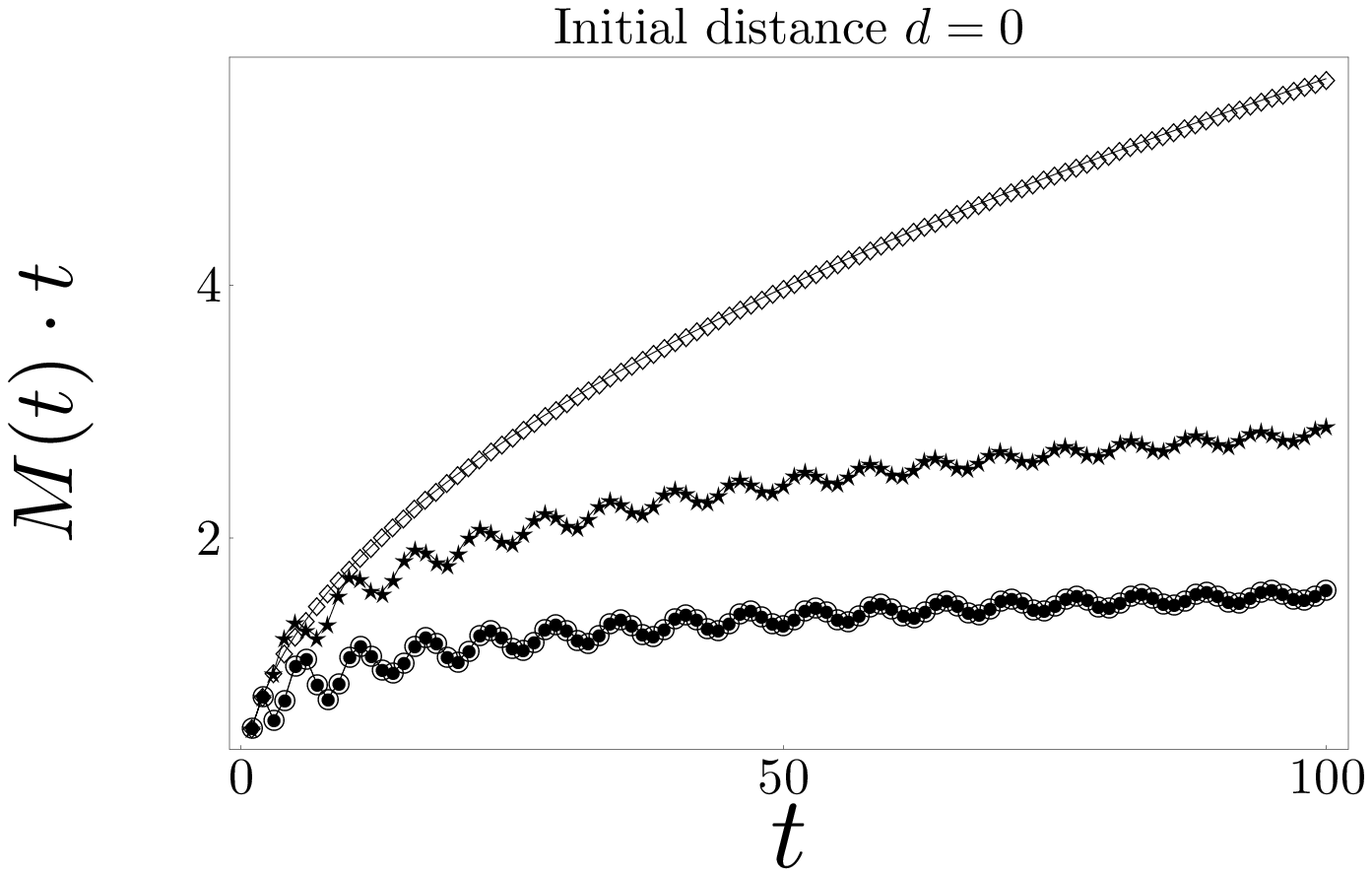}\vspace{24pt}
\includegraphics[width=0.7\textwidth]{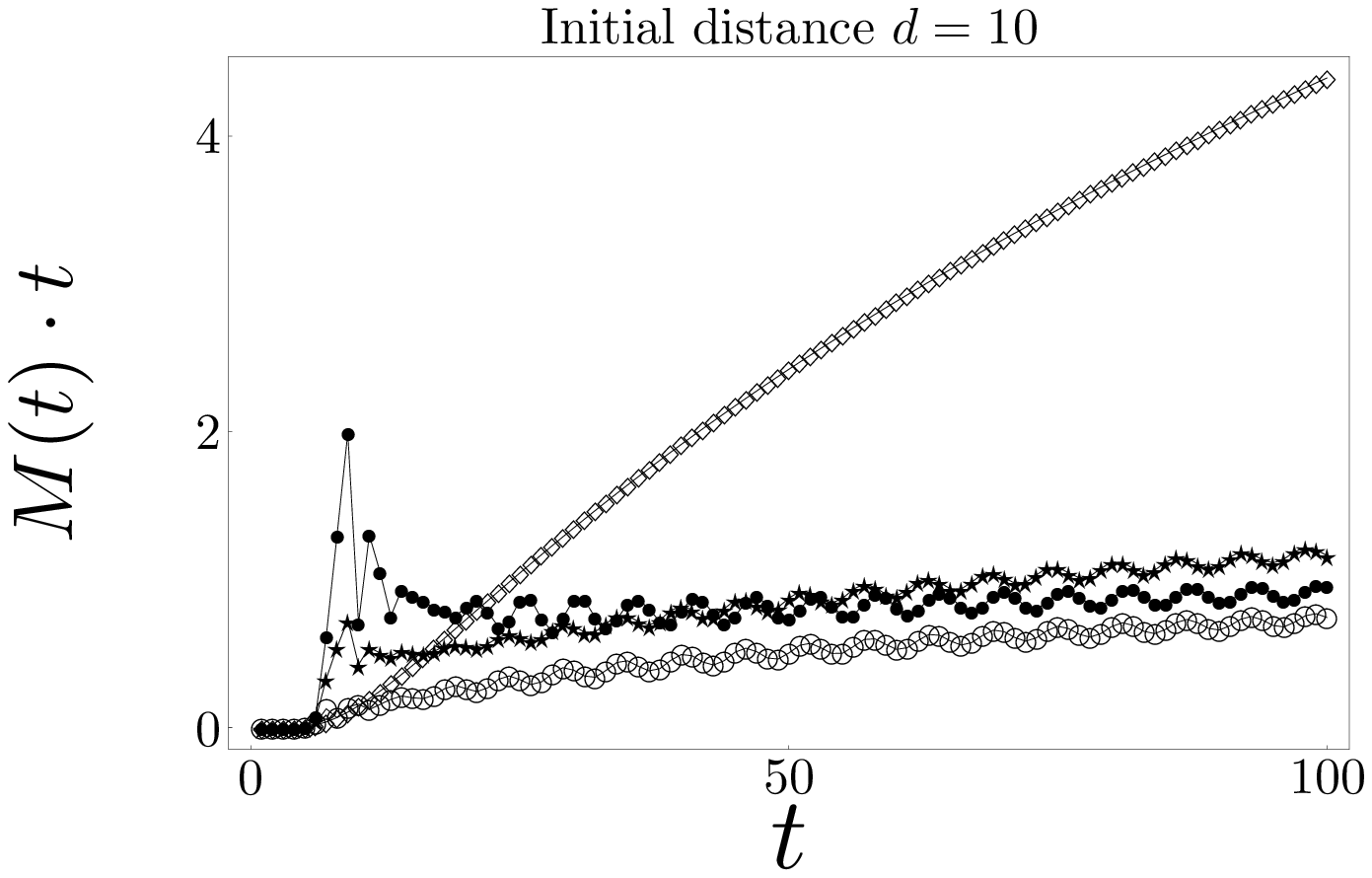}
\end{center}
\caption{Long-time behaviour of the meeting probability in the classical and quantum walk. In the upper plot both particles start the walk from the origin. In the lower plot the particles are initially separated by 10 lattice points. To highlight the asymptotic scaling of the meeting probability we plot the latter one multiplied by the number of steps. We can clearly see the difference between the classical and quantum walk. In the quantum case the re-scaled  meeting probability shows a logarithmic increase. On the other hand, the growth is much faster (with a square root of $t$) for the classical case. The numerical results are in good agreement with the analytical estimation of Appendix~\ref{app:d} which are summarized in (\ref{apq}).}
\label{fig:62}
\end{figure}

We illustrate these results in Figure~\ref{fig:62}. We plot the meeting probability multiplied by the number of steps to unravel the different scaling in the classical and quantum case. In the upper plot both particles start from the origin, whereas in the lower plot the initial distance is 10 lattice points. The numerical results are consistent with the analytical estimation of (\ref{apq}) and support the approximation (\ref{mq:d0}).

\begin{figure}[p]
\begin{center}
\includegraphics[width=0.7\textwidth]{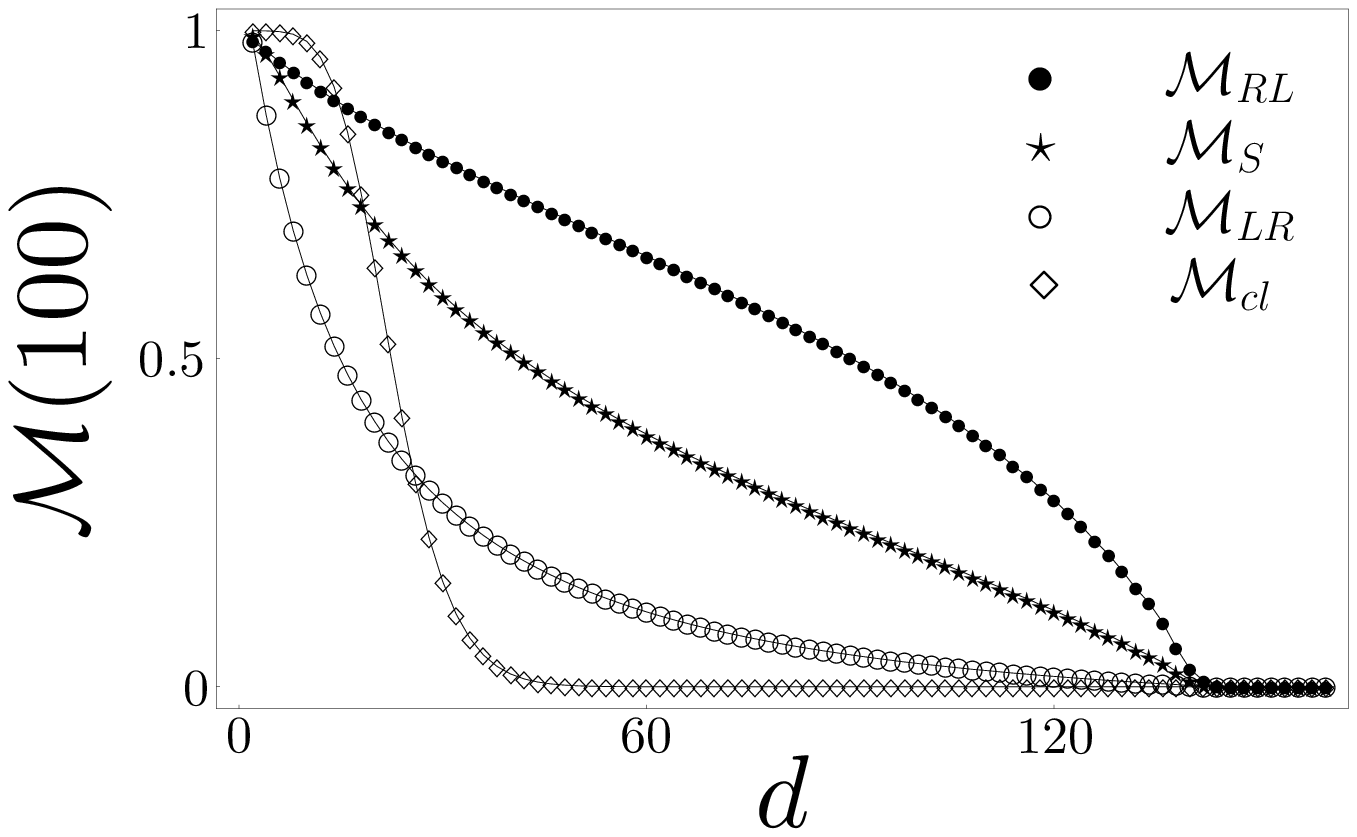}\vspace{24pt}
\includegraphics[width=0.7\textwidth]{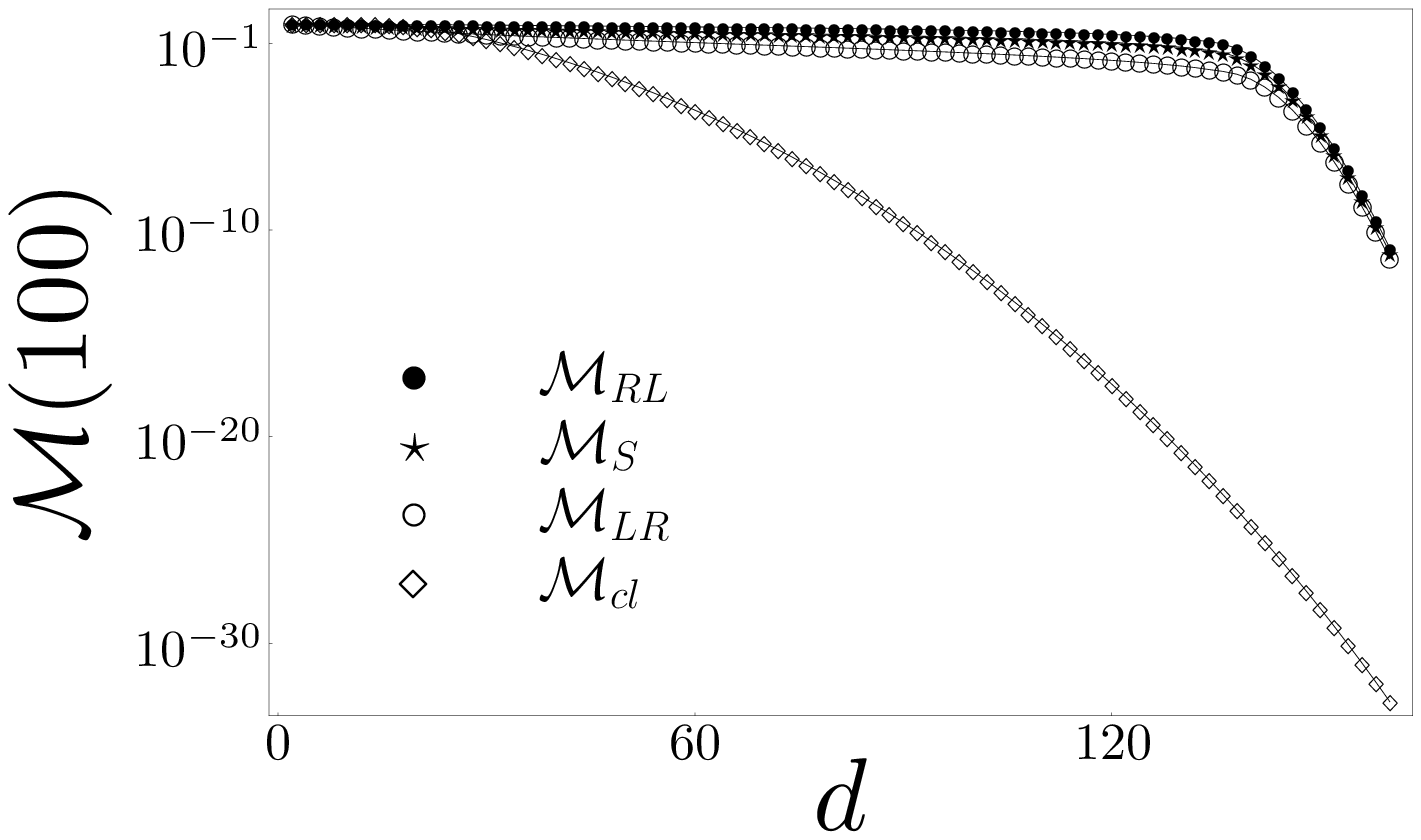}
\end{center}
\caption{The overall meeting probability for two distinguishable quantum and classical particle during first 100
steps as a function of the initial distance. The same plot on the logarithmic scale. Only the values for even points are plotted since for odd
initial distance the particles never meet.}
\label{fig:63}
\end{figure}

We focus on the overall meeting probability defined by (\ref{ov}). In Figure~\ref{fig:63} we plot the overall probability that the two
particles will meet during the first $T=100$ steps.

On the first plot we present the difference between the three studied quantum situations, whereas the second plot, where the
meeting probability is on the log scale, uncovers the difference between the quantum and the classical random walk. In the log
scale plot we see that the overall meeting probability decays slower in the quantum case then in the classical case, up to to
the initial distance of $\sqrt{2}T$. This can be understood by the shape and the time evolution of a single particle probability
distribution. After $t$ steps the maximums of the probability distribution are around the point $s\pm\frac{t}{\sqrt{2}}$, where
$s$ is the initial starting point of the quantum particle. For $t=100$ steps the peaks are around the points $s\pm 70$. When
the two particles are initially more then 140 points away, the peaks do not overlap, and the meeting probability is given by just
the tails of the single particle distributions, which have almost classical behavior.

Finally, we note that the overall meeting probability ${\cal M}(T,d)$ defined in (\ref{ov}) converges to one as $T$ approaches infinity for both classical and quantum walk independent of the initial distance. Indeed, to estimate ${\cal M}(T,d)$ we rewrite it in the form
\begin{equation}
{\cal M}(T,d) = 1 - \exp\left[\ln\left(\prod\limits_{t=1}^T\left(1-M(t,d)\right)\right)\right],
\label{appd:eq2}
\end{equation}
and estimate the exponent with the first order Taylor expansion
\begin{equation}
\ln\left(\prod_{t=1}^T\left(1-M(t,d)\right)\right)=\sum_{t=1}^T \ln{\left(1-M(t,d)\right)}\approx -\sum_{t=1}^T M(t,d).
\label{appd:eq3}
\end{equation}
The scaling of the meeting probability $M(t,d)$ both in the classical case (\ref{apcl}) and in the quantum case (\ref{apq}) is slow enough such that the sum in (\ref{appd:eq3}) diverges to $-\infty$ as $T$ grows. Consequently, the exponential in (\ref{appd:eq2}) vanishes as $T$ grows. Hence, the overall meeting probability converges to unity for both classical and quantum walk, i.e. the particles will meet with certainty during their time evolution.

\section{Effect of the entanglement}
\label{chap:6e}

\begin{figure}[p]
\begin{center}
\includegraphics[width=0.7\textwidth]{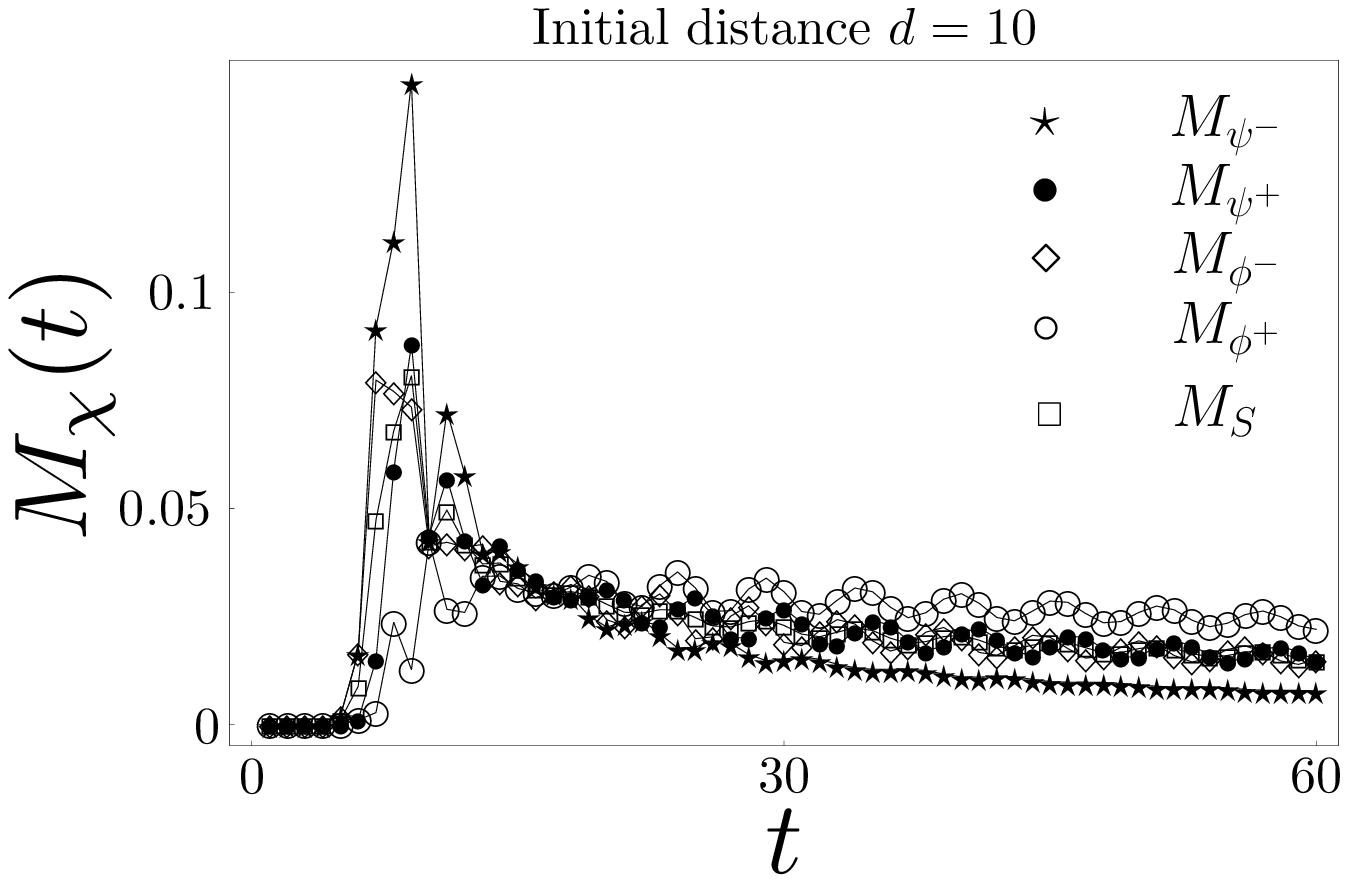}\vspace{24pt}
\includegraphics[width=0.7\textwidth]{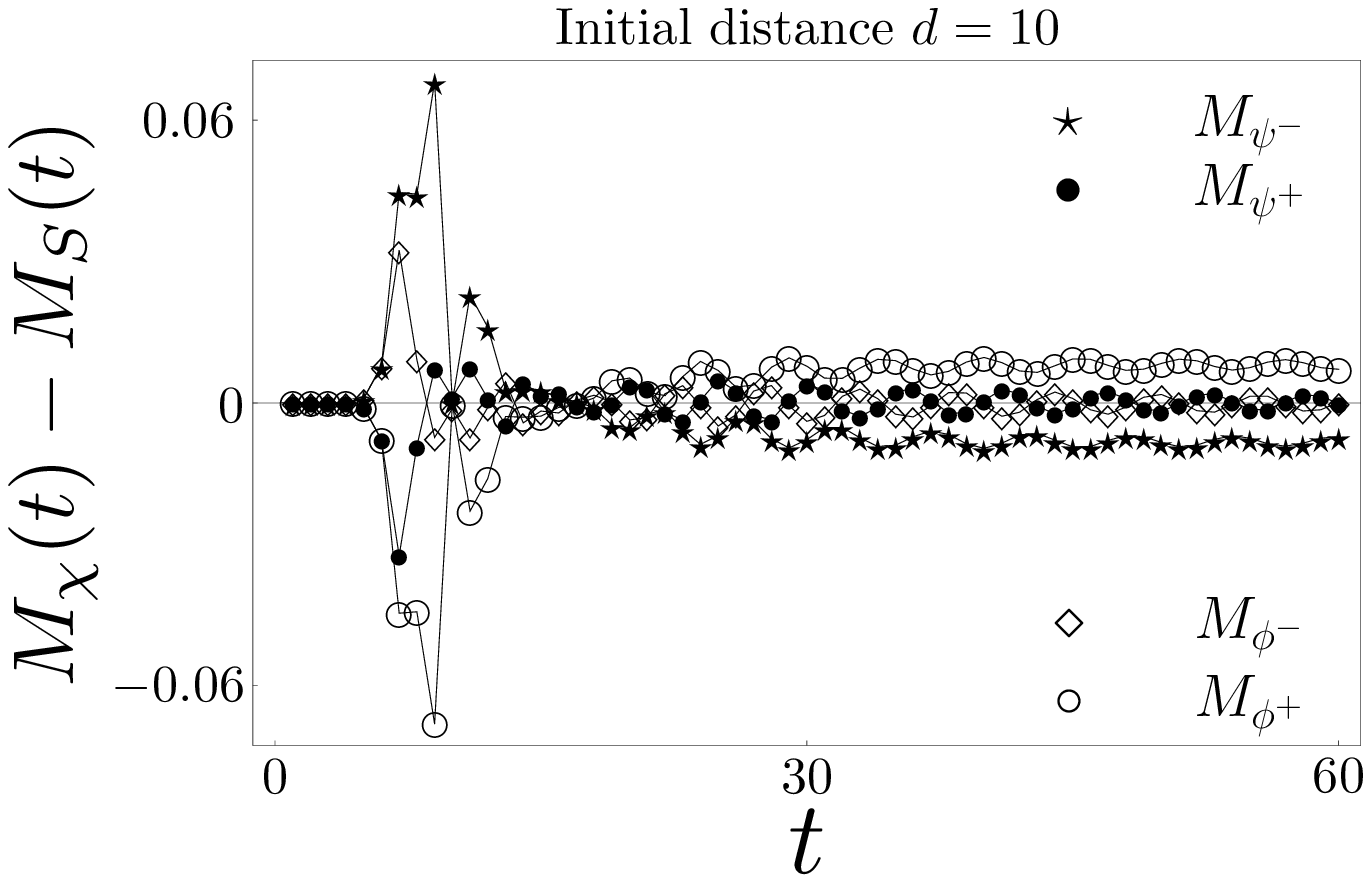}
\end{center}
\caption{Comparison of the meeting probability for the initially entangled coins and the symmetric case. The initial distance between the two particles is set to 10 points. As the initial coin states we choose the Bell states (\ref{bell}). We observe that the effect of the entangled coin state on the meeting probability can be both positive or negative. In the lower plot we show the difference in the meeting probability with respect to the symmetric case. We find that the effect of $|\psi^-\rangle$ is opposite to $|\phi^+\rangle$ and $|\phi^-\rangle$ is opposite to $|\psi^+\rangle$.}
\label{fig:65}
\end{figure}

We will consider the case when the two distinguishable particles
are initially entangled. According to (\ref{ment}) the meeting
probability is no longer given by a product of a single particle
probability distributions. However, it can be described using
single particle probability amplitudes. We consider the initial
state of the following form
$$
|\psi(0)\rangle = |0, 2d\rangle\otimes|\chi\rangle,
$$
where $|\chi\rangle$ is one of the Bell states
\begin{eqnarray}
\label{bell}
\nonumber |\psi^\pm\rangle &=&
\frac{1}{\sqrt{2}}\left(|LR\rangle\pm|RL\rangle\right),\\
|\phi^\pm\rangle & = &
\frac{1}{\sqrt{2}}\left(|LL\rangle\pm|RR\rangle\right).
\end{eqnarray}
The corresponding probability distributions resulting from such initial states have the form
\begin{eqnarray}
\label{pent}
\nonumber P_{\psi^\pm}(m,n,t) & = & \frac{1}{2} \sum_{i,j=L,R}\left|\psi_i^{(L)}(m,t)\psi_j^{(R)}(n-2d,t)\pm \psi_i^{(R)}(m,t)\psi_j^{(L)}(n-2d,t)\right|^2,\\
P_{\phi^\pm}(m,n,t) & = & \frac{1}{2}\sum_{i,j=L,R}\left|\psi_i^{(L)}(m,t)\psi_j^{(L)}(n-2d,t)\pm\psi_i^{(R)}(m,t)\psi_j^{(R)}(n-2d,t)\right|^2,
\end{eqnarray}
where $\psi^{L(R)}(m,t)$ are the probability amplitudes from (\ref{psi:LR}) which describe the state of a single particle after $t$ steps starting the quantum walk from the origin with the initial coin state $L(R)$. The meeting probabilities are given by the sum of the diagonal terms in (\ref{pent})
\begin{eqnarray}
\nonumber M_{\psi^\pm}(t,d) & = & \frac{1}{2}\sum_m \sum_{i,j=L,R}\left|\psi_i^{(L)}(m,t)\psi_j^{(R)}(m-2d,t)\pm\psi_i^{(R)}(m,t)\psi_j^{(L)}(m-2d,t)\right|^2,\\
\nonumber M_{\phi^\pm}(t,d) & = & \frac{1}{2}\sum_m \sum_{i,j=L,R}\left|\psi_i^{(L)}(m,t)\psi_j^{(L)}(m-2d,t)\pm\psi_i^{(R)}(m,t)\psi_j^{(R)}(m-2d,t)\right|^2.
\end{eqnarray}
The reduced density operators for both coins are maximally mixed for all four Bell states (\ref{bell}). From this fact follows that the reduced density operators of the particles are
\begin{eqnarray}
\nonumber \rho_1(t) & = & \frac{1}{2}\left(|\psi^{(L)}(t)\rangle\langle\psi^{(L)}(t)|+|\psi^{(R)}(t)\rangle\langle\psi^{(R)}(t)|\right)\\
\nonumber \rho_2(t) & = & \frac{1}{2}\left(|\psi^{(L)}_d(t)\rangle\langle\psi^{(L)}_d(t)|+|\psi^{(R)}_d(t)\rangle\langle\psi^{(R)}_d(t)|\right),
\end{eqnarray}
where the states $|\psi^{L(R)}_d(t)\rangle$ are analogous to $|\psi^{L(R)}(t)\rangle$ expressed in (\ref{psi:LR}) but with shifted starting point by $2d$, i.e.
$$
|\psi^{(L,R)}_d(t)\rangle = \sum_m\left(\psi^{(L,R)}_L(m-2d,t)|m,L\rangle+\psi^{(L,R)}_R(m-2d,t)|m,R\rangle\right).
$$
The reduced probabilities are therefore
\begin{eqnarray}
\label{red}
\nonumber P_1(m,t) & = & \frac{1}{2}(P^{(L)}(m,t)+P^{(R)}(m,t))\\
P_2(m,t) & = & P_1(m-2d,t),
\end{eqnarray}
which are symmetric and unbiased. Notice that the product of the reduced probabilities (\ref{red}) gives the probability distribution of a symmetric case studied in the previous section. Therefore to catch the interference effect in the meeting problem we compare the quantum walks with entangled coin states
(\ref{bell}) with the symmetric case $M_{S}$. Figure~\ref{fig:65} shows the meeting probabilities and the difference $M_\chi-M_{S}$, the
initial distance between the two particles was chosen to be 10 points.

\begin{figure}[h]
\begin{center}
\includegraphics[width=0.7\textwidth]{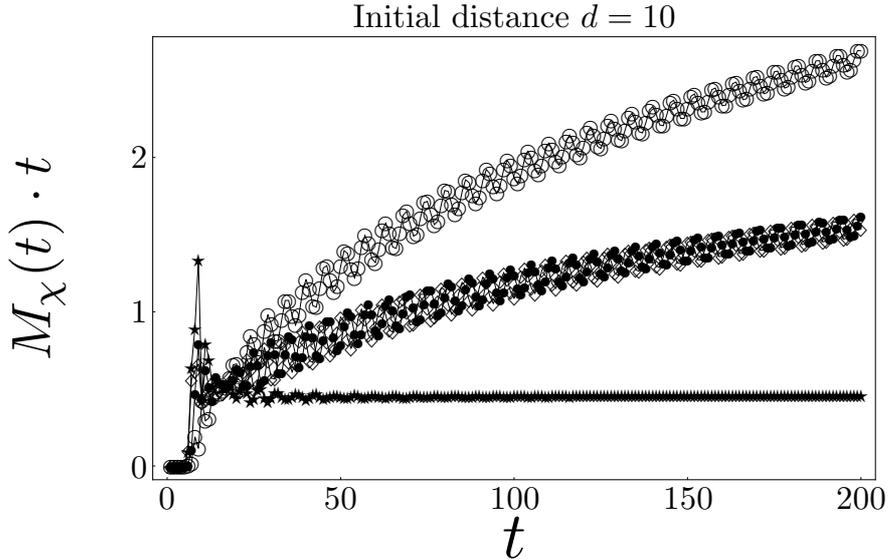}
\end{center}
\caption{Asymptotic behaviour of the meeting probability for the initially entangled coins. In order to unravel the asymptotic scaling of the meeting probability we multiply $M_\chi(t)$ by the number of steps $t$. We see that for the Bell states $|\psi^+\rangle$ (black dots) and $|\phi^{\pm}\rangle$ (open circles/diamonds) the rescaled meeting probability $M_\chi(t)\cdot t$ shows a logarithmic increase with $t$, while for $|\psi^-\rangle$ (stars) the value of $M_\chi(t)\cdot t$ levels. These results indicate that the asymptotic decay of the meeting probability is faster for the singlet state $|\psi^-\rangle$ compared to the other Bell states or factorized initial conditions.}
\label{fig:68}
\end{figure}

We see that the effect of the entanglement could be both positive or negative. Notice that
\begin{eqnarray}
\nonumber M_{\psi^-}(t,d)-M_S(t,d) & = & -\left(M_{\phi^+}(t,d)-M_S(t,d)\right)\\
\nonumber M_{\phi^-}(t,d)-M_S(t,d) & = & -\left(M_{\psi^+}(t,d)-M_S(t,d)\right),
\end{eqnarray}
so the effect of $|\psi^-\rangle$ is opposite to $|\phi^+\rangle$ and $|\phi^-\rangle$ is opposite to $|\psi^+\rangle$. The main
difference is around the point $t\approx\sqrt{2}d$, i.e., the point where for the factorized states the maximum of the meeting
probability is reached. The peak value is nearly doubled for $M_{\psi^-}$, but significantly reduced for $M_{\phi^+}$. On the long time scale,
however, the meeting probability $M_{\psi^-}$ decays faster than in the other situations. According to the numerical results presented in Figure~\ref{fig:68},
the meeting probabilities for $|\psi^+\rangle$ and $|\phi^\pm\rangle$ maintain the asymptotic behavior $\ln{t}/t$,
but for $|\psi^-\rangle$ it goes like
$$
M_{\psi^-}(t,d)\sim\frac{1}{t}.
$$
The initial entanglement between the particles influences the height of the peaks giving the maximum meeting probability and affects
also the meeting probability on the long time scale.

Let us briefly comment on the overall meeting probability. As we have discussed in the previous section the overall meeting
probability converges to one only if the decay of the meeting probability is not faster than $\frac{1}{t}$. As we have seen the
entanglement could speed-up the decay of the meeting probability but it is never faster than $\frac{1}{t}$. Therefore we conclude
that for the initially entangled particles the overall meeting probability converges to one.

\section{Meeting problem for indistinguishable particles}
\label{chap:6f}

We turn to the meeting problem for two indistinguishable particles. As an example, we
consider the initial state of the form $|1_{(0,R)}1_{(2d,L)}\rangle$, i.e. one particle starts at the site
zero with the right coin state and one starts at $2d$ with the left state. This corresponds to the case $M_{RL}$ for the
distinguishable particles. The meeting probabilities are according to (\ref{mb}), (\ref{mf}) given by
\begin{eqnarray}
\label{bfmp}
\nonumber M_B(t,d) & = & \sum_{m}\left(\frac{}{}2|\psi_L^{(R)}(m,t)|^2|\psi_L^{(L)}(m-2d,t)|^2 + 2|\psi_R^{(R)}(m,t)|^2|\psi_R^{(L)}(m-2d,t)|^2+\nonumber \right.\\
\nonumber & & \quad\quad \left.+|\psi_L^{(R)}(m,t)\psi_R^{(L)}(m-2d,t) + \psi_R^{(R)}(m,t)\psi_L^{(L)}(m-2d,t)|^2\frac{}{}\right),\\
\nonumber \\
M_F(t,d) & = & \sum_m\left(|\psi_L^{(R)}(m,t)\psi_R^{(L)}(m-2d,t)-\psi_R^{(R)}(m,t)\psi_L^{(L)}(m-2d,t)|^2\frac{}{}\right).
\end{eqnarray}

\begin{figure}[p]
\begin{center}
\includegraphics[width=0.7\textwidth]{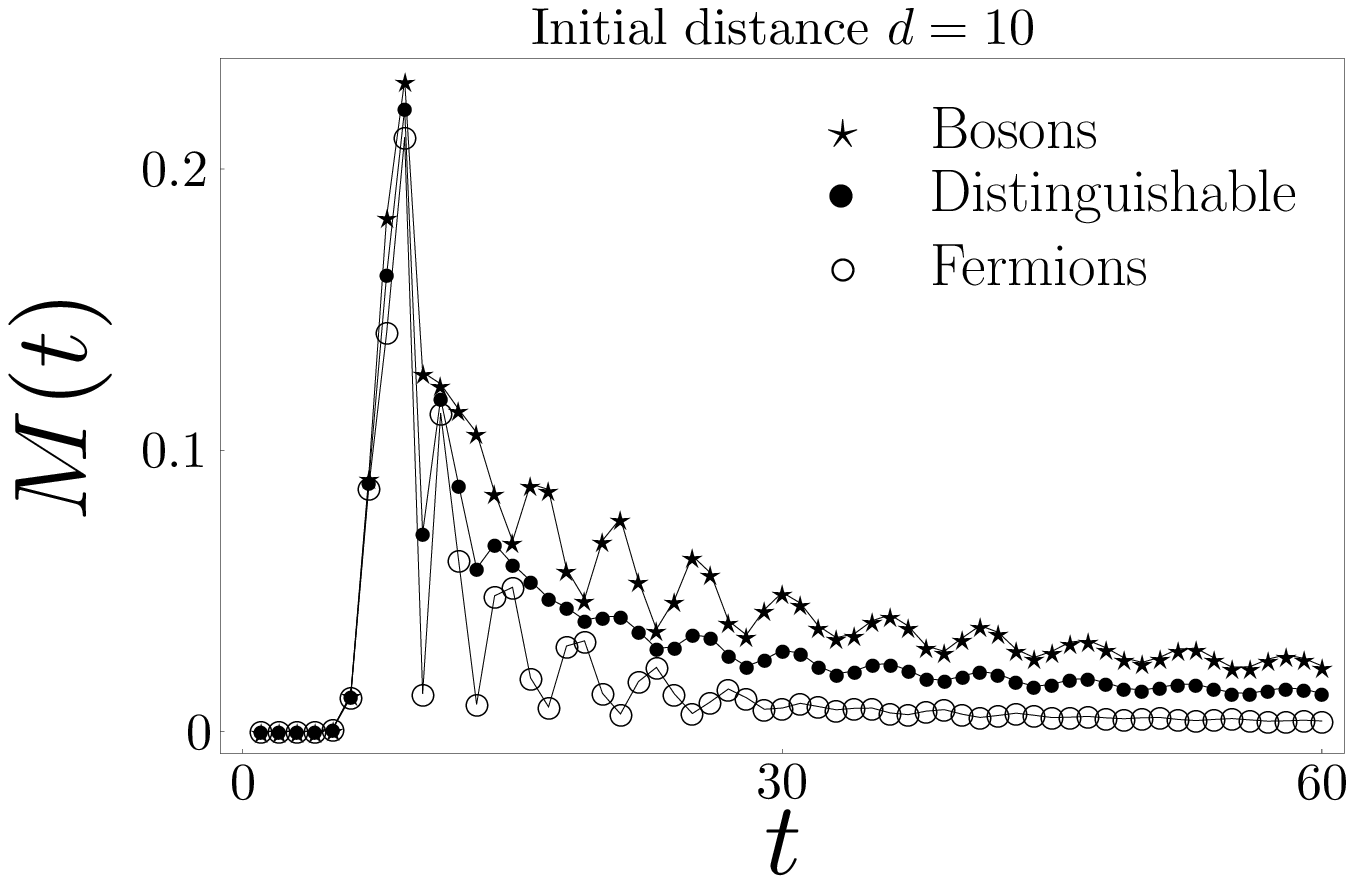}\vspace{24pt}
\includegraphics[width=0.7\textwidth]{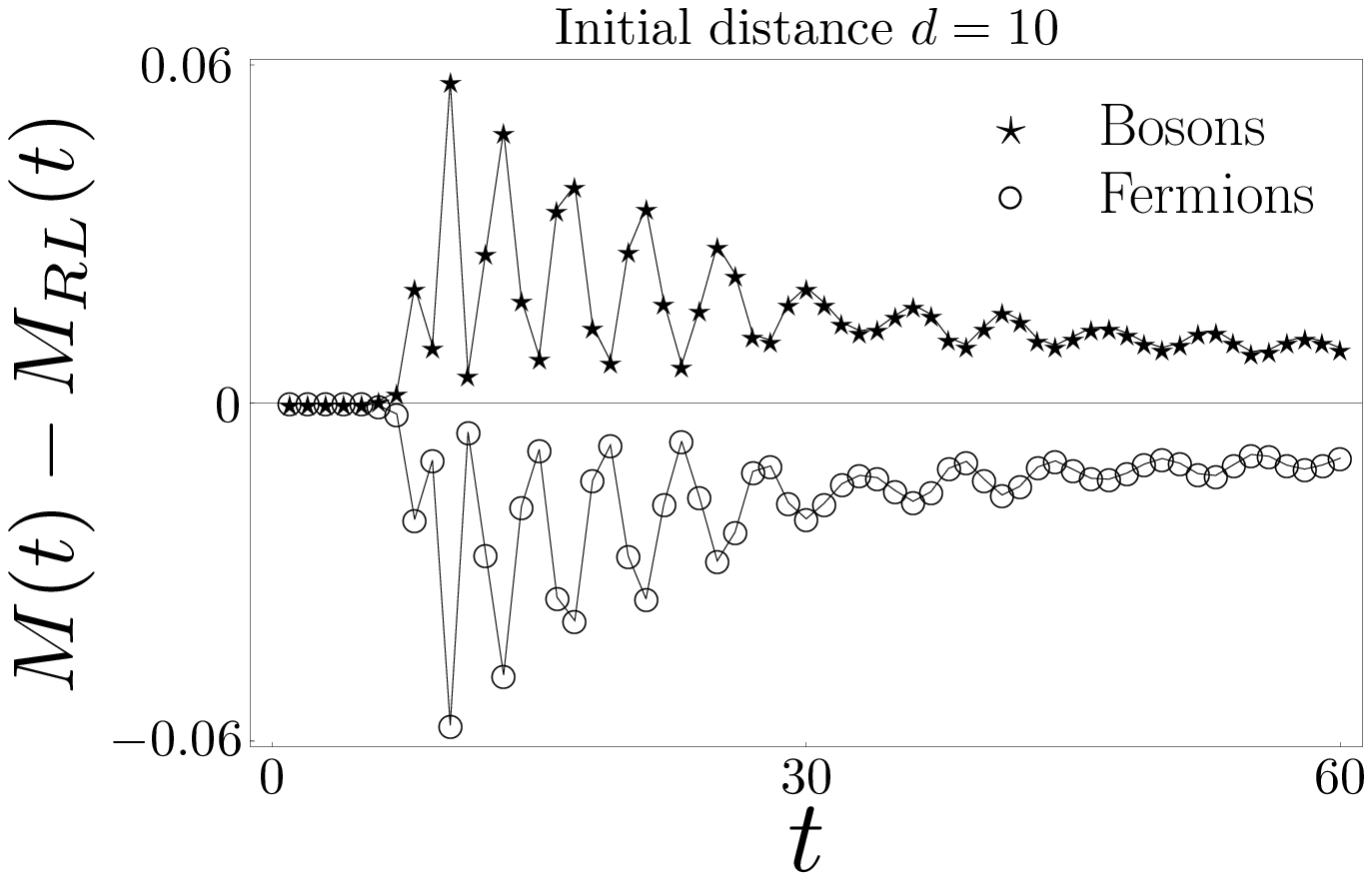}
\end{center}
\caption{Comparison of the meeting probability for bosons, fermions and distinguishable particles. The initial distance between the two particles is set to 10 points. We find that the maximum value of the meeting probability is almost unaffected. However, for longer times we observe an increase in the meeting probability for bosons and decrease for fermions. In the lower plot we show the difference in the meeting probability for bosons and fermions with respect to distinguishable particles. We find that the increase of the meeting probability for bosons is the same as the decrease for fermions.}
\label{fig:66}
\end{figure}

In Figure~\ref{fig:66} we plot the meeting probabilities and the difference $M_{B,F}-M_{RL}$.

\begin{figure}[h]
\begin{center}
\includegraphics[width=0.7\textwidth]{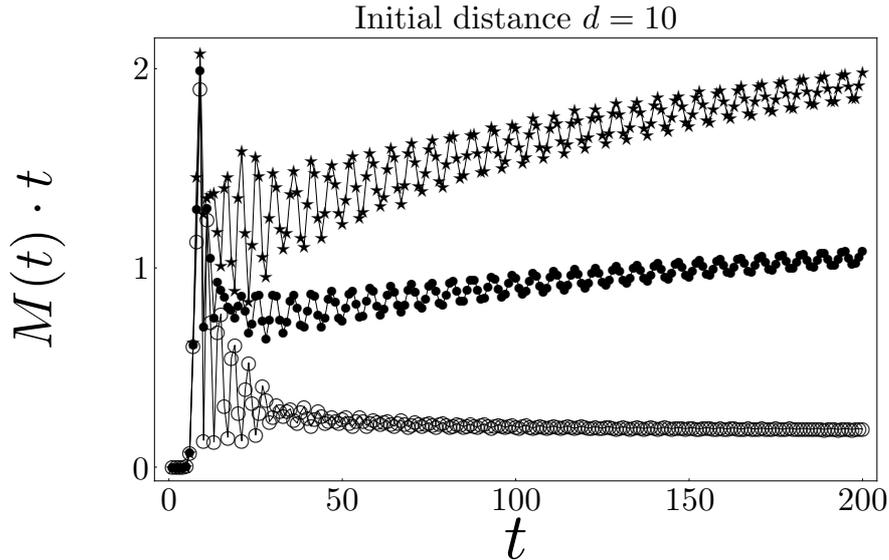}
\end{center}
\caption{Asymptotic behaviour of the meeting probability for bosons, fermions and distinguishable particles. In order to unravel the asymptotic scaling of the meeting probability we multiply $M_(t)$ by the number of steps $t$. We see that for bosons (stars) and distinguishable particles (black dots) the rescaled meeting probability $M(t)\cdot t$ shows a logarithmic increase with $t$, while for fermions (open circles) the value of $M(t)\cdot t$ levels. These results indicate that the meeting probability decays faster for fermions.}
\label{fig:69}
\end{figure}

From the figure we infer that the peak value is in this case only slightly changed. Significant differences appear on the long time
scale. The meeting probability is greater for bosons and smaller for fermions compared to the case of distinguishable particles. This
behavior can be understood by examining the asymptotic properties of the expressions (\ref{bfmp}). Numerical evidence presented in Figure~\ref{fig:69} indicates that the meeting probability for bosons has the asymptotic behavior of the form $\ln(t)/t$. However, for fermions the decay of the meeting probability
is faster having the form
$$
M_F(t,d)\sim\frac{1}{t} .
$$
The fermion exclusion principle simply works against an enhancement of the meeting probability.

For the overall meeting probability we can use the same arguments as in the previous section and conclude that it will converge to
one for both bosons and fermions.

\section{Conclusions}
\label{chap:6g}

We have defined and analyzed the meeting problem in the quantum walk on an infinite line with two quantum particles. For
distinguishable particles we have derived analytical formulas for the meeting probability. The asymptotic behavior following from
these results shows that the meeting probability decays faster but not quadratically faster than in the classical random walk. This
results in the slower convergency of the overall meeting probability, however it still converges to one. This is due to the
fact that the meeting probability does not decay faster than $\frac{1}{t}$. Such a situation might occur in higher dimensional
walks and could result in yet another difference between the classical and the quantum walks. We have studied the influence of
the entanglement and the indistinguishability of the particles on the meeting probability. The influence is particularly visible for
fermions and in the case of distinguishable particles for the case of initial entangled singlet state. Although the meeting
probability decays faster in these cases the overall meeting probability will still converge to one, as the decay is never
faster than the threshold $\frac{1}{t}$.


\nchapter{Conclusions}

Quantum walks are a specialized field on the border between quantum information theory and statistical physics which attracted a lot of interest in recent years. A number of novel effects have been found and are still under investigation. In the present thesis we contributed to these investigations.

In particular, we extended the concept of recurrence and P\'olya number to quantum walks. The particular measurement scheme employed in our definition preserves the effect of the additional degrees of freedom offered by quantum mechanics on the P\'olya number. We developed the tools needed for the analysis of the recurrence nature of quantum walks. The actual analysis revealed that quantum walks can be operated in physically different regimes. These regimes cover localization as well as ballistic spreading of the walker's wave packets. We found that the free parameters we have at hand in a coined quantum walk have a crucial impact on its dynamics and are capable of changing its behaviour from recurrent to transient. Striking diversity of quantum walks in contrast to classical random walks was pointed out. The present results prove the usefulness of the P\'olya number concept for quantum walks and support our expectation of its applicability in related domains.

The recurrence of quantum walks under the effect of bias was analyzed. For classical random walks breaking the symmetry results in immediate turnover from recurrence to transience. However, the ballistic nature of the quantum walk is able to compensate for the bias and the recurrence can be preserved. We identified the range of parameters for which the recurrence behaviour of biased quantum walks on a line diverse from classical random walks.

Finally, we considered quantum walks with two particles. This makes the additional properties offered by quantum mechanics like entanglement or indistinguishability accessible. We analyzed the effect of these non-classical features on the meeting probability and pointed out the difference from the classical random walk.

The presented results provide a step in the classification of coined quantum walks, in particular on higher-dimensional lattices. We have identified several extreme modes of the dynamics of quantum walks. Our next goal is to exploit the free parameters of the coin operator which will allow us to shed light on the connection between these different regimes.

Within our definition the recurrence of a quantum walk describes the revival of a particular quantity, namely the probability at the origin, rather than the revival of a quantum state. Nevertheless and quite surprisingly, even full revivals are possible in quantum walk settings. This effect is closely related to localization. Indeed, for localizing quantum walks the propagator has a non-empty point spectrum which allows for stationary and oscillating states. However, the point spectra of the presently known localizing quantum walks are rather simple leading only to oscillations with a period of two steps. Finding quantum walks with a broader point spectrum will lead to novel features including full and fractional revival dynamics.

Our definition of the P\'olya number of a quantum walk is connected to a specific measurement scheme. Needless to say, we can consider schemes where the measurements are performed in a different manner and define the P\'olya number accordingly. It is interesting to analyze the influence of various measurement schemes on the recurrence nature of the quantum walk. Preliminary results indicate that our definition gives an upper limit for the P\'olya number.

The meeting problem for two quantum walkers which we have studied presents a step towards quantum walks involving many particles. It is certainly worth to analyze other various quantities available in multi-particle settings, e.g. the angular correlations among the particle's positions. Moreover, up to date the particles performing quantum walk were considered non-interacting. To investigate the effect of interactions between the particles on the dynamics of quantum walks is one of our next goals.


\begin{appendices}

\chapter{Recurrence of Random Walks}
\label{app:a}

In this appendix we review the main results on the recurrence in classical random walks. First, we show how the recurrence is related to the probability at the origin. Then we discuss the recurrence of unbiased random walks on $d$-dimensional lattices. Finally, we analyze the recurrence of biased random walks on a line. For a more comprehensive reviews we refer to the literature \cite{hughes,revesz}.

We begin with the problem analyzed by P\'olya in 1921 \cite{polya}. Consider a particle performing a random walk on an infinite $d$-dimensional lattice. The particle is initially localized at the origin of the lattice. The probability $P$ that the particle returns to the origin during the time evolution is
called the P\'olya number of the walk. Random walks are classified as {\it recurrent} or {\it transient} depending on whether their P\'olya number equals to one, or is less than one, respectively. If the random walk is recurrent the particle returns to the origin with certainty. On the hand, for transient random walks there is a non-zero probability that the particle never returns to its starting point. In other words, there is a non-vanishing probability of escape.

The P\'olya number of a classical random walk can be defined in the following way \cite{revesz}. Let $q_0(t)$ be the probability that the particle returns to the origin for the {\it first time} after $t$ steps. Since these events are mutually exclusive we can add up their probabilities and the series
\begin{equation}
P\equiv\sum\limits_{t=1}^\infty q_0(t)
\label{polya:1}
\end{equation}
gives the probability that at least once the particle has returned to the origin, i.e. the P\'olya number. However, the definition (\ref{polya:1}) is not very practical for determining the recurrence nature of a random walk. We can express the P\'olya number in terms of the probability $p_0(t)$ that the particle can be found at the origin at any given time instant $t$. Indeed, it is easy to see that the probability at the origin $p_0(t)$ and the first return probability $q_0(t)$ fulfills the following relations
\begin{eqnarray}
\nonumber p_0(0) & = & 1\\
\nonumber p_0(1) & = & q_0(1)\\
\nonumber p_0(2) & = & q_0(2)+q_0(1)p_0(1)\\
\nonumber p_0(3) & = & q_0(3)+q_0(2)p_0(1)+q_0(1)p_0(2)\\
\nonumber & \vdots & \\
\nonumber p_0(n) & = & q_0(n)+q_0(n-1)p_0(1)+\ldots+q_0(1)p_0(n-1).
\end{eqnarray}
Simply adding all of these equations together might lead to a divergent series. Therefore, we first multiply the $n$-th equation by $z^n$ with $|z|<1$. Adding these modified equations we find the relation
\begin{equation}
F(z) =  1 + F(z)G(z),
\label{ap1:eq1}
\end{equation}
where we have defined the following functions
\begin{eqnarray}
\nonumber F(z) & = & \sum\limits_{n=0}^\infty p_0(n)z^n\\
\nonumber G(z) & = & \sum\limits_{n=1}^\infty q_0(n)z^n.
\end{eqnarray}
Both series are convergent for $|z|<1$. Moreover, the P\'olya number $P$ can be evaluated from the function $G(z)$ by taking the limit $z\rightarrow 1^-$
$$
P = \lim\limits_{z\rightarrow 1^-} G(z) = \sum\limits_{n=1}^\infty q_0(n).
$$
From the relation (\ref{ap1:eq1}) we express the function $G(z)$ in the form
$$
G(z) = 1-\frac{1}{F(z)}.
$$
Finally, we take the limit $z\rightarrow 1^-$ and find the formula
$$
P = 1-\frac{1}{\sum\limits_{t=0}^{+\infty}p_0(t)},
$$
which expresses the P\'olya number $P$ in terms of the probability at the origin $p_0(t)$.

The recurrence behaviour of a random walk is determined solely by the infinite sum
\begin{equation}
{\cal S} \equiv \sum_{t=0}^{\infty}p_0(t).
\label{ap1:eq2}
\end{equation}
We find that $P$ equals unity if and only if the series ${\cal S}$ diverges \cite{revesz}. In such a case the walk is recurrent. On the other hand, if the series $\cal S$ is convergent, the P\'olya number $P$ is strictly less than unity and the walk is transient. The convergence of the series $\cal S$ is determined by the asymptotic behaviour of the probability at the origin $p_0(t)$. Indeed, we find that if $p_0(t)$ decays faster than $t^{-1}$ the sum is finite, while if the decay of $p_0(t)$ is slower the sum is divergent. Hence we find the following criterion for recurrence of random walks --- the random walk is recurrent if and only if the probability at the origin decays like $t^{-1}$ or slower as $t$ approaches infinity.

In the following we use the above mentioned criterion to analyze the recurrence of biased and unbiased random walks.


\section{Unbiased random walks on $\mathds{Z}^d$}
\label{app:a1}

Let us begin with the unbiased random walk on a line. At each time step the particle has two possibilities --- it can move to the right or to the left by a unit distance with equal probability $1/2$. The probability distribution generated by such a random walk is easily found to be
$$
P(m,t) = \frac{1}{2^t} {t\choose \frac{t+m}{2}}.
$$
The probability at the origin is thus given by ( for even number of steps $2t$ )
$$
p_0(t) = \frac{1}{4^t} {2t\choose t}.
$$
Using the Stirling's formula
\begin{equation}
n! \approx \sqrt{2\pi n}\left(\frac{n}{e}\right)^n
\label{ap1:eq3}
\end{equation}
we find that the asymptotical behaviour of the probability at the origin is determined by
$$
p_0(t) \approx \frac{1}{\sqrt{\pi t}}.
$$
Hence, the series $\cal S$ defined in (\ref{ap1:eq2}) is divergent. Consequently, we find that the unbiased random walk on a line is recurrent.

Recurrence of unbiased random walks on higher-dimensional lattices can be analyzed in a similar way \cite{revesz}. One finds that the asymptotics of the probability at the origin is determined by the dimension of the lattice $d$ in the following form
$$
p_0(t) \sim t^{-\frac{d}{2}}.
$$
It follows that the series $\cal S$ determining the recurrence of a random walk is divergent only for the dimensions $d=1,2$ and convergent for $d\geq 3$. We conclude that the random walks on a line and in the plane are recurrent while higher-dimensional random walks are transient, the result originally found by P\'olya in 1921 \cite{polya}.

Concerning the value of the P\'olya number for the transient case Montroll \cite{montroll:1956} showed that for the dimensions $d>2$ the following relation holds
$$
P(d) = 1-\frac{1}{u(d)},
$$
where $u(d)$ can be expressed in terms of an integral of the modified Bessel function of the first kind \cite{abramowitzstegun}
$$
u(d) = \int\limits_0^\infty \left[I_0\left(\frac{t}{d}\right)\right]^d e^{-t} dt.
$$
However, the closed form of the function $u(d)$ is known only for $d=3$ due to the Watson's triple integral \cite{watson:1939} with the result
$$
u(3) = \frac{\sqrt{6}}{32\pi^3}\Gamma\left(\frac{1}{24}\right)\Gamma\left(\frac{5}{24}\right)\Gamma\left(\frac{7}{24}\right)\Gamma\left(\frac{11}{24}\right)\approx 1.516.
$$
For higher dimensions $d>3$ one has to evaluate the integral numerically. We present an overview of the numerical values of the P\'olya number \cite{montroll:1956} for a different dimensions $d$ in Table~\ref{app:a:tab}.

\begin{table}[h]
\begin{center}
\begin{tabular}{|c|c|}
  \hline
  \multirow{2}{*}{Dimension} & \multirow{2}{*}{P\'olya number}\\
  & \\ \hline
  \multirow{2}{*}{3} & \multirow{2}{*}{0.340537} \\
  & \\ \hline
  \multirow{2}{*}{4} & \multirow{2}{*}{0.193206} \\
  & \\ \hline
  \multirow{2}{*}{5} & \multirow{2}{*}{0.135178} \\
  & \\ \hline
  \multirow{2}{*}{6} & \multirow{2}{*}{0.104715} \\
  & \\
  \hline
\end{tabular}
\caption{P\'olya number of a random walk on $\mathds{Z}^d$ in dependence of the dimension $d$.}
\label{app:a:tab}
\end{center}
\end{table}


\section{Biased random walks on a line}
\label{app:a2}

\begin{figure}[h]
\begin{center}
\includegraphics[width=0.6\textwidth]{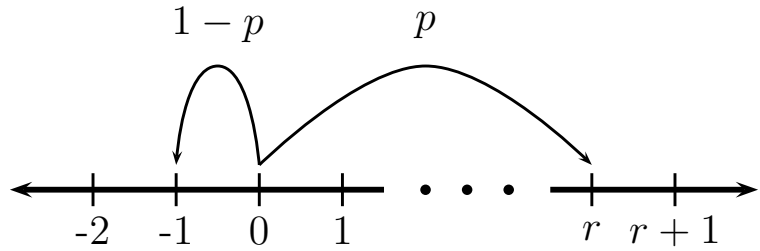}
\caption{Schematics of the biased random walk on a line. The particle can move to the right by a distance $r$ with the probability $p$.
The length of the step to the left is unity and the probability of this step is $1-p$.}
\label{fig5}
\end{center}
\end{figure}

Let us consider biased random walks on a line. The bias can be introduced in two ways --- the step in one direction is greater than in the other one and the probability of the step to the right is different from the probability of the step to the left (see \fig{fig5}).

Consider a random walk on a line such that the particle can make a
jump of length $r$ to the right with probability $p$ or make a unit
size step to the left with probability $1-p$. As we have discussed a random walk is
recurrent if and only if the probability to find the particle at the
origin at any time instant $t$ does not decays faster than $t^{-1}$.
This probability is easily found to be
$$
p_0(t) = (1-p)^{\frac{t r}{r+1}}p^{\frac{t}{r+1}}{t\choose \frac{t r}{r+1}}.
$$
With the help of the Stirling's formula (\ref{ap1:eq3}) we find the asymptotical behaviour of the probability at the origin
$$
p_0(t)\approx \frac{r+1}{\sqrt{2\pi r t}}\left[(1-p)^{\frac{r}{r+1}}p^{\frac{1}{r+1}}\frac{r+1}{r^\frac{r}{r+1}}\right]^t.
$$
The asymptotics of the probability $p_0(t)$ therefore depends on
the value of
$$
q = (1-p)^{\frac{r}{r+1}}p^{\frac{1}{r+1}}\frac{r+1}{r^\frac{r}{r+1}}.
$$
Since $q\leq 1$ the probability $p_0(t)$ decays exponentially
unless the inequality is saturated. Hence, the random walk is
recurrent if and only if $q$ equals unity. This condition is
satisfied for
\begin{equation}
\label{rw:cond}
p = \frac{1}{r+1},
\end{equation}
i.e. the probability of the step to the right has to be inversely
proportional to the length of the step.

\begin{figure}[h]
\begin{center}
\includegraphics[width=0.6\textwidth]{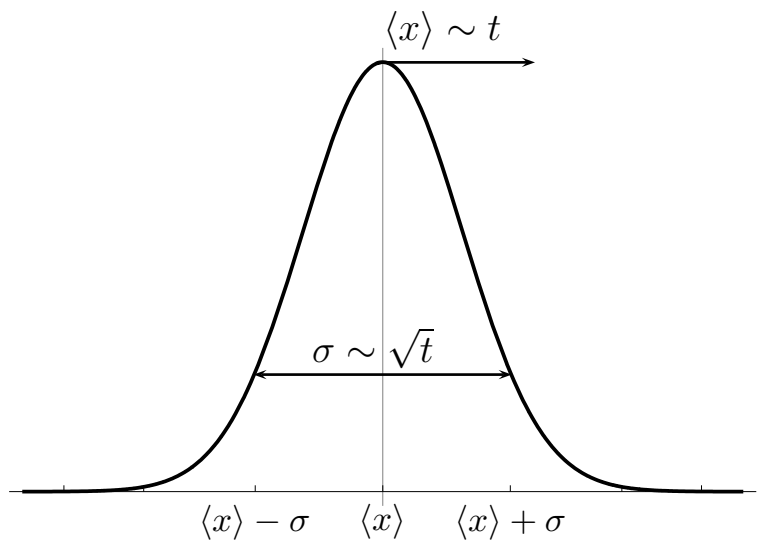}
\caption{Spreading of the probability distribution versus the motion
of the mean value of a biased classical random walk on a line. While the
spreading is diffusive ($\sigma\sim\sqrt{t}$) the mean value
propagates with a constant velocity ($\langle x\rangle\sim t$). The
probability inside the $\sigma$ neighborhood of the mean value
$\langle x\rangle$ behaves like $t^{-\frac{1}{2}}$. On the other
hand, outside the $\sigma$ neighborhood the decay is exponential.
Hence, if the mean value $\langle x\rangle$ does not vanish the
origin of the walk leaves the $\sigma$ neighborhood of the mean
value. In such a case the probability at the origin decays
exponentially and the walk is transient.}
\label{fig6}
\end{center}
\end{figure}

This result can be well understood from a different point of view, as we illustrate in \fig{fig6}.
The spreading of the probability distribution is diffusive, i.e.
$\sigma\sim\sqrt{t}$. The probability in the $\sigma$ neighborhood
of the mean value $\langle x\rangle$ behaves like $t^{-\frac{1}{2}}$
while outside this neighborhood the probability decays
exponentially. Therefore for the random walk to be recurrent the
origin must lie in this $\sigma$ neighborhood for all times $t$.
However, if the random walk is biased the mean value of the position
$\langle x\rangle$ varies linearly in time, thus it is a faster
process than the spreading of the probability distribution. In such
a case the origin would lie outside the $\sigma$ neighborhood of the mean
value after a finite number of steps leading to the exponential
asymptotic decay of the probability at the origin $p_0(t)$. Hence,
the random walk is recurrent if and only if the mean value of the
position equals zero. Since the individual steps are independent of
each other the mean value after $t$ steps is simply a $t$ multiple
of the mean value after single step, i.e.
$$
\langle x (t)\rangle = t \langle x(1)\rangle = t\left[p(r+1)-1\right].
$$
We find that the mean value equals zero if and only if the condition (\ref{rw:cond})
holds.

\chapter{Recurrence Criterion for Quantum Walks}
\label{app:b}

Let us prove that the recurrence criterion for quantum walks is the same as for random walks, i.e. the P\'olya number equals one if and only if the series
$$
{\cal S} \equiv \sum_{t=0}^{\infty}p_0(t)
$$
diverges.

According to the definition of the P\'olya number Eq. (\ref{polya:def}) for quantum walks we have to prove the equivalence
$$
\overline{P}\equiv\prod\limits_{t=1}^{+\infty}\left(1-p_0(t)\right) = 0 \Longleftrightarrow {\cal S}=+\infty.
$$
We note that the convergence of both the sum ${\cal S}$ and the product $\overline{P}$ is unaffected if we omit a finite number of terms.

Let us first consider the case when the sequence $p_0(t)$ converges to a non-zero value $0<a\leq 1$. Obviously, in such a case the series ${\cal S}$ is divergent. Since $p_0(t)$ converges to $a$ we can find for any $\varepsilon>0$ some $t_0$ such that for all $t>t_0$ the inequalities
$$
1-a-\varepsilon\leq 1-p_0(t)\leq 1-a+\varepsilon.
$$
hold. Hence, we can bound the infinite product
\begin{equation}
\lim\limits_{t\rightarrow +\infty}\left(1-a-\varepsilon\right)^t\leq\overline{P}\leq\lim\limits_{t\rightarrow +\infty} \left(1-a+\varepsilon\right)^t.
\label{app2:eq1}
\end{equation}
Since we can choose $\varepsilon$ such that
$$
\left|1-a\pm\varepsilon\right|<1,
$$
we find that limits both on the left-hand side and the right-hand side of Eq. (\ref{app2:eq1}) equals zero. Hence, the product $\overline{P}$ vanishes.

We turn to the case when $p_0(t)$ converges to zero. We denote the partial product
$$
\overline{P}_n = \prod\limits_{t=1}^n(1-p_0(t)).
$$
Since $1-p_0(t)>0$ for all $t\geq 1$ we can consider the logarithm
\begin{equation}
\ln{\overline{P}_n} = \sum\limits_{t=1}^n\ln\left(1-p_0(t)\right)
\label{app3}
\end{equation}
and rewrite the infinite product as a limit
\begin{equation}
\overline{P} = \lim\limits_{n\rightarrow +\infty}e^{\ln{\overline{P}_n}}
\label{app2}.
\end{equation}
Since $p_0(t)$ converges to zero we can find some $t_0$ such that for all $t>t_0$ the value of $p_0(t)$ is less or equal than $1/2$. With the help of the inequality
$$
-2x\leq\ln\left(1-x\right)\leq -x
$$
valid for $x\in\left[0,1/2\right]$ we find the following bounds
$$
-2\sum\limits_{t=1}^n p_0(t)\leq\ln\overline{P}_n\leq -\sum_{t=1}^n p_0(t).
$$
Hence, if the series ${\cal S}$ is divergent the limit of the sequence ${\left(\ln{\overline{P}_n}\right)}^\infty_{n=1}$ is $-\infty$ and according to Eq. (\ref{app2}) the product $\overline{P}$ vanishes. If, on the other hand, the series ${\cal S}$ converges the sequence ${\left(\ln{\overline{P}_n}\right)}^\infty_{n=1}$ is bounded. According to Eq. (\ref{app3}) the partial sums of the series $\sum\limits_{t=1}^{+\infty}\ln\left(1-p_0(t)\right)$ are bounded and since it is a series with strictly negative terms it converges to some negative value $b<0$. Consequently, the sequence ${\left(\ln{\overline{P}_n}\right)}^\infty_{n=1}$ converges to $b$ and according to Eq. (\ref{app2}) the product equals
$$
\overline{P} = e^b>0.
$$
This completes our proof.


\chapter{Method of Stationary Phase}
\label{app:c}

In order to determine the recurrence nature of a quantum walk one has to analyze the asymptotic behaviour of the probability at the origin. As we have shown in Section~\ref{chap:2d} the probability amplitude of the particle being at the origin of the quantum walk after $t$ steps is given by a sum of integrals of the form
\begin{equation}
I(t) = \int\limits_V e^{i\ \omega(\mathbf{k}) t} f(\mathbf{k})d\mathbf{k}.
\label{app:c:eq1}
\end{equation}
The recurrence of a quantum walk is determined by the asymptotics of such integrals. The method of stationary phase is a suitable tool for such analysis.

In the following we briefly review the main concepts of the method of stationary phase. First, we treat the one-dimensional integrals. Then we turn to the multivariate integrals. We find that the crucial contribution to the integral (\ref{app:c:eq1}) as $t$ approaches infinity arises from the {\it stationary points}, i.e. the points where the derivative of the phase $\omega(\mathbf{k})$ vanishes. We discuss how the amount of stationary points and the "flatness" of the phase at the stationary point influences the asymptotic behaviour of the integral $I(t)$. For a more comprehensive analysis we refer to the literature \cite{statphase2,statphase}.


\section{One-dimensional integrals}
\label{app:c1}

Let us begin with the one-dimensional integral of the form
\begin{equation}
I(t) = \int\limits_a^b e^{i\ \omega(k) t} f(k) dk,
\label{app:c:eq2}
\end{equation}
where $f$ and $\omega$ are smooth functions and $\omega$ is real-valued. We see that in the region of $k$ where $\omega(k)$ changes considerably the exponential $e^{i\ \omega(k) t}$ oscillates rapidly as $t$ approaches infinity. Assuming that the function $f$ is slowly varying compared to these rapid oscillations we find that this region of integration does not contribute significantly to the integral $I(t)$. Obviously, the most important contributions to the integral (\ref{app:c:eq2}) arise from the regions where the oscillations of the exponential are least rapid, which occur precisely at the stationary points $k_0$ of the phase $\omega$
$$
\omega'(k_0) = \left.\frac{d\omega}{dk}\right|_{k_0} = 0.
$$
The "flatness" of the phase at the stationary point determines the order of this contribution --- the more derivatives of the phase vanishes at the stationary point the slower the contribution decays as $t$ approaches infinity. Here we assume that the function $f$ is non-zero at the stationary point, otherwise the contribution to the integral $I(t)$ vanishes.

\subsection{No stationary points}

Let us first consider the case when the phase $\omega$ has no stationary points inside the integration domain. Then there exists $\varepsilon>0$ such that
$$
\left|\omega'(k)\right|>\varepsilon
$$
for all $k$. Performing the integration in (\ref{app:c:eq2}) per parts with
\begin{eqnarray}
\nonumber u(k) & = & \frac{f(k)}{i t \omega'{k}}, \quad v'(k) = i t \omega'(k) e^{i\ \omega(k) t},\\
\nonumber u'(k) & = & \frac{1}{i t} \frac{f'(k)\omega'(k)-f(k)\omega''(k)}{\omega'(k)^2}, \quad v(k) = e^{i\ \omega(k) t},
\end{eqnarray}
we find that $I(t)$ can be expressed in the form
\begin{equation}
I(t) =  \frac{1}{i t}\left[\frac{f(k)}{\omega'(k)}e^{i\ \omega(k)t}\right]_a^b-\frac{1}{it}\int\limits_a^b e^{i\ \omega(k)t}\frac{f'(k)\omega'(k)-f(k)\omega''(k)}{\omega'(k)^2}dk.
\label{app:c:eq3}
\end{equation}
We see that $I(t)$ decays at least like $t^{-1}$ as $t$ approaches infinity. Moreover, the second term in (\ref{app:c:eq3}) has the same form as the original integral (\ref{app:c:eq2}). Hence, if in addition the first term in (\ref{app:c:eq3}) vanishes, e.g. if the function $f$ equals zero at the boundaries of the integration domain, we find by repeated integration per parts that $I(t)$ decays faster than any inverse polynomial in $t$.

\subsection{First-order stationary points}

We turn to the case of $\omega$ having a single stationary point coinciding with the left endpoint of the interval $k_0=a$ ( any integral where the phase has more than one stationary point can be decomposed into a sum of such integrals ) and assume that the stationary point is of the first order, i.e. $\omega'(a)=0$ but $\omega''(a)\neq 0$. We then expand the phase into a Taylor series
$$
\omega(k) \simeq \omega(a) + \frac{\omega''(a)}{2}(k-a)^2
$$
around the stationary point $k_0=a$. Since we assume that the function $f$ is slowly varying we put it equal to its value at the stationary point $f(k)\approx f(a)$. With these estimations we find
\begin{equation}
I(t) \simeq f(a)e^{i\ \omega(a)t}\int\limits_a^b e^{i\frac{\omega''(a)}{2}(k-a)^2 t} dk.
\label{app:c:eq4}
\end{equation}
Let us estimate the remaining integral. We substitute for $y=k-a$ and extend the integration domain to $[0,+\infty)$
$$
\int\limits_a^b e^{i\frac{\omega''(a)}{2}(k-a)^2 t} dk = \int\limits_{0}^{b-a} e^{i\frac{\omega''(a)}{2} y^2 t} dy \approx \int\limits_{0}^{+\infty} e^{i\frac{\omega''(a)}{2} y^2 t} dy.
$$
This is the familiar Fresnel integral
\begin{equation}
F(\gamma) = \int\limits_{0}^{+\infty} e^{i\gamma y^2} dy = \frac{\Gamma\left(\frac{1}{2}\right)}{2}|\gamma|^{-\frac{1}{2}} e^{i {\rm sign}\gamma\frac{\pi}{4}}.
\label{app:c:fresnel}
\end{equation}
Inserting this result into the expression (\ref{app:c:eq4}) we finally arrive at the formula
\begin{equation}
I(t) \simeq \left[f(a)\frac{\Gamma\left(\frac{1}{2}\right)}{2}e^{i\ \omega(a)t\pm i\frac{\pi}{4}}\left(\frac{2}{|\omega''(a)|}\right)^\frac{1}{2}\right] t^{-\frac{1}{2}}
\label{app:c:eq5}
\end{equation}
describing the behaviour of the integral $I(t)$ for large values of $t$. The plus (minus) sign in the exponential in (\ref{app:c:eq5}) corresponds to the second derivative of the phase at the stationary point $\omega''(a)$ being positive (negative). To conclude, we find that if the phase $\omega(k)$ has a stationary point of the first order the integral $I(t)$ decays like $t^{-1/2}$ as $t$ approaches infinity.

\subsection{Higher-order stationary points}

We close this section by the analysis of the integral $I(t)$ when the phase $\omega$ has a stationary point $k_0=a$ of the order of $p-1$, i.e.
$$
\omega'(a) = \omega''(a) = \ldots = \omega^{(p-1)}(a) = 0,\quad \omega^{(p)}(a)\neq 0.
$$
In such a case the Taylor expansion of the phase reads
$$
\omega(k) \simeq \omega(a) + \frac{\omega^{(p)}(a)}{p!}(k-a)^p.
$$
Performing similar approximations as above we find
\begin{equation}
I(t) \simeq f(a)e^{i\ \omega(a)t}\int\limits_a^b e^{i\frac{\omega^{(p)}(a)}{p!}(k-a)^p t} dk.
\label{app:c:eq6}
\end{equation}
In the remaining integral we substitute for $y = k-a$ and extend the upper limit of the integration to $+\infty$
$$
\int\limits_a^b e^{i\frac{\omega^{(p)}(a)}{p!}(k-a)^p t} dk = \int\limits_{0}^{b-a} e^{i\frac{\omega^{(p)}(a)}{p!}y^p t} dy \approx \int\limits_{0}^{+\infty} e^{i\frac{\omega^{(p)}(a)}{p!}y^p t} dy.
$$
We find a generalization of the Fresnel integral (\ref{app:c:fresnel}) which is readily evaluated
\begin{equation}
F_p(\gamma) = \int\limits_{0}^{+\infty} e^{i\gamma y^p} dy = \frac{\Gamma\left(\frac{1}{p}\right)}{p}|\gamma|^{-\frac{1}{p}} e^{i {\rm sign}\gamma\frac{\pi}{2p}}.
\label{app:c:fresnel:p}
\end{equation}
Finally, inserting this result into the Eq. (\ref{app:c:eq6}) we arrive at the estimation
\begin{equation}
I(t) \simeq \left[f(a)\frac{\Gamma\left(\frac{1}{p}\right)}{p}e^{i\ \omega(a)t\pm i\frac{\pi}{2p}}\left(\frac{p!}{|\omega^{(p)}(a)|}\right)^\frac{1}{p}\right] t^{-\frac{1}{p}},
\label{app:c:eq7}
\end{equation}
where the plus (minus) sign corresponds to positive (negative) value of $\omega^{(p)}(a)$. From (\ref{app:c:eq7}) we find that the contribution of the stationary point of the order $p-1$ to the integral $I(t)$ behaves like $t^{-1/p}$ as $t$ approaches infinity. The flatness of the phase at the stationary point reduces the rate at which the integral $I(t)$ decays.


\section{Multivariate integrals}
\label{app:c2}

We turn to the asymptotic analysis of the multidimensional integrals of the form
\begin{equation}
I(t) = \int\limits_V e^{i\ \omega(\mathbf{k}) t} f(\mathbf{k})d\mathbf{k}.
\label{app:c2:eq1}
\end{equation}
We assume that both functions $\omega(\mathbf{k})$ and $f(\mathbf{k})$ are smooth and $\omega$ is real-valued. Similarly to the one-dimensional case, the main contribution to the integral arise from the stationary points of the phase $\omega$, i.e. points $\mathbf{k}_0$ where the gradient of $\omega$ vanishes
$$
\left.\nabla \omega(\mathbf{k})\right|_{\mathbf{k}=\mathbf{k}_0} = \mathbf{0}.
$$
As in the previous Section we approximate the phase around the stationary point by the Taylor expansion. In addition, we have to change the coordinates in such a way that the resulting integral factorizes into a product of one-dimensional integrals. Each of the 1-D integrals can be estimated by means provided in the previous Section.

In the following we review the main results of the asymptotics of (\ref{app:c2:eq1}) in dependence on the properties of the phase $\omega(\mathbf{k})$. For a more detailed analysis we refer to the literature \cite{statphase2,statphase}.

\subsection{No stationary points}
\label{app:c2a}

Let us begin with the case when the gradient of $\omega$ is non-vanishing inside the integration domain $V$. From the divergence theorem we find
\begin{equation}
I(t) = -\frac{i}{t}\int\limits_{\partial V} (\mathbf{u}\cdot\mathbf{n})e^{i\ \omega t}ds + \frac{i}{t}\int\limits_V (\nabla\cdot\mathbf{u})e^{i\ \omega t}d\mathbf{k},
\label{app:c2a:eq1}
\end{equation}
where $\partial V$ is the boundary of $V$, $\mathbf{n}$ is the unit vector normal to the boundary and the vector function $\mathbf{u}(\mathbf{k})$ is given by
$$
\mathbf{u}(\mathbf{k}) = \frac{\nabla\omega(\mathbf{k})}{|\nabla\omega(\mathbf{k})|^2} f(\mathbf{k}).
$$
The expression (\ref{app:c2a:eq1}) indicates that $I(t)$ decays at least like $t^{-1}$. Suppose that the function $f(\mathbf{k})$ vanishes smoothly on the boundary of $V$. In such a case the contour integral in (\ref{app:c2a:eq1}) equals zero. The remaining volume integral in (\ref{app:c2a:eq1}) is of the same kind as the original integral $I(t)$. Hence, by repeating the same procedure as above we find that the integral $I(t)$ decays faster than any inverse polynomial in $t$.

\subsection{Non-degenerate stationary points}
\label{app:c2b}

We turn to the case when the phase $\omega(\mathbf{k})$ has a single stationary point $\mathbf{k}_0$ inside the integration domain. We assume that $\mathbf{k}_0$ is non-degenerate, i.e. the Hessian matrix evaluated at the stationary point
\begin{equation}
H_{ij}(\mathbf{k_0}) = \left.\left(\frac{\partial^2\omega}{\partial k_i\partial k_j}\right)\right|_{\mathbf{k}=\mathbf{k}_0}
\label{app:c2b:hessian}
\end{equation}
is regular. We expand the phase around the stationary point into the second order
$$
\omega(\mathbf{k}) \simeq \omega(\mathbf{k}_0) + \frac{1}{2}\sum_{i,j} \left(k_i - {k_0}_i\right) H_{i,j}(\mathbf{k}_0) \left(k_j - {k_0}_j\right).
$$
Assuming that $f(\mathbf{k})$ is slowly varying we can evaluate it at the stationary point and extract it from the integral. Substituting for
$$
\bm{\kappa} = \mathbf{k}-\mathbf{k}_0
$$
and extending the integration from the finite volume $V$ to $\mathds{R}^n$ we arrive at the following estimation of the integral (\ref{app:c2:eq1})
\begin{equation}
I(t) \simeq f(\mathbf{k}_0) e^{i\ \omega(\mathbf{k}_0) t} \int\limits_{\mathds{R}^n} \exp\left(\frac{i}{2}\sum_{i,j} \kappa_i H_{ij}(\mathbf{k}_0) \kappa_j t \right) d\bm{\kappa}.
\label{app:c2b:eq1}
\end{equation}
The integral in (\ref{app:c2b:eq1}) can be reduced into the product of $n$ one-dimensional Fresnel integrals (\ref{app:c:fresnel}). Indeed, the Hessian matrix (\ref{app:c2b:hessian}) is real and symmetric since we assumed $\omega(\mathbf{k})$ to be smooth. Hence, it can be diagonalized with the help of the orthogonal matrix $O$. In the new coordinate system
\begin{equation}
\mu_i = \sum_j O_{ij}\kappa_j
\label{app:c2b:coor}
\end{equation}
the bilinear form in (\ref{app:c2b:eq1}) is given by the sum of purely quadratic terms
$$
\sum_{i,j} \kappa_i H_{ij}(\mathbf{k}_0) \kappa_j = \sum_i \lambda_i(\mathbf{k}_0) \mu_i^2,
$$
where $\lambda_i(\mathbf{k}_0)$ are eigenvalues of the Hessian matrix (\ref{app:c2b:hessian}) at the stationary point $\mathbf{k}_0$. Since the matrix $O$ is orthogonal the change of coordinates (\ref{app:c2b:coor}) has a unit Jacobian. Hence, using the substitution (\ref{app:c2b:coor}) we decompose the integral in (\ref{app:c2b:eq1}) into the product of one-dimensional Fresnel integrals
$$
\int\limits_{\mathds{R}^n} \exp\left(\frac{i}{2}\sum_{i,j} \kappa_i H_{ij}(\mathbf{k}_0) \kappa_j t \right) d\bm{\kappa} = \prod\limits_{j=1}^n\int\limits_\mathds{R} \exp\left(\frac{i}{2}\lambda_j(\mathbf{k}_0) \mu_j^2 t\right) d\mu_j,
$$
which are readily evaluated with the help of (\ref{app:c:fresnel}). Finally, we arrive at the following approximation of the integral (\ref{app:c2:eq1})
\begin{equation}
I(t) \simeq \left[f(\mathbf{k}_0) e^{i\ \omega(\mathbf{k}_0) t + i \nu(\mathbf{k}_0)\frac{\pi}{4}} \sqrt{\frac{(2\pi)^n}{\left|\det H_{ij}(\mathbf{k}_0)\right|}}\right] t^{-\frac{n}{2}},
\label{app:c2b:eq2}
\end{equation}
where $\nu(\mathbf{k}_0)$ is the sum of the signs of the eigenvalues of the Hessian matrix
$$
\nu(\mathbf{k}_0) = \sum_j {\rm sign} \lambda_j(\mathbf{k}_0).
$$
We find that contribution from the non-degenerate stationary points to the $n$-dimensional integral (\ref{app:c2:eq1}) is of the order of $t^{-n/2}$.

\subsection{Continuum of stationary points}
\label{app:c2c}

We close this Appendix by briefly discussing the asymptotic scaling of the integral (\ref{app:c2:eq1}) when the phase $\omega(\mathbf{k})$ has a curve of stationary points $\gamma$, i.e.
$$
\forall\mathbf{k}\in\gamma\qquad \nabla\omega(\mathbf{k}) = \mathbf{0}.
$$
Without loss of generality we assume that $\omega(\mathbf{k})=0$ at the stationary curve $\gamma$ which is considered to be smooth and without any loops. Moreover, we restrict ourselves to two-dimensional integrals, i.e. $n = 2$. As shown in \cite{statphase}, Chapter VIII.9, the main contribution of the continuum of stationary points to the asymptotic expansion of the integral (\ref{app:c2:eq1}) is
$$
I(t) \simeq \left[\sqrt{2\pi}e^{i\frac{\pi}{4}}\int\limits_\gamma \frac{f\left(k_1(s),k_2(s)\right)}{\sqrt{\frac{\partial^2\omega}{\partial k_1^2}+\frac{\partial^2\omega}{\partial k_2^2}}}ds\right] t^{-\frac{1}{2}},
$$
where $s$ is the parametrization of the curve $\gamma$. We find that in comparison with the case of the isolated non-degenerate stationary point analyzed in Section~\ref{app:c2b} the continuum of stationary points has reduced the decay of the integral $I(t)$ by a factor of square-root.


\chapter{Meeting Problem}
\label{app:d}

In this Appendix we analyze the meeting problem in classical and quantum walk. We derive analytical formulas for the asymptotic behaviour of the meeting probability.

\section{Meeting problem in the classical random walk}
\label{app:d1}

Let us define the meeting problem on the classical level. We
assume two particles which in each step of the process can perform
randomly a step to the left or to the right on a one dimensional
lattice labelled by integers. Initial distance between the two
particles is $2d$, because for odd initial distance the two particles
never meet, due to the transitional invariance we can assume that
one particle starts from the origin and the other one in
the vertex $2d$. We assume complete randomness, i.e. the
probabilities for the step right or left are equal. We ask for the
probability that the two particles meet again after $t$ steps
either at a certain position $m$ or we might ask for the total
probability to meet (the sum of probabilities at all of the
possible positions). A simple analysis reveals that the probability
to meet at a certain position $m$ equals
$$
M_{cl}(t,m,d) = \frac{1}{2^{2t}} {t\choose \frac{t+m}{2}}
{t\choose \frac{t+m-2d}{2}}.
$$
The total probability that the two particles are reunited after
$t$ steps reads
$$
M_{cl}(t,d) = \sum\limits^t_{m=2d-t}\frac{1}{2^{2t}} {t\choose
\frac{t+m}{2}} {t\choose \frac{t+m-2d}{2}},
$$
which simplifies to
\begin{equation}
\label{a1}
M_{cl}(t,d) = \frac{1}{2^{2t}}{2t\choose m+d}.
\end{equation}
To obtain the asymptotic behavior of the meeting probability we
approximate the single particle probability distribution by a
gaussian
$$
P_{cl}(x,t,d) = \frac{1}{\sqrt{\pi t}}\exp\left(-\frac{(x-2d)^2}{2t}\right),
$$
which leads to the following estimate on the meeting probability
$$
M_{cl}(t,d) \approx \int\limits_{-\infty}^{+\infty}P_{cl}(x,t,0)P_{cl}(x,t,d)dx =
\frac{1}{\sqrt{\pi t}} \exp\left(-\frac{d^2}{t}\right).
$$
Finally, for a fixed initial distance $d$ we get the long-time approximation for $t>d^2$
$$
M_{cl}(t,d)\approx \frac{1}{\sqrt{\pi t}}\left(1-\frac{d^2}{t}\right).
$$

\section{Meeting problem in the quantum walk}
\label{app:d2}

Let us derive analytical formulas for the meeting probabilities in the quantum case. We consider the following initial states:

({\it i}) right for the first particle and left for
the second
$$
\psi_{RL}(0,2d,0) = 1 ,
$$

({\it ii}) symmetric initial conditions
$1/\sqrt{2}(|L\rangle+i|R\rangle)$ for both
$$
\psi(0,2d,0) = \frac{1}{2}\left( \begin{array}{c}
  1 \\
  i \\
  i \\
  -1
\end{array}\right) ,
$$

({\it iii}) left for the first particle and right for the second
$$
\psi_{LR}(0,2d,0) = 1.
$$

For $t\geq\sqrt{2}d$ we consider the slowly varying part of the single particle probability distribution derived in \cite{nayak} which has the form
$$
P^{(L,R)}_{slow}(x,t) = \frac{2}{\pi t\left(1\pm\frac{x}{t}\right)\sqrt{1-\frac{2x^2}{t^2}}},
$$
if the initial coin state was $|L\rangle$ or $|R\rangle$, while for the symmetric initial condition it reads
$$
P^{(S)}_{slow}(x,t) = \frac{1}{2}\left(P^{(L)}_{slow}(x,t)+P^{(R)}_{slow}(x,t)\right) = \frac{2}{\pi t\left(1-\frac{x^2}{t^2}\right)\sqrt{1-\frac{2x^2}{t^2}}}.
$$
We then estimate the sums in (\ref{mq}) defining the meeting probabilities by integrals
\begin{eqnarray}
\label{M1}
\nonumber M_{RL}(t,d) & \approx & \frac{2}{\pi^2 t^2}\int\limits_{2d-\frac{t}{\sqrt{2}}}^{\frac{t}{\sqrt{2}}}\frac{dx}{(1-\frac{x}{t})(1+\frac{x-2d}{t})\sqrt{1-2\frac{x^2}{t^2}}\sqrt{1-2\frac{(x-2d)^2}{t^2}}}\\\nonumber\\
\nonumber M_{S}(t,d) & \approx & \frac{2}{\pi^2 t^2}\int\limits_{2d-\frac{t}{\sqrt{2}}}^{\frac{t}{\sqrt{2}}}\frac{dx}{(1-\frac{x^2}{t^2})(1-\frac{(x-2d)^2}{t^2})\sqrt{1-2\frac{x^2}{t^2}}\sqrt{1-2\frac{(x-2d)^2}{t^2}}}\\\nonumber\\
M_{LR}(t,d) & \approx & \frac{2}{\pi^2
t^2}\int\limits_{2d-\frac{t}{\sqrt{2}}}^{\frac{t}{\sqrt{2}}}\frac{dx}{(1+\frac{x}{t})(1-\frac{x-2d}{t})\sqrt{1-2\frac{x^2}{t^2}}\sqrt{1-2\frac{(x-2d)^2}{t^2}}}
\end{eqnarray}
which can be evaluated in terms of elliptic integrals. Notice that
the integrals diverge for $d=0$, i.e. for the case when the two
particles start at the same point. For now we suppose that $d>0$. The formulas
(\ref{M1}) can be expressed in the form
\begin{eqnarray}
\label{Mq2}
\nonumber M_{RL}(t,d) & \approx & F_+\left\{2(t-d)(t-(4-2\sqrt{2})d)K(a)+\frac{}{}\right.\\
\nonumber & & \left.+\sqrt{2}\left((t-(4+2\sqrt{2})d)(t-(4-2\sqrt{2})d)\Pi(b_+|a)-t^2\Pi(c_+|a)\frac{}{}\right)\right\}\\
\nonumber \\
\nonumber M_{S}(t,d) & \approx & \frac{ \pi^2 F_+F_-}{4}\left\{
16d(t^2-d^2)(t+(4+2\sqrt{2})d)(t-(4-2\sqrt{2})d)K(a)+\frac{}{}\right.\\
\nonumber & & +\sqrt{2}(t+(4+2\sqrt{2})d)(t-(4+2\sqrt{2})d)(t+(4-2\sqrt{2})d)\times \\
\nonumber & & \times(t-(4-2\sqrt{2})d)\left((t+d)\Pi(b_+|a)+(t-d)\Pi(b_-|a)\frac{}{}\right)-\\
\nonumber & & -\sqrt{2}t^2\left((t+d)(t+(4+2\sqrt{2})d)(t+(4-2\sqrt{2})d)\Pi(c_+|a)+\frac{}{}\right.\\
\nonumber & & \left.\left.+(t-d)(t-(4+2\sqrt{2})d)(t-(4-2\sqrt{2})d)\Pi(c_-|a)\frac{}{}\right)\right\}\\
\nonumber \\
\nonumber M_{LR}(t,d) & \approx &
F_-\left\{2(t+d)(t+(4+2\sqrt{2})d)K(a)-\frac{}{}\right.\\
\nonumber & & \left.-\sqrt{2}\left((t+(4+2\sqrt{2})d)(t+(4-2\sqrt{2})d)\Pi(b_-|a)-t^2\Pi(c_-|a)\frac{}{}\right)\right\}.\\
\end{eqnarray}
Here $K(a)$ is the complete elliptic integral of the first kind and $\Pi(x|a),\Pi(y|a)$ are the complete elliptic integrals of the
third kind (see e.g. \cite{abramowitzstegun}, chapter 17). The coefficients $a,b_\pm,c_\pm$ and $F_\pm$ are given by
\begin{eqnarray}
\nonumber F_\pm & = & \frac{2t}{\pi^2 d(t\mp
d)(t(2+\sqrt{2})\mp 4d)(t(2-\sqrt{2})\mp 4d))} \\
\nonumber a & = & i\sqrt{\frac{t^2}{2d^2}-1} \\
\nonumber b_\pm & = & \frac{(1\pm \sqrt{2})(t-\sqrt{2}d)}{d(\sqrt{2}\mp 2)}\\
\nonumber c_\pm & = & \frac{(t(\sqrt{2}\mp
2)+4d)(t-\sqrt{2}d)}{\sqrt{2}d(t(\sqrt{2}\pm 2)-4d)}.
\end{eqnarray}

Let us analyze the asymptotic behavior of the meeting probability. We begin with the observation that the
coefficients at the highest power of $t$ with the elliptic
integrals of the third kind are the same but with the opposite
signs for $\Pi(b|a)$ and $\Pi(c|a)$. Moreover, $b_\pm$ and $c_\pm$
goes like $-t$ as $t$ approaches infinity, and thus all of the
$\Pi$ functions have the same asymptotic behavior. Due to the
opposite sign for $\Pi(b|a)$ and $\Pi(c|a)$ the leading order
terms cancel and the contribution from this part to the meeting
probability is of higher order of $1/t$ compared to the
contribution from the complete elliptic integral of the first kind
$K(a)$. The asymptotic of the function $K(a)$ is given by
$$
K(a)\approx
\frac{d\sqrt{2}\ln{\left(\frac{2\sqrt{2}t}{d}\right)}}{t}.
$$
Inserting this approximation into (\ref{Mq2}) we find that the leading order term of the meeting probability in all the three studied
situations is given by
$$
M_{D}(t,d) \sim \frac{\ln{\left(\frac{2\sqrt{2}t}{d}\right)}}{t}.
$$

\end{appendices}


\part{Factorization with Exponential Sums}
\label{part:2}

\nchapter{Introduction}
\label{chap8}

Factorization of integers is a famous NP problem \cite{Wegener,Mertens} and the difficulty to decompose a number into prime factors lies at the heart of several encryption schemes \cite{RSA,Menezes}. However, Peter Shor found \cite{shor} that a quantum computer is capable of finding factors of a given number efficiently. The fundamental advantage of the Shor's algorithm compared to the classical algorithms is the massive use of quantum parallelism and entanglement. On the other hand, the physical realizations of the Shor's algorithm are very challenging and are so far limited to a proof of principle experiment \cite{vandersypen}.

Recently, several schemes for integer factorization based on Gauss sums \cite{lang:1970,davenport:1980,schleich:2005:primes} were proposed
\cite{clauser:1996,harter:2001,harter:2001b,mack:2002,mack:proc,merkel:ijmpb:2006,merkel:FP,rangelov}. For a review see e.g. \cite{zubairy:science}. In contrast to the Shor's algorithm, factorization using Gauss sums consists of a feasible factor test based on a classical interference scheme. The proposals employ the so-called normalized truncated Gauss sum
$$
{\cal A}_N^{(M)}(\ell) = \frac{1}{M+1}\sum\limits_{m=0}^{M}\exp\left(2\pi i\, m^2 \frac{N}{\ell}\right).
$$
Here $N$ is the number to be factored, $\ell$ is a trial factor and $M$ is the truncation parameter. The capability of Gauss sums to factor numbers stem from the fact that the {\it signal} - the absolute value of the Gauss sum which is measured in the experiment, attains the maximal value only for a factor. For non-factors destructive interference yields a small signal. In the most elementary approach we have to perform this factor test for every number smaller than $\sqrt{N}$. As a consequence the method scales as $\sqrt{N}$ and is therefore exponential.On the other hand, the physical realizations of Gauss sums are less demanding than the implementations of the Shor's algorithm. Indeed, recent experiments based on NMR \cite{mehring:NMR:2006,Suter,suter2}, cold atoms \cite{rasel}, ultra-short pulses \cite{girard,girard2,girard3} and Bose-Einstein condensate \cite{sadgrove} have successfully demonstrated the possibility to find the prime factors of up to 17-digit numbers.

In the NMR settings \cite{mehring:NMR:2006,Suter,suter2} a sequence of RF pulses with linearly increasing relative phase shifts is applied to the ensemble of nuclear spins. After each pulse the echo, i.e. the polarization of the spins, is measured. Finally, all echoes are summed and for the proper choice of relative phase shifts of the RF pulses the resulting signal has the form of the Gauss sum.

The experiment with cold atoms presented in \cite{rasel} employs two long-living hyperfine ground states of rubidium. The atoms are launched by a system of magneto-optical traps and prepared in the atomic ground state by appropriate pulse sequence. After the preparation the atoms interact with a sequence of Raman pulses driving a transition between the hyperfine states. Similarly to the NMR experiments the individual pulses have to be properly phase shifted. Finally, after all pulses are applied a fluorescence detection measures the populations in both hyperfine states. The sum of these interference signals determines the Gauss sum.

In \cite{girard,girard2,girard3} the Gauss sum is implemented by a sequence of shaped femtosecond laser pulses. Individual laser pulses are properly phase shifted by a complex spectral mask. The interference produced by the pulse train is analyzed with a spectrometer. Due to the temporal Talbot effect the frequency component of the electric field is determined by a Gauss sum.

The experiment \cite{sadgrove} uses diffraction of the BEC on an optical lattice. One of the beams which creates the optical lattice is designed with a specific phase jumps. The pulse separates the atoms in the BEC into different momentum orders. In the absorption image a diffraction pattern determined by the Gauss sum is observed in which high-momentum atoms represent factors and low-momentum atoms represent non-factors.

As we have mentioned, the signal for a non-factor is suppressed and its value depends on the number of terms in the Gauss sum. In the experiment we have to take into account the limited resolution of the measured signal. Hence, to be able to distinguish factors from non-factors we have to add a sufficient number of terms in the Gauss sum. However, in all experiments performed so-far the individual contributions to the Gauss sums are created by individual pulses. Hence, the total number of terms in the Gauss sum is limited by the decoherence time of the system used in the experiment. Because of these two antagonistic effects we have to find conditions under which the algorithm based on Gauss sums successfully finds the factors of a given number $N$. We answer these questions in the following Chapters.

Chapter~\ref{chap9} deals with truncated Gauss sums and is based on \cite{opttrunc}. We find that the truncated Gauss sums offer good discrimination of factors from non-factors since the gap between their corresponding signals can reach a value of almost $30\%$. Moreover, we show that to reach such a gap the number of terms in the Gauss sum $M$ we have to add, i.e. the number of laser pulses we have to apply in the experiment, has to be of the order of the fourth-root of $N$. The total number of the resources needed for the success of the factorization scheme based on the truncated Gauss sum is thus
$$
{\cal R} \sim \sqrt[4]{N}\cdot \sqrt{N} = N^\frac{3}{4}.
$$

In Chapter~\ref{chap10} which is based on \cite{stef:exp:sum} we extend the idea of factorization of integers from Gauss sums to exponential sums of the form
$$
{\cal A}_N^{(M,j)}(\ell) \equiv \frac{1}{M+1} \sum_{m=0}^M \exp\left[2\pi i\, m^j\frac{N}{\ell}\right].
$$
Here the power of the phase is no longer quadratic like in the case of the Gauss sum but is given by a positive integer $j$. The faster growth of the phase results in the reduction of the number of terms $M$ that has to be added to the $2j$-th root of $N$. The total number of resources necessary to factorize a number $N$ using exponential sum is given by
$$
{\cal R}_j \sim \sqrt[2j]{N}\cdot \sqrt{N} = N^\frac{j+1}{2j}.
$$
Hence, we can save experimental resources by applying exponential sums with larger value of $j$. On the other hand, the gap between the signals of factors and non-factors shrinks as the power of the phase $j$ increases. This can make the experimental data inconclusive, unless a sufficient resolution is guaranteed. We summarize our results in the Conclusions.


\chapter{Factorization with Gauss sums}
\label{chap9}

\nsection{Introduction}

Gauss sums \cite{lang:1970,davenport:1980,schleich:2005:primes} play an important role in many phenomena of physics ranging from the Talbot effect of classical optics \cite{talbot:1836} via the curlicues emerging in the context of the semi-classical limit of quantum mechanics \cite{berry:curlicues:1:1988,berry:curlicues:2:1988}, fractional revivals \cite{leichtle:PRL:1996,leichtle:PRA:1996} and quantum carpets \cite{schleich:2001} to Josephson junctions \cite{schopohl}. Moreover, they build a bridge to number theory, especially to the topic of factorization. Indeed, they can be viewed as a discrimination function of factors versus non-factors for a given natural number. The essential tool of this factorization scheme \cite{merkel:FP} is the periodicity of the Gauss sum.

Usually Gauss sums extend over some period which leads to the {\it complete Gauss sum}. However, recent experiments based on NMR \cite{mehring:NMR:2006,Suter,suter2}, cold atoms \cite{rasel}, ultra-short pulses \cite{girard,girard2,girard3} and Bose-Einstein condensate \cite{sadgrove} have demonstrated the possibility of factoring numbers using a {\it truncated Gauss sum}, where the number of terms in the sum is much smaller than the period. Therefore, factorization with truncated Gauss sums offers enormous experimental advantages since the number of terms is limited by the decoherence time of the system. In the present Chapter we address the dependence of the number of terms needed in order to factor a given number. In particular, we find an optimal number of terms which preserves the discrimination property and at the same time minimizes the number of terms in the sum.

In order to factor a number $N$ we analyze the {\it signal}, i.e. the absolute value of the Gauss sum, for integer arguments $\ell=1,\ldots,\lfloor\sqrt{N}\rfloor$. We call the graphical representation of the signal data {\it factorization interference pattern}. In order to gain information about the factors of $N$ we analyze the factorization interference pattern:
Whenever the argument $\ell$ corresponds to a factor of $N$ we observe the maximal signal value of unity. For most non-factor arguments this signal value is significantly below unity. However, for {\it ghost factors} we observe signal values close to unity even though these arguments do not correspond to an actual factor of $N$. Thus ghost factors spoil the discrimination of factors from non-factors in such a factorization interference pattern. Fortunately, ghost factors can be suppressed below a given threshold by extending the upper limit of the summation in the Gauss sum. This goal of completely suppressing all ghost factors provides us with an upper bound on the truncation parameter. This upper bound represent a sufficient condition for the success of our Gauss sum factorization scheme. The analysis of the number of ghost factors evaluated by the {\it ghost factor counting function} $g(N,M)$, which depends on the number to be factorized $N$ and the truncation parameter $M$, reveals that this upper bound on the truncation parameter is also a necessary condition for the success of our Gauss sum factorization scheme.

The Chapter is organized as follows:
We first briefly review in Section~\ref{chap9:1} the central idea of the factorization scheme based on the Gauss sums. In particular, we introduce complete and truncated Gauss sums and compare the resources necessary to factor a given number $N$. We find the first traces of ghost factors in the factorization interference pattern based on the truncated Gauss sum. Since the truncation of the Gauss sum weakens the discrimination of the factors from non-factors, we dedicate Section~\ref{chap9:2} to deriving a deeper understanding of this feature. We find four distinct classes of arguments $\ell$ which result in utterly different behaviours of the truncated Gauss sum. Rewriting the truncated Gauss sum in terms of the curlicue sum allows us to identify the class of problematic arguments - the ghost factors. Moreover, we identify a natural threshold which separates factors from non-factors. For a rigorous argument we refer to Appendix~\ref{appendA}. In Section~\ref{chap9:3} we obtain an upper bound on the truncation parameter of the Gauss sum needed to suppress the signal of all ghost factors below the natural threshold. Ghost factors appear whenever the ratio of the number to be factored and a trial factor is close to an integer. This fact allows us to replace the Gauss sum by an appropriate Fresnel integral. From this expression we find the scaling law $M\sim\sqrt[4]{N}$ for the truncation parameter $M$, which represents the sufficient condition for the success of our Gauss sum factorization scheme. We discuss the applicability of the Fresnel approximation in the Appendix~\ref{appendB}. Finally, we analyze the ghost factor counting function in Section~\ref{chap9:4} and show that the fourth-root law is also necessary for the success of our factorization scheme, even if we relax the threshold value or allow limited error tolerance. We conclude in Section~\ref{chap9:5}.


\section{Factorization based on Gauss sums: appearance of ghost factors}
\label{chap9:1}

\begin{figure}[p]
\begin{center}
\includegraphics[width=0.65\textwidth]{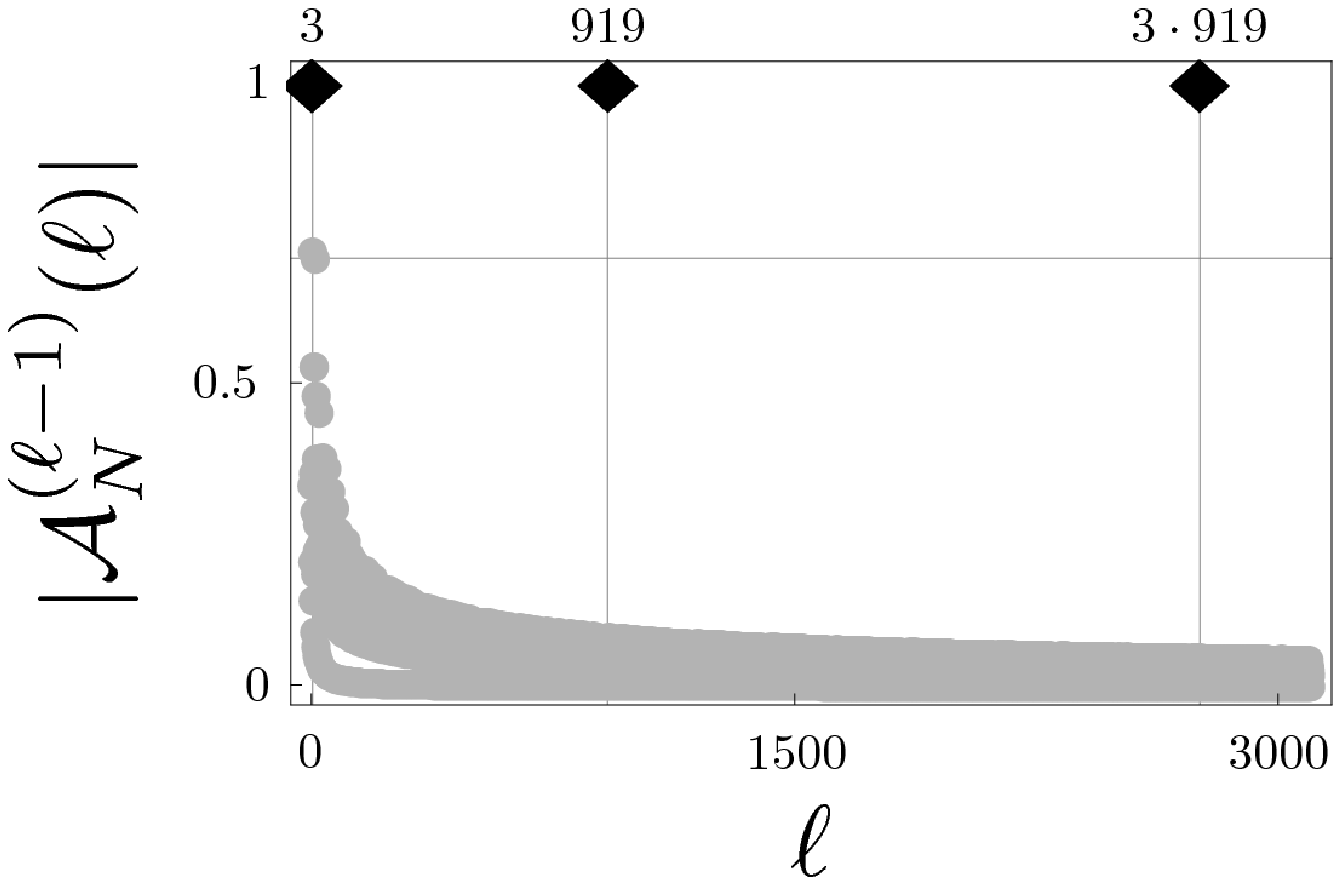}\vspace{24pt}
\includegraphics[width=0.65\textwidth]{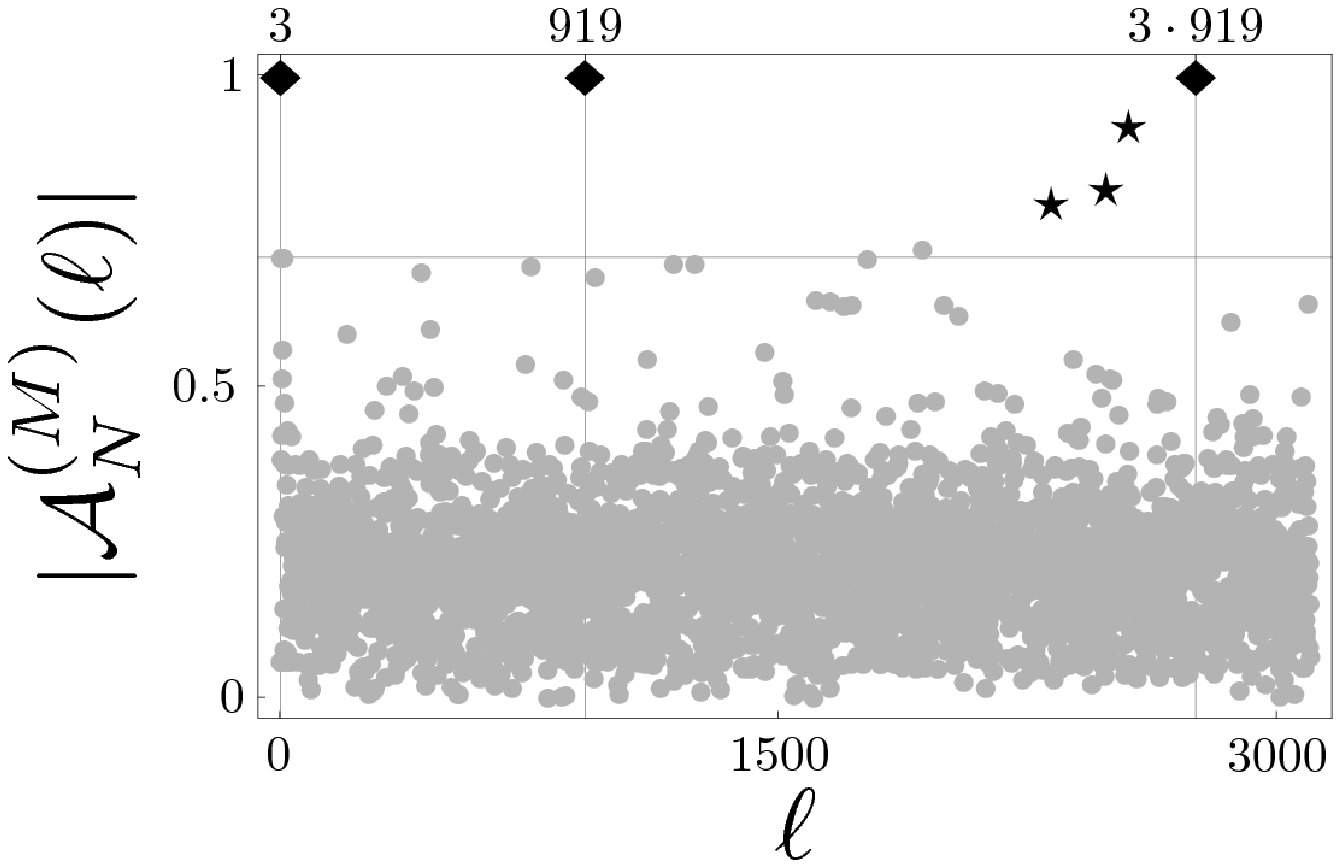}
\caption[Truncating the summation range in the Gauss sum]
{Influence of the truncation parameter $M$ of the incomplete Gauss sum ${\cal A}_{N}^{(M)}(\ell)$ defined in (\ref{gauss2}) on the contrast of the factorization interference pattern for the example $N=9624687=3\cdot 919 \cdot 3491$. The upper picture shows the corresponding pattern for the complete Gauss sum defined in (\ref{complete:GS}) where $M= \ell-1$. For the lower plot we have truncated the Gauss sum after $M=\ln N=16$ terms. At factors of $N$ indicated by vertical lines the Gauss sum assumes the value of unity marked by black diamonds.
The complete Gauss sum enjoys an impressive contrast due to a
suppressed signal value at all non-factors. However, also the truncated Gauss
sum with a relatively small number of terms allows to discriminate
factors from non-factors. However, we also observe several ghost factors marked by stars.}
\label{truncated}
\end{center}
\end{figure}

To start our analysis we first consider the complete normalized quadratic Gauss sum
\begin{equation}
\label{complete:GS}
{\cal A}_N^{(\ell-1)}(\ell) = \frac{1}{\ell}\sum\limits_{m=0}^{\ell-1}\exp\left(2\pi
i\, m^2 \frac{N}{\ell}\right),
\end{equation}
which is frequently used in number theory. Here $N$ is the integer to be factorized and the integer argument $\ell$ scans through all numbers from 1 to $\lfloor\sqrt{N}\rfloor$ for factors of $N$. If $\ell$ is a factor then all terms in the sum contribute with a value of unity and thus the resulting  signal value $|{\cal A}_N^{(\ell-1)}(\ell)|$  is one. However, for non-factor arguments the signal value is suppressed considerably as illustrated on the left in Figure~\ref{truncated}. Thus the absolute value of the Gauss sum allows one to discriminate between factors from non-factors.

Factorization based on the complete Gauss sum (\ref{complete:GS}) has several disadvantages. First of all, the limit of the sum depends on the trial factor $\ell$. Thus the number of terms in the sum increases with $\ell$ up to $\sqrt{N}$. Hence, to obtain a complete factorization interference pattern in total
\begin{equation}
\sum_{\ell=1}^{\sqrt{N}} \ell=\frac{1}{2}\sqrt{N}(\sqrt{N}-1)\approx \frac{1}{2}N
\label{resource:complete}
\end{equation}
terms have to be added.

In the recent experimental demonstrations \cite{mehring:NMR:2006,Suter,suter2,rasel,girard,girard2,girard3,sadgrove} of our Gauss sum factorization scheme the number of terms in the sum translates directly into the number of pulses applied onto the system, or the number of interfering light fields. Due to the decoherence it is favorable to use as few pulses as possible. Hence the experiments employ a constant number $M$ of pulses for each argument $\ell$ to be tested. Thus the resulting signal is of the form of a truncated Gauss sum
\begin{equation}
{\cal A}_N^{(M)}(\ell) = \frac{1}{M+1}\sum\limits_{m=0}^{M}\exp\left(2\pi
i\, m^2 \frac{N}{\ell}\right),
\label{gauss2}
\end{equation}
rather than a complete Gauss sum of (\ref{complete:GS}). Hence we have to add
\begin{equation}
\sum_{\ell=1}^{\sqrt{N}} M = M\cdot\sqrt{N}
\label{resource}
\end{equation}
terms to obtain the factorization pattern with the truncated Gauss sum. With this fact in mind we treat the number of terms in the Gauss sum as a resource for this factorization scheme.

The experiments impressively demonstrate that the truncated Gauss sums are also well suited to discriminate in the factorization interference pattern between factors from non-factors, even though the summation range does not cover a full period. As a drawback we find that the signal value at non-factor arguments is not suppressed as well as in the case of the complete Gauss sum.

In order to illustrate the effect of truncating the Gauss sum we compare in Figure~\ref{truncated} the factorization interference patterns for the complete Gauss sum ${\cal A}_N^{(\ell-1)}(\ell)$ (left) and for the truncated Gauss sum ${\cal A}_N^{(M)}(\ell)$ (right). In a first guess we chose the truncation parameter $M=\ln N$ to depend logarithmically on the number to be factorized. It is remarkable that the small number  $M=16$ of terms in the truncated Gauss sum is sufficient to reveal the factors of a seven-digit number like $N=9624687$. On the other hand we observe a number of data-points with signal values close to one (stars), for example at the argument $\ell=2555$. In an experiment such points might lead us to wrong conclusions in the interpretation of a factorization interference pattern. Thus we call arguments $\ell$ resulting in such critical values of the signal {\it ghost factors}.

\section{Classification of trial factors}
\label{chap9:2}

The frequency of appearance of ghost factors is the central question of our study. Indeed, how many terms in the truncated Gauss sum are needed in order to suppress the occurrence of ghost factors. However, we first need to identify the class of arguments which results in ghost factors.

Figure \ref{truncated} already indicates that there are different classes of trial factors: (i) factors with constant value of $|{\cal A}_N^{(M)}(\ell)|$, (ii) typical non-factors at which already few terms $M$ are sufficient to suppress the value of $|{\cal A}_N^{(M)}(\ell)|$ considerably, (iii) ghost factors at which a larger summation range is needed to suppress the value of $|{\cal A}_N^{(M)}(\ell)|$, and finally (iv) threshold non-factors at which the value of $|{\cal A}_N^{(M)}(\ell)|$ {\it cannot} be suppressed by increasing $M$.

\begin{figure}[h]
\begin{center}
\includegraphics[width=0.6\textwidth]{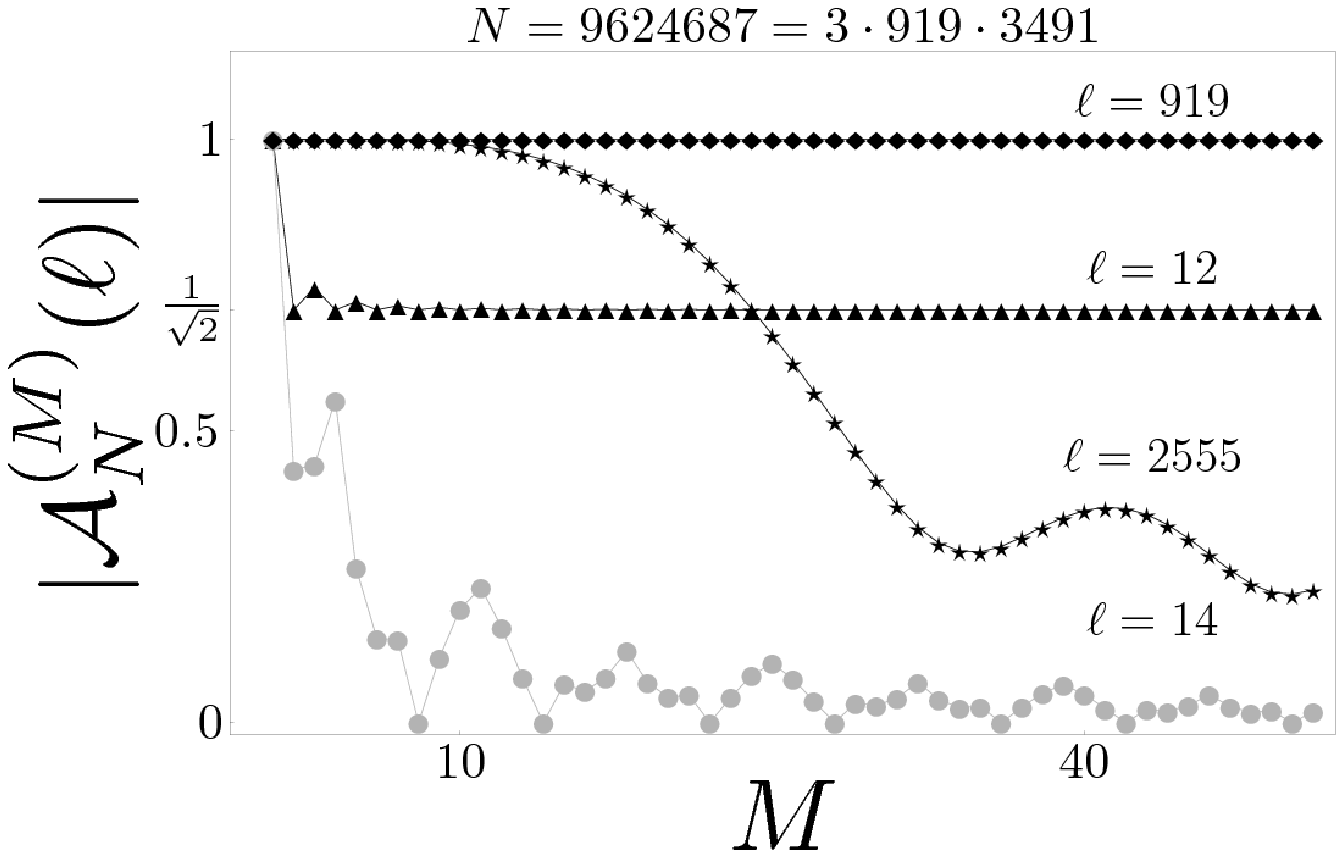}
\caption
{Emergence of four different classes of arguments $\ell$ of the truncated Gauss sum of (\ref{gauss2})  from its dependence on its truncation parameter $M$ for $N=9624687=3\cdot 919\cdot 3491$. We show the signal value $|{\cal A}_N^{(M)}(\ell)|$ for four arguments $\ell$ as a function of the truncation parameter $M$. For factors of $N$, such as $\ell=919$ depicted by the black diamonds, the signal is constant and equal to unity. For typical non-factors, such as $\ell=14$ depicted by the gray dots, the signal is suppressed considerably already for small values of the truncation parameter $M$. However, for ghost factors, such as $\ell=2555$ depicted by stars, much more terms in the sum (\ref{gauss2}) are needed to suppress the signal. For arguments  such as $\ell=12$ depicted by the black triangles, the signal levels off at a non-vanishing threshold determined by $1/\sqrt{2}$ and it is impossible to suppress it by further by increasing the truncation parameter $M$.}
\label{arguments}
\end{center}
\end{figure}

To illustrate this we plot in \fig{arguments} the truncated Gauss sum of (\ref{gauss2}) for $N=9624687$ and various arguments $\ell$ characteristic for each one of the class (i-iv) as a function of the truncation parameter $M$.

To which class of arguments (i-iv) the given $\ell$ belongs is determined by the relation between the argument $\ell$ and the number we are factorizing $N$, namely on the value of the fraction $2N/\ell$ which enters the Gauss sum (\ref{gauss2}). Indeed, for the number $N=9624687$ and the arguments $\ell$ used in \fig{arguments} we find the following: (i) for a factor $\ell=919$ the fraction $2N/\ell$ is an even integer, (ii) for a typical non-factor $\ell=14$ the fraction $2N/\ell$ is close to an odd integer, (iii) for a ghost factor $\ell=2555$ the fraction $2N/\ell$ is close to an even integer, (iv) for a threshold non-factor $\ell=12$ the fraction $2N/\ell$ is an even integer plus one-half. Thus we see that the class of $\ell$ is given by the fractional part of the fraction $2N/\ell$. Hence, in order to bring out these classes most clearly, we represent the truncated Gauss sum (\ref{gauss2}) in a different form. For any argument $\ell$ we decompose the fraction $2N/\ell$ into the closest even integer $2k$ and the fractional part $\rho(N,\ell)=p/q$ with $|\rho|<1$ and $p$, $q$ being coprime, i.e.
\begin{equation}
\rho(N,\ell)=\frac{2N}{\ell}-2k.
\label{fracpart}
\end{equation}
Since $\exp\left(2\pi i\, m^2 \cdot k\right)=1$ the Gauss sum (\ref{gauss2}) reads
\begin{equation}
\label{gauss3}
{\cal A}_N^{(M)}(\ell)= s_M\left(\rho(N,\ell)\right)
\end{equation}
where we have introduced the normalized curlicue function \cite{berry:curlicues:1:1988},\cite{berry:curlicues:2:1988}
\begin{equation}
s_M(\tau)\equiv\frac{1}{M+1}\sum\limits_{m=0}^M \exp\left(i \pi\, m^2 \tau \right)
\label{curl}
\end{equation}
which we consider for a real argument $\tau$ with $-1\leq\tau\leq 1$.

The connection (\ref{gauss3}) between the truncated Gauss sum ${\cal A}_N^{(M)}(\ell)$ defined in  (\ref{gauss2}) and the normalized curlicue sum $s_M(\tau)$ for a given $N$ is established by the fractional part $\rho(N,\ell)$ of the fraction $2N/\ell$. Indeed, factors of $N$ correspond to $\rho=0$. All other values of $\rho$ correspond to non-factors. In particular, ghost factors have $\rho$ values close to zero.

\begin{figure}
\centering
\includegraphics[width=0.8\textwidth]{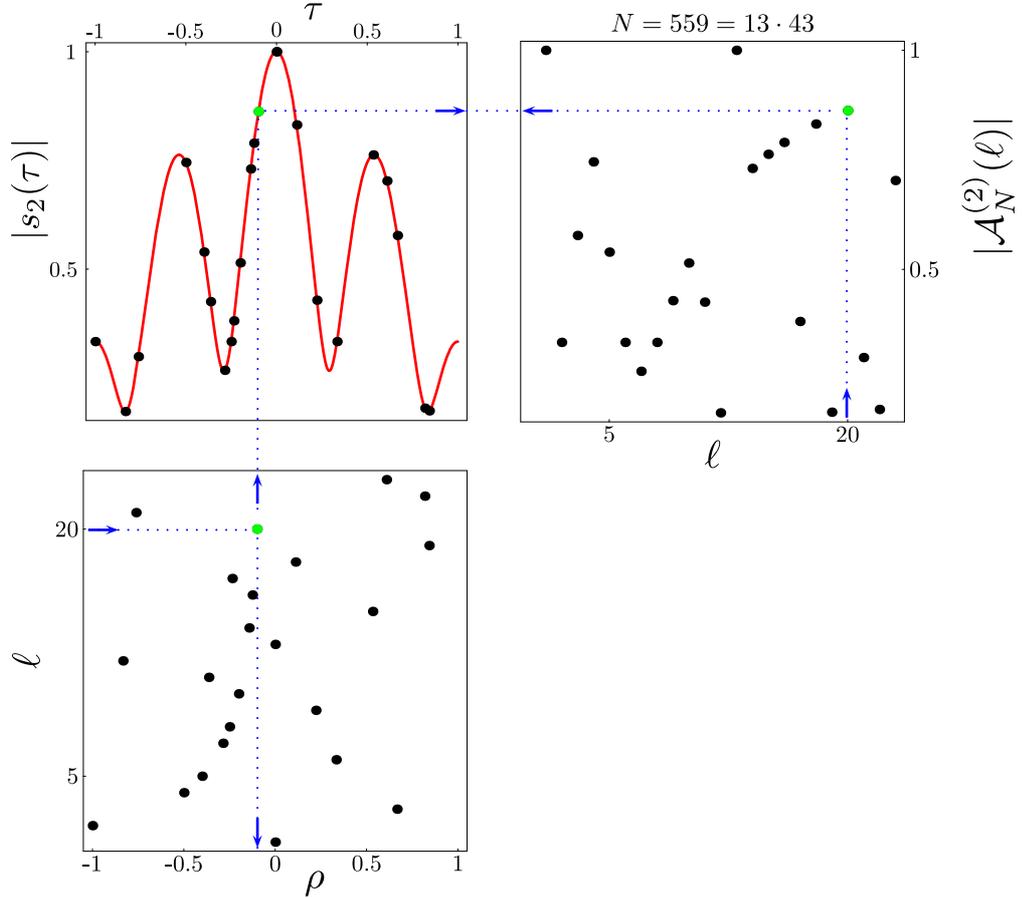}
\caption{Connection between the truncated Gauss sum ${\cal A}_N^{(M)}(\ell)$ and the normalized curlicue sum $s_M(\tau)$ established by the fractional part $\rho(N,\ell)$ of the fraction $2N/\ell$. Here the number $N$ to be factorized is $N=559=13\cdot 43$ with the truncation parameter $M=2$. In the lower-left plot we assign to every value of $\ell$ the fractional part $\rho(N,\ell)$ defined by (\ref{fracpart}) for the number $N=559$ as exemplified by $\ell=20$ and the green dot. In the upper-left plot we show the master curve $|s_2(\tau)|$ indicated by the red curve and given by the absolute value of the normalized curlicue sum (\ref{curl}). This curve is an even function with respect to $\tau$  and attains the values above $1/\sqrt{2}$ only in the narrow peak located at $\tau=0$. The factorization interference pattern for $N=559$ shown in the upper-right corner follows from the dots in the upper-left plot in a two step process going through the master curve: from $\ell$ we find the fractional part $\rho(N,\ell)$ which determines through the master curve the signal value as indicated by the arrows.}
\label{master}
\end{figure}

We depict the connection of ${\cal A}_N^{(M)}(\ell)$ with $s_M(\tau)$ via $\rho(N,\ell)$  in \fig{master} for the number to be factorized $N=559=13\cdot 43$ and the truncation parameter $M=2$. The upper-left plot represents the master curve $|s_2(\tau)|$ (blue line) given by the absolute value of the normalized curlicue sum (\ref{curl}). The function $|s_M(\tau)|$ is even with respect to $\tau$, since
$$
s_M(-\tau)=s_M^*(\tau).
$$
Hence, it depends only on the absolute value of $\tau$. Moreover, we note two characteristic domains of $|s_M(\tau)|$: (i) the function starts at unity for $\tau=0$ and decays for increasing $\tau$. This central peak around $\tau=0$ is the origin of the ghost factors. (ii) After this initial decay oscillations set in whose amplitudes seem to be bound. Indeed, in the Appendix~\ref{appendA} we show that in the limit of large $M$ the absolute value of the normalized curlicue sum $|s_M(\tau)|$ evaluated at non-zero rational $\tau$ is bounded from above by $1/\sqrt{2}$.

The lower-left plot shows the distribution of the fractional parts $\rho(N,\ell)$ given by (\ref{fracpart}). The dots in the upper-left plot arise from the projection of the fractional parts (\ref{fracpart}) of the lower-left plot onto the master curve. Those data points represent the factorization interference pattern for $N=559$, as depicted on the right.

\section{Upper bound on the truncation by complete suppression of ghost factors}
\label{chap9:3}

\begin{figure}
\centering
\includegraphics[width=0.7\textwidth]{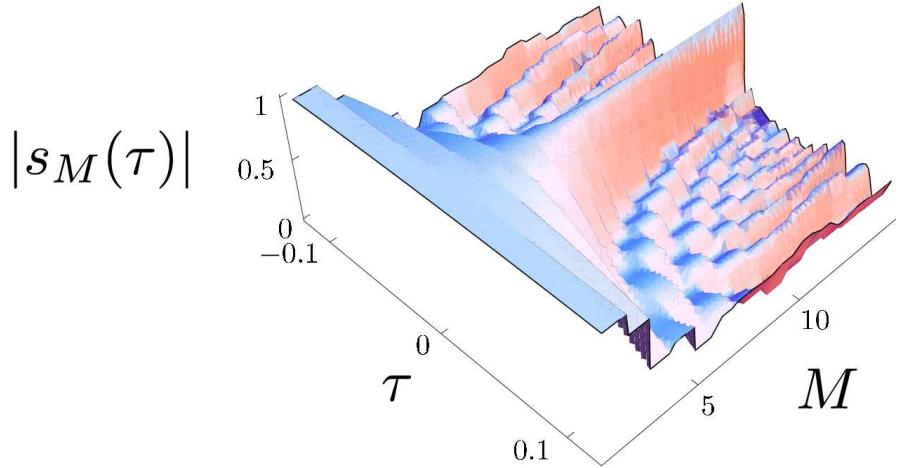}
\caption{Absolute value $|s_M(\tau)|$ of the normalized curlicue function in the neighborhood of $\tau=0$ in its dependence on the fractional part $\tau$ and the truncation parameter $M$. The function starts at unity for $\tau=0$ and decays for increasing $\tau$. This decay becomes faster as we increase $M$. This behaviour is at the heart of the suppression of the ghost factors.}
\label{gs3d}
\end{figure}

Ghost factors emerge from the central peak of the absolute value of the normalized curlicue function. Our goal is to suppress these ghost factors by increasing the truncation parameter $M$. For this purpose we display in \fig{gs3d} the normalized curlicue sum (\ref{curl}) in the neighborhood of $\tau=0$ in its dependence on $\tau$ and $M$. Indeed, we find a narrowing of the central peak with increasing $M$. In this way we can suppress the ghost factors below a natural threshold.

As shown in Appendix~\ref{appendA} for non-zero positive rational $\tau=p/q$ the absolute value of the normalized curlicue sum is asymptotically bounded from above by $1/\sqrt{2}$. Due to the connection (\ref{gauss3}) between the normalized curlicue sum $s_M(\tau)$ and the Gauss sum ${\cal A}_N^{(M)}(\ell)$ it is natural to use this bound as a natural threshold between factors and non-factors. This observation allows us to define the ghost factor properly: ghost factors $\ell$ of a number $N$ arise when the fractional part $\rho(N,\ell)$ of $2N/\ell$ leads to a value of the normalized curlicue function $|s_M(\rho)|$ in the domain between $1/\sqrt{2}$ and unity.

We determine the truncation parameter $M_0$ such that we can push the absolute value of the Gauss sum for all ghost factors below the natural threshold of $1/\sqrt{2}$. Ghost factors appear for small values of $\tau$. This fact allows us to replace the Gauss sum by an integral which leads us to an estimate for the truncation parameter $M_0$.

Indeed, with the substitution $u=\sqrt{2\tau}m$ we can approximate the normalized curlicue function
\begin{equation}
\label{fresnel}
s_M(\tau)
\approx \frac{1}{M}\int\limits_0^{M} du\,
\exp\left(i\, \pi m^2\tau\right)= \frac{F(M\sqrt{2\tau})}{M\sqrt{2\tau}}
\end{equation}
with the Fresnel integral \cite{abramowitzstegun}
$$
F(x)=\int\limits_0^x du\,\exp\left(i\, \frac{\pi}{2} u^2\right)
$$
familiar from the diffraction from a wedge \cite{born:wolf}. We note that in the continuous approximation the normalized curlicue function depends only on the product $M\cdot \sqrt{2\tau}$.

\begin{figure}
\centering
\includegraphics[width=0.7\textwidth]{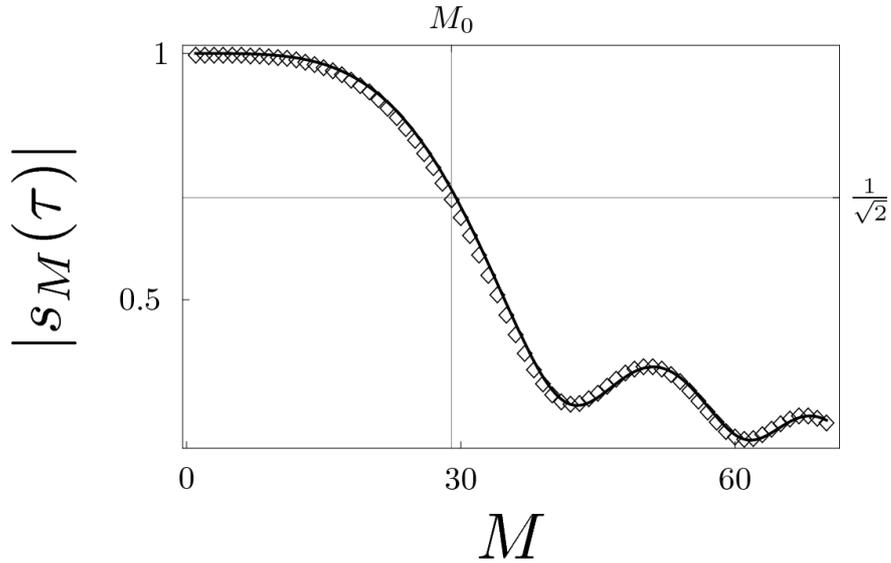}
\caption[Suppression of ghost factors]
{Comparison between the exact discrete normalized curlicue sum (\ref{curl}) shown by diamonds and its approximation (\ref{fresnel}) by the continuous Fresnel integral depicted by the black curve. We display the absolute value $|s_M(\tau)|$ as a function
of the number $M$ of terms contributing to the curlicue sum (\ref{curl}) for $\tau=10^{-3}$. The horizontal line marks the threshold $1/\sqrt{2}$ of the signal and the vertical line indicates the upper bound $M_0$ (\ref{M-N}) required to suppress a ghost factor corresponding to $\tau=10^{-3}$.}
\label{gf:M}
\end{figure}

In Figure~\ref{gf:M} we compare the absolute value of the discrete curlicue sum $s_M(\tau)$ and the continuous Fresnel integral
$F(M\sqrt{2\tau})/(M\sqrt{2\tau})$ at small value $\tau = 10^{-3}$. This approximation impressively models the results of the discrete curlicue sum.

We are looking for the truncation parameter $M_0$ such that for a given fractional part $\tau$ the absolute value of the integral (\ref{fresnel}) is equal to $\frac{1}{\sqrt{2}}$. We denote $\alpha(\xi)$ as the solution of the transcendental equation
$$
\frac{|F(\alpha)|}{\alpha}=\xi.
$$
In particular, for the natural threshold $\xi=1/\sqrt{2}$ defining the ghost factors we find the numerical value of $\alpha(\xi)\approx 1.318$. From the fact that $F$ depends only on the product of $M\cdot\sqrt{2\tau}$ it follows that
\begin{equation}
\label{alpha}
\alpha(\xi) = M_0 \sqrt{2\tau}.
\end{equation}

For the factorization of the number $N$ the argument $\ell$ is varied within the interval $[1,\sqrt{N}]$. Consequently, the minimal fractional part
$$
\rho_{\rm min}(N)\sim\frac{2}{\sqrt{N}}.
$$
arises from the ratio $2N/\ell$ when the denominator takes on its maximum value $\ell=\sqrt{N}$.

Finally, we arrive at
\begin{equation}
\label{M-N}
M_0\approx \frac{\alpha(\xi)}{\sqrt{2 \rho_{\rm min}(N)}}\approx \frac{\alpha(\xi)}{2}\sqrt[4]{N}.
\end{equation}
Hence, $M_0$ represents an upper bound for the number of terms in the truncated Gauss sum (\ref{gauss2}) required to push {\it all} non-factors below the threshold of $\xi$. In particular, we find that to suppress all ghost factors below the natural threshold $\xi=1/\sqrt{2}$ we need $M_0\approx 0.659\sqrt[4]{N}$ terms in the truncated Gauss sum. However, we point out that the power-law (\ref{M-N}) arises from the fact that we use quadratic phases and will be unchanged by relaxing the threshold value $\xi$, as the change of this threshold will only change the prefactor $\alpha(\xi)$.

We conclude this section by noting that the scaling law rests on approximating the normalized curlicue sum by the Fresnel integral. In Appendix~\ref{appendB} we analyze the range of applicability of the Fresnel integral approximation (\ref{fresnel}) and find that our results lie within the validity of the approximation.


\section{Ghost factor counting function: inevitable scaling law}
\label{chap9:4}

In the preceding section we have derived a scaling law between the number $M$ of terms of the truncated Gauss sum to factor a given number $N$. This estimate is a {\it sufficient} condition for the success of the Gauss sum factorization scheme. In the present section we show that it is also a {\it necessary} condition. In order to illustrate this feature we first choose logarithmic truncation $M=\ln N$ and show that at the end of our factorization scheme we will be left with too many candidate factors, most of them being a ghost factor. Moreover, we show that we cannot achieve a more favorable scaling than the fourth-root dependence, (\ref{M-N}), even if we tolerate a limited number of ghost factors.

To answer these questions we introduce the ghost factor counting function
\begin{equation}
g(N,M)\equiv \#\left\{\ell=1,\ldots,\lfloor\sqrt{N}\rfloor\ \rm{with}\ \frac{1}{\sqrt{2}}<|{\cal A}_N^{(M)}(\ell)|<1\right\}
\label{gf:count}
\end{equation}
which yields the number of data-points with critical values of the signal in the factorization interference pattern for a given $N$ and a chosen truncation $M$.

\begin{figure}
\begin{center}
\includegraphics[width=0.6\textwidth]{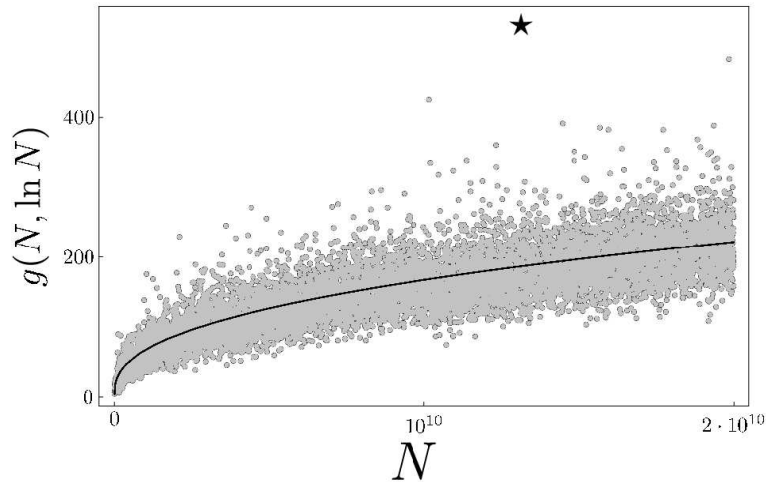}
\caption
{A logarithmic dependence of the truncation parameter $M$ on $N$ is not sufficient to suppress ghost factors. The ghost factor counting function $g(N,M)$ calculated for 10000 random odd numbers $N$ out of the interval $[1,2\cdot 10^{10}]$ with $M=\ln N$ grows faster than the logarithm of $N$. The black line describes the general trend given by (\ref{gf:fit}). We observe strong deviations, as exemplified by $N=13064029441$ highlighted by the star and discussed in Section~\ref{non-uniform}.}
\label{gfcount}
\end{center}
\end{figure}

In order to study the behaviour of the ghost factor counting function $g(N,M)$ on a broad range of numbers $N$ we show in Figure~\ref{gfcount} $g(N,M=\ln N)$ for 10000 random numbers $N$ out of the interval $[1,2\cdot 10^{10}]$. Here we choose the truncation parameter to depend logarithmically $M\approx \ln N$ on $N$. This result shows that the number of ghost factors $g(N,M)$ for $M\approx\ln N$ grows faster than the logarithm of $N$. Hence the logarithmic truncation $M\approx\ln N$ is not sufficient for the success of our Gauss sum factorization scheme. We provide an explanation for the general trend observable in Figure~\ref{gfcount} in Section~\ref{uniform} and discuss the deviations in Section~\ref{non-uniform}.

In evaluating the number of ghost factors we proceed in two steps. First, we make use of the connection (\ref{gauss3}) between the truncated Gauss sum
${\cal A}_N^{(M)}(\ell)$ and the normalized curlicue sum $s_M(\tau)$. As already pointed out in Section~\ref{chap9:2} the ghost factors appear only for $\tau$ values lying in the small interval $[-\tau_0,\tau_0]$ around zero. The Fresnel integral approximation from Section~\ref{chap9:3} allows us to determine the fractional part $\tau_0$ where the normalized curlicue sum assumes the value $1/\sqrt{2}$. In the second step we relate the number of ghost factors $g(N,M)$ to $\tau_0$ by a density argument.

We determine $\tau_0$ with the help of the continuous approximation of the curlicue sum. From (\ref{alpha}) we obtain
\begin{equation}
\label{tau0}
\tau_0= \tau_0(M) \approx \frac{\alpha^2}{2 M^2}
\end{equation}
and we thus arrive at the total width
$2\tau_0\approx\alpha^2/M^2$ of the interval of fractional parts resulting in signal values larger than $1/\sqrt{2}$.

We relate the number of ghost factors $g(N,M)$ to the width of the interval $2\tau_0$ via the distribution of fractional parts $\tau$ for a given $N$. First, we consider a uniform distribution. Here we derive an analytical estimation for $g(N,M)$ which explains the general trend in Figure~\ref{gfcount}. Second, we discuss the case of numbers $N$ where the distribution of fractional parts cannot be approximated as uniform. Finally, we analyze a trade-off between a smaller truncation parameter at the expense of more ghost factors. We show that this approach will not change the power-law (\ref{M-N}).

\subsection{Uniform distribution of fractional parts}
\label{uniform}

Let us first assume for simplicity that the distribution of the fractional parts $\tau$ is uniform for a given number $N$. Here the number of ghost factors $g(N,M)$ is directly proportional
$$
\frac{g(N,M)}{\sqrt{N}}\approx\frac{2 \tau_0}{2}.
$$
to the width $2\tau_0$ of the interval of the fractional parts which lead to ghost factors.

Recalling the dependence of $\tau_0$ on $M$ (\ref{tau0}) we conclude that the number of ghost factors
\begin{equation}
\label{nbghost}
g(N,M)\approx \frac{1}{2}\left(\frac{\alpha}{M}\right)^2\sqrt{N}.
\end{equation}
depends via an inverse power-law on the truncation parameter $M$.

In Figure~\ref{gfcount} we already found indications that $g(N,M=\ln N)$ grows faster than the logarithm of $N$. Indeed, from (\ref{nbghost}) we obtain
\begin{equation}
g(N,\ln N)
\approx\frac{1}{2}\left(\frac{\alpha}{\ln N}\right)^2\sqrt{N}.
\label{gf:fit}
\end{equation}
which implies that $g(N,\ln N)$ behaves like $\sqrt{N}$.
In Figure~\ref{gfcount} we display a fit according to (\ref{gf:fit}). We find that this fit well describes the general trend over a large range of numbers $N$. However, we also observe strong variations around this general trend.
The deviations indicate that the distribution of fractional parts is not uniform for certain numbers $N$. We analyze such numbers in Section~\ref{non-uniform}.


\subsection{Non-uniform distribution of the fractional parts}
\label{non-uniform}

In Figure~\ref{gfcount} we find that for certain numbers the actual number of ghost factors $g(N,M)$ considerably deviates from our estimation (\ref{gf:fit}). In the following we show that for such numbers the distribution of the fractional parts cannot be treated as uniform.

This unfavorable case occurs when $N$ has only few divisors, but another number $N'=N+k$ close to $N$ has a lot of divisors (with $|k|\ll N$). For example for the number
$$
N=13064029441=21647\cdot 603503
$$
highlighted in \fig{gfcount} by the circle we find that
$$
N'=N-1=2^8\cdot 3\cdot 5\cdot 11\cdot 17\cdot 23\cdot 113
$$
obviously has a lot of divisors.

Let us consider $\ell'$ which is a divisor of $N'=N+k$ but not of $N$. It follows that if $\ell'>2k$ the fractional part of $2N/\ell'$ is equal to
\begin{equation}
\rho(N,\ell')=-\frac{2k}{\ell'}.
\label{hyp}
\end{equation}
If we consider a plot of the fractional part $\rho(N,\ell)$ of $2N/\ell$ as a function of $\ell$ we will find that for divisors $\ell'$ of $N'$ the resulting fractional parts are aligned on the hyperbola (\ref{hyp}) and are attracted to zero. Hence for $N$ the distribution of fractional parts $\rho(N,\ell)$ is not uniform.

In the factorization interference pattern of $N$ data-points associated with arguments $\ell'$ corresponding to divisors of $N'$ are also aligned on the curve
\begin{equation}
\gamma_k^{(M)}(l) \equiv \left|s_M\left(\frac{2k}{\ell}\right)\right|.
\label{gamma}
\end{equation}
As for large values of $\ell'$ the associated fractional part $-2k/\ell'$ tends to zero the resulting signal values $|{\cal A}^{(M)}_N(\ell')|$ approaches unity. Hence the divisors of $N'$ become ghost factors of $N$.

\begin{figure}
\begin{center}
\includegraphics[width=0.45\textwidth]{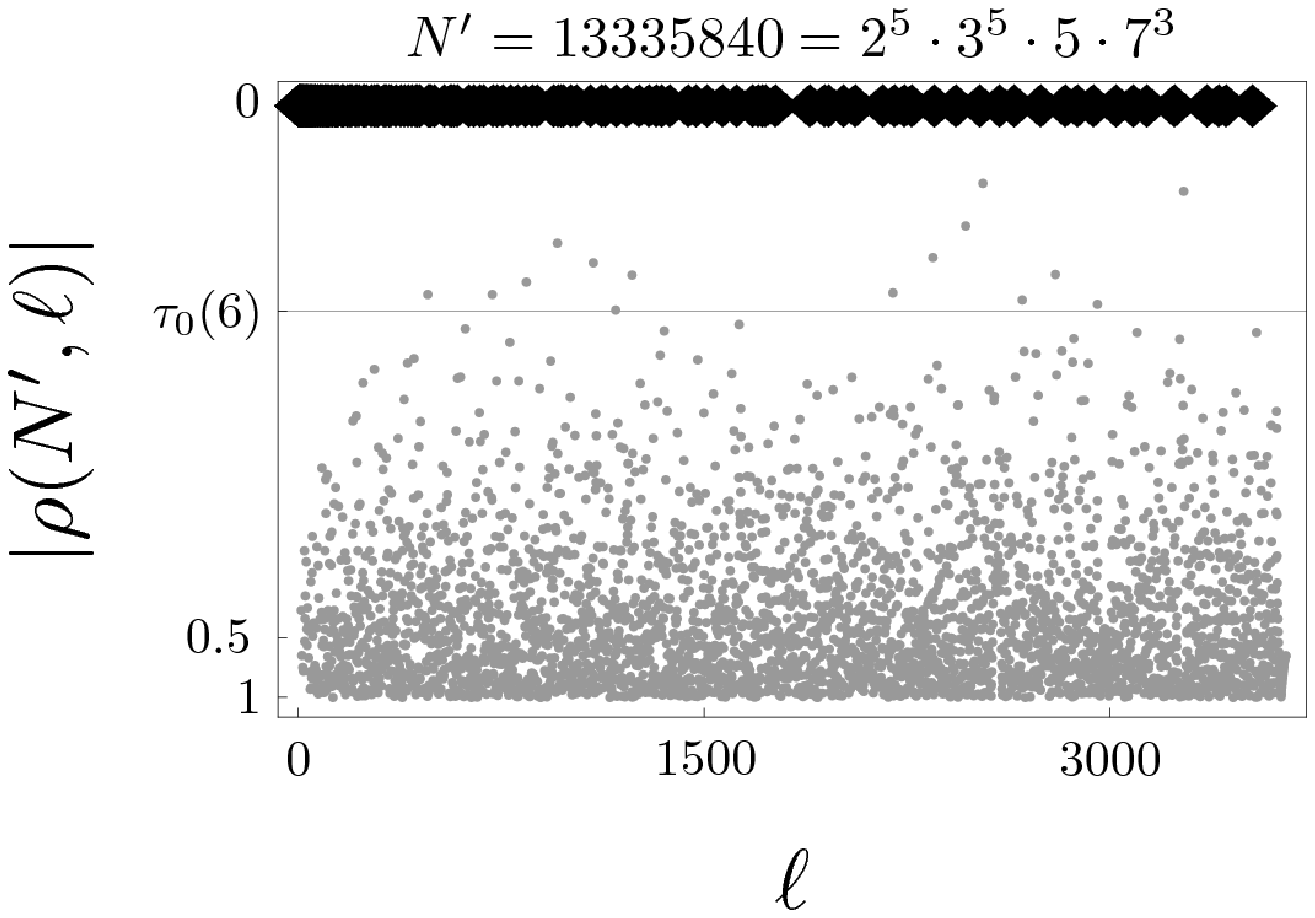}\hfil
\includegraphics[width=0.45\textwidth]{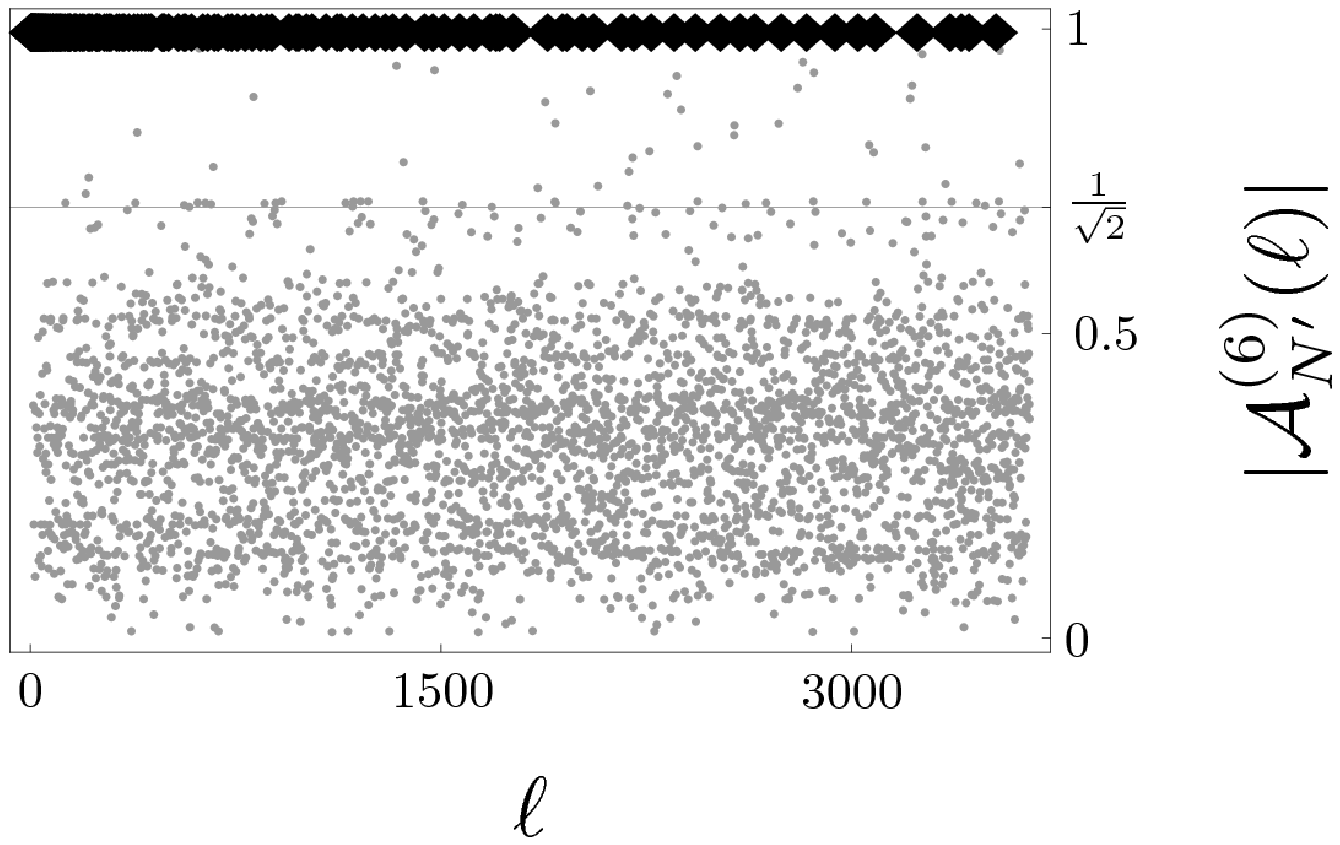}
\includegraphics[width=0.45\textwidth]{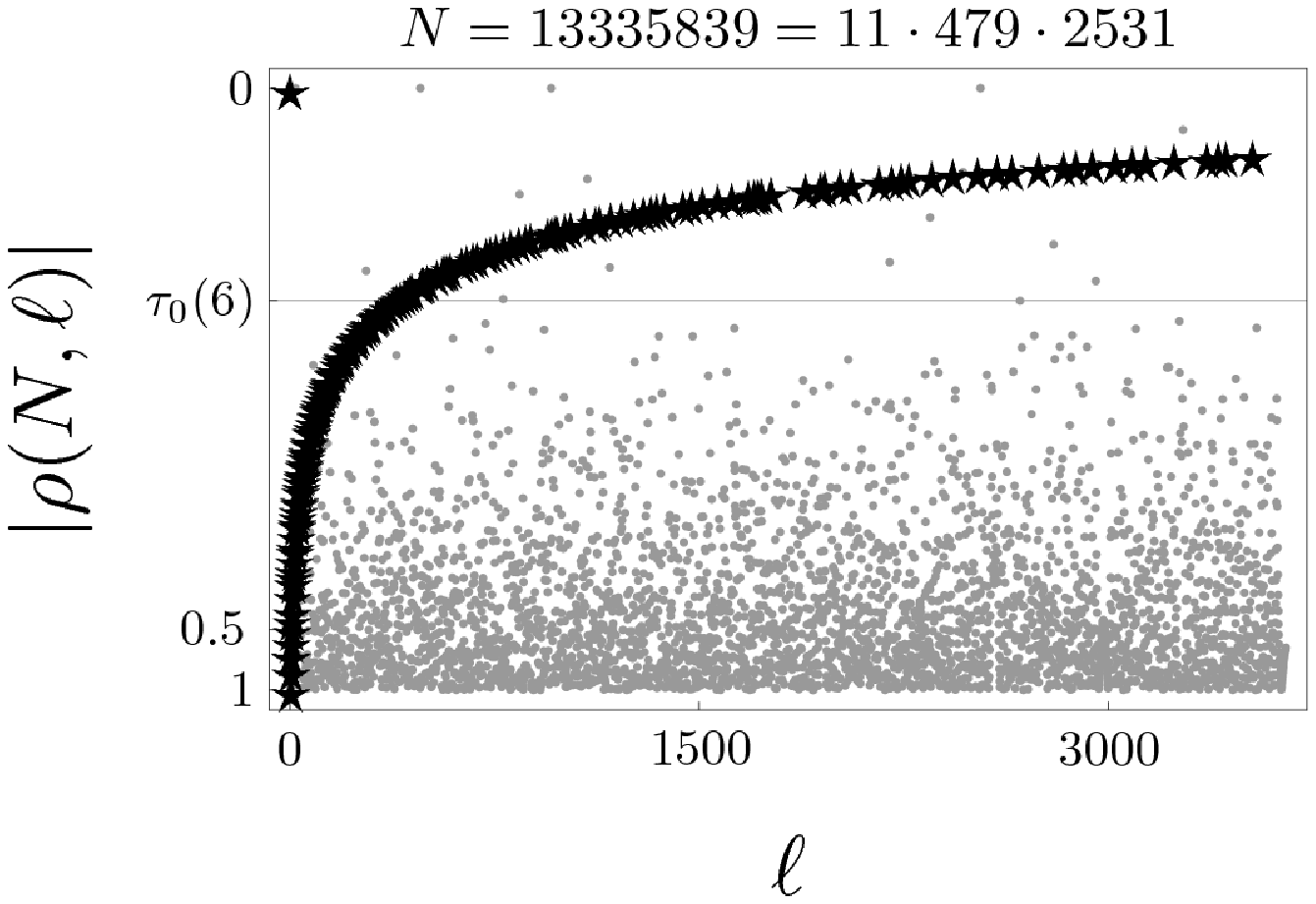}\hfil
\includegraphics[width=0.45\textwidth]{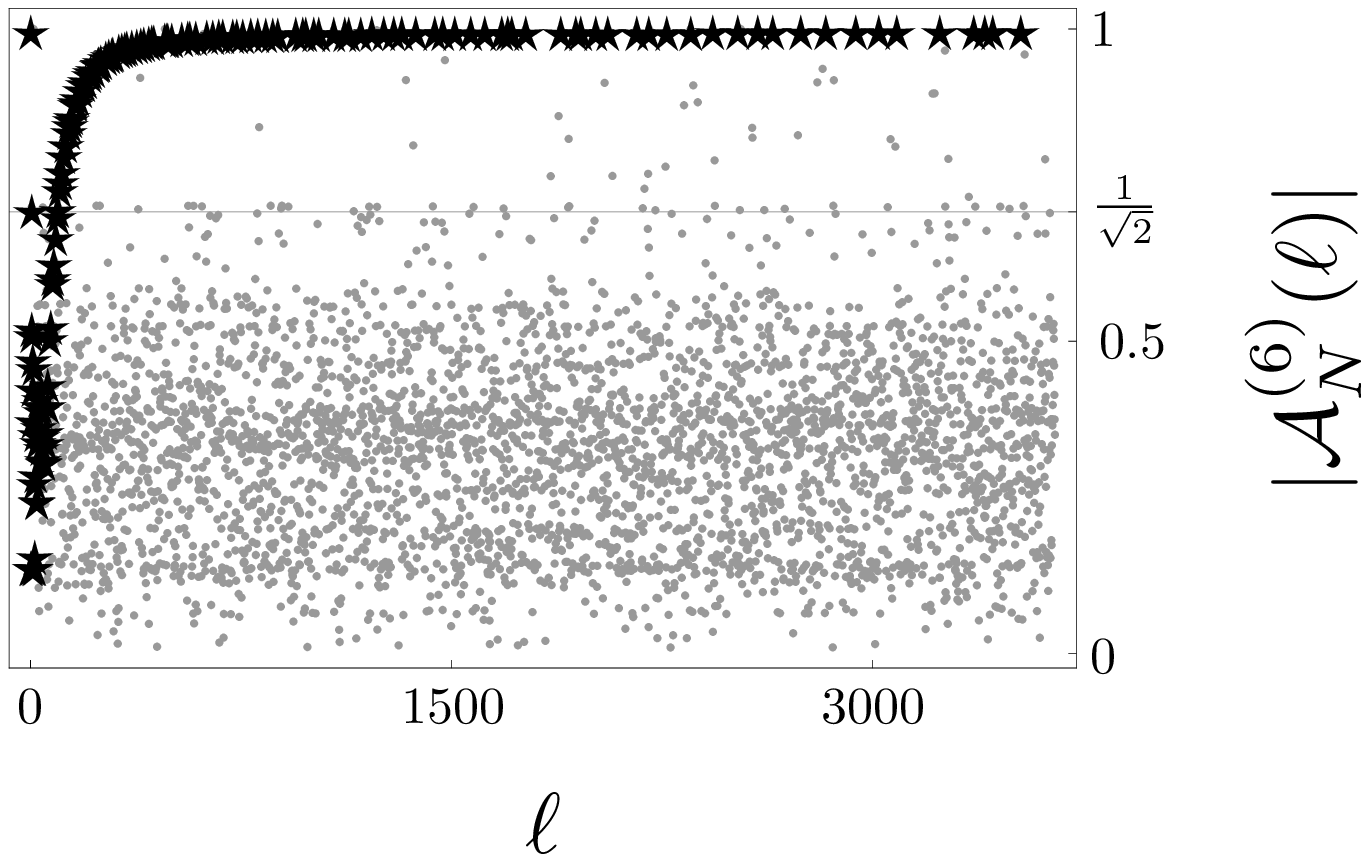}
\caption
{Emergence of ghost factors of $N$ from factors of $N'$. We display the distributions of the fractional parts (left column) and the factorization interference patterns (right column) for the numbers  $N'=13335840=2^5\cdot 3^5\cdot 5\cdot 7^3$ which is rich in factors and $N=N'-1=13335839=11\cdot 479\cdot 2531$ which is rich in ghost factors. To emphasize the region of fractional parts which lead to ghost factors we use a logarithmic scale for $|\rho|$ on the vertical axes. The number $N'$ has a lot of divisors, as depicted on the upper-left plot by the straight line of black diamonds. In the factorization interference pattern shown on the right these divisors correspond to a straight line of signals equal to unity. However, the divisors of $N'$ are non-factors for $N=N'-1$. Moreover, they are aligned on a hyperbola (\ref{hyp}) and attracted to zero as shown in the lower-left plot where we can clearly identify the hyperbola of stars. Consequently, in the factorization interference pattern shown on the right this hyperbola translates into the curve of ghost factors.}
\label{signal:tau}
\end{center}
\end{figure}

We illustrate this fact in Figure~\ref{signal:tau} where we plot the distribution of the fractional parts and the factorization interference pattern for two numbers: $N'$ rich in factors and $N=N'-1$ rich in ghost factors. To emphasize the region of fractional parts which lead to ghost factors we use the logarithmic scale. Here we have chosen $N'=13335840=2^5\cdot 3^5\cdot 5\cdot 7^3$ which obviously has a lot of divisors, as depicted on the upper-left plot by the straight line of black diamonds. In the factorization interference pattern shown on the right these divisors correspond to a straight line of signals equal to unity. However, the divisors of $N'$ are non-factors of $N=N'-1=13335839=11\cdot 479\cdot 2531$. Moreover, they are aligned on a hyperbola (\ref{hyp}) and attracted to zero as shown in the lower-left plot where we can clearly identify the hyperbola of stars. Consequently, in the factorization interference pattern plotted on the right this hyperbola of arguments with small fractional parts (\ref{hyp}) translates into the curve of ghost factors, as depicted on the right.


\subsection{Optimality of the fourth-root law}

In Section~\ref{chap9:3} we have derived the fourth-root law (\ref{M-N}) as an upper bound on the truncation parameter. We will show that it is also necessary for the success of our factorization scheme.

The analysis of $g(N,M)$ revealed that it behaves similarly to the inverse power in $M$ (\ref{nbghost}). The closer the distribution of the fractional parts for a given $N$ the better the estimation (\ref{nbghost}) fits the actual data.

\begin{figure}
\begin{center}
\includegraphics[width=0.6\textwidth]{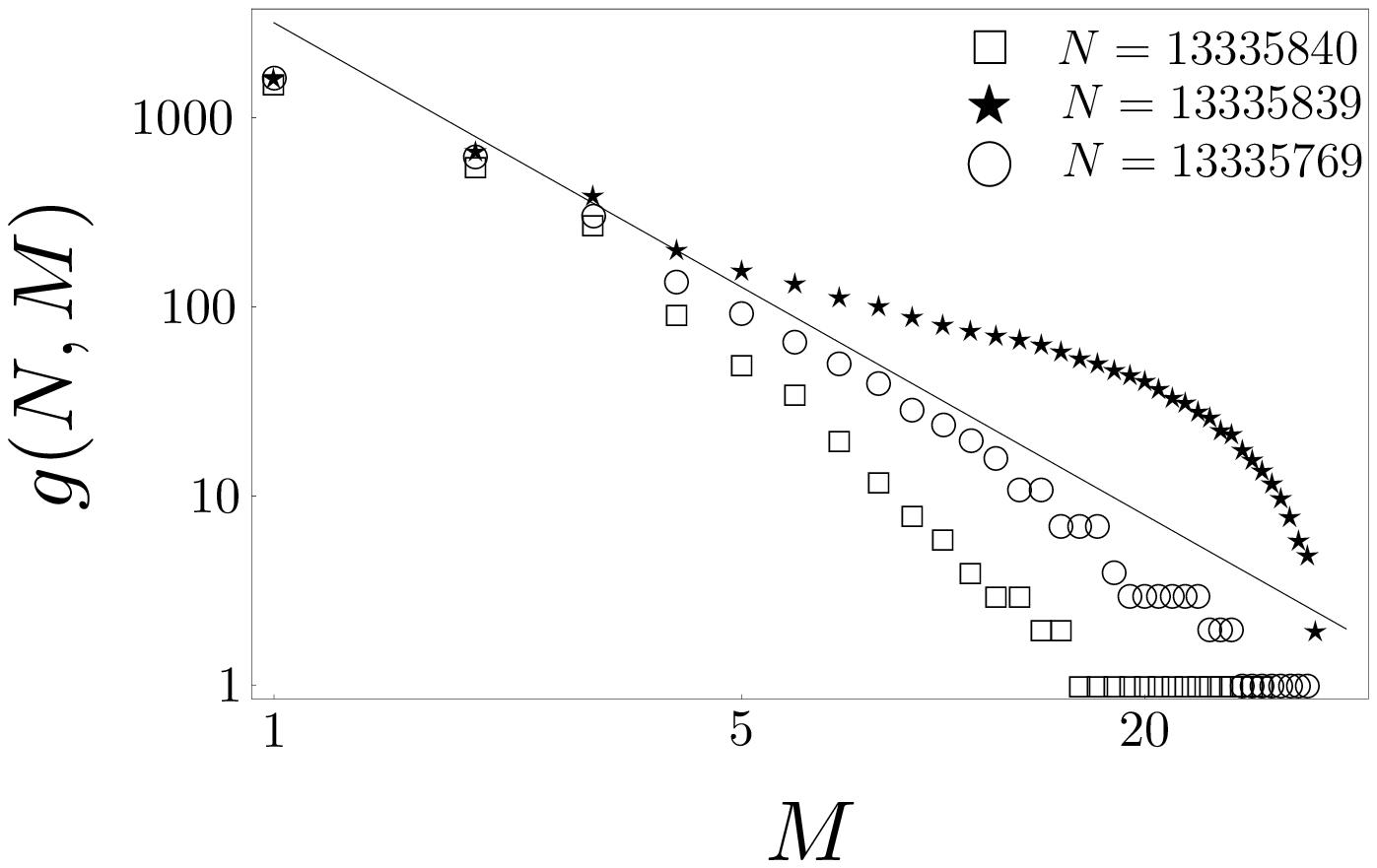}
\caption
{The number $g(N,M)$ of ghost factors expressed by the ghost factor counting function (\ref{gf:count}) as a function of the truncation parameter $M$ for three characteristic examples. We use a log-log plot to bring out the scaling of $g(N,M)$ with $M$. For the number $N=13335769$ the scaling $g(N,M)\sim M^{-2}$ predicted by (\ref{nbghost}) with the help of the Fresnel integral is satisfied. In contrast for $N=13335839$ which is rich in ghost factors we see a strong deviation. Finally, for $N=13335840$ which is poor in ghost factors due to the fact it has many divisors the ghost factor counting function $g(N,M)$ decays even faster than the estimation (\ref{nbghost}) predicts.}
\label{plaw}
\end{center}
\end{figure}

In Figure~\ref{plaw} we present the log-log plot of $g(N,M)$ as a function of the truncation parameter $M$ for three characteristic examples.  First, for the number $N=13335769$ which has the fractional parts $\rho(N,\ell)$ of $2N/\ell$ distributed almost uniformly we find that scaling $\sim M^{-2}$ predicted by (\ref{nbghost}) is obeyed. For $N=13335839$ we find strong deviations for larger values of $M$ due to the fact that the actual distribution of fractional parts is not uniform. Finally, for $N=13335840$ the ghost factor counting function $g(N,M)$ decays even faster than the estimation (\ref{nbghost}) predicts. Nevertheless, in all three cases the number of ghost factors drops down rapidly in the beginning.

The inverse power-law (\ref{nbghost}) suggests an alternative truncation of the Gauss sum when we tolerate a limited number of ghost factors, say $K$. Indeed, the power-law reduces the number of ghost factors considerably for small values of $M$. On the other hand, it has a long tail, which implies that we have to include many more terms in the Gauss sum in order to discriminate the last few ghost factors. However, this approach will not change the power law dependence of $M$, (\ref{M-N}), as the equation~(\ref{nbghost}) yields that
$$
M_K \approx \frac{\alpha}{\sqrt{2K}}\sqrt[4]{N}
$$
terms are required to achieve this goal. Let us point out that this results holds if we can approximate the distribution of the fractional parts by uniform distribution. However, as we have seen in \fig{plaw}, if this simplification is not feasible such $M_K$ might be even greater. Therefore we cannot achieve a better scaling on $N$ than  $\sqrt[4]{N}$, even if we tolerate a limited number of ghost factors.

We conclude that the scaling $M_0\sim\sqrt[4]{N}$ of the upper limit of the Gauss sum ${\cal A}_N^{(M)}$ provides both {\it sufficient} and {\it necessary} condition for the success of our factorization scheme. Using $M_0$ terms in the Gauss sum we can suppress {\it all} ghost factors for {\it any} number $N$. From the relation (\ref{resource}) we see that we need to add
$$
{\cal R}\sim \sqrt[4]{N}\cdot\sqrt{N}=N^{\frac{3}{4}}
$$
terms for the success of the factorization scheme based on the truncated Gauss sum. In comparison with the value of ${\cal R}\sim N$ of terms required for the complete Gauss sum (\ref{resource:complete}) we have gained a factor of fourth-root. We emphasize that we cannot reduce the amount of resources further.


\section{Conclusions}
\label{chap9:5}

We have analyzed the conditions required for the success of the factorization algorithm based on the truncated Gauss sums. Four distinct classes of candidate factors $\ell$ with respect to the number to be factorized $N$ have been identified. In particular, with the help of the normalized curlicue sum we have found a simple criterion for the most problematic class of ghost factors. The natural threshold of the signal value of the Gauss sum which can be employed to discriminate factors from non-factors was identified. We have derived the scaling law $M_0\sim\sqrt[4]{N}$ for the upper limit of the Gauss sum which guarantees that all ghost factors are suppressed, i.e. the signal values for all non-factors lie below the natural threshold. Unfortunately, we cannot achieve a more favorable scaling even if we change the threshold value or tolerate a limited amount of non-factors.

However, a generalization of Gauss sums to sums with phases of the form $m^j$ with $2<j$ might offer a way out of the fourth-root scaling law. Indeed, such a naive approach suggests the scaling law $M_0\sim\sqrt[2j]{N}$. For an exponential phase dependence $m^m$ we would finally achieve a logarithmic scaling law. However, these new phases bring in new thresholds and a more detailed analysis is needed. The answer to these questions is presented in the following Chapter~\ref{chap10}.

Moreover, the analysis of the non-uniform distribution of the fractional parts provides us with a new perspective on the ghost factors. So far we have treated them as problematic trial factors which might spoil the identification of factors from the factorization interference pattern. However, the fact that the ghost factors of $N$ are factors of numbers close to $N$ offers an interesting possibility -- by factorizing $N$ we can find candidate factors of numbers close to $N$. Indeed, as we have found in (\ref{gamma}) the factors of $N\pm k$ align on the curve $\gamma_k^{(M)}(\ell)$ in the factorization interference pattern of $N$. Hence, if we identify the data points lying on these curves we find candidate factors of $N\pm k$. However, to take advantage of this positive aspect of ghost factors we need a very good resolution of the experimental signal data.

\begin{figure}
\begin{center}
\includegraphics[width=0.7\textwidth]{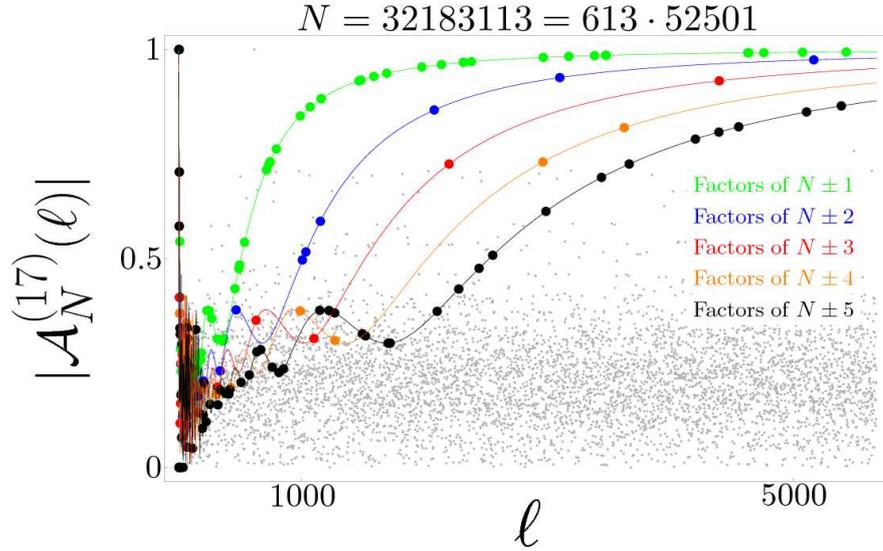}
\caption
{Factors of $N\pm k$ obtained from the ghost factors of the factorization interference pattern of $N = 32183113 = 613\cdot 52501$ with the truncation parameter $M = 17 \approx \ln{N}$. Such a choice of $M$ is clearly not sufficient to suppress all ghost factors. However, the remaining ghost factors can be fitted to the curves $\gamma_k^{(17)}(\ell)$ for $k=1,\ldots, 5$. Hence, we can identify candidate factors of numbers close to $N$, in our case up to $N\pm 5$.}
\label{usefulghost}
\end{center}
\end{figure}

We illustrate this feature in \fig{usefulghost} on the factorization interference pattern of $N = 32183113 = 613\cdot 52501$. Here we have chosen the truncation parameter according to $M\approx\ln{N}\approx 17$ which leads to an interference pattern with several ghost factors. However, we can clearly fit the ghost factors to curves $\gamma_k^{(17)}(\ell)$ for $k=1,\ldots,5$. Hence, by factorizing $N$ we also find candidate factors of $N\pm k$ with $k=1,\ldots,5$.

\chapter{Factorization with Exponential sums}
\label{chap10}

\nsection{Introduction}
\label{sec10:1}

In the present Chapter we extend the idea of factorization with the help
of Gauss sums by considering exponential sums. Here the phase is
proportional to $m^j$ where $m$ is the summation index and $j$ is an
integer. We show that in such a case the truncation depends on the
inverse of this function, i.e. $M\sim\sqrt[2j]{N}$. Hence, we can save experimental resources by employing rapidly increasing phase
functions. The extreme limit of an exponential sum where the phase
varies exponentially with the summation index, i.e. $m^m$, should
then be the optimal choice. We briefly address this case and demonstrate
by a numerical analysis that the truncation parameter depends only
logarithmically on the number to be factored.

It is interesting to note that recently an experiment \cite{suter2}
based on NMR has used an exponential sum with $j=5$ to factor
a 17-digit number consisting of two prime factors of the same order.
In this experiment $\pi$-pulses \cite{sargent} drive a two-level atom. By choosing the phases of the pulses appropriately we can achieve
a situation in which the resulting polarization is
determined by a truncated exponential sum with a particular choice of $j$.
Moreover, even the extreme case of an exponential phase $m^m$ can be realized
in this way.

We introduce exponential sums in Section \ref{sec10:2} and show that they allow us to
discriminate between factors and non-factors. In particular, we demonstrate by a numerical example that phases which increase as
$m^3$ suppress ghost factors more effectively than Gauss sums which have phases proportional to $m^2$. This feature is our motivation to
study the factorization properties of exponential sums. In Section~\ref{sec10:3} we have shown that for truncated Gauss sums the
influence of the truncation parameter $M$ depends crucially on the choice of trial factors. We have identified four classes: (i)
factors, which are not influenced by $M$, (ii) threshold trial factors, which are also independent of $M$, (iii) typical
non-factors, which decay very quickly, and (iv) ghost factors, which decay slowly. In Section \ref{sec10:3} we perform a similar analysis
for exponential sums. The numerical calculations of Section \ref{sec10:2} are confirmed in Section \ref{sec10:4} by an analytic argument. We show that the
number of terms which have to be summed in order to suppress the signal of all ghost factors depends on the $2j$-th root of the
number to be factored. For all exponential sums except the Fourier sum there exist non-factors for which the signal cannot be suppressed below certain thresholds by further increasing the truncation parameter. The values of these thresholds are determined by the power $j$ and can be close to the maximal signal of unity corresponding to a factor. In such a case we cannot achieve a sufficient contrast between the signals of factors and non-factors. We discuss the restrictions imposed by this fact on our factorization scheme in Section~\ref{sec10:5}. Our analysis indicates that rapidly increasing phases suppress ghost
factors most effectively. This feature suggests to consider the extreme case with the phase $m^m$. We briefly address this case in Section \ref{sec10:6} where we present numerical simulations indicating that the resources scale only logarithmically. However, in contrast to sums involving a fixed exponent, we no longer have the tools of number theory at hand to prove perfect discrimination of factors from non-factors. Nevertheless, in the Appendix~\ref{appendC} we demonstrate that the sum actually discriminates factors from non-factors. We summarize our results in the conclusions of Section \ref{sec10:7}.


\section{Factorization with exponential sums}
\label{sec10:2}

For our purpose to factorize numbers
we use truncated and normalized exponential sums of the type
\begin{equation}
{\cal A}_N^{(M,j)}(\ell) \equiv \frac{1}{M+1} \sum_{m=0}^M
\exp\left[2\pi i\, m^j\frac{N}{\ell}\right],
\label{eqn:Gausslike}
\end{equation}
where the phases are determined by the integer power $j$. Here $N$
is the number to be factored and $\ell$ is a trial factor which
scans through all integers between $1$ and $\lfloor\sqrt{N}\rfloor$.
In the experiments performed so far the upper bound $M$ in the sum
is equal to the number of pulses applied.

In the case of $j=1$ the exponential sum reduces to a Fourier sum.
For $j=2$ we find the truncated Gauss sum
\begin{equation}
{\cal A}_N^{(M)}(\ell)\equiv{\cal A}_N^{(M,2)}(\ell) = \frac{1}{M+1}
\sum_{m=0}^M \exp\left[2\pi i\, m^2\frac{N}{\ell}\right].
\label{eqn:Gauss}
\end{equation}
In the case of $j=3$ the sum
\begin{equation}
{\cal A}_N^{(M,3)}(\ell) = \frac{1}{M+1} \sum_{m=0}^M \exp\left[2\pi i\,
m^3\frac{N}{\ell}\right]
\label{kummer}
\end{equation}
is the truncated version of the {\it Kummer sum} named after the mathematician Ernst Kummer
(1810-1893).

The capability of the exponential sums, \eq{eqn:Gausslike}, to
factor numbers stems from the fact that for an integer factor $q$ of
$N$ with $N=q \cdot r$ all phases in ${\cal A}_N^{(M,j)}$ are
integer multiples of $2\pi$. Consequently, the terms add up
constructively and yield ${\cal A}_N^{(M,j)}(q)=1$. When $\ell$ is
not a factor the phases oscillate with $m$ and the signal $|{\cal
A}_N^{(M,j)}(\ell)|$ takes on small values. In order to factor a
number $N$ we analyze $|{\cal A}_N^{(M,j)}(\ell)|$ for arguments
$\ell$ out of the interval $[1,\sqrt{N}]$. We refer to the graphical
representation of the signal data as {\it factorization interference
pattern}.

\begin{figure}[p]
\centering
\includegraphics[width=0.65\textwidth]{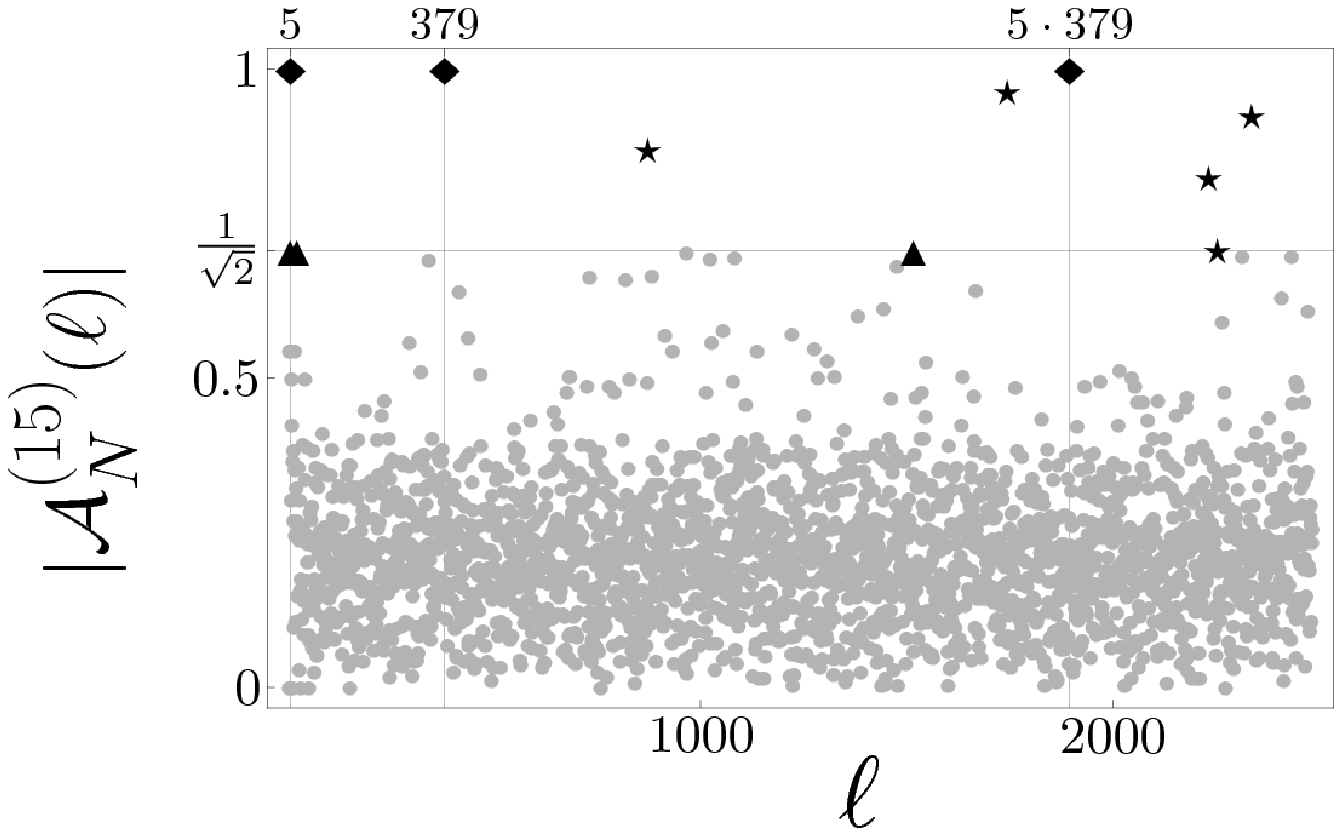}\vspace{24pt}
\includegraphics[width=0.65\textwidth]{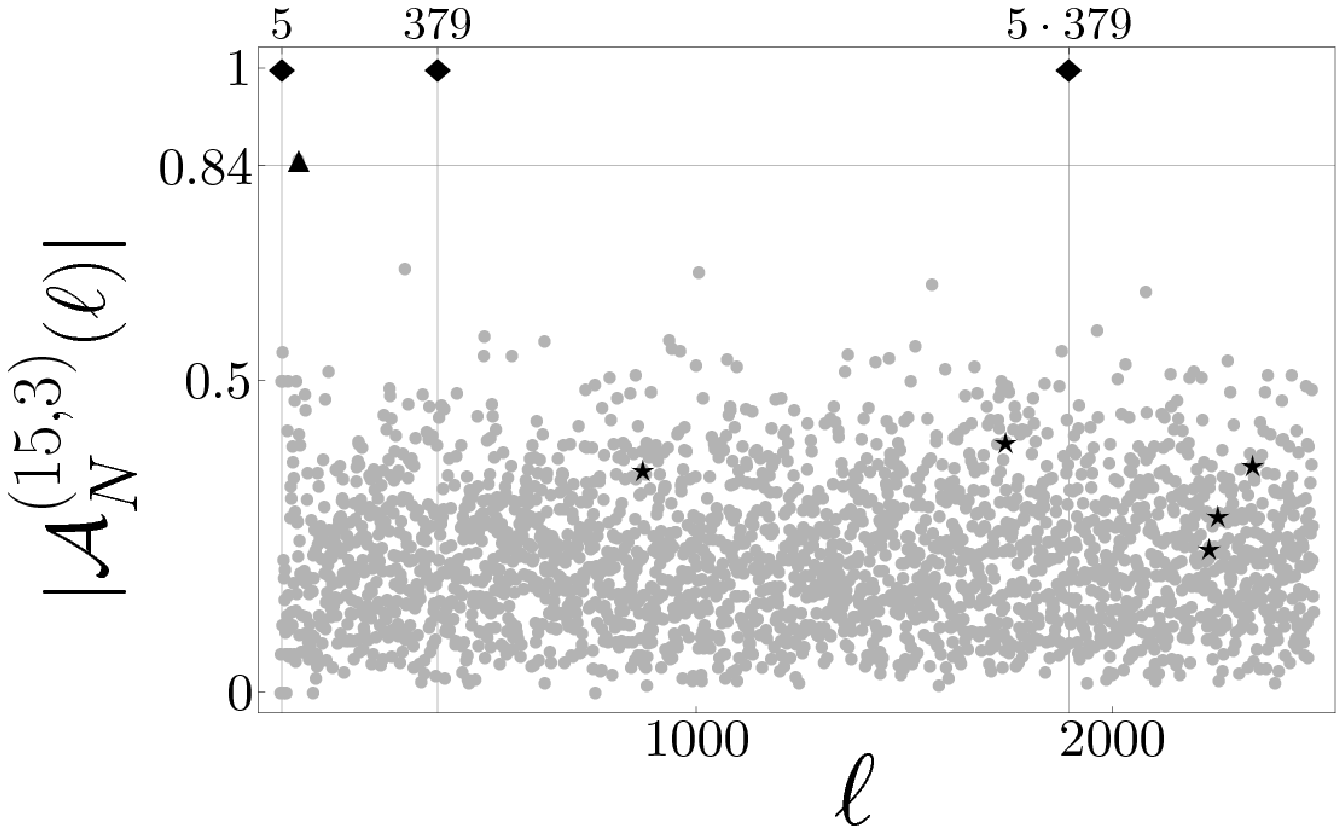}
\caption{ Factorization interference patterns of the number $N
=6172015 = 5\cdot 379\cdot 3257$ resulting from the Gauss sum (upper plot)
and the Kummer sum (lower plot). Here we have chosen the truncation
parameter $M\approx\ln{N}\approx 15$. The factors of $N$, depicted
by black diamonds, correspond to the signal value of unity. For most
of the non-factors, depicted by gray dots, the signal value is well
suppressed. However, in the case of Gauss sum we note that for a few
non-factors, depicted by stars, the signal is close to that of a
factor. Since such arguments can be misinterpreted as factors of $N$
we call them ghost factors. The presence of ghost factors in the
factorization interference pattern indicates that the choice of the
truncation parameter $M\approx\ln{N}$ is not sufficient for the
Gauss sum. However, the cubic phases in the Kummer sum grow faster
than the quadratic phases in the Gauss sum. As a result, the
truncation parameter $M=15$ is sufficient to suppress all ghost
factor. Moreover, some trial factors result in a threshold value of
the signal depicted by black triangles which cannot be suppressed by
further increasing the truncation parameter $M$. In the case of the
Gauss sum the threshold is $1/\sqrt{2}$ whereas for the Kummer sum
it has the value $~0.844$.}
\label{factpat}
\end{figure}

In \fig{factpat} we show the factorization interference patterns of
the number $N = 6172015 = 5\cdot 379\cdot 3257$ resulting from the
Gauss sum (left) and from the Kummer sum (right) for the choice of
the truncation parameter $M = 15\approx\ln{N}$. In both cases the
factors of $N$ lead to the maximal signal of unity depicted by black
diamonds. In contrast for most of the non-factors the signal
represented by gray dots is well suppressed. However, for the Gauss
sum there appear some non-factors, the so-called {\it ghost
factors}, where the signal indicated by black stars is still close
to that of a factor. We recognize that the corresponding
factorization pattern resulting from the Kummer sum does not display
any ghost factors. The origin of this positive feature lies in the
fact that the cubic phase of the Kummer sum shows a stronger
increase than the quadratic variation of the Gauss sum.

\section{Classification of trial factors}
\label{sec10:3}

In the preceding section we have shown using numerical examples that the influence of the truncation parameter of the exponential sums depends crucially on the choice of the trial factors. In the present section we analyze this feature in more detail and identify four classes of trial factors.

For this purpose we start from the decomposition of the fraction $N/\ell$ into an integer $k$ and the fractional part
$$
\rho(N,\ell) = \frac{N}{\ell}-k
$$
with $|\rho|\leq 1/2$. Indeed, the integer part contributes only as the multiplication by unity in \eq{eqn:Gausslike} and we find
$$
{\cal A}_N^{(M,j)}(\ell) = {\cal S}_j^{(M)}\left(\rho(N,\ell)\right)
$$
where we have introduced the sum
$$
{\cal S}_j^{(M)}(\rho)\equiv\frac{1}{M+1}\sum\limits_{m=0}^M \exp\left(2\pi i\, m^j\rho\right)
$$
This elementary analysis allows us to identify four classes of the fractional part. Indeed, we find in complete analogy to the Gauss sums \cite{opttrunc} : ({\it i}) for $\rho(N,\ell)=0$ the trial factor $\ell$ is a factor of $N$, ({\it ii}) for $|\rho(N,\ell)| = t_j$ the trial factor $\ell$ results in a threshold value $T_j$ of the exponential sum, where the values of $t_j$ and $T_j$ are determined by the power $j$, ({\it iii}) for $\rho(N,\ell)$ appropriately away from the origin the trial factor $\ell$ is a typical non-factor of $N$, ({\it iv}) for $\rho(N,\ell)\sim 0$ the trial factor $\ell$ is a ghost factor of $N$.

\begin{figure}[h]
\centering
\includegraphics[width=0.65\textwidth]{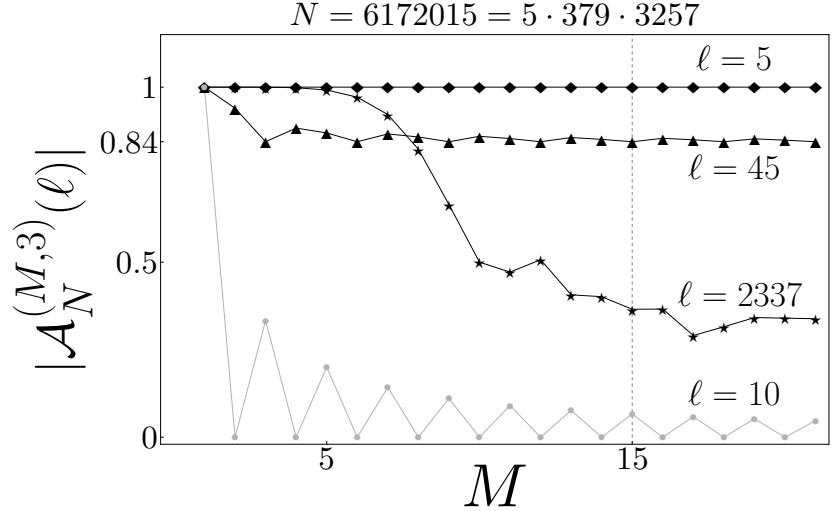}
\caption{Four classes of trial factors $\ell$ illustrated by the dependence of the Kummer sum $|{\cal A}_N^{(M,3)}(\ell)|$ on the truncation parameter $M$. In order to compare with \fig{factpat} where $M=15$ as indicated by a vertical dashed line we have chosen again $N=6172015 = 5\cdot 379\cdot 3257$. For factors of $N$, such as $\ell=5$ depicted by black diamonds, the signal is constant and equals to unity. For typical non-factors, such as $\ell=10$ depicted by gray dots, the signal is suppressed considerably already for small values of the truncation parameter $M$. However, for ghost factors, such as $\ell=2337$ depicted by black stars, more terms in the sum (\ref{kummer}) are needed to suppress the signal. Finally, for certain arguments, such as $\ell=45$ depicted by black triangles, the signal levels at non-vanishing threshold and it is impossible to suppress it further by increasing the truncation parameter $M$.}
\label{fig:arguments}
\end{figure}

We illustrate the different dependence of representatives of these classes on the truncation parameter $M$ in \fig{fig:arguments} using the example of the truncated Kummer sum (\ref{kummer}). We find signals which are independent of $M$ and equal to unity. They indicate factors. Moreover, we note, a rapid suppression of the signal for a typical non-factor. However, for a ghost factor the signal is close to that of a factor and we have to include more terms in the sum \eq{kummer} in order to suppress it. Moreover, we find that for certain trial factors $\ell$ the signal levels off at a non-zero threshold value and thus cannot be reduced at all.

\section{Scaling law of the truncation parameter}
\label{sec10:4}

In Section~\ref{sec10:2} we have shown that the ghost factors spoil the discrimination of factors from non-factors. Fortunately, we can suppress the signal of a ghost factor by increasing the truncation parameter $M$. In this context the truncated Gauss sums were analyzed in \cite{opttrunc} and it was shown that one needs $M\sim\sqrt[4]{N}$ terms in the sum in order to suppress the signal of all ghost factors considerably. We derive the corresponding scaling law $M_j\sim\sqrt[2j]{N}$ of an exponential sum ${\cal A}_N^{(M,j)}$. In \cite{opttrunc} the upper bound for the truncated Gauss sum \eq{eqn:Gauss} was obtained by approximating the Gauss sum by the Fresnel integral. We perform a similar analysis for the exponential sums.

Since ghost factors result from small values of the fractional part
$\rho\equiv N/\ell-k$ we replace the exponential sum by an integral,
i.e.
$$
{\cal A}_N^{(M,j)}(\ell)={\cal S}_j^{(M)}(\rho)\approx \frac{1}{M}\int\limits_0^M e^{2\pi i m^j\rho}dm.
$$
This approximation is justified by the van der Corput method
\cite{Kowalski} approximating sums by sums of shifted integrals.

With the help of the substitution $m^j\rho \equiv u^j$ and
$dm=du/\sqrt[j]{\rho}$ we find
$$
{\cal A}_N^{(M,j)}(\ell)\approx F_j(M\cdot\sqrt[j]{\rho})
$$
where
$$
F_j(x)\equiv\frac{1}{x}\int\limits_0^{x} e^{2\pi i u^j}du \: .
$$
This analysis brings out most clearly that for small fractional parts
$\rho$ the truncation parameter $M$ and $\rho$ appear in the
exponential sum only as the product $M\cdot\sqrt[j]{\rho}$.

In order to suppress the absolute value $|{\cal A}_N^{(M,j)}(\ell)|$
below a given value $\xi$ we have to choose the upper bound $M$
according to
$$
M\cdot\sqrt[j]{\rho}=\alpha
$$
where $\alpha$ is the solution of the integral equation
$$
|F_j(\alpha)| = \xi
$$
which leads us to the relation
$$
M=\alpha(\xi)\rho^{-\frac{1}{j}}.
$$

This result shows that the smaller the fractional part
$\rho(N,\ell)$ of the ghost factor $\ell$ the more terms are
required. Since the largest trial factor is of the order of
$\sqrt{N}$ the smallest attainable fractional part
$$
\rho_{\rm min}(N) \sim \frac{1}{\sqrt{N}}
$$
gives an upper bound
\begin{equation}
M_j\approx\alpha(\xi)\rho_{\rm min}^{-\frac{1}{j}} \approx \alpha(\xi)\sqrt[2j]{N}
\label{rule}
\end{equation}
on the truncation parameter
$M$.

Hence, in order to suppress all ghost factors of $N$  we require an order of $\sqrt[2j]{N}$ terms in the exponential sum ${\cal A}_N^{(M,j)}$. We point out that the scaling law (\ref{rule}) is inherent in the exponential sum since the change of $\xi$ only modifies the pre-factor $\alpha(\xi)$.

\begin{figure}[htbp]
\centering
\includegraphics[width=0.65\textwidth]{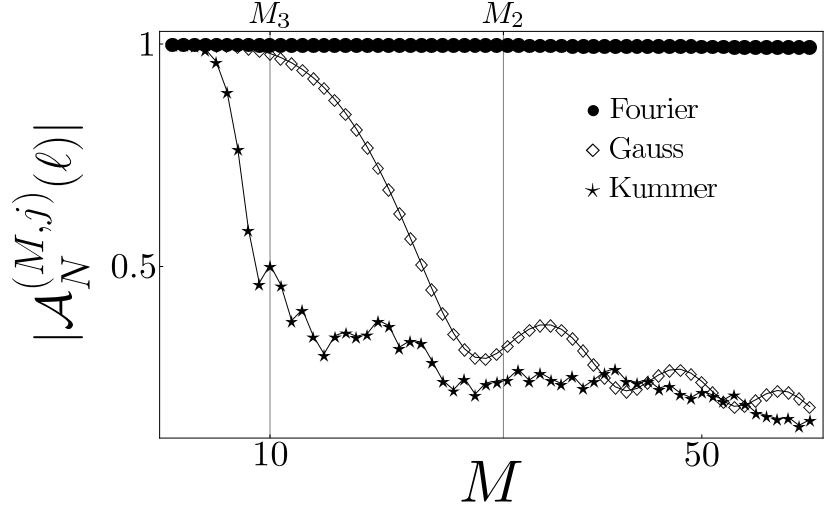}
\caption{Decay of the signal $|{\cal A}_N^{(M,j)}(\ell)|$ for increasing truncation parameter $M$ exemplified by the Fourier $(j=1)$, Gauss $(j=2)$ and Kummer $(j=3)$ sum. Here we have chosen $N=10^6+1$ and $\ell=10^3$ resulting in the fractional part $\rho(N,\ell)=10^{-3}\approx 1/\sqrt{N}$. For the Fourier sum (black dots) we find an extremely slow decay of the signal. On the other hand, for the Gauss sum (diamonds) already $M_2\sim\sqrt[4]{N}\approx 32$ terms are  sufficient to suppress the signal considerably. This requirement is further reduced for the Kummer sum (stars) to $M_3\sim\sqrt[6]{N}\approx10$. We find that our numerical results are in good agreement with the analytical estimate \eq{rule}.}
\label{fig:supp}
\end{figure}

In \fig{fig:supp} we illustrate the behaviour of $|{\cal A}_N^{(M,j)}(\ell)|$ for $N=10^6+1$ and $\ell=10^3$ resulting in the fractional part $\rho(N,\ell)=10^{-3}\approx 1/\sqrt{N}$ as a function of the truncation parameter $M$. We visualize the effect of the power $j$ on the suppression of $|{\cal A}_N^{(M,j)}(\ell)|$ by presenting three different curves: ({\it i}) black dots correspond to the Fourier sum with linear phases, ({\it ii}) diamonds represent the Gauss sum, and finally ({\it iii}) stars result from the Kummer sum with cubic phases. We find that for the Fourier sum the suppression of the signal is extremely slow. Indeed, according to the estimate \eq{rule} we need $M_1\sim\sqrt{N}\approx 10^3$ terms in order to suppress the signal considerably. On the other hand, for the Gauss sum already $M_2\sim\sqrt[4]{N}\approx 32$ terms suffice to reduce the signal, in agreement with \eq{rule}. Finally, for the Kummer sum the decay of the signal is even faster. We find that $M_3\sim\sqrt[6]{N}\approx10$ terms are sufficient to suppress the signal, in agreement with \eq{rule}.

\begin{figure}[htbp]
\centering
\includegraphics[width=0.65\textwidth]{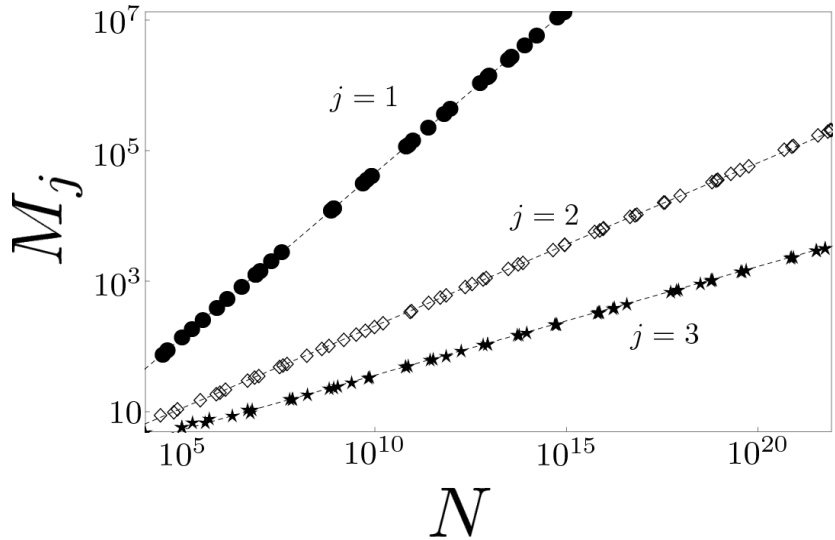}
\caption{Number $M_j$ of terms needed to suppress the signal of all ghost factors of $N$ below the value $0.7$ for the Fourier sum (black dots), Gauss sum (open diamonds) and Kummer sum (stars). To unravel the scaling of $M_j$ with $N$ we use a log-log scale. The dashed lines follow from the estimate $M_j\sim\sqrt[2j]{N}$ given by (\ref{rule}).}
\label{fig:supp2}
\end{figure}

In order to verify the scaling law (\ref{rule}) for a broad range of $N$ we have calculated numerically the truncation parameter $M_j$ needed to suppress all ghost factors of $N$ below the value $\xi$. We have chosen $N$ randomly from the interval $\left[10^4,10^{20}\right]$ and considered $\xi=0.7$. In \fig{fig:supp2} we present the results for the Fourier sum (black dots), Gauss sum (open diamonds) and Kummer sum (stars). To unravel the scaling law we use a logarithmic scale for both $N-$ and $M-$ axes. The numerical results are in excellent agreement with the estimates (\ref{rule}) indicated by the dashed lines.

\section{Threshold}
\label{sec10:5}

An experiment must also take into account the limited measurement accuracy. Thus for the success of our factorization scheme we need a good contrast between the signals of factor and non-factors, i.e. we require that the signals of all non-factors are suppressed below the estimated measurement error. However, due to the existence of the thresholds discussed in Section~\ref{sec10:3} this suppression might be impossible for certain powers $j$. In such a case we might misinterpret the signal arising from a non-factor as that of a factor. Hence, such exponential sums ${\cal A}_N^{(M,j)}$ are not suitable for integer factorization.

Relation (\ref{rule}) shows that the faster the phase grows the less terms in the exponential sum are needed in order to suppress the signal of a ghost factor argument $\ell$. However, the suppression of the signal might be impossible for all arguments $\ell$, as we have seen already in \fig{fig:arguments}. This feature is closely related to the power $j$ determining the phase.

The absolute value $|{\cal A}_N^{(M,j)}(\ell)|$ depends on how many different roots of unity we find in the sum. These roots of unity are given by
\begin{equation}
\exp{\left(2\pi i m^j \frac{N}{\ell}\right)}=\exp{\left(2\pi i m^j \rho(N,\ell)\right)}=\exp{\left(2\pi i m^j \frac{p}{q}\right)}
\label{phase:iden}
\end{equation}
where $p/q$ is the coprime rational representation of $\rho(N,\ell)$. This is equivalent to
$$
m^j\frac{N}{\ell} q \equiv 0,\ 1,\ \ldots,\ q-1\ \textrm{mod}\ q,
$$
i.e. the terms in the exponential sum $|{\cal A}_N^{(M,j)}(\ell)|$ attain at most $q$ different values.

For the Fourier sum we find all $q$ different roots $\exp{\left(2\pi i m/q\right)}$ with $m=0,\ldots,\ q-1$ of unity. Moreover, since they are distributed symmetrically on the unit circle they cancel each other out. Hence, for the Fourier sum we can suppress the signal $|{\cal A}_N^{(M,1)}|$ of any non-factor $\ell$ below any given value by extending the summation range $M$.

However, for exponential sums ${\cal A}_N^{(M,j)}$ with powers $2\leq j$ we are not guaranteed to find all different roots of unity. Moreover, since $j\neq 1$ the corresponding roots of unity $\exp{\left(2\pi i m^j p/q\right)}$ are not necessarily distributed symmetrically on a unit circle. Hence, they do not cancel themselves completely. In such a case the signal $|{\cal A}_N^{(M,j)}(\ell)|$ has a non-zero limit as $M$ tends to infinity. This limit value determines the threshold and depends on how many different roots of unity we find in the sum and their distribution on the unit circle. If we find only few different roots of unity which are moreover close to each other on the unit circle the signal $|{\cal A}_N^{(M,j)}(\ell)|$ attains values close to unity and cannot be suppressed further by increasing the truncation parameter $M$, even though $\ell$ does not correspond to a factor of $N$.

\begin{figure}[htbp]
\centering
\includegraphics[width=0.4\textwidth]{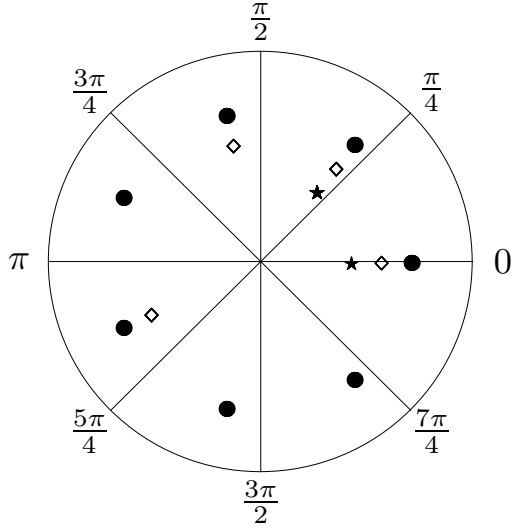}
\caption{The roots of unity contained in the exponential sums ${\cal A}_N^{(M,j)}(\ell)$ exemplified by the Fourier sum (j=1, black dots), the Gauss sum (j=2, open diamonds) and a higher order exponential sum (j=6, black stars). Here we have chosen $N=99$ and $\ell=7$ which leads to $\rho(N,\ell)=p/q=1/7$. For the Fourier sum we find all seven different roots of unity.  However, in the Gauss sum only four different roots of unity appear. This number is further reduced to just two different roots of unity in the higher order exponential sum with power $j=6$.}
\label{fig:phases}
\end{figure}


The fewest possible terms in the sum ${\cal A}_N^{(M,j)}$ for a non-factor $\ell$ occur if $j+1$ is the prime number $q$ from the rational representation of $\rho(N,\ell)$. In such a case we find from the Euler's Theorem (see e.g. Chapter 3 in \cite{Rosen})
$$
m^j \equiv
\left\lbrace{
\begin{tabular}{ccl}
$1$ & if & $q$ is not a divisor of $m$ \\
$0$ & if & $q$ is a divisor of $m$
\end{tabular}}\right.
$$
so $m^j\cdot p$ is either congruent to $p$ or $0$ mod $q$. With the help of the periodicity $m^j\cdot p\equiv (m+q)^j\cdot p\ \textrm{mod}\ q$ and the relation (\ref{phase:iden}) we obtain for $M+1$ being a multiple of $q$
\begin{eqnarray}
\nonumber {\cal A}_N^{(M,j)}(\ell) &  = & \frac{1}{M+1}\sum_{m=0}^M e^{2\pi i m^j\frac{N}{\ell}} = \frac{1}{q}\sum_{m=0}^{q-1} e^{2\pi im^j\frac{p}{q}}\\
\nonumber & = & \frac{1}{q}\left(1+(q-1)e^{2\pi i\frac{p}{q}}\right).
\end{eqnarray}
Hence we find for the absolute value squared
$$
|{\cal A}_N^{(M,j)}(\ell)|^2=\frac{1}{q^2}((1+(q-1)\cos(\frac{2\pi p}{q}))^2+(q-1)^2\sin^2(\frac{2\pi p}{q})).
$$
Substituting $q=j+1$ we find for $p=1$ the threshold value of the sum ${\cal A}_N^{(M,j)}$
$$
T_1(j) = \frac{1}{j+1}\sqrt{j^2+1+2j\cos{\left(\frac{2\pi}{j+1}\right)}}.
$$
For $p>1$ or for more than two different terms in the sum ${\cal A}_N^{(M,j)}$ the threshold will always be smaller.

\begin{figure}[htbp]
\centering
\includegraphics[width=0.7\textwidth]{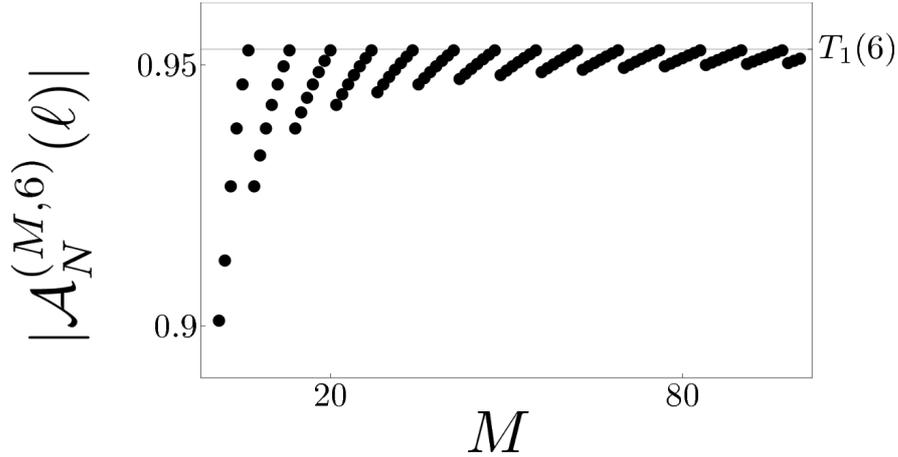}
\caption{Emergence of the threshold for the exponential sum ${\cal A}^{(M,j)}_N$ with the power $j=6$ for increasing truncation parameter $M$. We have chosen $N=99$ and $\ell=7$ resulting in $\rho(N,\ell)=p/q=1/7$. The signal converges to the value of $T_1(6)\approx 0.953$ and cannot be suppressed by a further increase of $M$.}
\label{fig:thresh}
\end{figure}

To illustrate this we plot in \fig{fig:thresh} the behaviour of the signal $|{\cal A}_N^{(M,6)}(\ell)|$ as a function of the truncation parameter $M$. Here we have chosen $N=99$ and $\ell=7$ resulting in $\rho(N,\ell)=p/q=1/7$. Hence, $q=7=1\cdot 6+1$ and we find that the signal converges to the threshold value $T_1(6)\approx 0.953$.

More generally, for prime denominator $q=k\cdot j+1$ the sum ${\cal A}_N^{(M,j)}$ contains at most $k+1$ different terms. For the case of $k=2$ an analogous calculation results in the threshold value
$$
T_2(j) = \frac{1}{2j+1}\left(1+2j\cos{\left(\frac{2\pi}{2j+1}\right)}\right).
$$
Obviously, for large powers $j$ the values of $T_{1,2}(j)$ are very close to one.

The above derived results indicate that the exponential sums ${\cal A}_N^{(M,j)}$ with powers $j$ larger than two can be used for integer factorization only when the experimental data are sufficiently precise. For the Fourier sum the signal for any non-factor can be suppressed below any given value. However, according to \eq{rule} we have to include a number of terms of the order of the square-root of $N$ to achieve this suppression. The quadratic Gauss sum of \eq{eqn:Gauss} provides a reasonable compromise between the number of terms needed and the non-factor discrimination. The gap between the signal of a factor and the greatest threshold is approximately 30$\%$ which should be sufficient for the experimental realization. The number of terms in the sum needed is according to \cite{opttrunc} reduced to the fourth-root of $N$.


\section{Factorization with an exponential phase}
\label{sec10:6}

One way to improve the scaling law might be offered by an exponential sum where the phase is not governed by a polynomial as in (\ref{eqn:Gausslike}) but by an exponential function. This idea leads to the sum
$$
{\cal E}_N^{(M)}(\ell) \equiv\frac{1}{M+1}\sum_{m=0}^M \exp\left[2\pi i m^m \frac{N}{\ell}\right].
$$

\begin{figure}[h]
\centering
\includegraphics[width=0.4\textwidth]{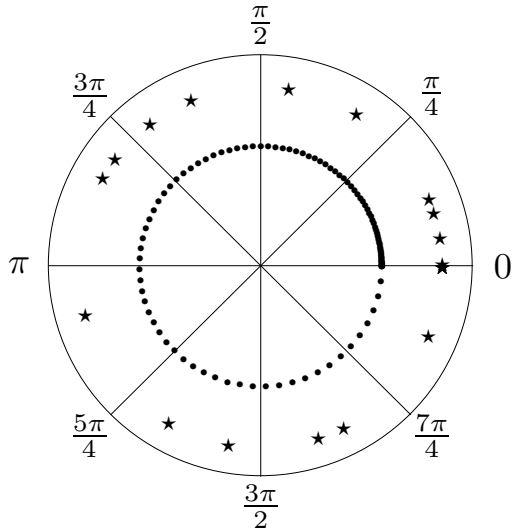}
\caption{Distribution of the roots $e^{2\pi i m^2 p/q}$ (dots) and $e^{2\pi i m^m p/q}$ (stars) of unity for quadratic and exponential phase, respectively. Here we have chosen $p=1$ and $q=10^4$. Since the fraction $p/q$ is small we observe an accumulation of the roots for small values of $m$ in the case of the quadratic phase.}
\label{fig:phases2}
\end{figure}

We present a numerical analysis which confirms a logarithmic scaling law. In Section \ref{sec10:4} we have found that the number of $M_j$ terms needed to suppress all ghost factors for the exponential sum ${\cal A}_N^{(M,j)}$ scales like $M_j\sim\sqrt[2j]{N}$, i.e. $M_j$ is determined by the inverse function of the phase evaluated at $\sqrt{N}$. This feature arises from the fact that the rising exponent prevents the function from accumulating values near unity for small arguments $m$, as we illustrate in \fig{fig:phases2}.

\begin{figure}[h]
\centering
\includegraphics[width=0.7\textwidth]{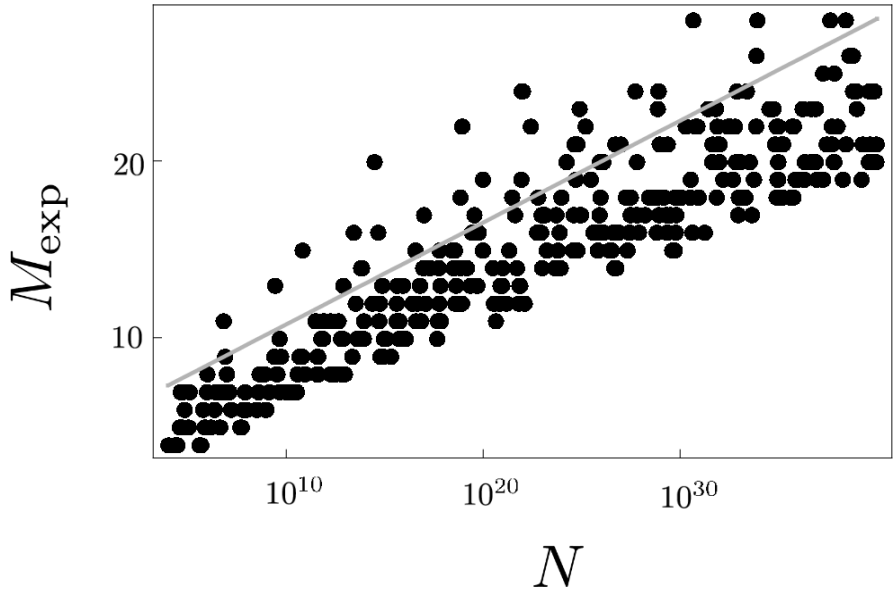}
\caption{Number $M_{\rm{exp}}$ of terms needed to suppress the signal $|{\cal E}_N^{(M)}|$ of all non-factors of $N$ below the value $0.7$. To unravel the scaling of $M$ we use a logarithmic scale for $N$. The gray line represents the estimate $M\sim\ln\sqrt{N}$. The plot indicates that already an order of $\ln{\sqrt{N}}$ terms in the exponential sum ${\cal E}_N^{(M)}$ is sufficient to find all factors of $N$.}
\label{scaling:exp}
\end{figure}

This result suggests that for the exponential sum ${\cal E}_N^{(M)}$ already a logarithmic number of terms $M_{\rm{exp}}\sim\ln \sqrt{N}$ should be sufficient to eliminate all ghost factors. Moreover, our numerical analysis summarized in \fig{scaling:exp} indicates that the largest threshold for ${\cal E}_N^{(M)}$ occurs around the value $0.5$. Hence, we can achieve perfect discrimination of factors from non-factors.

However, in contrast to sums involving a fixed exponent, we no longer have the tools of number theory at hand to prove perfect discrimination of factors from non-factors. Moreover, since the derivative of $m^m$ grows faster then $m^m$ itself, standard techniques to approximate these exponential sums by integrals cannot be applied. Nevertheless, in the Appendix~\ref{appendC} we demonstrate that it is still possible to show that the sum actually discriminates factors from non-factors by methods of elementary number theory (see \cite{Rosen} for example).


\section{Conclusions}
\label{sec10:7}

In the present Chapter we have extended the idea of factorization with Gauss sums to exponential sums where the phase is governed by a power $j$ of the summation index. These sums are also capable of non-factor discrimination in complete analogy to Gauss sums. However, the truncation parameter $M_j$ needed to achieve a significant suppression of ghost factors of the number $N$ scales like $M_j\sim \sqrt[2j]{N}$ . Hence, we can save experimental resources by employing exponential sums with large powers $j$. On the other hand the gap between the signal of a factor and the greatest threshold value shrinks as $j$ grows. Therefore, exponential sums with large values of $j$ can be used for integer factorization only if the expected imperfections in the experiment are smaller than this gap.

We have also presented numerical simulations of factoring numbers using an exponential sum with exponentially increasing phases. Here the resources scale only logarithmically. Moreover, our results indicate that the gap survives.

Our results also show a connection to two recent experiments \cite{suter2,girard3} which factored a 13-digit and a 17-digit number using a Monte-Carlo sampling technique of a complete Gauss sum. This method accepts a small fraction of ghost factors and achieves a logarithmic scaling very much in the spirit of the exponential phase.

It is interesting to compare and contrast these two approaches. Ghost factors arise from the addition of neighbouring phase factors which only deviate slightly from each other. However, when many terms are added the phase factors are distributed homogeneously on the unit circle. The Monte-Carlo technique does not add up consecutive terms but tries to collect those terms which almost cancel each other. On the other hand, the exponential phase guarantees that neighbouring phase factors deviate significantly from each other and no ghost factors can arise. This feature leads to the logarithmic scaling.


\nchapter{Conclusions}

Various schemes for factorization of numbers based on exponential sums have been developed recently. Their relative simplicity when compared to the celebrated Shor's algorithm results in several advantages for the experimental realizations. First of all, exponential sums can be easily implemented in various physical systems. Moreover, thanks to the sufficiently long coherence times larger numbers can be factorized.

In the previous Chapters we presented the necessary conditions for the success of the factorization schemes based on exponential sums. We found that the number of terms in the sum, which directly translates to the number of pulses in the experiment, needed for the suppression of all ghost factors is determined by the inverse of the function determining the growth of the phase evaluated at the square-root of the number to be factorized. The exponential sums with rapidly growing phases are therefore more suitable for the suppression of ghost factors. On the other hand, the non-factors resulting in the threshold signal values become a significant problem. In general, with the faster growing phase the thresholds appear closer to the maximal signal of unity corresponding to the factor. Hence, for the successful physical implementation of the exponential sums algorithm for factorization of numbers one has to guarantee sufficient resolution of the measured signal. The quadratic Gauss sum analyzed in Chapter~\ref{chap9} provides a reasonable compromise between the number of terms needed and the non-factor discrimination. Most of the experiments performed to date benefited from this fact.

Needless to say, the simplicity of the analyzed schemes follows from the fact that they do not employ entanglement which is the key for the exponential speed-up of the Shor's algorithm over the classical ones. Indeed, factorization of numbers based on exponential sums relies only on interference. The resources scale exponentially like in the case of all known classical algorithms. To improve this scaling law by involving entanglement is our next goal.


\begin{appendices}

\chapter{Determination of Threshold}
\label{appendA}

In this Appendix we show that for non-zero positive rational $\tau=p/q$ the absolute value of the normalized curlicue sum is asymptotically bound from above by $1/\sqrt{2}$. This property follows immediately from \cite{berry:curlicues:1:1988}. Indeed, as shown in \cite{berry:curlicues:1:1988} the asymptotic behaviour of the curlicue sum
$$
{\cal C}_M(\tau) = \sum\limits_{m=0}^M \exp\left(i \pi\, m^2 \tau \right)
$$
for rational $\tau=p/q$ depends on the product $p\cdot q$. We find that for $p\cdot q$ being odd the curlicue is bounded. In such a case the absolute value of the normalized curlicue sum $s_M(\tau)$ decays with increasing $M$ like $|s_M(\tau)|\sim M^{-1}$. On the other hand for $p\cdot q$ being even the curlicue is unbounded and its growth can be approximated by
$$
|C_M(\tau)|\approx (M +1) (\tau_0\cdot\tau_1\cdot\ldots\cdot\tau_{\mu-1})^{1/2}
$$
where
\begin{equation}
\tau_j = (1/\tau_{j-1}) \rm{\ mod\ } 1\quad  \rm{if}\quad  \tau_{j-1}\neq 0
\label{appendA:rec}
\end{equation}
belongs to the $j$-th step in the repeating curlicue pattern \cite{berry:curlicues:1:1988} with $\tau_0=\tau$. Consequently, the limit of the absolute value of the normalized curlicue sum is non-vanishing and for large $M$ we can approximate $|s_M(\tau)|$ by the finite product
$$
|s_M(\tau)|\approx (\tau_0\cdot\tau_1\cdot\ldots\cdot\tau_{\mu-1})^{1/2}
$$

\begin{figure}
\begin{center}
\includegraphics[width=0.7\textwidth]{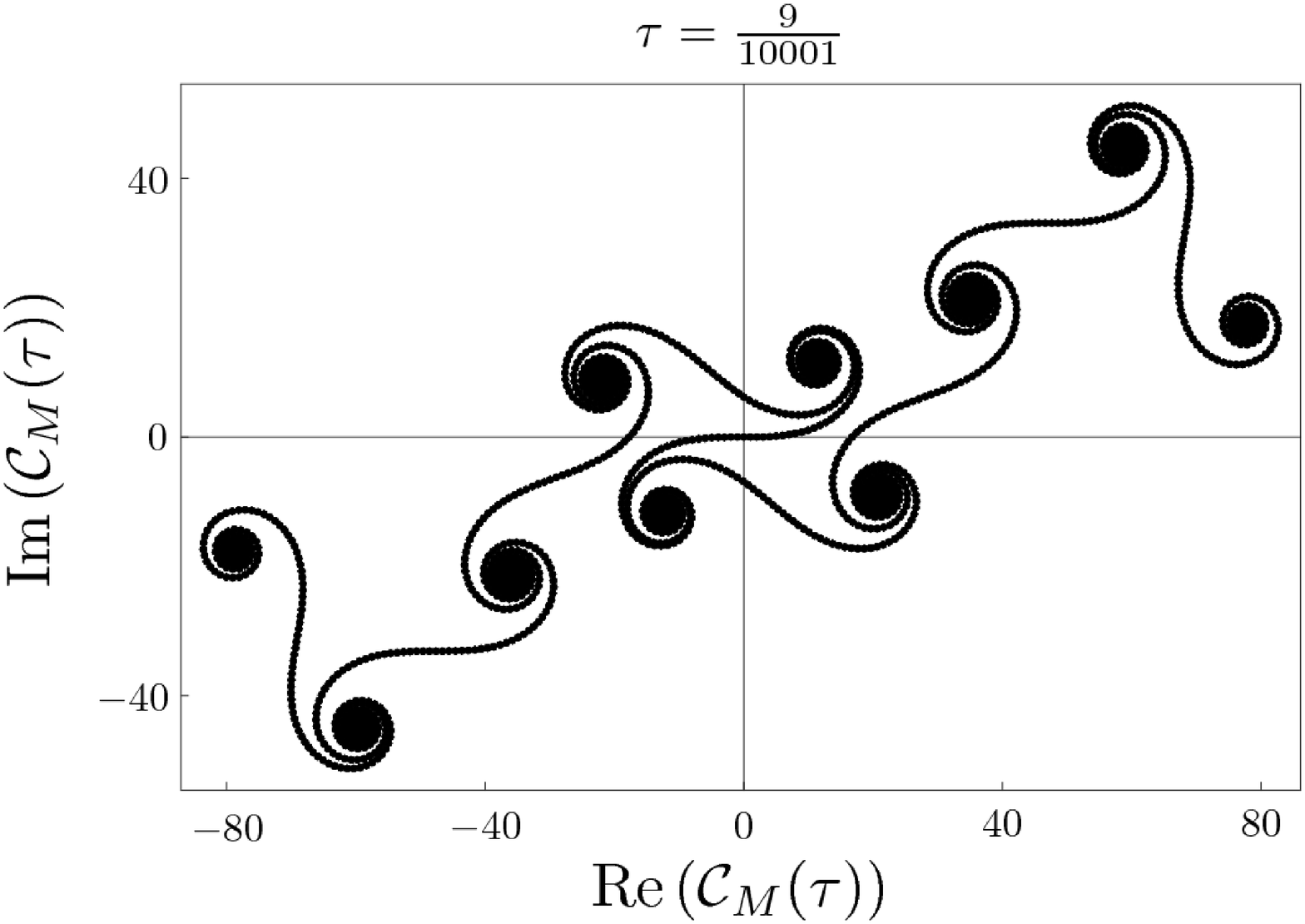}\vspace{60pt}
\includegraphics[width=0.8\textwidth]{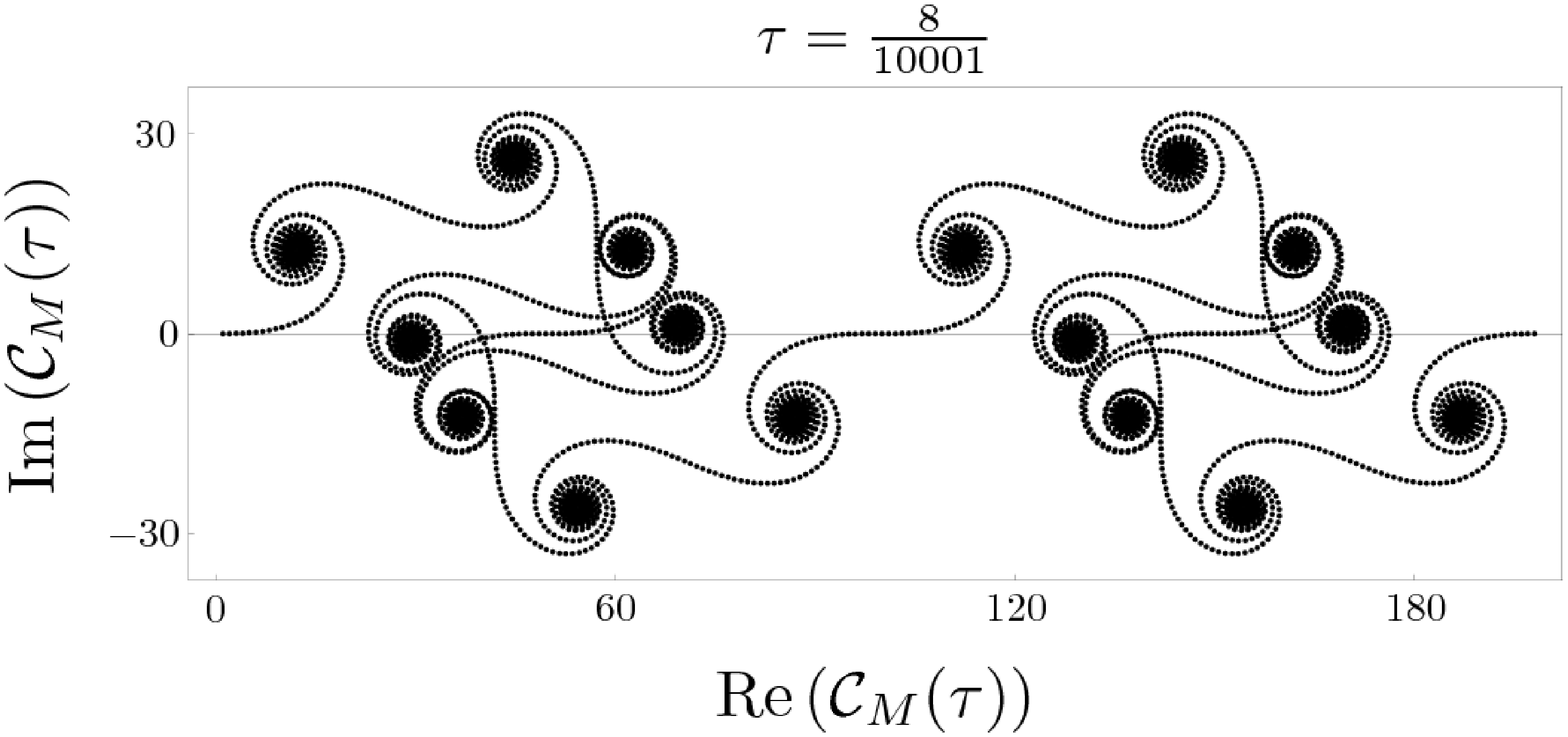}
\caption
{The behaviour of the curlicue sum ${\cal C}_M(p/q)$ in dependence on the parity of the product $p\cdot q$. In the upper plot we choose $\tau = p/q = 9/10001$ for which the product $p\cdot q$ is odd. We find that the curlicue repeats itself and is bounded. As a second example we choose $\tau = p/q = 8/10001$ where the product $p\cdot q$ is even. In such a case the curlicue expands, as depicted in the lower plot.}
\label{appendA:curl}
\end{center}
\end{figure}

We illustrate this feature in \fig{appendA:curl} where we show two different curlicues $C_M(\tau)$. In the upper plot we choose $\tau = \frac{9}{10001}$ for which the product $p\cdot q$ is odd. In such a case the function $C_M(\tau)$ is periodic in $M$ and the curlicue is bounded. On the other hand for $\tau = \frac{8}{10001}$ the curlicue depicted in the lower plot is unbounded and its ultimate growth is linear.

Let us determine the asymptotic bound of $|s_M(\tau)|$. The recursion (\ref{appendA:rec}) terminates \cite{berry:curlicues:1:1988} at $\tau_\mu=0$ which implies $\tau_{\mu-1}=1/b$ where $b$ is a natural number. Since $\tau_j< 1$ we find the estimate
$$
|s_M(\tau)|\leq\sqrt{\tau_{\mu-1}}=1/\sqrt{b}.
$$
The case $b=1$ cannot be produced by the recursion formula since all $\tau_j$ are strictly less than one. As a consequence the absolute value of the normalized curlicue sum $|s_M(\tau)|$ is asymptotically bound from above by $1/\sqrt{2}$.


\chapter{Applicability of the Fresnel approximation}
\label{appendB}

Let us comment the range of applicability of the continuous approximation (\ref{fresnel}). The scaling law $M_0\sim \sqrt[4]{N}$  connecting the number to be factored with the truncation parameter $M_0$ necessary to push all ghost factors below the threshold $1/\sqrt{2}$ relies on the approximation of the normalized curlicue function by the Fresnel integral. For large values of $N$ the scaling law requires large values of $M_0$. However, for large $M$ the continuous approximation might not hold any more.

For the continuous approximation to hold the phase difference
$$
\pi  \left((m+1)^2-m^2\right)\tau=\pi (2m+1)\tau
$$
of two successive terms in the sum (\ref{curl}) should at most be of the order of $\pi$. Together with the fact that the maximal phase difference appears for $m=M$ we arrive at the inequality
$$
\tau (2M+1) < 1.
$$
Indeed, this condition is violated for sufficiently large $M$.

When we recall that for a given $N$ the smallest fractional part is $\tau_{\rm min}=1/\sqrt{N}$ we arrive at the constraint
$$
M_c \approx \frac{1}{4}\sqrt{N}.
$$
on the maximal value $M_c$ of the truncation parameter for a given $N$. Thus $M_c\sim\sqrt{N}$ provides an upper bound on the validity of the Fresnel approximation, (\ref{fresnel}). Since $M_0\sim\sqrt[4]{N}$ the Fresnel approximation is valid.


\chapter{Discrimination Property for Variable Exponents}
\label{appendC}

In this Appendix we prove that the exponential sums with exponential phase allows us to distinguish factors from non-factors of a given number. The discrimination property of the exponential sums with a fixed exponent rests on the fact that only
for integer values of $l$ which are factors of $N$, the sum can take the value unity.
There is a number theoretical argument supporting this fact, as long as the exponent $j$
in the sum (\ref{eqn:Gausslike}) is fixed. This feature comes from the distribution of the values $\exp(2\pi i m^j \frac{N}{\ell})$
on the unit circle. For fixed $j$, it is impossible to hit the same point twice as $m$ increases provided we use a truncation parameter $M$ below $\sqrt[2j]{N}$. However, for a variable power $m^m$ that is an exponential phase, this non-recurrence property
is not obvious. In this case we need to prove the discrimination property explicitly.

The value $\exp(2\pi i m^m\frac{N}{\ell})$ depends on the fractional part of $m^m\frac{N}{\ell}$ only.
We hit the same point twice for different values $m$ and $n$ if and only if
\begin{equation}
m^m\frac{N}{\ell}-n^n\frac{N}{\ell}=k
\label{rec}
\end{equation}
where $k$ is an integer.

As in (\ref{phase:iden}) we make use of the coprime rational representation of $\rho(N,\ell)=p/q$ and find that the phase factor
$$
\exp\left(2\pi i m^m\frac{N}{\ell}\right) = \exp\left(2\pi im^m\rho(N,\ell)\right) = \exp\left(2\pi i \frac{pm^m}{q}\right)
$$
is a $q$-th root of unity. In particular, it is the $(pm^m)$-th one if we enumerate them
counter-clockwise starting from the zeroth root $1=\exp(2\pi i \frac{0}{q})$.
Note that $c$-th and $d$-th roots coincide if and only if $q$ is a divisor of $c-d$.

So the discrimination property depends on the fact, that there are values $c$ and $d$
such that
$$
q\;\textrm{is not a divisor of}\; pc^c-pd^d \: .
$$
The discrimination threshold does not depend only on the number of such pairs, but also on the position of the corresponding roots
of unity. Opposite roots of unity eliminate themselves in the sum, so the worst case
occurs if these roots accumulate on the same position.

We consider two cases: for large $q$, the first numbers in the sequence $pm^m$ will be below $q$,
so any pair chosen from the beginning of the sequence cannot fulfill the recurrence condition (\ref{rec}),
so they correspond to pairwise distinct roots. As a consequence, the absolute value of the sum cannot
assume the value unity.

For small $q$, we show that the $p$-th root
$\exp(2\pi i \frac{p}{q})$ and its conjugate $\exp(-2\pi i \frac{p}{q})$
appear in the sum, which leads to the elimination of their imaginary parts.
According to Euler's Theorem \cite{Rosen}
there is an even $m$ such that $pm^m$ corresponds to the first root $\exp(2\pi i \frac{1}{q})$ and
$j=m/2$ gives $pj^j$, which corresponds to the
$(-1)$-root $\exp(-2\pi i \frac{1}{q})=\exp(2\pi i \frac{q-1}{q})$.
The sum of this conjugate pair is a real number strictly below unity.

\end{appendices}





\begin{thebibliography}{x}

\addcontentsline{toc}{chapter}{References} \setlinespacing{1.2}

\nsection{List of Author's Publications}

\bibitem[I]{stef:prl}
M. \v Stefa\v n\'ak, I. Jex and T. Kiss, Phys. Rev. Lett. \textbf{100}, 020501 (2008)

\bibitem[II]{kiss:recurrence}
T. Kiss, L. Kecsk\'es, M. \v Stefa\v n\'ak and I. Jex, Phys. Scripta T \textbf{135}, 014055 (2009)

\bibitem[III]{stef:pra}
M. \v Stefa\v n\'ak, T. Kiss and I. Jex, Phys. Rev. A \textbf{78}, 032306 (2008)

\bibitem[IV]{stef:njp}
M. \v Stefa\v n\'ak, T. Kiss and I. Jex, New J. Phys. \textbf{11}, 043027 (2009)

\bibitem[V]{stef:meeting}
M. \v Stefa\v n\'ak, T. Kiss, I. Jex and B. Mohring, J. Phys. A \textbf{39}, 14965 (2006)

\bibitem[VI]{opttrunc}
M. \v Stefa\v n\'ak, W. Merkel, W. P. Schleich, D. Haase and H. Maier, New J. Phys. \textbf{9}, 370 (2007)

\bibitem[VII]{stef:exp:sum}
M. \v Stefa\v n\'ak, D. Haase, W. Merkel, M. S. Zubairy and W. P. Schleich, J. Phys. A \textbf{41}, 304024 (2008)

\newpage


\nsection{References}

\bibitem{pearson}
K. Pearson, Nature \textbf{72}, 294 (1905)

\bibitem{brown}
R. Brown, Phil. Mag. \textbf{4}, 161 (1828)

\bibitem{einstein}
A. Einstein, Ann. Phys. (Leipzig) 17, 549 (1905); 19, 371 (1906)

\bibitem{smoluchowski}
M. Smoluchowski, Ann. Phys. (Leipzig) 21, 756 (1906)

\bibitem{overview}
N. Guillotin-Plantard and R. Schott, {\it Dynamic Random Walks: Theory and Application}, Elsevier, Amsterdam (2006)

\bibitem{rw:compsc1}
C. Papadimitriou, {\it Computational Complexity}, Addison Wesley, Reading (1994)

\bibitem{rw:compsc2}
R. Motwani and P. Raghavan, {\it Randomized Algorithms}, Cambridge University Press, Cambridge (1995)

\bibitem{graph:connect}
A. Sinclair, {\it Algorithms for Random Generation and Counting, a Markov Chain Approach}, Birkhauser Press, Boston (1993)

\bibitem{3-sat}
U. Sch\"oning, 40th Annual Symposium on Foundations of Computer Science, IEEE, New York, 17 (1999)

\bibitem{matrix:perm}
M. Jerrum, A. Sinclair and E. Vigoda, in Proceedings of the 33th STOC, New York, 712 (2001)

\bibitem{aharonov}
Y. Aharonov, L. Davidovich and N. Zagury, Phys. Rev. A \textbf{48}, 1687 (1993)

\bibitem{meyer1}
D. Meyer, J. Stat. Phys. \textbf{85}, 551 (1996)

\bibitem{meyer2}
D. Meyer, Phys. Lett. A \textbf{223}, 337 (1996)

\bibitem{watrous}
J. Watrous, J. Comput. Syst. Sci. \textbf{62}, 376 (2001)

\bibitem{farhi}
E. Farhi and S Gutmann, Phys. Rev. A \textbf{58}, 915 (1998)

\bibitem{childs}
A. Childs, E. Farhi and S. Gutmann, Quantum Inf. Process. \textbf{1}, 35 (2002)

\bibitem{feynman}
R. P. Feynman and A. R. Hibbs, {\em Quantum Mechanics and Path Integrals}, International Series in Pure and Applied Physics, McGraw-Hill, New York (1965)

\bibitem{birula}
I. Bialynicki-Birula, Phys. Rev. D \textbf{49}, 6920 (1994)

\bibitem{hillery:2003}
M. Hillery, J. Bergou and E. Feldman, Phys. Rev. A \textbf{68}, 032314 (2003)

\bibitem{hillery:2004}
E. Feldman and M. Hillery, Phys. Lett. A \textbf{324}, 277 (2004)

\bibitem{kosik:2005}
J. Ko\v s\'{\i}k and V. Bu\v zek, Phys. Rev. A \textbf{71}, 012306 (2005)

\bibitem{hillery:2007}
E. Feldman and M. Hillery, J. Phys. A \textbf{40}, 11343 (2007)

\bibitem{strauch}
F. W. Strauch, Phys. Rev. A \textbf{74}, 030301 (2006)

\bibitem{chandra:08}
C. M. Chandrashekar, Phys. Rev. A \textbf{78}, 052309 (2008)

\bibitem{childs:09}
A. M. Childs, Phys. Rev. Lett. \textbf{102}, 180501 (2009)

\bibitem{lovett:09}
N. B. Lovett, S. Cooper, M. Everitt, M. Trevers and V. Kendon, {\it pre-print} arXiv:0910.1024 (2009)

\bibitem{bruss:leuchs}
D. Bru\ss\  and G. Leuchs (Eds.), {\it Lectures on Quantum Information}, Wiley-VCH, Berlin (2006)

\bibitem{kempe:ovw}
J. Kempe, Contemp. Phys. \textbf{44}, 307 (2003)

\bibitem{Venegas-Andraca}
S. E. Venegas-Andraca, {\em Quantum Walks for Computer Scientists}, Morgan and Claypool (2008)

\bibitem{konno:book}
N. Konno, {\em Quantum Walks}, in {\em Quantum Potential Theory}, Eds. U. Franz and M. Sch\"urmann, Lecture Notes in Mathematics \textbf{1954}, pp. 309-452, Springer-Verlag, Heidelberg (2008)

\bibitem{muelken:prl}
O. M\"ulken, A. Blumen, T. Amthor, Ch. Giese, M. Reetz-Lamour and M Weidemueller, Phys. Rev. Lett. \textbf{99}, 090601 (2007)

\bibitem{muelken:pre1}
O. M\"ulken O, V. Bierbaum and A. Blumen, Phys. Rev. E \textbf{75}, 031121 (2007)

\bibitem{engel}
G. S. Engel, T. R. Calhoun, E. L. Read, T. K. Ahn, T. Man\v cal, Y. C. Cheng, R. E.  Blankenship and G. R. Fleming, Nature \textbf{446}, 782 (2007)

\bibitem{mohseni}
M. Mohseni, P. Rebentrost, S. Lloyd and A. Aspuru-Guzik, J. Chem. Phys. \textbf{129}, 174106 (2008)

\bibitem{caruso}
F. Caruso, A. W. Chin, A. Datta, S. F. Huelga and M. B. Plenio, J. Chem. Phys. \textbf{131}, 105106 (2009)

\bibitem{kempe}
D. Aharonov, A. Ambainis, J. Kempe and U. Vazirani, in Proceedings of the 33th STOC, ACM, New York, 50 (2001)

\bibitem{ambainis}
A. Ambainis, E. Bach, A. Nayak, A. Vishwanath and J. Watrous, Proceedings of the 33th STOC, ACM, New York, 60 (2001)

\bibitem{shenvi:2003}
N. Shenvi, J. Kempe and K. B. Whaley, Phys. Rev. A \textbf{67}, 052307 (2003)

\bibitem{ambainis:2003}
A. Ambainis, SIAM J. Comput., \textbf{37}, 210 (2007)

\bibitem{childs:04}
A. M. Childs, J. Goldstone, Phys. Rev A \textbf{70}, 022314 (2004)

\bibitem{kendon:2006}
V. Kendon, Phil. Trans. R. Soc. A 364, 3407 (2006)

\bibitem{magniez}
F. Magniez, A. Nayak, J. Roland and M. Santha, in Proceedings of the 33th STOC, ACM, New York, 575 (2007)

\bibitem{aurel:2007}
A. Gabris, T. Kiss and I. Jex, Phys. Rev. A \textbf{76}, 062315 (2007)

\bibitem{vasek}
V. Poto\v cek, A. Gabris, T. Kiss and I. Jex, Phys. Rev. A \textbf{79}, 012325 (2009)

\bibitem{2dw1}
B. Tregenna, W. Flanagan, R. Maile and V. Kendon, New J. Phys. \textbf{5}, 83.1 (2003)

\bibitem{miyazaki}
T. Miyazaki, M. Katori and N. Konno, Phys. Rev. A \textbf{76}, 012332 (2007)

\bibitem{chandrashekar:2007}
C. M. Chandrashekar, R. Srikanth and R. Laflamme, Phys. Rev. A \textbf{77}, 032326 (2008)

\bibitem{bach:2004}
E. Bach, S. Coppersmith, M. P. Goldschen, R. Joynt and J. Watrous, J. Comput. Syst. Sci. \textbf{69}, 562 (2004)

\bibitem{kempe:2005}
J. Kempe, Prob. Th. Rel. Fields \textbf{133} (2), 215 (2005)

\bibitem{krovi:2006a}
H. Krovi and T. A. Brun, Phys. Rev. A \textbf{73}, 032341 (2006)

\bibitem{krovi:2006b}
H. Krovi and T. A. Brun, Phys. Rev. A \textbf{74}, 042334 (2006)

\bibitem{kendon:2006b}
V. Kendon, Math. Struct. in Comp. Sci \textbf{17}(6), 1169 (2006)

\bibitem{varbanov:2008}
M. Varbanov, H. Krovi and T. A. Brun, Phys. Rev. A \textbf{78}, 022324 (2008)

\bibitem{nayak}
A. Nayak and A. Vishwanath, {\it pre-print} arXiv:quant-ph/0010117v1 (2001)

\bibitem{konno:2002}
N. Konno, Quantum Inform. Compu. \textbf{2}, 578 (2002)

\bibitem{konno:2005b}
N. Konno, J. Math. Soc. Japan \textbf{57}, 1179 (2005)

\bibitem{carteret}
H. A. Carteret, M. E. H. Ismail and B. Richmond, J. Phys. A \textbf{36}, 8775 (2003)

\bibitem{Grimmett}
G. Grimmett, S. Janson and P. F. Scudo, Phys. Rev. E \textbf{69}, 026119 (2004)

\bibitem{2dqw}
T. D. Mackay, S. D. Bartlett, L. T. Stephenson and B. C. Sanders, J. Phys. A \textbf{35}, 2745 (2002)

\bibitem{localization}
N. Inui, Y. Konishi and N. Konno, Phys. Rev. A \textbf{69}, 052323 (2004)

\bibitem{1dloc}
N. Inui, N. Konno and E. Segawa,  Phys. Rev. E \textbf{72}, 056112 (2005)

\bibitem{sato:2008}
M. Sato, N. Kobayashi, M. Katori and N. Konno, {\it pre-print} arXiv:0802.1997v1 (2008)

\bibitem{sanders}
B. C. Sanders, S. D. Bartlett, B. Tregenna and P. L. Knight, Phys. Rev. A \textbf{67}, 042305 (2003)

\bibitem{jeong}
H. Jeong, M. Paternostro and M. S. Kim, Phys. Rev. A \textbf{69}, 012310 (2004)

\bibitem{pathak}
P. K. Pathak and G. S. Agarwal, Phys. Rev. A \textbf{75}, 032351 (2007)

\bibitem{dur}
W. D\"ur, R. Raussendorf, V.M. Kendon and H.-J. Briegel, Phys. Rev. A \textbf{66}, 052319 (2002)

\bibitem{eckert}
K. Eckert, J. Mompart, G. Birkl and M. Lewenstein, Phys. Rev. A \textbf{72}, 012327 (2005)

\bibitem{chandrashekar:2006}
C.M. Chandrashekar, Phys. Rev. A \textbf{74}, 032307 (2006)

\bibitem{Orsolya}
O. K\'alm\'an, T. Kiss and P. F\"oldi, Phys. Rev. B \textbf{80}, 035327 (2009)

\bibitem{karski}
M. Karski, L. Förster, J. Choi, A. Steffen, W. Alt, D. Meschede and A. Widera, Science \textbf{325}, 174 (2009)

\bibitem{schmitz}
H. Schmitz, R. Matjeschk, Ch. Schneider, J. Glueckert, M. Enderlein, T. Huber and T. Schaetz, Phys. Rev. Lett. \textbf{103}, 090504 (2009)

\bibitem{Schreiber}
A. Schreiber, K. N. Cassemiro, V. Poto\v cek, A. Gabris, P. Mosley, E. Andersson, I. Jex and Ch. Silberhorn, {\it pre-print} arXiv:0910.2197 (2009)



\bibitem{polya}
G. P\'olya, Mathematische Annalen \textbf{84}, 149 (1921)

\bibitem{montroll:1956}
E. W. Montroll, J. SIAM 4, 241 (1956)

\bibitem{domb:1954}
C. Domb, Proc. Cambridge Philos. Soc. \textbf{50}, 586 (1954)

\bibitem{hughes}
B. D. Hughes, {\it Random walks and random environments, Vol. 1: Random walks}, Oxford University Press, Oxford (1995)

\bibitem{montroll:1964}
E.W. Montroll, in {\it Random Walks on Lattices}, edited by R. Bellman (American Mathematical Society, Providence, RI), Vol. \textbf{16}, 193 (1964)

\bibitem{revesz}
P. R\'ev\'esz, {\it Random walk in random and non-random environments}, World Scientific, Singapore (1990)

\bibitem{jarnik}
V. Jarn\'{\i}k, {\it Diferenci\'aln\'{\i} po\v cet II}, Academia, Prague, 121 (1976)

\bibitem{statphase}
R. Wong, {\it Asymptotic Approximations of Integrals}, SIAM, Philadelphia (2001)

\bibitem{abramowitzstegun}
M. Abramowitz and I. A. Stegun, {\em Handbook of Mathematical Functions
with Formulas, Graphs, and Mathematical Tables}, Dover Publications (1972)

\bibitem{watson:1939}
G. N. Watson, Quart. J. Math., Oxford Ser. 2 10, 266 (1939)










\bibitem{statphase2}
N. Bleistein and R. A. Handelsman, {\it Asymptotic Expansions of Integrals}, Holt, Rinehart and Winston, New York, (1975)



























\bibitem{Wegener}
I. Wegener, {\em Complexity Theory}, Springer-Verlag, Berlin (2005)

\bibitem{Mertens}
S. Mertens and C. Moore, {\em The Nature of Computation}, Oxford University Press, Oxford (2007)

\bibitem{RSA}
R. L. Rivest, A. Shamir and L. Adleman, Communications of the ACM \textbf{21}, 120 (1978)

\bibitem{Menezes}
A. J. Menezes, P. C. van Oorschot and S. A. Vanstone, {\em Handbook of Applied Cryptography}, CRC Press (1996)

\bibitem{shor}
P. Shor, SIAM J. Comput. \textbf{26}, 1484 (1997)

\bibitem{vandersypen}
L. M. K. Vandersypen, M. Steffen, G. Breyta, C. S. Yannoni, M. H. Sherwood and I. L. Chuang, Nature (London) \textbf{414}, 883 (2001)

\bibitem{lang:1970}
S. Lang, {\em Algebraic Number Theory}, Addison Wesley, New York (1970)

\bibitem{davenport:1980}
H. Davenport, {\em Multiplicative Number Theory}, Springer, New York (1980)

\bibitem{schleich:2005:primes}
H. Maier and W. P. Schleich, {\em Prime Numbers 101: A Primer on Number Theory}, Wiley-VCH, New York (2008)

\bibitem{clauser:1996}
J. F. Clauser and J. P. Dowling, Phys. Rev. A \textbf{53}, 4587 (1996)

\bibitem{harter:2001}
W. G. Harter, Phys. Rev. A \textbf{64}, 012312 (2001)

\bibitem{harter:2001b}
W. G. Harter, J. Mol. Spec. \textbf{210}, 166 (2001)

\bibitem{mack:2002}
H. Mack, M. Bienert, F. Haug, M. Freyberger and W. P. Schleich, Phys. Stat. Sol. (b) \textbf{233}, 408 (2002)

\bibitem{mack:proc}
H. Mack H, M. Bienert, F. Haug, F. S. Straub, M. Freyberger and W. P. Schleich, in {\em Experimental Quantum Computation}, Eds. P. Mataloni and F. De Martini, Elsevier, Amsterdam (2002)

\bibitem{merkel:ijmpb:2006}
W. Merkel, O. Crasser, F. Haug, E. Lutz, H. Mack, M. Freyberger, W. P. Schleich, I. Sh. Averbukh, M. Bienert, B. Girard, H. Maier and G. G. Paulus, Int.~J.~of~Mod.~Phys. B \textbf{20}, 1893 (2006)

\bibitem{merkel:FP}
W. Merkel, I. Sh. Averbukh, B. Girard, G. G. Paulus and W. P. Schleich, Fortschr. Phys. \textbf{54}, 856 (2006)

\bibitem{rangelov}
A. A. Rangelov, J. Phys. B \textbf{42}, 021002 (2009)

\bibitem{zubairy:science}
M. S. Zubairy, Science \textbf{316}, 554 (2007)

\bibitem{mehring:NMR:2006}
M. Mehring, K. M\"uller, I. Sh. Averbukh, W. Merkel and W. P. Schleich, Phys. Rev. Lett. \textbf{98}, 120502 (2007)

\bibitem{Suter}
T.S. Mahesh, N. Rajendran, X. Peng and D. Suter, Phys. Rev. A \textbf{75}, 062303 (2007)

\bibitem{suter2}
X. Peng and D. Suter, Euro. Phys. Lett. \textbf{84}, 40006 (2008)

\bibitem{rasel}
M. Gilowski, T. Wendrich, T. M\"uller, Ch. Jentsch, W. Ertmer, E. M. Rasel and W. P. Schleich, Phys. Rev. Lett. \textbf{100}, 030201 (2008)

\bibitem{girard}
D. Bigourd, B. Chatel, W. P. Schleich and B. Girard, Phys. Rev. Lett. \textbf{100}, 030202 (2008)

\bibitem{girard2}
S. Weber, B. Chatel and B. Girard, Euro. Phys. Lett. \textbf{83}, 34008 (2008)

\bibitem{girard3}
S. Weber, B. Chatel and B. Girard, in {\it Conference On Lasers And Electro-Optics Quantum Electronics And Laser Science Conference}, 3002 (2008)

\bibitem{sadgrove}
M. Sadgrove, S. Kumar and K. Nakagawa, Phys. Rev. Lett. \textbf{101}, 180502 (2008)



\bibitem{talbot:1836}
H. F. Talbot, Phil. Mag. \textbf{9}, 401 (1836)

\bibitem{berry:curlicues:1:1988}
M. V. Berry and J. Goldberg, Nonlinearity \textbf{1}, 1 (1988)

\bibitem{berry:curlicues:2:1988}
M. V. Berry, Physica D \textbf{33}, 26 (1988)

\bibitem{leichtle:PRL:1996}
C. Leichtle, I. Sh. Averbukh and W. P. Schleich, Phys. Rev. Lett. \textbf{77}, 3999 (1996)

\bibitem{leichtle:PRA:1996}
C. Leichtle, I. Sh. Averbukh and W. P. Schleich, Phys. Rev. A \textbf{54}, 5299 (1996)

\bibitem{schleich:2001}
M. V. Berry, I. Marzoli and W. P. Schleich, Physics World \textbf{14}, 39 (2001)

\bibitem{schopohl}
J. Oppenl\"ander, Ch. H\"aussler and N. Schopohl, Phys. Rev. B \textbf{63}, 024511 (2000)

\bibitem{born:wolf}
M. Born and E. Wolf, {\em Principles of Optics}, Pergamon Press, Oxford (1993)

%
%

\bibitem{sargent}
M. Sargent, M. O. Scully and W. E. Lamb, {\em Laser Physics}, Addison-Wesley, Reading (1974)

\bibitem{Kowalski}
H. Iwaniec and E. Kowalski, {\em Analytic Number Theory}, American
Mathematical Society, Providence (2004)

\bibitem{Rosen}
K. Ireland and M. Rosen, {\em A Classical Introduction to Modern Number Theory}, Springer-Verlag, Heidelberg (1990)

\end{thebibliography}
\end{document}